\documentclass[abstractpage,publicabstractpage,figures,tables,phd]{uithesis}

\usepackage{amsfonts}
\usepackage{amsmath}
\usepackage{amssymb}
\usepackage{array}
\usepackage{float}
\usepackage[toc,page,titletoc]{appendix}
\usepackage{longtable}
\usepackage{graphicx}
\usepackage[normalem]{ulem}
\usepackage{nicefrac}
\usepackage{units}
\usepackage{rotating}
\usepackage{afterpage}
\usepackage{caption}

\usepackage{mathtools}
\usepackage{mathrsfs}
\usepackage{tensor}
\usepackage{cite}
\usepackage{latexsym}
\usepackage{epsfig}
\usepackage{subfig}
\usepackage{graphics}
\usepackage{dsfont}
\usepackage{enumerate}

\usepackage[colorlinks=false, linktocpage=true]{hyperref}

\usepackage[all]{hypcap}

\usepackage{etoolbox}
\apptocmd{\thebibliography}{\interlinepenalty 10000\relax}{}{}

\usepackage{slashed}

\usepackage{enumitem}
\setlist{noitemsep,topsep=1pt, partopsep=4pt, parsep=2pt}
\setenumerate{noitemsep,topsep=1pt, partopsep=4pt, parsep=2pt}

\let\savedCaption=\caption
\renewcommand*{\caption}[2][\shortcaption]{%
  \def\shortcaption{#2}
  \normalsize\singlespace
  \savedCaption[#1]{#2}}

\makeatother 

\let\baraccent=\=

\renewcommand{\=}[1]{\stackrel{#1}{=}}

\newcommand{\al}[1]{\begin{align}#1\end{align}}
\newcommand{\bs}{\begin{split}}
\newcommand{\es}{\end{split}}

\def\root2{\sqrt{2}}

\newcommand{\mbb}[1]{\mathbb{#1}}
\def\a{\alpha}
\def\b{\beta}

\def\la{\lambda}

\def\G{\Gamma}
\def\g{\gamma}

\newcommand{\beq}{\begin{equation*}}
\newcommand{\eeq}{\end{equation*}}
\newcommand{\all}[1]{\begin{align*}#1\end{align*}}
\newcommand{\ld}{\mathscr{L}}

\usepackage{amsthm}

\theoremstyle{plain}
\newtheorem{thm}{Theorem}[section]

\title{THE DIFFEOMORPHISM FIELD}
\author{Delalcan Kilic}
\dept{Physics}
\setboolean{multipleSupervisors}{false}
\advisor{Professor Vincent G. J. Rodgers}
\memberOne{Vincent G. J. Rodgers}
\memberTwo{Yannick Meurice}
\memberThree{Wayne N. Polyzou}
\memberFour{Hao Fang}
\memberFive{Yasar Onel}
\submitdate{May 2018}
\copyrightyear{2018}

\setboolean{copyrightpage}{true} 

\dedication{dedication} 
\setboolean{dedicationpage}{true} 

\ackfile{thesisAck} 
\abstractfile{thesisAbstract} 
\publicabstractfile{publicAbstract} 

\setboolean{figures}{true} 
\setboolean{tables}{false}

\begin{document}

\frontmatter



\chapter{introduction}

The diffeomorphism field is introduced to the physics literature in  \cite{rairodgers90}. The authors obtained geometric actions by integrating the Kirillov form on the coadjoint orbits of Kac-Moody (KM)  and Virasoro algebras, two infinite-dimensional and centrally extended Lie algebras\footnote{A similar analysis was also done in \cite{alekseev}.}. These algebras are reviewed in Sections \ref{Kac-Moody Algebra} and  \ref{Virasoro Algebra}.

The parts of the geometric actions coming  from the centers of KM and  Virasoro algebras   are, respectively,  Wess-Zumino-Witten (WZW) action \cite{WZNW}  and  Polyakov's two dimensional quantum gravity (P2DG) action in lightcone gauge (LCG) \cite{polyakov2Dgravity}, describing  bosonization of the gauge and gravitational coupling of the chiral fermions in 2D. WZW and P2DG theories are reviewed in Sections \ref{WZNW Action} and \ref{Polyakov 2D Gravity}, and the construction of the geometric actions on  the coadjoint orbits of  KM and Virasoro algebras in Sections \ref{Kac-Moody Geometric Action} and  \ref{Virasoro Geometric Action}.

 The remaining terms in the geometric actions suggest the following. The non-central  part of the KM coadjoint element can be identified as a background Yang-Mills (YM) field coupled to the WZW field \cite{divecchia}, \cite{redlich87}.  Similarly, the non-central part of the Virasoro coadjoint element can be identified as a background rank-two field   coupled to the Polyakov field.    
 This rank-two field and its higher dimensional extensions are called the diffeomorphism field, or the diff field in short.  P2DG action in LCG is considered \cite{knizhnik} as the gravitational analog of  WZW action. Diff field is, in the same sense, the gravitational analog of  YM field, which is central to the Standard Model.

 In 2D, Einstein tensor identically vanishes so Einstein's theory of gravity does not provide dynamics for the spacetime metric.   Einstein-Hilbert action yields  Euler-characteristic, providing only topological information  about the spacetime. Therefore,  dynamics for gravity can arise only from quantum anomalies.  P2DG action is originally introduced as the effective action encoding the conformal anomaly \cite{polyakov81bosonic} and carries dynamical information. Since the background diff field couples to the Polyakov metric, it provides a source for cosmological constant and its dynamics would affect the spacetime. In particular, it may solve the dark energy puzzle. 
 
 If the 2D result, that the diff field is the gravitational analog of the YM field, holds in higher dimensions then the  graviton  may not be described by the spacetime metric or derivable from it, as has been thought. Alternatively, the diff field theory may be constructed in a way to encompass Einstein's theory to fix the quantization problem. These are currently speculative statements, yet suggesting the motivation of pursuing research on this subject. See, for instance, \cite{hendersonrajeev} for a quantum gravity theory on a circle suggested along a similar idea.

There are two distinct approaches for constructing  a dynamical theory for the diff field, leaving aside the most recent approach to be discussed at the end. In the first approach, Virasoro coadjoint action  is considered as the Lie derivative of a rank-two object. This rank-two object is not a  tensor due to the central term in its Lie derivative. Therefore, covariantization\footnote{By covariantization we mean lifting the space and time indices of tensors to spacetime indices, and lifting the partial spatial and temporal derivatives to covariant spacetime derivatives.}  can not yield a scalar under general coordinate transformations (GCTs) formed only from the diff field and its derivatives. One needs to introduce other objects, which also do transform inhomogeneously, into the theory in order to get a GCT-scalar Lagrangian.

Affine connection (metric or not) also transforms inhomogeneously. In fact, in 1D, it is easy to obtain a particular functional of connection coefficients that transform in the same way as a Virasoro coadjoint element (Section \ref{connectionVircoadjointelement}). However, extension of this relation to higher dimensions is highly nontrivial (Section \ref{HigherDimLiftofSigma}), so this approach has been mostly evaded. Two places, where this approach is held, are \cite{lanophd} and \cite{rodgers19942d}. The former claims to obtain a covariant action for the diff field. The latter is the theory of a $(1,3)$-field. We  restrict our attention to the examination of rank-two proposals for the diff field in this thesis.

In the second approach, one considers a rank-two tensor whose field theory yields a constraint equation such that this constraint equation reduces in 1D to the isotropy equation on Virasoro coadjoint orbits. This constraint is called the diff-Gauss law since the analogous constraint of YM theory is the Gauss law, which reduces in 1D to the isotropy equation on the KM coadjoint orbit.
Since the main field of the theory is  proposed to be a tensor, covariantization yields a GCT-scalar. This is the approach that has been followed the  most.

There are two subcases to consider in the second approach. One can introduce the diff-Gauss law as an implicit constraint i.e. one imposed on the phase space via an equivalence relation (invariance under the field lift of the Virasoro coadjoint transformation). In \cite{LR95} authors followed this approach using \cite{rajeev88} as a guide. They introduced the Virasoro analog of the Wilson loop, and obtained a finite reduction (i.e. a theory with a finite-dimensional phase space) of the diff field theory. 
We review \cite{rajeev88} in Section \ref{Rajeev88Summary}, and  \cite{LR95}  in Section \ref{ReviewDXGravity}. 

In the other subcase (\cite{LR96}, \cite{BLR97}, \cite{BLR00},  \cite{rodgers2007general}, \cite{takeshithesis} ) the diff-Gauss law is made explicit i.e. introduced into the action. This method is called the transverse action method.
 By this method one recovers the gauge-invariant YM action from the gauge-fixed contents of it.   We  do this construction in Section \ref{YMfromKM}.
An important part of this thesis, Chapters \ref{Transverse Actions}, \ref{Diff Field in 2D Minkowski} and \ref{2DMinkowskiConstraint}, is devoted to the examination of the tranverse formalism. Let us outline the transverse formalism procedure to obtain  the YM Lagrangian from the KM algebra, and the diff field Lagrangian from the Virasoro algebra.

It is known that the coadjoint action, $\text{ad}^*_{\Lambda} A$ of KM algebra is equivalent to the gauge transformation, $\delta_{\Lambda} A_1$ of a YM field in 1D. In 2D, it can  be identified as the residual gauge transformation of the spatial component of a YM field $A_{\mu}$, in the temporal gauge $A_0=0$. To build the transverse Lagrangian associated with the KM algebra one uses the latter identification.  Introducing a conjugate momentum $\pi^1$ to $A_1$ one can obtain the gauge transformation of $A_1$ through the Poisson bracket (PB) relation $\delta A_1 = \{A_1 , \int \Lambda G \}$. The generator $G$ of the transformation is the well-known Gauss law. 

Next, one introduces a Lagrangian formed from the symplectic term $\partial_0 A_1 \pi^1$, the Hamiltonian $\pi^1 \pi_1$ and  the Gauss law times  a Lagrange multiplier $\lambda$. Introducing\footnote{In the references listed above YM form of the momentum was directly assumed. Since in the diff field case we do not know the final result to be reached, we proposed an ansatz that could deduce the YM form.} an ansatz of the form $\pi^1 = \partial^0 A^1 + ...$  one recomputes the momentum from the constructed Lagrangian. This yields the YM momentum for $\lambda=A_0$, and  the YM Lagrangian in 2D, $\sim \pi^1 \pi_1= F^{01} F_{01}$.
 One can straightforwardly lift this  to higher dimensions, and covariantize. Note that upon covariantization  gauge structure of the YM theory is preserved. That is, $A_0$ is still nondynamical, and the Gauss law is still a first-class constraint generating time-independent gauge transformations of $A_i$. 

With the lead from KM-YM pair, 
one lifts the  transformation  of a Virasoro coadjoint element $D$ to the Lie derivative of a rank-two object $D_{\mu \nu}$, the diff field. This transformation has a third-order inhomogeneous term $\partial_{\mu} \partial_{\nu} \partial_{\lambda} \xi^{\lambda}$. Hence, $D_{\mu \nu}$ does not transform as a tensor at this point. In 2D, one can recover the transformation of $D$ as the Lie derivative of $D_{11}$ under spatial, time-independent coordinate transformations.  Introducing a conjugate momentum $X^{11}$ to $D_{11}$, one obtains the operator $G_1$ generating $\delta D_{11}$, namely, the diff-Gauss law. This operator is called the diff-Gauss law because it is introduced precisely in the same way as the ordinary Gauss law. Therefore, it is  expected, in the end, to be a first-class constraint generating the Lie derivative $\delta D_{ij}$  as a local symmetry of the theory just as the first-class constraint Gauss law generates the gauge symmetry $\delta A_i$. 

Using the corresponding ingredients, i.e. the symplectic term $\partial_0 D_{ij} X^{ij}$, the Hamiltonian $X^{ij} X_{ij}$, and  the diff-Gauss law $G_i$ times its Lagrange multiplier $D_0^{ \ i}$, one obtains a Lagrangian. Then one introduces the analogous ansatz\footnote{Note that in the references stated above, the momentum was taken as $X =\dot{D}$. We observed that with this choice one does not recover back the same momentum from the constructed Lagrangian. In fact the same kind of choice in the gauge theory case does not yield YM Lagrangian. 

The new ansatz leads to the momentum squared form of the diff Lagrangian, and one obtains three and four-point self-interaction terms of the diff field, again in analogy with the YM theory. To distinguish the modified and old theories in the analysis, we call the latter, the BLRY theory. Whenever we would like to exemplify a computational technique we use BLRY theory rather than the full theory for simplicity. } $X^{11} = \partial^0 D^{11} + \cdots$ for the momentum, inserts it in the Lagrangian, and recomputes the momentum from the constructed Lagrangian. With this momentum the Lagrangian attains the same form as in YM theory $\sim X^{ij} X_{ij}$. 

The following step is covariantization just as in the YM case. However, at the starting point we had a nontensor rank-two field. Covariant derivative is not defined on such an object and even if we blindly applied the covariant derivative formula of a rank-two tensor to it, such a derivative would preserve non-covariance of the object. Similarly, contraction of spacetime indices of such an object and its derivatives will not yield a GCT-scalar Lagrangian. Hence, at this point  diff field is regarded as a tensor. Moreover, upon covariantization the Lagrange multiplier $D_0^{\ i}$ becomes dynamical, and the diff-Gauss law is no more obtained as a constraint.  Hence, contrary to the YM case, at this step we lose connection to the origins of the theory. This is expected because in the diff field case the local symmetry itself is coordinate invariance.

We  construct the transverse action for the diff field in Section \ref{Diffaction}.    
Its supersymmetric extension is obtained in Section \ref{DiffSUSY} using \cite{GR01} as a guide. We  analyze the transverse diff theory in 2D Minkowski spacetime before covariantization in Section \ref{DXNTheoryPart1}, and the covariantized theory in Sections \ref{TransverseAfter} and \ref{BLRYTotalHamiltonian}.

Interactions of the diff field  is obtained by a prescription that emerges from examining the structure of the self-interaction of the diff field \cite{BLR00} in the transverse action. When this prescription is applied to the point particle and spinor interactions, the resulting expression suggests that the diff field is a perturbation to the spacetime metric. We  review interactions of the diff field in Section \ref{DiffInteractions}. Motivated by the coadjoint action of the semidirect product of Virasoro and KM algebras, we treat the diff field as transforming nontrivially under  gauge transformations\footnote{In \cite{lanophd} also the diff field is treated this way.}. We examine application of the interaction prescription to the spin-one coupling with this treatment. 

Note that even if we turn off covariantization, and treat diff field as a nontensor, the diff-Gauss law turns out to be inconsistent for the chosen standard kinetic term (Section \ref{ConstraintDXN}). In the Dirac-Hamiltonian analysis, new constraints arise, and these are all derivable from the kinetic term. The diff-Gauss law turns out to be second-class unless the kinetic term itself is a constraint. Hence, we turn the kinetic term  into a constraint. Then  the diff-Gauss law becomes a first-class constraint, and no new constraints arise.  In this case, however, dynamics is lost (Section \ref{FreezingDiffTheory}). We investigate an alternative kinetic term  in Section \ref{AlternativeTransverseTheoryTprimeG}. 

Note that although covariant transverse diff theory is  inconsistent with its own philosophy, it has mathematically consistent subcases i.e. gauge-fixed reductions without constraint inconsistencies. Namely, it is not a theory with local Virasoro symmetry, and the motivation coming from geometric actions is lost (i.e. diff field being the gravitational analog of the gauge field), but it still provides subcases with dynamical content (momenta, field equations and so on) that is related to the  Virasoro algebra in some way (e.g. appearance of the KdV equation and its variants). One such case  is in a gauge we call the chiral gauge. The diff-Gauss law does not arise, as the field equation of $D_0^{\ i}$ component reduces to $0=0$.  In 2D, in this gauge, covariant transverse theory reduces to a theory with two decoupled fields (one a function of time only and the other a function of space only) which seems to be related to the geometric action associated with the direct product of two Virasoro algebras (Section \ref{VirasoroDirectProduct}). We investigate this in Section \ref{TransverseAfter}.

In Section \ref{CoordGaugeTrfoftheDiffField}, we review the tranverse method, outline all its problems  and discuss how they are related. We decide that the most important issue in the theory is covariantization. We abandon covariantization and go back to the approach of treating the diff field as a nontensor. We look for alternative ways to recover covariance.  We investigate complementing\footnote{As we mentioned above, in \cite{lanophd}, a covariant theory of the diff field is proposed along similar lines i.e. by introducing connection coefficients into the action. We could not verify their result, but it would be interesting to investigate how this theory may be related to the transverse theory, or whether it actually fulfills our goal by providing a gauge theory of the diff field. For this one needs to check whether the theory provides the diff-Gauss law as a first-class constraint generating a coordinate transformation that reduces in 1D to the Virasoro coadjoint action. } the diff field with connection coefficients (using the results of Section \ref{HigherDimLiftofSigma}) and propose modifications which recover full covariance for the interactions of the diff field while keeping spatial covariance of the diff Lagrangian.

Problems of the transverse theory lead us to investigate alternative methods to obtain a theory of the diff field. One such method is  the Euler-Poincare formalism, an alternative Lagrangian formalism suited for Lie groups (Section \ref{DiffEulerPoincare}). Application of this formalism to diff field, however, leads to a dynamical theory within a coadjoint orbit, rather than producing dynamics with gauge degrees of freedom lying on the orbits i.e. with the diff-Gauss law being a first-class constraint generating the Virasoro coadjoint transformation as a local symmetry. 

Next, we extensively examine the analog of the Wilson loop for the diff field in Section \ref{diffWilson}.   For this we follow the references \cite{scherer88}, \cite{hendersonrajeev}. The latter claims to obtain the Virasoro analog of the theory in \cite{rajeev88}, just as \cite{LR95}, but we believe the Wilson loop to be used for such a theory should be  associated with the operator $\nabla^{(3)}$ (introduced in Section \ref{Virasorocovariantderivatives} ) rather than the Hill operator $\nabla^{(2)}$. Indeed, in \cite{LR95} a Wilson loop associated with $\nabla^{(3)}$ was introduced, but we believe its implementation was incomplete (Section \ref{DiffWilsonLR95}). We produce results associated with $\nabla^{(3)}$ that may be needed for future research on this project.

As the final part of the thesis we review an entirely different approach  proposed recently \cite{brensinger} to obtain a dynamical diff theory. Diff field is identified as part of a TW projective connection. The authors introduce a curvature-squared type action for the diff field based on this identification. We are going to provide a quick summary of this work in Section \ref{TWdiffdynamicaltheory}.  Our focus in this thesis is on the clarification of the relationship between  TW connections and  diff field. This is investigated in Section \ref{TWVirasororeln}.

Let us also briefly discuss the notation and the conventions used in the thesis. 
Throughout the thesis, summation convention is used both for algebraic and tensorial sums unless there is potential confusion. Derivative of a quantity with respect to a variable is frequently denoted by a subscript, e.g., $\partial/\partial \rho =: \partial_{\rho}$. In 1D,  derivative with respect to a single coordinate is  denoted by a prime unless it is a temporal parameter, in which case it is denoted by a dot. 
In 2D,  derivative with respect to time and space is also be denoted by a dot and a prime, respectively. 

The sign convention for the metric is $(+t , - x, -x, \cdots)$. To avoid culmination of negative signs,  in any dimensions we  write $\sqrt{g}$ for the metric determinant even when $g$ is negative; what is implied is $\sqrt{|g|}$.  Certain sections require additional notational and conventional changes, they are noted beforehand. 

Analogs of objects of the YM theory are named with "diff-..." in the case of the diff field e.g. diff-Gauss law, diff-Wilson loop etc.

\chapter{Preliminaries}
\section{Coadjoint Orbits and Kirillov Form}
\subsection{Adjoint Action}
Let $G$ be a Lie group and $\mathfrak{g}$ its Lie algebra. Consider the conjugation map by a fixed element $g \in G$ 
\al{  
C_g : G \rightarrow G : h \mapsto ghg^{-1}
}
and its pushforward
\al{
\text{Ad}_g \equiv (C_g)_* : \mathfrak{g} \rightarrow \mathfrak{g}
}
For $X \in \mathfrak{g}$ we can explicitly write
\al{
\text{Ad}_g X = \frac{d}{dt} (R_{g^{-1}} \circ L_g \circ \exp(tX) ) \bigg|_{t=0}
}
For matrix groups this simplifies to
\al{ \label{AdMatrixGroup}
\text{Ad}_g X = g X g^{-1} 
}

The map 
\al{
\text{Ad} : G \times \mathfrak{g} \rightarrow \mathfrak{g} : (g,X) \mapsto \text{Ad}_g X
}
defines an action of $G$ on its Lie algebra $\mathfrak{g}$ called the adjoint action. Any group action on a vector space defines a representation of the group; $\mathfrak{g}$ is a vector space. The representation for the adjoint action is defined by 
\al{
\text{Ad} : G \mapsto \mathfrak{gl}(\mathfrak{g}) : g \mapsto \text{Ad}_g 
}
and is called the adjoint representation. Using the properties of the pushforward and the group one can show that Ad indeed satisfies the properties of a representation
\al{
\text{Ad}_g \circ \text{Ad}_h = \text{Ad}_{gh} \hspace{0.3in} \text{and} \hspace{0.3in} (\text{Ad}_g)^{-1} = \text{Ad}_{g^{-1}}
}
Adjoint representation of $G$ induces a representation of its Lie algebra $\mathfrak{g}$ defined by
\al{
\text{ad} \equiv \text{Ad}_* : \mathfrak{g} \rightarrow \mathfrak{gl} (\mathfrak{g}) : X \mapsto \text{ad}_X
}
We shall call this the infinitesimal adjoint action and it can be explicitly written as 
\al{
\text{ad}_X = \frac{d}{dt} \bigg|_{t=0} \text{Ad}_{ \exp t X}
}
One can show using the flow of $X \in \mathfrak{g}$ that its action on $Y \in \mathfrak{g}$ yields
\al{
\text{ad}_X Y = \mathcal{L}_X Y = [X, Y ]
}
Using the Jacobi identity on $\mathfrak{g}$ one can show that ad is indeed a representation of $\mathfrak{g}$
\al{
\text{ad}_{[X,Y]} = [\text{ad}_X , \text{ad}_Y]
}

\subsection{Coadjoint Action} 
Let $V$ and $W$ be vector spaces, and let $A:V\rightarrow W$ be a linear map. The dual map $A^*:W^* \rightarrow V^*$ is defined by
\al{
(A^* b ) (v) \equiv b (A v)
}
where $b \in W^*$, $v$ $\in$ $V$. 
$\mathfrak{g}$ is a vector space and the adjoint action $\text{Ad}_g : \mathfrak{g} \rightarrow \mathfrak{g}$ is a linear map. Hence we can define its dual, $(\text{Ad}_g)^* : \mathfrak{g}^* \rightarrow \mathfrak{g}^*$ , as
\al{
[(\text{Ad}_g)^* b]  \ (u) \equiv b (\text{Ad}_g u)
} 
where  $b$ $\in$  $\mathfrak{g}^*$ and $u$ $\in$ $\mathfrak{g}$. 

 Vectors in $\mathfrak{g}$ are often called adjoint vectors  and ones in $\mathfrak{g}^*$ are called coadjoint vectors. Thus, a coadjoint vector is a linear functional on $\mathfrak{g}$. 
This is often written in the form of a  'pairing', a linear map $\left< \ | \ \right> : \mathfrak{g}^* \times \mathfrak{g} \rightarrow \mathbb{R}$ 
\al{
\left< b | u  \right> \equiv b (u)  
}

The coadjoint action  of $g$ $\in$ $G$ on $\mathfrak{g}^*$, denoted $Ad^*_g$,  is defined by 
\al{ \label{Coaddefinition}
\text{Ad}^*_g \equiv (\text{Ad}_{g^{-1}})^*
}
The reason for $g^{-1}$ on the right hand side is to make the pairing invariant under the action of the group. That is, if $b \in \mathfrak{g}^*$ and  $u \in \mathfrak{g}$ then we have
\al{ \label{PairingInvarianceGenericFinite}
\left< \text{Ad}^*_g b  |\text{Ad}_g u \right>   = \left< b | \text{Ad}_{g^{-1}} \text{Ad}_g u \right> =  \left< b| u\right> 
}
Practically one first introduces a pairing, then obtain the coadjoint action by declaring invariance of the pairing. 

The set $\mathfrak{g}^*$ is  a vector space just as $\mathfrak{g}$. Therefore, similar to the adjoint case, we can define the coadjoint representation of the group from the coadjoint action.

 The induced infinitesimal coadjoint  action  $\text{ad}^*_v : \mathfrak{g}^* \rightarrow T \mathfrak{g}^* \cong \mathfrak{g}^* $ is defined by
\al{
(\text{ad}^*_v b)u \equiv - b (\text{ad}_v u ) = - b ([v,u]) 
}
One can obtain this from the invariance condition \eqref{PairingInvarianceGenericFinite}  by considering the one-parameter subgroup generated by $v \in \mathfrak{g}$ i.e. $g = \exp(tv)$, differentiating with respect to $t$, and evaluating at $t=0$. 
Then the infinitesimal form of invariance   follows:
\al{ \label{infinitesimalinvariance}
v * \left< b | u \right>  =&  \left< v * b | u \right> + \left< b | v * u\right> \notag\\ =& \left< \text{ad}^*_v b | u \right> + \left< b | \text{ad}_v u \right> \notag\\ =& -b([v,u])+b([v,u]) = 0
}

The isotropy group $G_{b}$ of $b \in \mathfrak{g}^*$ under the  coadjoint action is defined by
\al{
G_{b} \equiv \{ g \in G \  | \  \text{Ad}^*_g  b = b \}
}
and is a subgroup of $G$. 
The isotropy algebra $\mathfrak{g}_{b}$ of $b$ is the Lie subalgebra of $\mathfrak{g}$ that generates the isotropy group $G_b$. It is given by 
\al{ \label{isotropyalgebrageneral}
\mathfrak{g}_{b} = \{ u \in \mathfrak{g} \ \ | \ \ \text{ad}^*_u  b = 0 \} 
}
The equation $\text{ad}^*_u b = 0$ is called the isotropy equation for the coadjoint element $b$. 
We will construct transverse actions in Chapter \ref{Transverse Actions} by lifting isotropy equations of algebras  to constraint equations of the corresponding field theory. 
\subsection{Coadjoint Orbits and Kirillov Form} 
The coadjoint orbit of  $b_0 \in \mathfrak{g}^*$ is defined by
\al{
\text{Orb}(b_0) \equiv \{ b \in \mathfrak{g}^*  \ | \ \exists g \in G \ \ \text{st} \ \  b = \text{Ad}^*_g b_0 \} 
}
and is a subspace of $\mathfrak{g}^*$. Kirillov \cite{kirillovorbit} showed that every coadjoint orbit of a Lie group $G$ is naturally equipped with a symplectic structure $\Omega$ (called the Kirillov form) that is invariant under the action of $G$.  A symplectic structure is a two-form   that is non-degenerate
and closed. 

$\Omega$ is defined as follows. The (coadjoint) action of two adjoint vectors $u, u' \in \mathfrak{g}$ on $b\in \mathfrak{g}^*$  yield two coadjoint vectors $a , a'$ that are tangent to the orbit at $b$ : 
\al{
a \equiv \text{ad}^*_u b \hspace{0.3in} \text{and} \hspace{0.3in} a' \equiv \text{ad}^*_{u'} b
}
Then the Kirillov form $\Omega$ is defined as
\al{ \label{Kirillovform}
\Omega ( a, a' ) \equiv \left< b | [u, u'] \right> 
}
$\Omega$ is antisymmetric because of the commutator on the right. The pairing is $G$-invariant by definition, so $\Omega$ is $G$-invariant :
\al{
\Omega ( a, a' ) = \left< b | [u, u'] \right> = \left< b_g | [u,u']_g \right> = \left< b_g | [ u_g , u'_g] \right> = \Omega ( a_g , a'_g) 
} 
where $b_g \equiv  \text{Ad}^*_g b$ ,  $u_g \equiv \text{Ad}_g u$ , $a_g \equiv  \text{ad}^*_{u_g} b_g$ and $g, g' \in G$.
 If $a=\text{ad}^*_u b$ is a nonzero coadjoint vector then $b$ is nonzero by linearity of $\text{ad}^*$. Then  there must be an adjoint vector $v$ (that does not commute with $u$) such that $(\text{ad}^*_u b) v = - \left<b | [ u,v] \right> \neq 0$ so that $\Omega$ is nondegenerate. 
 
In order to prove\footnote{The  proof here is from \cite{witten88}. For a rigorous proof see e.g. \cite{marsdenratiu} Chapter 14.}  closure of $\Omega$ we use the invariant formula for exterior derivatives. For a two-form $\lambda$ and vector fields $u,v,w$ on a manifold $M$ it reads
\al{
d \lambda ( u, v, w ) & = u \cdot \nabla  (\lambda (v, w ) ) + v \cdot \nabla  (\lambda (w, u)) + w \cdot \nabla  (\lambda (u,v)) \notag\\ & \ \ \ - \lambda ( [u,v], w) - \lambda([w, u],v) - \lambda ( [v,w],u)
}
The adjoint vectors $u, v, w \in \mathfrak{g}$ define the coadjoint tangent vectors $b_u$, $b_v$, $b_w$ on the orbit, where 
$b_u \equiv \text{ad}^*_u b \in T_b \mathfrak{g}^* \cong \mathfrak{g}^*$. The Kirillov  form $\Omega \in \Lambda^2 ( \text{Orb}(b) )$ acts on a pair of (tangent) coadjoint vectors on the orbit, and $d \Omega$ on  three of them. The invariant formula in this case reads 
\al{ \label{dOmegabubvbw}
d \Omega ( b_u, b_v , b_w) & = b_u \cdot \nabla ( \Omega (b_v, b_w ) ) 
+ b_v \cdot \nabla ( \Omega (b_w, b_u ) ) 
+ b_w \cdot \nabla ( \Omega (b_u, b_v ) )  \notag\\ & \ \ - \Omega ( [ b_u, b_v] , b_w )  - \Omega ( [ b_w, b_u] , b_v )  - \Omega ( [ b_v, b_w] , b_u ) 
}
Consider the first term on the right, $b_u \cdot \nabla ( \Omega (b_v, b_w))$. It  represents the change of $\Omega (b_v, b_w) = \left< b | [ v,w] \right> $ in $b_u$ direction, so is equal to the action of the adjoint element $u$ on the pairing,
which is zero by invariance of the pairing. 
Thus the first line on the right in \eqref{dOmegabubvbw} vanishes. 

Now, consider the first term in the second line. Since $\text{ad}^*$ is a representation of $\mathfrak{g}$ we have 
\al{
\Omega ( [b_u, b_v] , b_w ) & = \Omega ( [\text{ad}^*_u b , \text{ad}^*_v b] , \text{ad}^*_w b ) \notag\\ & = \Omega ( \text{ad}^*_{[u,v] }b , \text{ad}^*_w b ) = \Omega ( b_{[u,v]} , b_w ) = b( [ [u,v],w] ) 
}
Thus  the terms in the second line add up to zero by Jacobi identity on $\mathfrak{g}$ and linearity of $b$. This completes the proof of $d\Omega = 0$. 
\section{Construction of Geometric Actions on Coadjoint Orbits}
\label{GeoActonCoadOrbNonExact}
\subsection{Mechanics on Space of Paths in Phase Space} \label{NonExactMech}
Symplectic structure $\Omega$ is the main ingredient of Hamiltonian mechanics. Hamilton's equations describing the dynamics of a physical system can be written  as
\al{
\iota_{X_H} \Omega =  d H
}
where $H$ is the Hamiltonian, $X_H$ is the Hamiltonian vector field whose flow describes the evolution. Let $\Gamma = \{ \xi^i \}$ be the phase space. Then we can explicitly write 
\al{
\Omega_{ij} \dot{\xi}^j = \partial_{\xi^i} H
}
In most cases, the symplectic structure is not only closed but also (globally) exact  so that we can write it as the exterior derivative of the so called "canonical one-form", denoted by $\theta$ , i.e. $\Omega= d \theta$. Then the action can be written as  
\al{
 S = \int \theta - H dt
}

Consider a simple example, that of a two-dimensional phase space $\Gamma= \{ (p ,q )\}$ with $\omega = dp \wedge dq$ and $\theta = p  dq$. The action then reads
\al{
S = \int ( p dq - H dt) = \int dt \ (p \dot{q} - H )  = \int dt \ L
}

For coadjoint orbits, however, we do not, in general, enjoy this simplification. Balachandran et al \cite{zaccaria} \cite{balagauge} discusses an extension of symplectic mechanics when symplectic structure is not exact. Below is the outline.

 Instead of the phase space $\Gamma$ we consider the space of paths on $\Gamma$, denoted $P \Gamma$. The points on $P\Gamma$ can be defined by fixing a point $P_0$ in $\Gamma$ . Then an element of $P\Gamma$ is a path from $P_0$ to some other point $\xi$ in $\Gamma$. We may parametrize these paths as 
\al{
\gamma \in \{ \gamma ( \lambda) \ | \ 0 \leq \lambda \leq 1 , \gamma(0) = P_0 , \gamma(1) = \xi \}
}

\begin{figure}[H]
\begin{center}
\includegraphics[scale=.6]{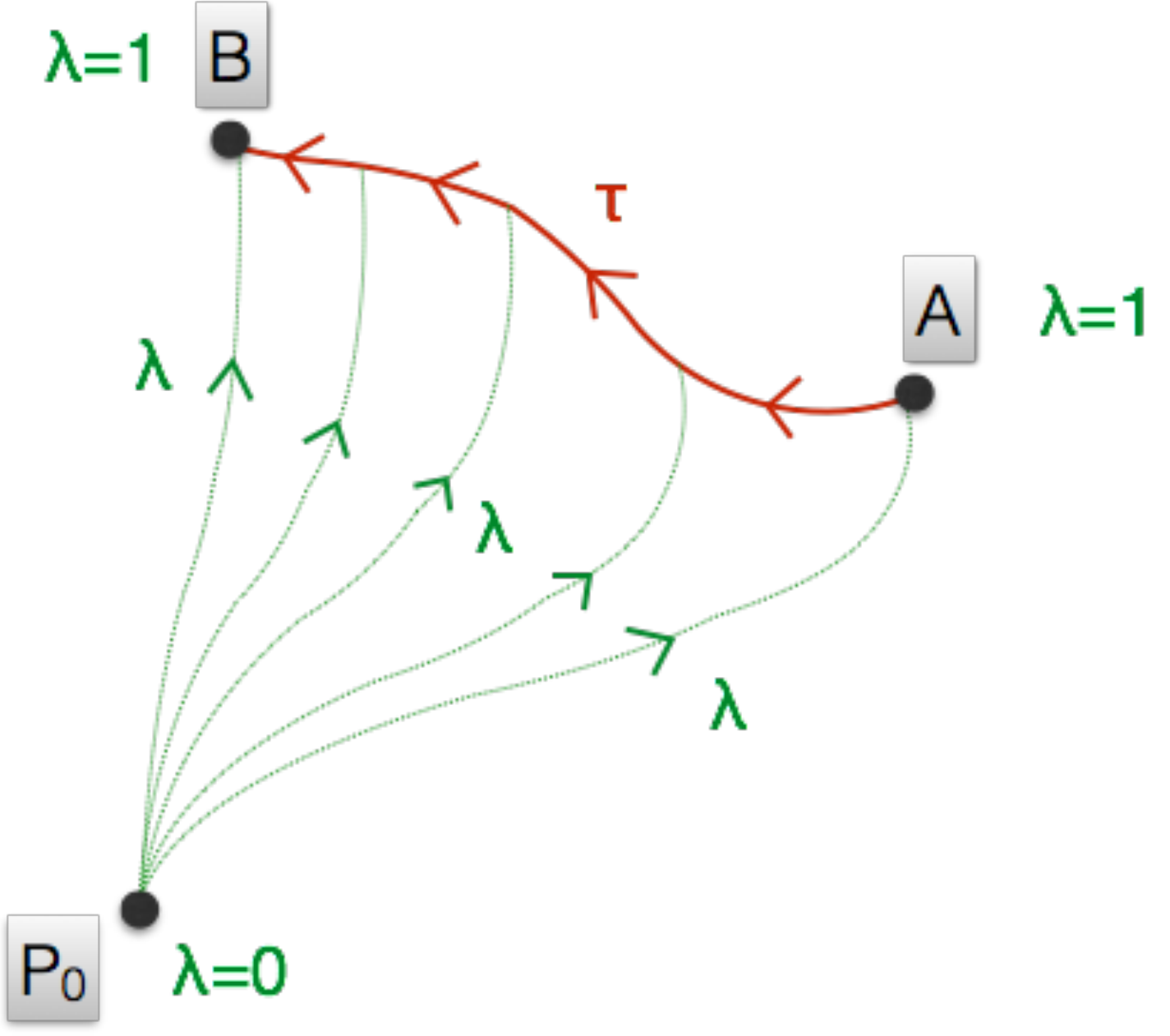}
\end{center}
  \caption{The Two-surface $m$ Traced by the Paths}
  \label{twosurface}
\end{figure}

Introducing also the time coordinate $\tau$ we get time-dependent paths, $\gamma ( \lambda , \tau)$ where $\gamma ( \lambda = 0 , \tau) = P_0$ and $\gamma ( \lambda = 1 , \tau ) = \xi(\tau)$. Here, $\{\xi(\tau)\}$ is a possible trajectory to be followed by the system. That is, we would like to obtain an action functional whose extremization yields equations only on $\{\xi(\tau)\}$. As $\lambda$ and $\tau$ vary,  the paths $\gamma(\lambda, \tau)$ sweep out a two-surface $m$ in $\Gamma$  (See Figure \ref{twosurface}). Its boundary is given by
\al{
\partial m = \partial m_1 \cup \partial m_2 \cup \partial m_3
}
where 
\al{
\partial m_1 & = \{ \xi(\tau) \ | \ \tau_i \leq \tau \leq \tau_f \} = AB \notag\\ \partial m_2  & = \{ \gamma(\lambda , \tau_i ) \ | \ 0 \leq \lambda \leq 1 \} = P_0A\notag\\ \partial m_3  & = \{ \gamma(\lambda , \tau_f ) \ | \ 0 \leq \lambda \leq 1 \} =  P_0B
}
The Hamiltonian $H$ is lifted to a functional $\tilde{H}$ on paths, as
\al{
\int_0^1 d \lambda \ \tilde{H} [ \gamma ( \lambda , \tau ) ] = H [ \gamma (1, \tau)] = H [\xi ( \tau)]
}
Then the action functional can be defined as 
\al{ \label{ZacAct}
S = \int_m ( \Omega - \tilde{H} \ d \lambda \wedge d \tau ) 
}
or, in coordinates, as
\al{
S =  \int \Omega_{ij} \partial_{\lambda} \gamma^i \partial_{\tau} \gamma^j d \lambda d \tau - \int_{\partial m_1 } H d \tau
}
Under variations, the point $P_0$ and the end paths $\partial m_2$ and $\partial m_3$ are to be held fixed. Equations of motion derived by varying the paths $\gamma$ then become
\al{
0 = \delta S = \int_{\partial m_1} \Omega_{ij} d \gamma^i \delta \gamma^j - \int_{\partial m_1} \partial_{\gamma^i} H \delta \gamma^i d \tau  
}
where $d \Omega = 0$ is used. This recovers  Hamilton's equations on $\partial m_1 = \{\xi(\tau) \}= AB$
\al{
\Omega_{ij} \dot{\gamma}^i = \partial_{\gamma^i} H 
}
\subsection{Geometric Actions on Coadjoint Orbits}

We choose to consider theories with vanishing Hamiltonian. The symplectic structure is the Kirillov form on coadjoint orbits of the infinite-dimensional Lie algebras, Kac-Moody and Virasoro. The Kirillov form is non-exact in each case, so we will employ the results of the previous section. Then according to \eqref{ZacAct}, the action functional (called the geometric action) is given simply by the integral of the Kirillov form on an orbit
\al{
S = \int_{\text{Orb}} \Omega
}

The orbit is parametrized as a two-surface $\{ (\lambda, \tau)\}$ so that we need to construct adjoint vectors $u_{\tau}, u_{\lambda}$ and coadjoint (tangent) vectors $b_{\tau} =\text{ad}^*_{u_{\tau}} (b)  , \ b_{\lambda}=\text{ad}^*_{u_{\lambda}} (b) $ describing changes in $\tau$ and $\lambda$ directions for a suitably chosen coadjoint vector $b=b(\tau,\lambda)$. Then using \eqref{Kirillovform} the action can  be explicitly written as 
\al{
S = \int_{\text{Orb}(b)} d \lambda \ d\tau \ \Omega ( b_{\tau} , b_{\lambda} ) =\int_{\text{Orb}(b)} d \lambda \ d\tau \ \left< b | [u_{\tau} , u_{\lambda}]\right> 
}

This will be done in the next chapter.
 The central part of the constructed geometric action will turn out to be the  WZW action in the KM case and P2DG action in LCG in the Virasoro case.

\section{Kac-Moody Algebra }
\label{Kac-Moody Algebra}
For our purposes Kac-Moody (KM) algebra and the geometric action on its coadjoint orbits play secondary roles. Therefore, we will not get into detail as much as we do for the Virasoro algebra. 
The main references for this section are  \cite{goddardolive86}, \cite{delius90} and \cite{WZNW}.

\subsection{Loop Group, Loop Algebra and Its Central Extension }
\label{Loop Group}

Let $G$ be a compact, connected, semi-simple Lie group. Then its Lie algebra $\mathfrak{g}$ is semi-simple with  Killing form $\delta^{ab}$ so that the structure constants with fully upper  indices $f^{abc}$ are defined and are fully antisymmetric.  The commutation relations for $\mathfrak{g}$ can then be written in a basis $\{T^a\}$ as  
\al{ \label{basealgebra}
[T^a , T^b ] = if^{abc} T^c
}
Since $G$ is connected, any element $g$ of $G$ can be obtained by exponentiation of an algebra element, i.e. $g = \exp(- i T^a \theta_a)$ with parameters $\theta_a$.
 
A smooth map $\gamma$ from circle $S^1=\{ z \in \mathbb{C} \, : \, |z| =1 \}$ to $G$ is called a loop in $G$. The set of loops forms a Lie group, called the loop group of $G$, denoted $LG$,  with the group multiplication defined by
\al{
(\gamma_1 \cdot \gamma_2 ) (z) \equiv \gamma_1 (z) \gamma_2 (z) 
}
On the right,  group multiplication of $G$ is implied. 

To obtain the Lie algebra $L\mathfrak{g}$ of $LG$ consider its  connected component consisting of maps $\gamma : S^1 \rightarrow G$ that can be continuously deformed to the constant map $\gamma(z)=1$. Then any element of this subset of $LG$ can be obtained using  functions $\theta_a(z)$ defined on the unit circle  as $\gamma(z) = \exp ( - i T^a \theta_a(z) )$.

For elements near the identity map we have $
\gamma(z) \approx 1 - i T^a \theta_a (z)$. 
Making a Laurent expansion, $ 
\theta_a (z) =  \theta_a^{n} z^n$, we see that the composite objects 
\al{ \label{loopgroupgenerators}
J^a_n \equiv T^a z^n
} 
are generators for the loop group. Indeed for elements near the identity we have $\gamma(z) \approx 1 - i  J^a_{n} \theta^n_{a}$.  
Using  \eqref{basealgebra} and \eqref{loopgroupgenerators} we get the commutation relations 
\al{
[J^a_m , J^b_n] = i f^{abc} J^c_{m+n}
}
for the loop algebra $L\mathfrak{g}$. Note that $\{J^a_0\}$ generate a subgroup of $LG$ isomorphic to its base group $G$. 

Since $G$ is  compact,  picking up a Hermitian basis of $G$-generators,  $T^{a\dagger} = T^a$,   the loop generators satisfy 
$J^{a\dagger}_{n} = J^a_{-n}$, 
where $z^* = z^{-1}$ is used. Such a representation of loop algebra basis generates unitary loops $\gamma(z)$ for real $\theta^n_a$ and $|z|=1$. 

The Kac-Moody algebra (or, more explicitly, the untwisted affine Kac-Moody algebra) associated with a compact finite-dimensional Lie algebra $\mathfrak{g}$ is the central extension of the loop algebra $L\mathfrak{g}$, defined by the commutation relations 
\al{ \label{KacMoodyalgebra1}
[J^a_m , J^b_n ] = i f^{abc} J^c_{m+n} + km \delta^{ab} \delta_{m+n}
}
For the detailed arguments leading to this form of the central extension see \cite{goddardolive86}. For a mathematically more precise way of expressing central extension, see \eqref{DeliusKMcommutation}.

\subsection{Current Algebra }
 \label{Current Algebra}
The loop algebra and its central extension appear in a variety of physical theories. Here, we  analyze one such theory\footnote{Later we are going to see the same current algebra appearing  in the WZW theory.}, namely, that of free massless quarks  in 2D. These quarks can be represented by $N$ massless Majorana fermions $\psi^i$ ($i=1, \cdots , N$). The action reads
\al{ \label{Nchiralfermionaction}
S = \frac{i}{2} \int d^2 x \ \overline{\psi}_k \gamma^{\mu} \partial_{\mu} \psi^k
}
where $\overline{\psi} = \psi^T \gamma^0$. This action leads to the Dirac equation 
\al{
\gamma^{\mu} \partial_{\mu} \psi^k = 0
}
Using the chirality matrix\footnote{See Section \ref{Lightcone Coordinates} for the conventions in 2D.} $\gamma_c$ we can decompose $\psi=(\psi_- \ \psi_+)^T$ with $\gamma_c \psi_{\pm} = \mp \psi_{\pm}$. Then the Dirac equation yields the Weyl equations for the chiral components
\al{ \label{Weylequations}
\partial_+ \psi_- = 0=\partial_- \psi_+ 
}
Therefore, we have $\psi_-= \psi_- (x^-)$ and $\psi_+=\psi_+(x^+)$, and the equations for $\psi_-$ and $\psi_+$ are decoupled in the massless case. 

Upon quantization we get the anticommutation relations  which can be written in terms of the  chiral components as
\al{
\{ \psi^i_{\pm} (x) , \psi^j_{\pm} (y) \} & = \hbar \delta (x-y) \delta^{ij} \\ \{ \psi^i_+(x) , \psi^j_- (y) \} & = 0
}

Since dynamics of the chiral components are decoupled in the massless case, the theory described has an internal O$(N)\times$O$(N)$ symmetry at the classical level.
Generators of O$(N)$ can be taken as $T^a = i M^a$, where $M^a$ are $N \times N$ real, antisymmetric matrices satisfying
\al{ \label{ONbasisMa}
[M^a , M^b]=f^{abc} M^c
}
Associated with this symmetry are the (classically) conserved  chiral currents
\al{
J^a_{\pm} = \frac{1}{2} \psi_{\pm}^T T^a \psi_{\pm}
} 
Their conservation read
\al{
\partial_- J_+ = 0 = \partial_+ J_-
}
so that $J^a_+$ ($J^a_-$) is a function of $x^+$ ($x^-$) only, as implied by \eqref{Weylequations}. 

 Quantization yields the following commutation relations
 \begin{subequations}
 \label{currentalgebra}
\al{ 
[J^a_{\pm} (x) , J^b_{\pm} (y) ] & = i \hbar f^{abc} J^c_{\pm} (x) \delta (x- y) + \frac{i \hbar^2}{2 \pi} k \delta^{ab} \delta' (x-y) \\
[J^a_{\pm} (x) , J^b_{\mp} (y) ] & = 0 
}
\end{subequations}
This algebra is none other than the KM algebra \eqref{KacMoodyalgebra1}. The c-number term, appearing upon quantization, is also known as the Schwinger term. Presence of the $\hbar^2$ coefficient  emphasizes that this is a second-order quantum effect (i.e. corresponds to a 1-loop Feynman diagram).

The energy momentum tensor for this theory is given by
\al{
T^{\mu \nu} = \frac{i}{4} \left( \overline{\psi} \gamma^{\mu}  \overset{\leftrightarrow}{\partial}_{\nu} \psi + \overline{\psi} \gamma^{\nu} \overset{\leftrightarrow}{\partial}_{\mu} \psi   \right)
}
It is traceless so the theory is classically conformally invariant.  Upon quantization its Laurent modes $\{L_n\}$ satisfy 
\al{
[L_m , L_n ] = (m-n) L_{m+n} +\frac{c}{12} m (m^2 -1) \delta_{m , -n}
}
with $c = N /2$. This is the Virasoro algebra. We will discuss the Virasoro algebra in detail in Section \ref{Virasoro Algebra}.

\subsection{Coadjoint Action of Kac-Moody Algebra }
For the purposes of construction of the geometric action on the coadjoint orbits of the Kac-Moody (KM) algebra we are going to use the conventions set in \cite{delius90}. 

The KM algebra is defined by
\al{ \label{DeliusKMcommutation}
[J^a_m,J^b_n]& =f^{abc} J^c_{m+n} + km \delta_{m+n} \delta^{ab} I \notag\\ [J^a_m, I] & = [I,I]=0
}
where $f^{abc}$ are the structure constants of the semi-simple Lie algebra $\mathfrak{g}$ underlying KM algebra, $\delta^{ab}$ is the Killing metric, $I$ is the generator for the central charge, and $k$ is a constant. Structure constants are taken real and $(J^a_m)^{\dagger}=-J^a_{-m}$, so that $k$ is also real. (See Section \ref{Loop Group} and for more details \cite{goddardolive86}.)

A general Lie algebra element is written as
\al{
\Lambda +\alpha kI & =  \Lambda^n_a J^a_n + \alpha k I   = \oint \frac{dz}{2 \pi i} \ \text{Tr} \ \Lambda(z) J(z) + \alpha k I  \notag\\ & \equiv (\Lambda(z), \alpha)
}  
where $\Lambda(z) = \Lambda^a(z) T^a$ , $J(z) =  J^a(z) T^a$ , $\text{Tr} \ T^a T^b = \delta^{ab}$ , $\Lambda^a(z)= z^n \Lambda^a_n$ and $J^a(z) =  z^{-n-1} J^a_n$. 
The contour integral is around the origin in the complex $z-$plane and the following  is  used:
\al{
\oint \frac{dz}{2 \pi i} \ z^{p-1} = \delta_p
}

The commutator between two general elements with central charges becomes 
\al{ \label{KMad}
[\Lambda + \alpha k I & , \Sigma + \beta k I ] \notag\\ & = \oint \frac{dz}{2 \pi i} \ \text{Tr}  \ [ \Lambda(z),\Sigma(z)]  J(z) + k \oint \frac{dz}{2 \pi i} \ \text{Tr} \ \partial_z \Lambda(z) \Sigma(z)  I \notag\\ & \equiv \left( [\Lambda ,\Sigma ] \ , \ \oint \frac{dz}{2 \pi i} \ \text{Tr} \ \partial_z \Lambda(z) \Sigma(z)  \right)
}
This defines the action of KM algebra on itself, i.e., the ad-action. The center of the commutator is  defined through the so-called two-cocyle 
\al{ \label{KMcocycle}
\omega ((\Lambda, \alpha) , (\Sigma, \beta)) \equiv \oint  \frac{dz}{2 \pi i} \ \text{Tr} \ \partial_z \Lambda(z) \Sigma(z) 
}

A finite group element, 
\al{
g = \exp ( \Sigma + \beta kI ) = \exp \left( \oint \frac{dz}{2 \pi i} \ \text{Tr} \ \Sigma(z) J(z) + \beta k I \right) \equiv \exp \Sigma(z),
}
acts on the algebra as
$( \Lambda(z),\alpha) \mapsto (\Lambda_g (z) , \alpha_g)\equiv \text{Ad}_g ( \Lambda(z),\alpha)  $ where
\al{ \label{AdjgKM}
 (\Lambda_g (z) , \alpha_g) = \left( g(z) \Lambda(z) g^{-1} (z) , \alpha + \oint \frac{dz}{2 \pi i} \ \text{Tr} \ \partial_z g(z) \Lambda(z) g^{-1} (z) \right)
}
Infinitesimal reduction of \eqref{AdjgKM} yields back \eqref{KMad}.

We will denote a coadjoint vector by $(A(z), a)$. The pairing $\left< \ | \ \right> : \mathcal{G}^* \times \mathcal{G} \rightarrow \mathbb{R}$ is chosen as
\al{ \label{KMpairing}
\left< ( A(z),a) | (\Lambda(z),\alpha) \right> = \oint \frac{dz}{2 \pi i} \ \text{Tr} \ A(z) \Lambda(z) + a \alpha
}
The coadjoint action $(A (z)  , a ) \mapsto ( A_g(z),a_g) \equiv \text{Ad}^*_g (A (z)  , a ) $ is then defined by  invariance of the pairing under the action of  $g$ :
\al{
\left< ( A_g(z),a_g) | (\Lambda_g(z),\alpha_g) \right> \overset{!}{=} \left< ( A(z),a) | (\Lambda(z),\alpha) \right>
}
Using \eqref{AdjgKM} and \eqref{KMpairing}, this condition yields
\al{ \label{CoadKMfiniteact}
 A_g(z) & = g(z) A(z) g^{-1} (z) - a \partial_z g(z) g^{-1} (z) \notag\\  a_g & = a
}
Notice that the coadjoint element $A$ can be identified as a gauge field in 1D. Alternatively, \eqref{CoadKMfiniteact} can be identified as a time-independent gauge transformation of $A_1 = A$ component of a Yang-Mills field $A_{\mu}$ in 2D.

\section{Virasoro Algebra }
\label{Virasoro Algebra}
\subsection{Diffemorphism Algebra in 1D }
In any dimensions the  Lie derivative of a vector field $\eta$ along another vector field $\xi$ can be written as
\al{
\ld_{\xi} \eta^a =  \xi^b \partial_b \eta^a - \eta^b \partial_b \xi^a \equiv (\xi \circ \eta)^a
}
and it satisfies
\al{ \label{Liebracket}
[ \ld_{\xi} , \ld_{\eta} ] = \ld_{\xi \circ \eta} 
}
This defines  the diffeomorphism algebra \cite{courant}. 

In 1D we can write the bracket above, explicitly,  as\footnote{One often uses the shorthand notation, 
$[\xi , \eta] = \xi \eta' - \eta \xi' $.}
\al{ \label{1Ddiffeoalgebra}
\left[\xi\,\frac{d}{d\theta} , \eta \,\frac{d}{d\theta} \right] = (\xi \eta' - \xi' \eta ) \frac{d}{d\theta} 
}

The Witt algebra (whose central extension is the Virasoro algebra) is a realization of 1D diffeomorphism algebra. Indeed on a circle, the realizations,
\al{ \label{adjointrealizationsVir} 
& \xi  = i e^{im \theta}\partial_{\theta} = L_m && \eta = i e^{i n \theta} \partial_{\theta} =L_n \\ 
& \xi   = - z^{m+1} \partial_z = L_m && \eta   = - z^{n+1} \partial_z = L_n
} 
yield 
\al{
[L_m , L_n ] = (m-n) L_{m+n}
}
Two copies of Witt algebra, with bases $\{L_m \}$ and $\{\bar{L}_n \}$ such that, for all $m , n \in \mathbb{Z}$, $[L_m , \bar{L}_n ]=0$, generate conformal symmetry in 2D at the classical level. 
\subsection{Virasoro Algebra}
We can centrally extend the 1D algebra by introducing a coordinate-invariant  cocycle $c( \ , \ ) : \mathfrak{g} \times \mathfrak{g} \rightarrow C$ with which the bracket \eqref{Liebracket} is modified to
\al{
[(\ld_{\xi} ; a) , (\ld_{\eta} ; b)] = (\ld_{\xi \circ \eta} ; c(\xi , \eta))
}
In order that the Jacobi identity is satisfied,  cocyle must be antisymmetric and must satisfy the  condition
\al{
( [ \xi , \eta ] , \zeta ) + ( [ \eta , \zeta ] , \xi ) + ( [ \zeta , \xi ] , \eta ) = 0
}

There are two commonly used conventions for the  central extension of the Virasoro algebra. The Gelfand-Fuchs cocyle is defined by \cite{gelfand} 
\al{ \label{GelfandFuchscocyle}
c(\xi,\eta) \equiv \int \frac{d\theta}{2 \pi} \xi'(\theta) \eta''(\theta)  = -\frac{1}{2} \int \frac{d\theta}{2 \pi} (\xi(\theta)  \eta'''(\theta) -\xi'''(\theta) \eta(\theta) )
}
where the second equality follows by partial integrations and using the fact that $\xi$ and $\eta$ are smooth vector fields on the circle. With this choice we get 
\al{ \label{GelfandFuchsVirasoro}
\left[\xi \frac{d}{d\theta} , \eta \frac{d}{d\theta} \right] = (\xi \eta' - \xi' \eta) \frac{d}{d\theta} - \frac{ic}{48 \pi} \int_0^{2 \pi}  d\theta \ (\xi \eta''' - \xi''' \eta) 
} 
where the additional factor of $ic/12$ is introduced for conventional purposes \cite{witten88}. Then in terms of the basis elements $\{L_m\}$ in  \eqref{adjointrealizationsVir} the commutation relations become
\al{ \label{GelfandVirasorocommutation}
[L_m , L_n ] = (m-n) L_{m+n} + \frac{c}{12} m^3 \delta_{m+n}
}

In string theory (see e.g. \cite{beckers} Section 2.4), the commonly used convention  differs   by adding a constant to $L_0$  to replace  \eqref{GelfandVirasorocommutation} by 
\al{ \label{stringVirasorocommutation}
[L_m , L_n ] = (m-n) L_{m+n} + \frac{c}{12} (m^3-m) \delta_{m+n}
}
 Then the subset $\{L_{-1} , L_0 , L_1 \}$ is "preserved", i.e. does not receive a contribution from the central extension. This subset generates the Lie group SL$(2, \mathbb{R})$ or SU$(1,1)$. These are $2 \times 2$ real matrices with unit determinant.

\subsection{Coadjoint Action of Virasoro Algebra }

In this section, we follow the conventions set in \cite{delius90}. 
Smooth vector fields $\xi \in \text{Vect}(S^1)$ on a circle generate  orientation preserving diffeomorphisms $F \in \text{Diff}(S^1)$.  The adjoint action of  $F$ on a smooth vector field  $\xi(\theta)$ reads 
\al{
F :  \xi(\theta) \mapsto \text{Ad}_F \xi \equiv \xi_F(\theta) 
}
such that 
\al{
\xi_F (F(\theta)) = F'(\theta) \xi(\theta) 
}
Indeed, for an infinitesimal diffeomorphism $F(\theta) =  \theta - \eta(\theta)$ this reduces to \eqref{1Ddiffeoalgebra}
\al{
\delta \xi \equiv \xi_{\theta-\eta(\theta)} - \xi_{\theta} = \eta \xi' - \eta'\xi = [\eta,\xi] = \text{ad}_\eta \xi
} 

The central charge transforms as 
\al{ \label{Adcentertrf}
F : a \mapsto a_F = a + \int \frac{d\theta}{2 \pi} S(\theta,F) \xi(\theta) 
}
where $S(\theta, F) \equiv SF(\theta)$ is the Schwarzian derivative \eqref{TheSchwarzian}.  
Indeed under an infinitesimal transformation $F(\theta) = \theta - \eta(\theta)$ we can compute
\al{ \label{Schwarzianinfinitesimal}
S(\theta , \theta - \eta(\theta) ) =- \eta'''(\theta)
}
 so that \eqref{Adcentertrf} reduces to \eqref{GelfandFuchscocyle}
\al{
\delta a = a_{\theta - \eta} - a_{\theta} = - \int \frac{d\theta}{2 \pi} \, \eta''' \, \xi = \int \frac{d\theta}{2 \pi} \, \eta \, \xi'''
}

 The coadjoint action is introduced by the invariant pairing\footnote{Note that the dual space $\mathcal{G}^*$ considered here is not the set of all linear functionals on $\mathcal{G}$. This is  sometimes called the regular dual \cite{ovsienkobook} or the smooth dual \cite{segal}. }, $\left< \ | \ \right> : \mathcal{G}^* \times \mathcal{G} \rightarrow \mathbb{R}$ chosen to be
\al{ \label{Virasoropairing}
\left< (u,b^*) | (\xi,a) \right> = b^*a + \int \frac{d\theta}{2 \pi} \, u(\theta)\, \xi(\theta) 
}
Invariance of the pairing under the action of a diffeomorphism $F$ means that
\al{
\left< (u_F,b_F^*) | (\xi_F,a_F) \right> = \left< (u ,b^*) | (\xi,a) \right> 
}
Then 
 the coadjoint action of $F$ on a coadjoint vector $(u,b^*)$ is obtained as
\al{ \label{coadpassive}
u_F (F(\theta))\equiv \text{Ad}^*_F u  & = (F'(\theta))^{-2} ( u(\theta) - b^* S(\theta , F))
\\ 
b_F^* & = b^*
}
Indeed using the given adjoint and coadjoint transformations we can verify the invariance condition  
\al{
\left< (u_F , b^*_F) | (\xi_F , a_F) \right> & = \int \frac{df}{2 \pi} u_F(f) \xi_F(f) + a_F b^*_F 
\notag\\ 
& = \int \frac{df}{2 \pi} (F'(\theta))^{-1} (u(\theta) - b^* SF(\theta) ) \xi(\theta) + b^* \int \frac{d\theta}{2 \pi} SF(\theta) \xi(\theta)  + b^* a 
\notag\\ 
& = \int \frac{d\theta}{2 \pi} u(\theta) \xi(\theta) - b^* \int \frac{d\theta}{2 \pi} SF(\theta) \xi(\theta) +b^* \int \frac{d\theta}{2 \pi} SF(\theta) \xi(\theta) + b^* a \notag\\ & = b^* a + \int \frac{d\theta}{2 \pi} u(\theta) \xi(\theta)  \notag\\ & = \left< (u ,b^*) , (\xi,a) \right> 
}
where $f \equiv F(\theta)$.

For an infinitesimal diffeomorphism $F(\theta) = \theta - \xi(\theta)$ the coadjoint action \eqref{coadpassive} reduces to 
\al{ \label{coadinfinitesimal}
\text{ad}^*_{\xi} \equiv \delta u \equiv u_{\theta - \xi} - u_{\theta} = \xi u' + 2 \xi' u + b^* \xi'''
}

Note that the transformation \eqref{coadpassive} corresponds to a passive  transformation. The active version can be obtained by inverting \eqref{coadpassive} using $\theta = F^{-1} (f)$ and using the Schwarzian identity \eqref{SchwarzianInverseIdentity}. 
This yields
\al{ \label{Virasoroactivecoad}
u_F(\theta) =  (F'(\theta))^{2}  u(F(\theta)) + b^* S(\theta , F)
}

\subsection{Coadjoint Element $\Sigma$ Formed from Affine Connection }
\label{connectionVircoadjointelement}

Under a coordinate transformation $x \mapsto \bar{x}(x)$, affine connection coefficients $\Gamma_{ab}^{c}$ transform  as
\al{\label{gamma trf}
\bar{\Gamma}^{c}_{ab} (\bar{x})= \frac{\partial \bar{x}^{c}}{\partial x^{d}}  \frac{\partial x^{e}}{\partial \bar{x}^{a}}  \frac{\partial x^{f}}{\partial \bar{x}^{b}} \Gamma^{d}_{e f} (x) -  \frac{\partial x^{d}}{\partial \bar{x}^{a}}  \frac{\partial x^{e}}{\partial \bar{x}^{b}}  \left( \frac{\partial^2 \bar{x}^{c}}{\partial x^{d} \partial x^{e}} \right)    
}
This deviates from the transformation of a (1,2)-tensor by the last term.  In 1D it reduces to
\al{\label{gammatrf1d}
\bar{\Gamma} (\bar{x} )= \frac{\partial x}{\partial \bar{x}} \Gamma (x) - \left( \frac{\partial x}{\partial \bar{x}} \right)^2 \frac{\partial^2 \bar{x}}{\partial x^2}
}
For an infinitesimal coordinate transformation $\bar{x} = x - \xi(x)$ we have $\partial \bar{x} / \partial x = 1- \xi'$ and $\partial x / \partial \bar{x} = 1+ \xi'$ (to first order in $\xi$).
We shall use the convention that if the argument of a field is suppressed, it is $x$ i.e. the original coordinate. 
Plugging these into \eqref{gammatrf1d} we get 
\al{
\bar{\Gamma} (\bar{x})   = \Gamma  + \Gamma \xi' +\xi''
}
On the other hand, we also have (by Taylor expansion)
\al{
\bar{\Gamma}(\bar{x}) = \bar{\Gamma} ( x- \xi)  = \bar{\Gamma}  - \xi \bar{\Gamma}' = \bar{\Gamma} - \xi \Gamma'
}
Combining the two expressions we get  
\al{ 
 \delta \Gamma  : = \bar{\Gamma}  - \Gamma   =  \xi \Gamma' + \Gamma \xi' +\xi'' 
}
where $\delta \Gamma (x)$ is the Lie variation of $\Gamma(x)$ with respect to the vector field $\xi(x)$. 

Using $[\partial, \delta]=0$ we can compute
\al{
\delta \Gamma' = (\delta \Gamma)' =  \xi \Gamma'' + 2 \xi' \Gamma' + \xi'' \Gamma + \xi'''
} 
Since $\delta$ is a derivation we also have
\al{ 
\delta (\Gamma^2/2) =  \Gamma \delta \Gamma
}
Combining the two results we can compute
\al{ \label{GammaVirasoro}
\delta (  \Gamma' - \Gamma^2/2 ) & = ( \xi \Gamma'' + 2 \xi' \Gamma' +  \xi'' \Gamma +  \xi''') -   \Gamma ( \xi \Gamma' + \Gamma \xi' + \xi'' ) \notag\\ & = \xi (  \Gamma' - \Gamma^2/2 )' + 2 \xi' ( \Gamma' - \Gamma^2/2 ) + \xi'''
}
In other words, the object
\al{ 
\Sigma \equiv \Gamma' - \Gamma^2/2
}
 transforms  as a Virasoro coadjoint element with central charge one
\al{\label{Sigmainfinitesimalcoad}
\delta \Sigma= \xi \Sigma' + 2 \xi' \Sigma + \xi'''
}

Let us define the  object $\Sigma_k \equiv \Gamma' + k \Gamma^2$. Then we can compute 
\al{
\delta \Sigma_k = \xi \Sigma_k' +2 \xi' \Sigma_k + \xi'' \Gamma (1+2k) + \xi''' 
}
So  for $k \neq -1/2$ the object $\Sigma_k$ almost transforms as a Virasoro coadjoint element, but there is an additional $\xi''$ center which breaks the invariance of the  pairing \eqref{Virasoropairing}.

Now consider the object $c\Sigma$.
Using \eqref{GammaVirasoro} it is easy to see that $c\Sigma$ transforms as a Virasoro coadjoint element of central charge $c$.

Next consider a rank-two tensor $S_{ab}$. It transforms under $x\mapsto \bar{x}(x)$ as 
\al{
\bar{S}_{ab} (\bar{x}) = \frac{\partial x^{c}}{\partial \bar{x}^{a}} \frac{\partial x^{d}}{\partial \bar{x}^{b}} S_{cd} (x)
}
Following a similar analysis as in above, in 1D, we get 
\al{
\delta S \equiv \bar{S}  - S =  \xi S' + 2 \xi' S
}

Adding a rank-two tensor to a rank-two object that transform as a Virasoro coadjoint element does yield another object that transform as a Virasoro coadjoint element with the same central charge. Indeed for an object $D$ defined by,
\al{
D = S+ c\Sigma
}we get the following  transformation 
\al{
\delta D & = \delta S + c \delta ( \Gamma' - \Gamma^2/2)  \notag\\ & =  \xi [ S' +   c  (\Gamma'' - \Gamma \Gamma' ) ] + 2 \xi' [ S + c(  \Gamma' - \Gamma^2/2)] +  c \xi ''' \notag\\ & = \xi D' + 2 \xi' D + c \xi'''    \label{Coad1Dtrf}
}
This calculation shows that we can use $c\Sigma = c(\Gamma'-\Gamma^2/2)$ as a core to build arbitrary Virasoro coadjoint elements of central charge $c$ by adding arbitrary rank-two tensors to it. In particular, we can build one from the spacetime metric $g$, $D_g \equiv g+c(\Gamma'-\Gamma^2/2)$. 

Finally consider two objects that transform as Virasoro coadjoint elements, $D_1 , D_2$ with the same central charge. Then we can compute
\al{ \label{Virasorocoadjointdifference}
\delta (D_1 - D_2) = \xi (D_1 -D_2)' + 2 \xi' (D_1 - D_2) 
}
Thus, the difference transforms as a rank-two tensor. This shows that using a multiple of $(\Gamma'-\Gamma^2/2)$ we  can always extract a  rank-two tensor out of a Virasoro coadjoint element.

\subsection{Higher Dimensional Lift of $\Sigma$}
\label{HigherDimLiftofSigma}

Although an affine connection  does not transform as a tensor, its  Lie derivative does. To see this one first computes the pullback of the connection coefficients (i.e. the usual coordinate transformation) and applies the formal definition of the Lie derivative to get  
\al{ \label{higherdimLiederofGamma}
\mathcal{L}_{\xi} \Gamma^{\mu}_{\nu \lambda} = \xi^{\rho} \partial_{\rho} \Gamma^{\mu}_{\nu \lambda} - \partial_{\rho} \xi^{\mu} \Gamma^{\rho}_{\nu \lambda} + \partial_{\nu} \xi^{\rho} \Gamma^{\mu}_{\rho \lambda} + \partial_{\lambda} \xi^{\rho} \Gamma^{\mu}_{\nu \rho} + \partial_{\lambda} \partial_{\nu} \xi^{\mu} 
}
The first four terms are what you would expect from a (1,2)-tensor and the last term is the inhomogeneous term representing the nontensoriality of $\Gamma$. The last term can be rewritten as part of $\nabla_{\lambda} \nabla_{\nu} \xi^{\mu}$, then one can show that  
\al{
\mathcal{L}_{\xi} \Gamma^{\mu}_{\nu \lambda} = \nabla_{\lambda} \nabla_{\nu} \xi^{\mu} - \xi^{\rho} R^{\mu}_{\ \ \nu \lambda \rho} 
}
The expression on the right is a tensor, so the Lie derivative of the connection coefficients form a tensor.

As discussed in the previous section the object $\Sigma=\Gamma'-\Gamma^2/2$ in 1D is a Virasoro coadjoint element of central charge one, so we can extract a pure tensor out of a diff field $D$ of central charge $c$ as $D - c\Sigma$. 
We lift the diff field  to a rank-two object $D_{\mu \nu}$ in higher dimensions since this is the most natural lift that follows from the coadjoint action\footnote{It is also possible to lift the diff field to pseudotensor densities with appropriate weight. See for instance \cite{rodgers19942d}.}.  Then $\Gamma'$ and $\Gamma^2$ should also have two free indices. 

For the $\Gamma'$ the possible lifts are
$\Gamma'  \sim \partial_{\lambda} \Gamma^{\lambda}_{\mu \nu} , \partial_{\mu} \Gamma^{\lambda}_{\lambda \nu}$ and $\partial_{\nu} \Gamma^{\lambda}_{\mu \lambda}$,
whereas for the $\Gamma^2$ we have $\Gamma^{\lambda}_{\mu \nu} \Gamma^{\sigma}_{\sigma \lambda} , \Gamma^{\lambda}_{\mu \lambda} \Gamma^{\sigma}_{\nu \sigma}$ and $\Gamma^{\lambda}_{\mu \sigma} \Gamma^{\sigma}_{\nu \lambda}$. Hence we can  form a general combination 
\al{ \label{higherdimcoadfield}
\Sigma_{\mu \nu}  \equiv a \partial_{\lambda} \Gamma^{\lambda}_{\mu \nu} +b \partial_{\mu} \Gamma^{\lambda}_{\lambda \nu}  + c  \partial_{\nu} \Gamma^{\lambda}_{\mu \lambda} + d \Gamma^{\lambda}_{\mu \nu} \Gamma^{\sigma}_{\sigma \lambda}+ e \Gamma^{\lambda}_{\mu \sigma} \Gamma^{\sigma}_{\nu \lambda}  + f \Gamma^{\lambda}_{\mu \lambda} \Gamma^{\sigma}_{\nu \sigma} 
}
and subject it to the condition 
\al{ \label{higherdimcentralchargecond}
q \equiv a+b+c = -2 (d+e+f)
}
This yields a higher-dimensional, rank-two lift of a Virasoro coadjoint element of central charge $q$. 

We can show by  direct computation
\al{ \label{lambda2esitlanbda3}
\partial_{\mu} \Gamma^{\lambda}_{\lambda \nu} =  \partial_{\nu} \Gamma^{\lambda}_{\mu \lambda} 
}
So there is a redundancy in  \eqref{higherdimcoadfield}. However, we intentionally introduced these two terms to keep  symmetry manifest. Also note that these are related to the metric determinant by 
\al{
\partial_{\mu} \Gamma^{\lambda}_{\lambda \nu} = \partial_{\nu} \left( \frac{1}{\sqrt{g}} \partial_{\mu} \sqrt{g} \right)
}

The most natural lift of the Virasoro coadjoint transformation (with linear center term ignored) \eqref{Coad1Dtrf}
reads 
\al{ \label{CoadNDtrf}
\delta D_{\mu \nu} = \xi^{\lambda} \partial_{\lambda} D_{\mu \nu} + \partial_{\mu} \xi^{\lambda} D_{\lambda \nu} + \partial_{\nu} \xi^{\lambda} D_{\mu \lambda} + q \partial_{\mu} \partial_{\nu} \partial_{\lambda} \xi^{\lambda}
}
which we use for building the transverse action for the diff field in Section \ref{Diffaction}. 
So the question is "Does the object $\Sigma_{\mu \nu}$ defined in \eqref{higherdimcoadfield} satisfy 
\al{ \label{deltaLambdaansatz}
\delta \Sigma_{\mu \nu} = \xi^{\lambda} \partial_{\lambda} \Sigma_{\mu \nu} + \partial_{\mu} \xi^{\lambda} \Sigma_{\lambda \nu} + \partial_{\nu} \xi^{\lambda} \Sigma_{\mu \lambda} + q \partial_{\mu} \partial_{\nu} \partial_{\lambda} \xi^{\lambda}
}
given the Lie derivative $\delta \Gamma^{\mu}_{\nu \lambda}$ \eqref{higherdimLiederofGamma}?" The answer turns out to be negative. Here are the results : We are going to denote the true Lie variation, i.e. one obtained using \eqref{higherdimLiederofGamma} by $\delta_{\text{tr}} \Sigma_{\mu \nu}$,  and the Lie variation obtained from the ansatz \eqref{deltaLambdaansatz} by $\delta_{\text{an}} \Sigma_{\mu \nu}$. Then we define the difference
\al{
\Delta_{\mu \nu} \equiv \delta_{\text{tr}} \Sigma_{\mu \nu} - \delta_{\text{an}} \Sigma_{\mu \nu}
} 
We find 
\al{
\Delta_{\mu \nu} & = (-a +d) \Gamma^{\rho}_{\mu \nu} \partial_{\rho} \partial_{\sigma} \xi^{\sigma} + (a+e) ( \Gamma^{\rho}_{\sigma \nu}  \partial_{\mu} \partial_{\rho} \xi^{\sigma} +  \Gamma^{\rho}_{\sigma \mu} \partial_{\rho} \partial_{\nu} \xi^{\sigma}) \notag\\ &  + f ( \Gamma^{\rho}_{\rho \nu} \partial_{\mu} \partial_{\sigma} \xi^{\sigma} + \Gamma^{\rho}_{\rho \mu} \partial_{\nu} \partial_{\sigma} \xi^{\sigma}) + (b+c+d) \Gamma^{\rho}_{\rho \sigma} \partial_{\mu} \partial_{\nu} \xi^{\sigma} 
}
Notice that the difference is only made up of terms of order $\xi''$, and  1D reduction of $\Delta$ vanishes with the condition \eqref{higherdimcentralchargecond} as expected. 

The next question is whether a subcase (with some of the terms set to zero) subject to condition \eqref{higherdimcentralchargecond} yields a vanishing $\Delta_{\mu \nu}$. The answer turns out to be negative again. There are simple cases which come close to the goal. For instance for the case $b=1=c=-d$ and $a=0=e=f$ we get
\al{ \label{simplecaseSigmamunu}
\Sigma_{\mu \nu} =   \partial_{\mu} \Gamma^{\lambda}_{\lambda \nu}  +   \partial_{\nu} \Gamma^{\lambda}_{\mu \lambda}  - \Gamma^{\lambda}_{\mu \nu} \Gamma^{\sigma}_{\sigma \lambda}
}
and 
\al{
\Delta_{\mu \nu} =   -  \Gamma^{\rho}_{\mu \nu} \partial_{\sigma} \partial_{\rho} \xi^{\sigma} + \Gamma^{\rho}_{\rho \sigma} \partial_{\mu} \partial_{\nu} \xi^{\sigma}
}

In Section \ref{Diffaction}, we are going to obtain \eqref{Coad1Dtrf} from \eqref{CoadNDtrf} by some gauge fixing arguments instead of the direct dimensional  lifting.

\subsection{Covariant Cocyle }
\label{Covariantcocyle}
Consider the Gelfand-Fuchs cocyle introduced before\footnote{For simplicity, $c$ is scaled by $2\pi$. }
\al{
c(\xi, \eta) =  -\frac{1}{2}\int d\theta\  (\xi \eta''' - \xi''' \eta)
}
If we place an affine connection $\nabla$ on circle this cocyle can be extended to 
\al{ \label{covcocyle}
c_{\Gamma}(\xi,\eta) = \frac{1}{2}\int dx^a  (\xi^b \nabla_a \nabla_b \nabla_c \eta^c ) - (\xi\leftrightarrow \eta)
}
covariantly, in higher dimensions. Expanding the  derivatives, in 1D, we get 
\al{
c_{\Gamma} (\xi,\eta) & = \frac{1}{2}\int dx \ \xi (\partial-\Gamma) \partial (\partial+\Gamma) \eta - (\xi \leftrightarrow \eta) \notag\\ & =  \frac{1}{2}\int dx  \  \xi (\eta''' + (2 \Gamma' - \Gamma^2) \eta') - ( \xi \leftrightarrow \eta)
}
Therefore we obtain \cite{courant}
\al{
c_{\Gamma}(\xi, \eta) = \int dx\  \frac{1}{2} (\xi \eta''' - \xi''' \eta) + \int dx \ (\xi \eta' - \xi' \eta) (\Gamma' - \Gamma^2/2)
}
We can rewrite this in terms of  the pairing \eqref{Virasoropairing} as
\al{
c_{\Gamma} (\xi, \eta) = c(\xi, \eta) + \left< \Gamma'-\Gamma^2/2 \, | \, [\xi, \eta] \right> 
}
Since we have shown that $\Sigma = \Gamma' -\Gamma^2/2$ transforms as a Virasoro coadjoint element of central charge one we can interpret the last term as the Kirillov form $\Omega$ on the  coadjoint orbit of $\Sigma$, evaluated on two tangent vectors $\Sigma_{\xi} , \Sigma_{\eta}$ obtained by the action of the adjoint vectors $\xi,\eta$ (equation \eqref{Kirillovform}). That is, 
\al{
 c_{\Gamma}(\xi,\eta) - c(\xi,\eta)&   = \Omega_{\Sigma} ( \Sigma_{\xi} , \Sigma_{\eta} ) \\ 
 \Sigma_{\xi}  \equiv \text{ad}^*_{\xi} \Sigma \  \ , \ \ & \Sigma_{\eta}  \equiv  \text{ad}^*_{\eta} \Sigma
}

\subsection{Chiral Splitting of Curvature } 
In this section we would like to investigate an interesting possibility related to the $N$D generalization of the diff field-affine connection relationship. Consider the Ricci curvature tensor\index{Ricci tensor}
\al{
R_{\alpha \beta} \equiv R^{\rho}_{ \ \ \alpha \rho \beta}  = \partial_{\rho} \Gamma^{\rho}_{\beta \alpha} - \partial_{\beta} \Gamma^{\rho}_{\rho \alpha} + \Gamma^{\rho}_{\rho \lambda} \Gamma^{\lambda}_{\beta \alpha}  - \Gamma^{\rho}_{\beta \lambda} \Gamma^{\lambda}_{ \rho \alpha} 
}
Adding and subtracting the term $a\partial_{\rho} \Gamma^{\rho}_{\beta \alpha}+(1-a)\partial_{\beta} \Gamma^{\rho}_{\rho \alpha}$ we can rewrite this as
\al{
R_{\alpha \beta}  &  = \Big( (1+a) \partial_{\rho} \Gamma^{\rho}_{\beta \alpha} +(1-a) \partial_{\alpha} \Gamma^{\rho}_{\rho \beta} - \Gamma^{\rho}_{\beta \lambda} \Gamma^{\lambda}_{ \rho \alpha}  \Big)  - \Big( a \partial_{\beta} \Gamma^{\rho}_{\rho \alpha} + (2-a) \partial_{\alpha} \Gamma^{\rho}_{\rho \beta} - \Gamma^{\rho}_{\rho \lambda} \Gamma^{\lambda}_{\beta \alpha}  \Big) \notag\\ & \equiv R^+_{\alpha \beta} - R^-_{\alpha \beta}
}
The point of this definition is that the 1D reduction of $R^{\pm}_{\alpha \beta}$ are each given by
\al{ \label{Rplusminus1D}
R^{\pm} = 2 \Gamma' - \Gamma^2
}
Each  transforms as a Virasoro coadjoint element with central charge two. Note that although the Ricci tensor vanishes in 1D, $R^{\pm}$ do not. Note also that we have $\partial_{\beta} \Gamma^{\rho}_{\rho \alpha}= \partial_{\alpha} \Gamma^{\rho}_{\rho \beta}$ so that $R^{\pm}_{\alpha \beta}$ are each symmetric. We could also add and subtract  possible $\Gamma^2$ terms to obtain a more generic splitting as long as we keep the ratio of coefficients of $\Gamma'$ and $\Gamma^2$ as in \eqref{Rplusminus1D}. 

Now, if we introduce two copies of the diff field, $D^{\pm}$, each with central charge $c$, it is possible to obtain two pure tensors out of the diff field in 1D : 
\al{
c\tilde{R}^{\pm} \equiv D^{\pm} -  cR^{\pm} 
}
Here $R^{\pm}$  are postulated to represent two "chiral components" of the Ricci tensor and $\tilde{R}^{\pm}$  are the tensorial chiral components of the diff-corrected curvature tensor. Explicitly we have  
\al{
c\tilde{R} & = cR -  (D^+ - D^-)  \notag\\ & = (cR^+ -  D^+ ) - (cR^- -  D^-) \notag\\ & = c\tilde{R}^+ - c\tilde{R}^- 
}

Therefore, at least in 1D, we can construct  a curvature tensor $\tilde{R}$  whose chiral components $\tilde{R}^{\pm}$ are tensorial with the chiral diff corrections. Whether this result would be extended to $N$D is an interesting mathematical quest to pursue.

\subsection{Schwarzian Chain }
\label{schwarzianchain}
Consider the active transformation of a Virasoro coadjoint element under the action of a diffeomorphism $g$,
\al{
\tilde{u}(x) = \left( \frac{dg}{dx} \right)^2 u(g(x)) + c \, Sg(x)
}
For a coadjoint element made of central charge $c$ only, this implies
\al{
(0,c) \overset{g}{\mapsto} (c\, Sg(x),c)
}
That is, for $u=0$ we have $\tilde{u}(x) = c \, Sg(x)$ under the map $\text{Ad}_g^*$. Then using the  identity \eqref{Schwarzian id},
under the action of a second diffeomorphism $h$ we get 
\al{ \label{Schwarzianchaingh}
(0,c) \overset{g}{\mapsto} (c\, Sg(x),c) \overset{h}{\mapsto} (c \, S(g\circ h)(x) , c)
}
Therefore, the Schwarzian derivative operator is an invariant of the Virasoro coadjoint action connecting a zero element to an infinite chain of nonzero elements obtained by diffeomorphisms. 

Now use the identity \eqref{Schwarzian id} again, but with the second transformation made infinitesimal\footnote{The computation here is in the active picture, so the transformation is taken to be $x \mapsto x+\xi$.}, $x \mapsto h(x) = x + \xi(x)$,
\al{
S (g \circ (x+\xi)) (x) & = Sg(x+\xi) (1+ 2 \xi') + \xi'''\notag\\ 
(Sg)_{\xi} (x) & = (1+ 2 \xi') Sg(x) + \xi (Sg)'(x) + \xi''' \notag\\ 
(Sg)_{\xi} (x) & = Sg(x) + \xi (Sg)' + 2 \xi' (Sg) + \xi'''
}
where we defined $S(g\circ(x+\xi)) = S(g + g \circ \xi) \equiv (Sg)_{\xi}$. Hence, we get 
\al{
\delta (Sg) (x) \equiv (Sg)_{\xi}(x) - (Sg)(x) =  \xi (Sg)' + 2 \xi' (Sg) + \xi''' 
}
Therefore, we see that the Schwarzian derivative of a  diffeomorphism transforms infinitesimally as a Virasoro coadjoint element of central charge one.

\section{Semi-direct Product of Virasoro and Kac-Moody Algebras  } \label{semidirect} 

Commutation relations of the semi-direct product of Virasoro and Kac-Moody algebras are given by
\al{
[L_m , L_n] & = (m-n) L_{m+n} + (c m^3 + h m ) \delta_{m+n} I_{\text{Vir}} \notag\\ [J_m^{a} , J_n^{b} ] & = i f^{abc} J_{m+n}^{c} + k m \delta_{m+n} \delta^{ab}  I_{\text{KM}} \notag\\  [L_m , J^{a}_n ] & = - n J^{a}_{m+n}  \notag\\ [I_{\text{Vir}}, \text{all} ] & = 0 = [I_{\text{KM}}, \text{all}] \label{semidirectproductalgebra}
}
where we introduced generators of centers $I_{\text{Vir}} , I_{\text{KM}}$ for each of the algebras, and we did not fix the Virasoro cocyle as in \eqref{stringVirasorocommutation}.  
Realization of the basis elements in angular and complex coordinates are given by 
\al{
L_m(\theta)  & = i e^{im\theta} \partial_{\theta} \hspace{0.2in} &&J^{a}_m(\theta) = T^{a} e^{im\theta} &&& \text{angular} \\ L_m(z) & = - z^{m+1} \partial_z  &&J^{a}_m(z) = T^{a} z^m &&&\text{complex} 
}
These are related by $z = e^{i\theta}$. Note that these realizations satisfy only the non-central part of \eqref{semidirectproductalgebra}. 

Dual elements will be denoted with tildes $\tilde{L}_m$ and $\tilde{J}^a_m$. They are defined through the  individual invariant pairings of the algebras without central extension
\al{
\left< \tilde{L}_m \Big| L_n \right> = \delta_{mn} \ \ \ , \ \ \  \left< \tilde{J}^a_m \Big| J^b_n \right> = \delta_{mn} \delta^{ab}
}

For the semi-direct product algebra we form an adjoint basis element $(L_m , J_n^a , \mu)$ and a coadjoint basis element $(\tilde{L}_m , \tilde{J}_n^a , \tilde{\mu})$. Then we introduce the invariant pairing
\al{ \label{semidirectpairing}
\left< (\tilde{L}_m , \tilde{J}^a_n , \tilde{\mu} ) \Big| (L_{m'} , J^{a'}_{n'} , \mu \right> = \delta_{mm'} + \delta_{nn'}\delta^{aa'} + \tilde{\mu} \mu
}
Recall the infinitesimal form of  invariance of the pairing :  if $u, v$ denote two adjoint  elements and $\alpha$ denote a coadjoint element then invariance reads 
\al{ \label{pairinginvariancegeneric}
0 = u * \left< \alpha | v \right> = \left< u * \alpha | v \right> + \left< \alpha | u*v \right> 
}
The commutation relations \eqref{semidirectproductalgebra} yield $u * v$ so that using \eqref{semidirectpairing} one can compute $\left< \alpha | u*v \right> $. Then again using \eqref{semidirectpairing} one can deduce $u*\alpha$, namely, the infinitesimal coadjoint acton for the semi-direct product algebra. We will state the result for generic adjoint and coadjoint elements below.

From the basis elements of the Virasoro and KM algebras and their duals  we can construct generic adjoint and coadjoint elements of the algebras as\footnote{The negative sign in the Virasoro adjoint element is introduced to avoid negative signs in the Lie derivative by switching the passive $x\mapsto x-\xi$ and active $x \mapsto x+\xi$ transformations.}
\al{ \label{semidirectgeneric}
& \xi(\theta) =  - \xi^n L_n (\theta)  &&\Lambda(\theta)= \Lambda_{a}^n J^{a}_n (\theta) &&&\text{adjoint} \\  & D(\theta) =  D^n \tilde{L}_n (\theta) &&A (\theta)=  A_{a}^n \tilde{J}^{a}_n (\theta) &&&\text{coadjoint}
}
Then generic  elements of the semi-direct product algebra become
\al{
\text{adjoint:} \hspace{0.2in} & \mathcal{F} = (\xi(\theta),\Lambda(\theta),a) \\ \text{coadjoint:} \hspace{0.2in} & B=(D(\theta),A(\theta),\mu)
}

Finally, we can write the (infinitesimal) coadjoint action as
\al{
\delta_{\mathcal{F}} B \equiv ad^*_{\mathcal{F}} B = (\delta D(\theta) , \delta A(\theta) , 0 ) 
}
By the procedure described following equation \eqref{pairinginvariancegeneric}, one can compute \cite{lano92}
\begin{subequations}
\label{semidirectiso}
\al{ \label{semidirectiso1}
\delta D (\theta) & = 2 \xi' D + D' \xi + \frac{c \mu}{2 \pi} \xi''' + \frac{ h \mu}{2 \pi} \xi' - \text{Tr}\  (A \Lambda') \\ \delta A(\theta) & = A' \xi + \xi' A - [\Lambda , A ] + k \mu \Lambda' \label{semidirectiso2}
}  
\end{subequations}
Setting $\delta D$ and $\delta A$ to zero we get the isotropy equations for the semi-direct product algebra.

\chapter{GEOMETRIC ACTIONS}

\section{WZW Action }
\label{WZNW Action}
Wess-Zumino-Witten (WZW) model arises in a variety of phenomena in 2D field theories.
First, we review its motivation.  Namely, it is  a closed form solution in 2D to the Wess-Zumino (WZ) functional, describing the low-energy effective action of QCD, and encoding the chiral anomaly. Then we  discuss  bosonization, namely the equivalence between  WZW model and the theory of 2D chiral fermions. Finally we  review Polyakov and Wiegmann's treatment which further clarifies its relation to chiral anomaly and motivates the correspondence between WZW theory and the P2DG theory in LCG. 
In the following we mainly follow \cite{divecchia}, \cite{WZNW},  \cite{polyakovwiegmann} and \cite{polyakovwiegmann84}. 

\subsection{WZ Functional in 2D }

Reconsider the massless, free theory of fermions in 2D having a U$(N) \times \text{U}(N)$ chiral flavor symmetry,
introduced in section \ref{Current Algebra}. If we couple this theory to a background gauge field (i.e. its dynamics can be ignored for the discussion)  $A_{\mu}$ the chiral symmetry is broken by the axial anomaly \cite{adleraxial}, \cite{belljackiwaxial}.  With the gauge coupling the action reads 
\al{  \label{2Dchiralfermiongaugecoupled}
S_F[ \psi , \overline{\psi} , A_{\mu}] = \frac{i}{2} \int d^2 x \ \overline{\psi}_k \slashed{D} \psi^k
}
where $\slashed{D} = i \gamma^{\mu} (\partial_{\mu} + A_{\mu} )$ and $A_{\mu} = v_{\mu} + \gamma_5 a_{\mu}$. Here $v_{\mu}$ is the vector gauge field and $a_{\mu}$ is the axial gauge field. 
 One way to express the anomaly is through the  effective action\footnote{There is a closely related but distinct notion of effective action, denoted by $\Gamma$ in the literature. For the distinction and the relationship between the two, see \cite{bilalanomaly} Section 3.6.} which is obtained by path integration over the fermionic degrees of freedom. The result is formally written as  
\al{ \label{2DFermionDeterminant}
W[A_{\mu} ] = \log \text{Det} \slashed{D}=\text{Tr} \log \slashed{D} 
}
It is more convenient to work with the chiral components of the gauge field, $A^L_{\mu}$, $A^R_{\mu}$ which transform under  the action of $(g_L , g_R ) \in$ U$(N) \times$U$(N)$ as  
\al{ \label{DiVec2}
A^L_{\mu}  \rightarrow g^{-1}_L (\partial_{\mu} + A_{\mu}^L) g_L \ ,  \ \ 
A^R_{\mu}  \rightarrow g^{-1}_R (\partial_{\mu} + A_{\mu}^R) g_R
}

The anomaly is exposed through the evaluation of the formal fermion determinant by a choice of regularization. The problem is in the measure of the path integral \cite{fujikawa}. There one sees that there is no regulator that preserves both the vector symmetry and the axial symmetry simultaneously. Thus, one is forced to choose preserving one of the symmetries, losing the other. Anomalous gauge symmetry is catastrophic for a theory since it leads to nonrenormalizability  and states of negative norm, thereby to the violation of unitarity\cite{bilalanomaly}. Hence, one evaluates the determinant by a regulator preserving the vector symmetry, which forms a subgroup of the gauge group, sacrificing the axial symmetry. In other words, under a vector transformation $(g_L, g_R)$ with $g_L = g_R$, $W[A_{\mu}]^{\text{reg}}$ is invariant whereas under a chiral transformation, i.e. $(g_L, g_R)$ with $g_R = g_L^{-1} \equiv g$ it changes by 
\al{\label{chiralWZtrf}
W[A^g_{\mu}]^{\text{reg}} = W[A_{\mu}]^{\text{reg}} + \text{WZ}(g^2 , A_{\mu} ) 
} 
This defines the Wess-Zumino functional, WZ, encoding the chiral anomaly. 

The WZ functional has been evaluated explicitly in 2D by Witten \cite{WZNW}, taking the name WZW. For this purpose, consider the following complexified parametrization of the 2D gauge field : 
\al{
A_+ & = A_0 + i A_1 \equiv B^{-1} \partial_+ B  
\\ 
A_- & = A_0 - i A_1 \equiv C^{-1} \partial_- C
}
Here, $A_+ = A_-^{\dagger}$ implies $B^{-1} = C^{\dagger}$. Then the effective action is given by
\al{
W[A_{\mu}] = I[BC^{-1}]
}
where $I$ is the WZW functional given by
\al{
I[G] = \frac{1}{8 \pi} \int_{\partial Q} d^2 x \ \text{tr} (\partial_{\mu} G \partial_{\mu} G^{-1} ) - \frac{i}{12 \pi} \int_Q d^3 x \ \epsilon^{ABC} \ \text{tr} (G^{-1} \partial_A G G^{-1} \partial_B G G^{-1} \partial_C G)
}
Here $Q$ is a 3D hemisphere with compactified 2D space as its boundary. Alternatively $\partial Q = S^2$ and $Q$ is a three-ball \cite{WZNW}.  The map $G$ originally defined on the two-dimensional boundary has  topologically (more precisely homotopically) distinct possible extensions to the three-space $Q$. The last term  above can be evaluated to be   $2 \pi n$, where $n \in \mathbb{Z}$ where $n$ is the winding number specifying the homotopy class of the extended map\footnote{To make this distinction apparent, Witten \cite{WZNW} denotes the extended map by a hat, so that all the $G$'s in the second integrand are hatted. We will use the same symbol for the original and extended maps throughout for simplicity.}. For more details, see  \cite{WZNW}, or \cite{NairQFT} Section 17.6.  

Under a vector gauge  transformation $g \in \text{U}(N)$ we have $B \rightarrow Bg$ and $C \rightarrow Cg$ so that the effective action $I(B C^{-1})$ is vector gauge invariant as desired. On the other hand, under a chiral transformation, $C \rightarrow Cg$ and $B \rightarrow B g^{-1}$ we get 
\al{
W[A^g_{\mu}] = I[B g^2 C^{-1} ] = I[BC^{-1}] + \text{WZ} [g^2 , A_{\mu}] 
}
or 
\al{
\text{WZ}[U,A_{\mu} ] = I[B U C^{-1}] - I[BC^{-1}]
}
A straightforward calculation shows that 
\al{ \label{WZexplicit}
\text{WZ}[U,A_{\mu}] = I[U] + \frac{1}{4 \pi} \int d^2 x \ \text{tr} ( A_+ U \partial_- U^{-1} & + A_- U^{-1} \partial_+ U \notag\\ & + A_+ U A_- U^{-1} - A_+ A_-)
}

 Under a chiral transformation defined by 
\al{
U \rightarrow g^{-1} U g^{-1} \ \ , \ \ B \rightarrow Bg \ \ , \ \ C \rightarrow C g^{-1}
}
we get
\al{
- \text{WZ} [U^g, A^g_{\mu}] & = I [B g^2 C^{-1}] - I [ B U C^{-1} ] \notag\\ & = - \text{WZ}[U, A_{\mu}] + \text{WZ} [ g^2 , A_{\mu} ] \label{DiVec14}
}
Comparing with \eqref{chiralWZtrf} we see that $-\text{WZ}$ can be taken as the effective action $W$.  Next, we discuss the bosonization, namely the (quantum) equivalence of the bosonic $\text{WZ}[U,A_{\mu}]$ theory and the  original Fermi theory in the background field $A_{\mu}$.

\subsection{Bosonization of Chiral Fermion Theory  }
The action \eqref{2Dchiralfermiongaugecoupled} of chiral fermions coupled to a background gauge field $A_{\mu}$  can be rewritten as 
\al{
S_F [ \psi , \bar{\psi} , A_{\mu} ]  = \int d^2 x \ [ \bar{\psi} i \slashed{\partial} \psi + \text{tr} (J_+ A_- + J_- A_+)]
}
where $J_{\pm}$ are the chiral currents. Equivalence of the WZW functional to the fermion theory is an example of bosonization, and it can be formally expressed as 
\al{ \label{WZbosonization}
\int \mathcal{D} \psi \ \mathcal{D} \bar{\psi} \ \exp ( - S_F [ \psi, \bar{\psi} , A_{\mu}] ) = \text{const} \times \int \mathcal{D} U \ \exp ( - \text{WZ} [ U , A_{\mu}])
}
where the bosonic functional measure $\mathcal{D}U$ is formally a product of Haar measures on U$(N)$. Unlike the Fermi theory, the Bose theory is assumed not to have any anomalies. Instead the lack of chiral invariance is explicit  in the bosonic action, whereas the quantum measure $\mathcal{D} U$ is taken chirally invariant. Indeed one uses an identity obtained from the chiral invariance of the Haar measure to reach the bosonization result \eqref{WZbosonization}.

Then for the abelian case, $U = \exp(i \varphi)$ with a scalar field $\varphi$, we get $\mathcal{D} U = \mathcal{D} \varphi$, and using \eqref{WZexplicit} the bosonization formula \eqref{WZbosonization} simplifies to
\al{
\int \mathcal{D} & \psi \ \mathcal{D} \bar{\psi} \ \exp \left( \int d^2 x \ ( \bar{\psi} i \slashed{\partial} \psi + J_+ A_- + J_- A_+ ) \right) \notag\\ & = \text{const}  \int \mathcal{D} \varphi \ \exp \left( - \int d^2 x \ \left[ \frac{1}{8 \pi} \partial_{\mu} \varphi \partial^{\mu} \varphi + \frac{i}{4 \pi} ( A_+ \partial_- \varphi + A_- \partial_+ \varphi \right]\right)
} 
By taking functional derivatives with respect to $A_+$ and $A_-$ we see that 
\al{ \label{currentcorrelators}
& \left< J_+ (x_-) \cdots J_+(x_n) J_- (y_1) \cdots J_- (y_m) \right>_F \notag\\ & \hspace{0.2in} = \left( -\frac{i}{4 \pi} \right)^{n+m} \left<  \partial_+ \varphi(x_1) \cdots \partial_+ \varphi(x_n) \partial_- \varphi(y_1)   \cdots  \partial_- \varphi(y_m) \right>_B
}
where $\left< \cdots \right>_F$  and $\left< \cdots \right>_B$ denote expectation values in the fermionic and bosonic theory, respectively, thereby expressing the usual result of the bosonization prescription  i.e. $J_{\pm} \leftrightarrow \partial_{\pm} \varphi$ \cite{colemanbosonization}, \cite{mandelslambosonization}.

Similarly, in the nonabelian case, for a generic $U$ in \eqref{WZexplicit}, varying \eqref{WZbosonization} with  respect to $A_-$ and setting $A_{\pm}=0$ we get 
\al{
& \left< J_+^{mn}(x) J_+^{m'n'} (x') \cdots \right>_F \notag\\ & \hspace{0.3in} = \left<(-1/4\pi)(U^{-1} \partial_+ U)^{mn}(x) (-1/4\pi)(U^{-1} \partial_+ U)^{m'n'}(x')\cdots \right>_B 
}
Varying \eqref{WZbosonization} with  respect to $A_+$ and setting $A_{\pm}=0$ we get 
\al{
& \left< J_-^{mn}(x) J_-^{m'n'} (x') \cdots \right>_F \notag\\ & \hspace{0.3in}  = \left<(-1/4\pi)(U \partial_+ U^{-1})^{mn}(x) (-1/4\pi)(U \partial_+ U^{-1})^{m'n'}(x')\cdots \right>_B 
}
These verify\footnote{See \cite{divecchia} for the mixed correlators $\left< J_+ J_- \cdots \right>$. } the bosonization rules introduced by Witten \cite{WZNW} 
\al{
J^{mn}_+ \leftrightarrow (-1/4\pi) (U^{-1} \partial_+ U)^{mn} \ \ , \ \ J^{mn}_- \leftrightarrow (-1/4\pi) (U \partial_+ U^{-1})^{mn} 
}
These satisfy \cite{WZNW} 
\al{
[ \text{Tr} AJ_-(x) , \text{Tr} BJ_-(y)] & = 2i \delta(x-y) \text{Tr}[A,B] J_-(x) + \frac{iN}{\pi} \delta'(x-y) \text{Tr} AB \\
[ \text{Tr} AJ_+(x) , \text{Tr} BJ_+(y)] & = 2i \delta(x-y) \text{Tr}[A,B] J_+(x) - \frac{iN}{\pi} \delta'(x-y) \text{Tr} AB \\ 
[ \text{Tr} AJ_-(x) , \text{Tr} BJ_+(y)] & =0
}
where $A$ and $B$ are arbitrary antisymmetric matrices. These are equivalent to \eqref{currentalgebra}. To see this take $A=M^a$, $B=M^b$ with the $O(N)$ basis $\{ M^a \}$ satisfying \eqref{ONbasisMa} and the normalization condition $\text{Tr} (M^a, M^b) = 2 \delta^{ab}$.

\subsection{Polyakov-Wiegmann's treatment} 
Here is another calculation \cite{polyakovwiegmann}, \cite{polyakovwiegmann84} which better shows that the WZW functional is the integrated anomaly, and the correspondence between the WZW action and the P2DG action in LCG (to be discussed in the next section). 

Consider the formal Dirac determinant in 2D 
\al{
W[A_{\mu}] = \log \text{Det} (\gamma^{\mu} ( i \partial_{\mu} + A_{\mu}))
}
The quantum current can be defined through
\al{ \label{quantumcurrenteffectiveaction}
J_{\mu} = \frac{\delta W}{\delta A_{\mu}} 
}
One can choose a regularization such that the following quantum relations hold 
\begin{subequations} \label{chiralanomalycurrentoperatorequations}
\al{ 
 \partial_{\mu} J^{\mu} + [ A_{\mu} , J^{\mu}]  &= 0
\\
 \epsilon^{\mu \nu} (\partial_{\mu} J_{\nu} + [A_{\mu} , J_{\nu} ] )  &= \frac{1}{2 \pi} \epsilon^{\mu \nu} F_{\mu \nu}
}
\end{subequations}
The first equation simply states the conservation of the vector current $J^{\mu}$. 
Note that due to the identity $\overline{\psi} \gamma^{\mu} \gamma_5 \psi = \epsilon^{\mu \nu} \overline{\psi} \gamma_5 \psi$ the left-hand side of the second equation is equivalent to $\partial_{\mu} J_5^{\mu} + [A_{\mu} , J^{\mu}_5]$ so that this equation states the chiral anomaly i.e. the  nonconservation of the chiral current $J^{\mu}_5$ with the anomaly function given on the right-hand side.

If we switch to the LCC taken in this section as $x^{\pm} = x^0 \pm x^1$ and introduce the parametrizations
\al{  \label{WZNWPolyakovchiralansatz}
A_+ = g^{-1} \partial_+ g \ \ , \ \ A_- = h^{-1} \partial_- h
} 
for the gauge field we see that equations \eqref{chiralanomalycurrentoperatorequations} are solved for the chiral currents by 
\begin{subequations}
\begin{gather}
J_+  = g^{-1} \partial_+ g - h^{-1} \partial_+ h  \\
J_-  = h^{-1} \partial_- h - g^{-1} \partial_- g 
\end{gather}
\end{subequations}

Now let us  restrict our attention to the axial gauge $A_- =0$, $h=I$
 we get the following variation for the effective action 
\al{ \label{quantumcurrenteffectiveactionLCC}
\delta W & = \int d^2 x \ \text{Tr} (J_-\delta A_+)  \\ & = \int d^2 x \ \text{Tr} (\partial_- (g^{-1} \partial_+ g) \delta g g^{-1} )
}
The solution to this equation is none other than the WZW functional 
\al{
W[g]  & = \frac{1}{2} \int_{\partial Q = S^2} d^2 x\ \text{Tr} (\partial_{\mu} g^{-1} \partial^{\mu} g) \notag\\ & \ \ \ \ + \frac{i}{8 \pi^2} \int_Q d^3y \ \epsilon^{ABC} \, \text{Tr} (g^{-1} \partial_A g g^{-1} \partial_B g g^{-1} \partial_C g)
}

If the $A_-=0$ gauge is turned off then the effective action becomes \cite{polyakovwiegmann84} 
\al{ \label{effective action full}
W[A_{\mu}] = W_+ (A_+) + W_- (A_-) + \text{Tr} ( A_+ A_-)
}
where the last term is a dimensionless counterterm added to make $W[A_{\mu}]$ gauge invariant. In terms of the parametrizations \eqref{WZNWPolyakovchiralansatz} this reads 
\al{
W [gh^{-1}] = W[g] + W[h^{-1}] +  \int d^2 x \ \text{Tr} (   (g^{-1} \partial_+ g) (h^{-1} \partial_- h) )
}

\section{Polyakov's 2D Gravity }
\label{Polyakov 2D Gravity}
In this section we are going to introduce the Polyakov 2D quantum gravity action (P2DG) from a number of perspectives. In doing so, we are aiming to show in what sense it is a quantum gravity action in 2D, and  its relation to WZW theory and to chiral fermion theories.

In 2D, the classical theory of gravity described by the Einstein-Hilbert action does not provide any dynamics as the Einstein equations reduce to $0=0$.  The Einstein-Hilbert action reduces (using the Gauss-Bonnet theorem)  to $2 \pi \chi$ where $\chi$ is the Euler characteristic which is an invariant under homeomorphisms. Hence, in 2D, Einstein's gravity  provides only topological information about the spacetime. 

Upon quantization, however, theories can pick up contributions from anomalies (forming the one-loop quantum effective action) in case the symmetries of classical theory fails to hold in the quantum theory.  In particular, anomalies can provide dynamics to the spacetime metric. 
The main references for this section are \cite{polyakov81bosonic}, \cite{polyakov2Dgravity}, \cite{knizhnik}.

\subsection{P2DG as Integrated Conformal Anomaly}
P2DG action arises as the effective action for the conformal anomaly. It is introduced in  \cite{polyakov81bosonic},  in the context of bosonic string theory, but it is relevant to any 2D conformal field theory. Classical implication of conformal invariance is the vanishing of the trace of the energy momentum tensor. Hence, breaking of the conformal symmetry at the quantum level, namely the conformal anomaly, arises as the nonvanishing of the trace and turns out to be given by
\al{ \label{2Dconformalanomaly}
g^{ab} \frac{\delta W}{\delta g^{ab}} = g^{ab} \left< T_{ab} \right> = \frac{D}{48 \pi} (R + \text{const})
}
Here $D$ is a constant, $T_{ab}$ the energy momentum tensor of the theory and $R$ is the Ricci scalar of the 2D spacetime underlying the theory. 

Equation \eqref{2Dconformalanomaly} can be solved for the effective action $W$ in covariant but nonlocal form, 
\al{ \label{Polyakov2Dgravitynonlocal}
W[g_{ab}]  = & - \frac{D}{96 \pi} \int d^2 x \, d^2 x' \, \sqrt{g(x)}  \, R(x) K(x,x') R(x')\, \sqrt{g(x')} \notag\\ & + \text{const} \int d^2 x \, \sqrt{g}
}
where $K$ is the kernel of the Laplacian 
\al{
\partial_a ( \sqrt{g} g^{ab} \partial_b ) K(x,x') = \delta (x - x')
}

For computational purposes one chooses a gauge for the metric\footnote{Reparametrization invariance combined with the symmetry of the metric tensor, reduce its number of independent degrees of freedom to one.}, in which $W$ becomes local. In the same paper $W$ is introduced in the conformal gauge $g_{ab} = \rho \delta_{ab}$, $R= \rho^{-1} \partial^2 \rho$. In this case we get the 2D Liouville gravity 
\al{
W[\rho] = - \frac{D}{96 \pi} \int d^2 x \ \left[ \frac{1}{2} (\partial_a \log \rho)^2 + \mu^2 \rho \right] 
}

\subsection{P2DG in Lightcone Gauge}
In \cite{polyakov2Dgravity}, Polyakov chooses a different gauge for the metric to evaluate the effective action for the conformal anomaly, namely, the lightcone gauge (LCG)
\al{
ds^2 = dx^+ dx^- + h_{++} (x^+ , x^-) (dx^+)^2
}
The metric underlying this line element is given by 
\al{ \label{2DPolyakovmetric}
g^{(+ -)}_{\mu \nu} = \left( \begin{array}{cc} h_{++} & 1/2 \\ 1/2 & 0 \end{array} \right)
}
Instead of substituting this metric into  \eqref{Polyakov2Dgravitynonlocal} he introduces the conformal anomaly operator relations, analogous to \eqref{chiralanomalycurrentoperatorequations} and \eqref{quantumcurrenteffectiveactionLCC}, in LCC, namely, 
\al{
\delta W & = \int T_{--} \delta h_{++} \\
\nabla_+ T_{--} & \equiv \partial_+ T_{--} - h_{++} \partial_- T_{--} - 2 (\partial_- h_{++}) T_{--} = \frac{d}{24 \pi} \partial_- R
}
The Ricci scalar for \eqref{2DPolyakovmetric} becomes $R = \partial_-^2 h_{++}$. Recall the mentioned analogy between WZW theory and the P2DG theory in LCG.  In the stated analogy we have the correspondences $A_+ \leftrightarrow h_{++}$, $J_- \leftrightarrow T_{--}$. We will have more to say about it below.
   
 Polyakov states that it is possible to work out $W[h_{++}]$ perturbatively and even in closed form. However,  he chooses to work with  an action that is obtained from it by a field redefinition. Namely he introduces a field $f$ defined by 
\al{ \label{Polyakov2Dansatz}
\partial_+ f = h_{++} \partial_- f
}
He points out to the analogy between \eqref{Polyakov2Dansatz} and \eqref{WZNWPolyakovchiralansatz}. The former redefines the $h_{++}$ component  in terms of a field $f$ and the latter redefines the $A_+$ component in terms of a field $g$. That's the first reason why P2DG action in LCG is considered as the gravitational WZW model. 

The effective action can then be written as 
\al{ \label{P2DGLCCactionf}
W[f]  \propto    \int d^2 x \ \left[ \frac{(\partial^2_- f) ( \partial_+ \partial_- f )}{(\partial_- f)^2} - \frac{(\partial^2_- f)^2 (\partial_+ f)}{(\partial_- f)^3} \right] 
}   
   
In analogy with \eqref{effective action full}, if the $h_{--}=0$ gauge is turned off, we would have \cite{polyakovBook2D}
\al{
W[h_{++}, h_{--}] = W_+[h_{++}] + W_- [h_{--}] + \Lambda(h_{++} , h_{--})
}
where $\Lambda$ is a counterterm. In particular, if we introduce a Polyakov field $\bar{f}$ for $W_-$ as well, in terms of the fields $f$ and $\bar{f}$ the effective action would read 
\al{ \label{P2DGLCCFullActionWithCounterterm}
W[f,\bar{f}] = W_+ [f] + W_- [\bar{f}] + \Lambda ( f, \bar{f})
}
where $W_- [\bar{f}]$ is of the same form as \eqref{P2DGLCCactionf} with $f \leftrightarrow \bar{f}$, $x^+ \leftrightarrow x^-$.  
We can obtain $W_{\pm}$ as parts of  geometric action on the orbits of the direct product of two Virasoro algebras (Section \ref{VirasoroDirectProduct}). The counterterm may arise from quantization conditions entangling the diff field components $D_{++}$, $D_{--}$. 

 \subsection{Chiral Fermions Coupled to Gravity in 2D  }

The approach followed here to introduce the P2DG action in LCG is from \cite{knizhnik}. We differ in our LCC conventions (Section \ref{Lightcone Coordinates}).

The Dirac Lagrangian in a curved spacetime is defined as 
\al{
\mathcal{L} & =  \sqrt{-g} \ \overline{\psi} \gamma^{\mu} \nabla_{\mu} \psi \notag\\ & = -(\det v) \ \overline{\psi} \gamma^a e_{\ a}^{\mu} \left(  \partial_{\mu} - \frac{1}{2} \omega_{\mu a b} \gamma^a \gamma^b \right) \psi
}
Here $v^a_{\ \mu}$ are the vielbein components, $\det v$ is the determinant of the vielbein, $e^{\mu}_{\ a}$ are the inverse vielbein components and $\omega_{\mu a  b}$ are the spin connection components. 
In 2D the spin connection $\omega_{\mu}$ vanishes (see e.g. \cite{nakahara} Section 7.10.3 ) thus this reduces to 
\al{ \label{diraccurved}
\mathcal{L} = -(\det v) \ \overline{\psi} \gamma^a e_{\ a}^{\mu}  \partial_{\mu} \psi
}
We use different letters for the vielbein and its inverse to avoid confusion. 

Using the LCC toolbox developed in Section \ref{Lightcone Coordinates} we can do the sum 
\al{
\mathcal{L}/\sqrt{2} =   \psi_- ( v^-_{\ +} \partial_- - v^-_{\ -} \partial_+) \psi_- + \psi_+ ( - v^+_{\ +} \partial_-  -v^+_{\ -} \partial_+) \psi_+ 
}
Note that in obtaining this result we haven't used any gauge fixing conditions. 

Next, we introduce the field redefinitions $\phi_- \equiv  \sqrt{v^-_{\ +}} \psi_- $ and $\phi_+ \equiv  \sqrt{v^+_{\ -}} \psi_+$. Upon action of the derivatives we get terms of the form $\phi_- \phi_-$ and $\phi_+ \phi_+$. These vanish by the Grassmann nature of the  $\phi_{\pm}$. Therefore, in the end, we are left with 
\al{ \label{2DfermionLagbeforePol}
\mathcal{L}/\sqrt{2} =   \phi_- \left( \partial_- - \frac{v^-_{\ -}}{v^-_{\ +}} \partial_+ \right) \phi_-  \  +  \   \phi_+ \left( \partial_+ - \frac{v^+_{\ +}}{v^+_{\ -}} \partial_- \right) \phi_+ 
}

Polyakov chooses the following  vielbein gauge fixing conditions 
\al{ \label{Polyakovvielbein}
v^-_{\ -} = 0 \ \ \ , \ \ \ v^-_{\ +} = 1/v^+_{ \ -}
}
Now, under a coordinate transformation $x \mapsto x'$  metric transforms with the inverse Jacobian, $\partial x/ \partial x'$, i.e. 
\al{
\tilde{\mathbf{g}} = (J^{-1})^T \mathbf{g} J^{-1}
}
where we used boldface letter for the metric tensor matrix, to avoid confusion with the metric determinant $g$. In the special case of the transformation from a flat metric, the vielbein matrix is the same as the inverse Jacobian matrix. Therefore, the ligthcone flat metric  
\al{
\eta = \left( \begin{array}{cc} 0 & 1 \\ 1 & 0    \end{array} \right)
}
transforms to  (coordinates are in order of $(x^+ , x^-)$ )
\al{
\mathbf{g} = \left( \begin{array}{cc} 2 v^-_{ \ +} v^+_{\ +} & v^-_{\ -} v^+_{\ +} + v^-_{ \ +} v^+_{\ -} \\  v^-_{\ -} v^+_{\ +} + v^-_{ \ +} v^+_{\ -}  & 2 v^-_{ \ -} v^+_{\ +-}    \end{array} \right)
}

Under the conditions  \eqref{Polyakovvielbein} this yields the Polyakov metric 
\al{
\mathbf{g} = \left( \begin{array}{cc} \frac{2v^+_{\ +}}{ v^+_{ \ -}} & 1 \\ 1 & 0    \end{array} \right)
}

For the Polyakov metric, the Lagrangian \eqref{2DfermionLagbeforePol}  reduces to
\al{
\mathcal{L}/\sqrt{2} =   \phi_+ \left( \partial_+ - \frac{g_{++}}{2}  \partial_- \right) \phi_+  \ \ \ + \ \ \   \phi_-  \partial_-  \phi_-
}
Dropping the $-$ chiral mode  we get (ignoring the factor of $\sqrt{2}$) 
\al{
\mathcal{L}_+ =\phi_+ \left( \partial_+ - \frac{g_{++}}{2}  \partial_- \right) \phi_+ 
}
Integrating over the fermionic modes we get the P2DG action in LCG\footnote{Note that Polyakov \cite{polyakov2Dgravity}, \cite{knizhnik} used the LCC definitions $x^{\pm} = (t \pm x)$ so is off by a factor of $1/2$ from our conventions. As a result the corresponding Lagrangian reads $\mathcal{L}_+  \propto \phi_+ ( \partial_+ - g_{++} \partial_-)\phi_+$. 
}
\al{ \label{2Deffectivegravitational}
W[g_{++}] \propto  \log \text{Det} \left( \partial_+ - \frac{g_{++}}{2}  \partial_- \right) = \text{Tr} \log \left( \partial_+ - \frac{g_{++}}{2}  \partial_- \right)
}
Since WZW action arises as the Dirac determinant with gauge coupling \eqref{2DFermionDeterminant}, we again see that P2DG action in LCG is the gravitational analog of the WZW action. That's why it is also called the gravitational WZW model.

\section{Kac-Moody Geometric Action }
\label{Kac-Moody Geometric Action}
In this section we review the construction of the geometric action on Kac-Moody (KM) coadjoint orbits\cite{delius90}. 
Using the conventions set in Section \ref{Kac-Moody Algebra} we first need to construct  adjoint and coadjoint vectors that are suitable for $(  \tau, \lambda )$-parametrized coadjoint orbit $m$. 
For this purpose we introduce group elements $g(z, \tau, \lambda)$ i.e. for each point $(\tau , \lambda)$ on the orbit we have a group element $g(z)$.  Then as adjoint  vectors we can take 
\al{
u_{\tau} \equiv g(z, \tau , \lambda) \partial_{\tau} g^{-1} (z, \tau , \lambda) \hspace{0.2in} \text{and} \hspace{0.2in} u_{\lambda} \equiv g(z, \tau , \lambda) \partial_{\lambda} g^{-1} (z, \tau , \lambda)
}
To construct the geometric action we need the commutator of $u_{\tau}$ and $u_{\lambda}$
\al{
(\Lambda, a) \leftrightarrow  [(u_{\tau} , c_{\tau}) , (u_{\lambda},c_{\lambda}) ]  = \left( [u_{\tau} , u_{\lambda}] , \oint \frac{dz}{2 \pi i} \ \text{Tr} \ \partial_z u_{\tau} u_{\lambda} \right)
}
As a coadjoint vector on the orbit we pick a fixed coadjoint vector $(A(z) , a)$ and act on it by $g(z, \tau , \lambda)$ :
\al{
(A(z) , a)_g = \Big( g(z, \tau, \lambda)A(z) g^{-1}(z, \tau, \lambda) - a \partial_z g(z, \tau , \lambda) g^{-1} (z, \tau, \lambda) \, , \, a\Big)
}

Forming the pairing between the constructed coadjoint and adjoint vectors, and integrating it over the  orbit $m$, we get the  action
\al{
S_{\text{KM}} & \equiv \int_{m} \Omega \notag\\ & =  \int d \lambda\  d \tau \ \left< (A, a)_g \ | [(u_{\tau}, c_{\tau}), (u_{\lambda},c_{\lambda})] \right> \notag\\ & = k \int d \lambda d\tau \left( \oint \frac{dz}{2 \pi i} \ \text{Tr} (g A g^{-1} [ g \partial_{\tau} g^{-1} , g \partial_{\lambda} g^{-1} g]) \right.
\notag\\ & \hspace{1.3in}  -a\oint \frac{dz}{2 \pi i} \ \text{Tr} (  \partial_z g g^{-1} [ g \partial_{\tau} g^{-1} , g \partial_{\lambda} g^{-1} ] )  \notag\\ & \hspace{1.8in}  \left. + a \oint \frac{dz}{2 \pi i} \ \text{Tr} \ \partial_z (g \partial_{\tau} g^{-1} ) g \partial_{\lambda} g^{-1} \right) 
}
where $g=g(z, \tau, \lambda)$.  We first write this as far as possible in terms of total $z, \tau$ and $\lambda$ derivatives. Total $z$ derivative terms vanish. For $\tau$ and $\lambda$ dependence, 
we choose to impose boundary conditions such that total $\tau$ derivatives vanish, and $g$ is $\tau$-independent at $\lambda=0$. This is equivalent to the requirement that the $g(z, \tau, \lambda)$ describing embedding of a 3-ball into the group manifold with $\lambda$ as the radial coordinate and $z$ and $\tau$,  the coordinates on $S_2$. With these we arrive at the WZW action plus a background field $A(z)$ interacting with the WZW field $g(z, \tau, \lambda)$ :
\al{
S & =  \frac{k a}{2}  \int d \tau \oint \frac{dz}{2 \pi i} \ \text{Tr} \bigg( \partial_{\tau} g (\lambda=1) \, \partial_z g^{-1} (\lambda=1)  \notag\\ & \left. \hspace{1.6in} + \frac{1}{3} \int d \lambda \ \epsilon^{\alpha \beta \gamma} \ \partial_{\alpha} g g^{-1} \partial_{\beta} g g^{-1} \partial_{\gamma} g g^{-1}   \right) \notag\\ & \ \ \ \ \  + k \int d \tau \oint \frac{dz}{2 \pi i} \ \text{Tr} ( A \, g^{-1} (\lambda =1) \, \partial_{\tau} g ( \lambda = 1) ) 
}
where $\epsilon^{z \lambda \tau}=1$. Exact correspondence with the original form of WZW action (as provided in \cite{WZNW}) is achieved via 
$g \leftrightarrow g(z, \tau , \lambda=1)$ , 
$\bar{g}  \leftrightarrow g(z, \tau, \lambda)$   , 
$d^2 x  \leftrightarrow dz d \tau$ and 
$d^3 y  \leftrightarrow dz  d \tau  d \lambda$.

In the angular coordinate $\theta$ the same analysis leads to the action,  \cite{rairodgers90} : 
\al{ \label{KMgeometricactionangular}
S & = \frac{ka}{4 \pi } \int d \tau \ d \theta \ \text{Tr} \ \left(  \partial_{\tau} g  \partial_{\theta} g^{-1} \Big|_{\lambda=1} + \frac{1}{3 } \int d \lambda \ \epsilon^{\alpha \beta \gamma} \  \partial_{\alpha} g g^{-1} \partial_{\beta} g g^{-1} \partial_{\gamma} g g^{-1}   \right) \notag\\ & + \frac{k}{2 \pi} \int_{\lambda=1} d \tau \ d \theta  \ \text{Tr} \ ( A \, g^{-1}  \partial_{\tau} g  ) 
}
where $A= A (\theta)$ and $g= g( \theta, \tau , \lambda)$. 

The coupling of WZW field to a background gauge field $A_{\mu}$ is given by the second term of \eqref{WZexplicit} (with $U \rightarrow g, \, A_+ \rightarrow A_{\tau}, \, A_- \rightarrow A_{\theta}$),
\al{ \label{diVecchia1}
S_{gA_{\mu}} = \frac{1}{4 \pi} \text{Tr} \ \int d \tau \ d \theta \ ( A_{\tau} g \partial_{\theta} g^{-1} + A_{\theta} g^{-1} \partial_{\tau} g + A_{\tau} g A_{\theta} g^{-1} - A_{\tau} A_{\theta} )
}
In temporal gauge, $A_{\tau}=0$, this reduces to
\al{ \label{diVecchia2}
S_{gA_{\theta}} & = \frac{1}{4 \pi} \text{Tr} \ \int d \tau \ d \theta \ ( A_{\theta} g^{-1} \partial_{\tau} g  ) 
}
Comparing this with the last term of the geometric action  \eqref{KMgeometricactionangular} we see that the background field $A$ can be identified with the $A_{\theta}$ component of a YM field $A_{\mu}$ in temporal gauge. 
This identification will be further motivated  by analyzing the infinitesimal coadjoint action of the KM coadjoint element $A$. In fact, using this we are going to obtain YM action from KM algebra by the transverse prescription in Section \ref{YMfromKM}.

\section{Virasoro Geometric Action }
\label{Virasoro Geometric Action}
In this section, we review the geometric action on Virasoro coadjoint orbits \cite{delius90}. 
Again we first need to construct adjoint and coadjoint vectors from the group elements parametrized on the orbit  $m = \{(\lambda , \tau)\}$  i.e. from group elements of the form $s(z,\lambda, \tau) \equiv  s ( \lambda , \tau) \cdot z$ where $s(\lambda, \tau)$ is the group element, and $s(z, \lambda , \tau)$ is the diffeomorphism formed by its action on $z$.

Let us begin by the adjoint element in the direction of $\tau$. We will use the differential operator  representation\footnote{Here minus sign is needed for consistency with the commutation relations.}, $-u_{\tau} \partial_{\tau} = s \partial_{\tau} s^{-1}$. Using $s \cdot  z = s(z) \equiv \bar{z}$ we get  $u_{\tau} ( \bar{z} ) =  \partial_{\tau} s (z,\lambda, \tau)$  or $u_{\tau} (z) = \partial_{\tau} s (s^{-1}(z,\lambda, \tau),\lambda, \tau)$. The analogous expression can be found for $u_{\lambda}$. 
Next we evaluate the commutator $[(u_{\tau} , c_{\tau} ) , (u_{\lambda} , c_{\lambda})]$ which using chain rule becomes
\al{
\xi_s(\bar{z}) = [u_{\tau} , u_{\lambda} ] (\bar{z}) = \partial_{\tau} s(z,\lambda, \tau) \frac{\partial_{\lambda } \partial_z s(z,\lambda, \tau)}{\partial_z s(z,\lambda, \tau)}  - \partial_{\lambda} s(z,\lambda, \tau) \frac{\partial_{\tau } \partial_z s(z,\lambda, \tau)}{\partial_z s(z,\lambda, \tau))} 
}

For the center, as well, it is more convenient to do the computation at $\tilde{z}$
\al{
\mu_s \equiv c(u_{\tau} , u_{\lambda}) 
& = \oint \frac{dz}{2 \pi i} \ \partial^3_z u_{\tau} (z) u_{\lambda} (z)  
\notag\\ 
& = \oint    \frac{d \bar{z}}{2 \pi i} \  \partial_{\bar{z}}^3 u_{\tau} (\bar{z})  u_{\lambda} (\bar{z}) \notag\\ 
&= \oint \frac{dz}{2 \pi i} \ \partial_z s \left[ ((\partial_z s)^{-1} \partial_z)^3 \partial_{\tau} s \right] \partial_{\lambda} s
}
where $s=s(z,\lambda,\tau)$.

To get the coadjoint vector on the orbit $m$, we act on  a fixed element $(D(z),b^*)$ by a parametrized group element $s(z, \lambda , \tau)$, i.e. 
\al{
(D_s(\bar{z}) , b^*_s) & =  \Big( (D(z)  -b^* S(z, s) )\left( \partial_z s \right)^{-2} , b^* \Big) 
 }
 
With all these ingredients we obtain the action
\al{
S & = \int d \tau \, d \lambda  \oint \frac{d \bar{z}}{2 \pi i}  \left( D_s(\bar{z} ) \xi_s (\bar{z})  +  b^*_s \mu_s \right)  
\notag\\ 
& =  \int d \tau  \, d \lambda \oint \frac{dz}{2 \pi i} \,  ( \partial_z s)^{-2} ( D(\bar{z}) - b^* S(z, s)) ( \partial_z \partial_{\tau} s \, \partial_{\lambda} s - \partial_z \partial_{\lambda} s \, \partial_{\tau} s ) \notag\\ & \ \ \ \ +  b^* \int d \tau \, d\lambda \,  \oint \frac{dz}{2 \pi i} \ \partial_z s \left[ ((\partial_z s)^{-1} \partial_z)^3 \partial_{\tau} s \right] \partial_{\lambda} s
}
where $s=s(z, \tau , \lambda)$.  The same boundary conditions as in the case of KM are assumed. Namely, those that make  $z$ total derivatives vanish  and make $s$,  $\tau$-independent for $\lambda=0$. After a fairly long  calculation  one reaches the following action
\al{ \label{Virgeoact1}
S = \int d\tau \ \oint \frac{dz}{2 \pi i} \ \frac{\partial_{\tau} s}{\partial_z s} \ D (z)  - \frac{b^*}{2} \int d \tau \oint \frac{dz}{2 \pi i} \ F_{\tau} (z, \tau, \lambda=1)
}
where 
\al{
F_{\tau} = \frac{(\partial_z^2 s)^2 (\partial_z s)}{(\partial_z s)^3} - \frac{(\partial_z^2 s) ( \partial_z \partial_{\tau} s)}{(\partial_z s)^2}
}

If we change the notation as $z \rightarrow x^-$, $\tau \rightarrow x^+$, $s \rightarrow f$, the second term in the action \eqref{Virgeoact1} is identical to P2DG action in LCG \eqref{P2DGLCCactionf}. Explicitly, in Polyakov's notation we have
\al{ \label{SDPolyakovf}
S = \int d^2 x \ D(x^-) \ \frac{\partial_+f}{\partial_- f}   -\frac{b^*}{2} \int d^2 x \ \left[ \frac{(\partial^2_- f) ( \partial_+ \partial_- f )}{(\partial_- f)^2} - \frac{(\partial^2_- f)^2 (\partial_+ f)}{(\partial_- f)^3} \right] 
}
Here the noncentral Virasoro coadjoint element $D(x^-)$  couples as a background field to the lightcone Polyakov field $f(x^+,x^-)$ in analogy with a background YM field $A$ coupling to the WZW field $g$ in the case of KM geometric action.

Consider the first term in \eqref{SDPolyakovf}. The coadjoint field $D(x^-)$  corresponds to $D_{zz} \leftrightarrow D_{\theta \theta} \leftrightarrow D_{--}$ component of a rank-two object $D_{\mu \nu}$ called the diffeomorphism field or the "diff field" in short. Using \eqref{Polyakov2Dansatz} the interaction term can be written as
\al{ \label{Pol2Ddiffintterm}
S_{\text{int}}=\int d^2x \ D_{--} g_{++}
} 
We can rewrite the integrand in the covariant form 
\al{ \label{lightconediffmetric}
D_{\mu \nu} h^{\mu \nu} =  D_{-+} - D_{--} g_{++}
} 
given the temporal gauge $D_{-+}=0$ for the diff field accompanied with Polyakov's LCG \eqref{2DPolyakovmetric} for the metric. This shows that the first term in \eqref{SDPolyakovf} is the coupling of the diff field  to the metric.  We can't, on the other hand, say anything about $D_{++}$ component since in LCG we have $g_{--}=0$. Hence, $D_{++}$ is simply invisible on coadjoint orbits. 

The argument that the Virasoro coadjoint element $D$ can be identified with the space-space component of a rank-two object will be further supported in Section \ref{Diffaction}, where we will construct a covariant action governing the dynamics of the diff field. 

\section{Semi-direct Product Geometric Action }
\label{Semi-direct Product Geometric Action}
\subsection{Action}
We have previously constructed the geometric actions for Virasoro and KM algebras separately. In the KM sector we obtained the WZW action plus the interaction of the WZW field $g$ with a background YM field $A_{\mu}$ in temporal gauge. In the Virasoro sector we obtained the P2DG theory plus the interaction of the Polyakov field $s$ with a background diff field tensor $D_{\mu \nu}$ in temporal gauge. 

When we consider the semi-direct product algebra we get the sum of the previous geometric actions plus corrections in the $A_{\mu} - g$ interaction term involving the Polyakov field $s$. This correction follows from the nontrivial action of the Virasoro generators on the KM generators. Let us  review the main steps \cite{LR95}. 

The two cocyle of the semi-direct product algebra is the sum of the cocycles of the algebras given in \eqref{KMcocycle} and \eqref{GelfandFuchscocyle}
\al{ \label{semidirectcocyle}
\omega\Big( (\xi_1, \Lambda_1, \mu_1)& , (\xi_2, \Lambda_2, \mu_2) \Big) \notag\\ & = \frac{c}{48 \pi i} \int_0^{2 \pi} d\theta \ (\xi'''_1  \xi_2  - \xi_1  \xi'''_2 ) + \frac{k}{2 \pi} \int_0^{2 \pi} d \theta  \ \text{Tr} (\Lambda_1  \Lambda'_2  )
}
For convenience, we took the Gelfand-Fuchs cocyle on the Virasoro sector, i.e. we did not include the linear center. 

 As before ($z \leftrightarrow \theta$)  we denote the KM group element  by $g(\lambda ,\tau , \theta)$  and the Virasoro group element by $s(\lambda, \tau ,\theta)$. In analogy with the notation of the previous two sections,  adjoint elements that describe the changes in $\lambda$ and $\tau$ directions can be taken as 
\al{
U_{\lambda} & = ( \partial_{\lambda} s  , \partial_{\lambda} g g^{-1} , 0 )  \notag\\ U_{\tau} & = ( \partial_{\tau} s, \partial_{\tau} g g^{-1} , 0 ) 
}
Denoting a composite group element by $\tilde{g} \equiv ( g, s)$, the adjoint action $\text{Ad}_{\tilde{g}} U$ become
\al{
\tilde{U}_{\lambda} \equiv \tilde{g}^{-1} U_{\lambda} \tilde{g} & = ( \partial_{\lambda} s / \partial_{\theta} s , g^{-1} \partial_{\lambda} g , 0 ) \notag\\ \tilde{U}_{\tau} \equiv \tilde{g}^{-1} U_{\tau} \tilde{g} & = ( \partial_{\tau} s / \partial_{\theta} s , g^{-1} \partial_{\tau} g , 0 ) 
}
Let us also denote $\tilde{B}_{\lambda} \equiv \text{ad}_{\tilde{U}_{\lambda}} B$ where $B=(D(\theta),A(\theta),\mu)$ is a fixed coadjoint element  and $B_0=(D(\theta),A(\theta),0)$,  its non-central part. 
Then the geometric action becomes
\al{
S_B & = \int d \lambda  \ d \tau \ \Omega ( \tilde{B}_{\lambda} , \tilde{B}_{\tau} )   \notag\\ & = \int d \lambda  \ d \tau \ \left< B | [ \tilde{U}_{\lambda} , \tilde{U}_{\tau} ] \right> \notag\\ & = \int d \lambda \ d\tau \ \left( \left< B_0 | [ \tilde{U}_{\lambda} , \tilde{U}_{\tau} ] \right> + \mu \omega (\tilde{U}_{\lambda}, \tilde{U}_{\tau})\right) 
}
Using \eqref{semidirectproductalgebra}, \eqref{semidirectgeneric}, \eqref{semidirectpairing} and \eqref{semidirectcocyle}, and performing partial integrations with the same boundary conditions as in the individual geometric actions one reaches, 
\al{\label{geoactionsemidirect}
S & = \underbrace{ \frac{1}{2 \pi} \int d \tau \ d \theta  \left( \frac{\partial_{\tau} s}{\partial_{\theta} s} \right) D}_{\text{coupling to background diff field}} \notag\\ & \underbrace{+ \frac{1}{2 \pi} \int d \lambda \ d \theta \ d \tau \ \text{Tr} \left( A \left( \frac{\partial_{\lambda} s}{\partial_{\theta} s} \partial_{\theta} ( g^{-1} \partial_{\tau} g ) - \frac{\partial_{\tau} s}{\partial_{\theta} s} \partial_{\theta} ( g^{-1} \partial_{\lambda} g ) + [ g^{-1} \partial_{\lambda} g,g^{-1} \partial_{\tau} g]\right) \right)}_{\text{coupling to background gauge field}} \notag\\ & \underbrace{- \frac{\mu c}{48 \pi} \int d \tau \ d \theta \ \left( \frac{\partial^2_{\theta} s}{(\partial_{\theta} s)^2} \partial_{\tau} \partial_{\theta} s - \frac{(\partial^2_{\theta} s)^2}{(\partial_{\theta} s)^3} \partial_{\tau} s \right)}_{\text{Polyakov gravity}} \notag\\ & \underbrace{- \frac{\mu k}{4 \pi} \int d \tau \ d \theta \ \text{Tr} \ ( g^{-1} \partial_{\theta} g g^{-1} \partial_{\tau} g) + \frac{\mu k}{4 \pi} \int d \lambda \ d \tau \ d \theta \ \text{Tr} \ ( [ g^{-1} \partial_{\theta} g, g^{-1} \partial_{\lambda} g] g^{-1} \partial_{\tau} g)}_{\text{WZW}}
}
The KM and Virasoro geometric actions obtained in the previous two sections can be recovered from this action by setting $s(\theta)=0$ and $g(\theta)=0$, respectively. 

\subsection{Equations of Motion}

The equations of motion that follow \cite{LR95} from the geometric action  \eqref{geoactionsemidirect} are 
\al{ \label{isotropy1}
0 & = (\partial_{\theta} D)   \frac{\partial_{\tau} s }{ \partial_{\theta} s} + 2 D \ \partial_{\theta} \left( \frac{\partial_{\tau} s }{ \partial_{\theta} s} \right) +  \frac{c \mu}{24 \pi} \partial^3_{\theta} \left( \frac{\partial_{\tau} s }{ \partial_{\theta} s} \right) - \text{Tr} \{ A \ \partial_{\theta} ( g^{-1} \partial_{\tau} g ) \}  
\\ 0 & = A \ \partial_{\theta} \left( \frac{\partial_{\tau} s }{ \partial_{\theta} s} \right) + \left( \partial_{\theta} A \right) \frac{\partial_{\tau} s }{ \partial_{\theta} s}  - [ g^{-1} \partial_{\tau} g , A] + k \mu \ \partial_{\theta} ( g^{-1} \partial_{\tau} g ) \label{isotropy2}
}

Let us simplify the notation a bit. First we will denote $\tau$-derivative with a dot and $\theta$-derivative with a prime. We will also take $(\partial_{\tau} s / \partial_{\theta} s)\equiv \xi$,  $g^{-1} \partial_{\tau} g \equiv \Lambda$,  $ k \mu \equiv e^{-1}$ and $(c \mu / 24 \pi) \equiv q$. With all these   the equations  read
\al{ \label{isotropy1b}
0 & = D' \xi + 2 D \xi' + q \xi''' - \text{Tr} \{ A \Lambda' \} \\ 
0 & = \xi' A + A' \xi - [ \Lambda , A] + e^{-1} \Lambda'
\label{isotropy2b}
}
These are  equations \eqref{semidirectiso}, with the linear center of the Virasoro algebra ignored (Section \ref{vanishingbeta}), and with the left-hand sides set to zero, i.e. $\delta D = 0 = \delta A$. In other words, the equations of motions turn out to be the isotropy equations. 

In the absence of diffeomorphisms ($\xi=0$),  equation \eqref{isotropy2b} has solutions 
\al{ 
g(\theta , \tau) = L(\theta) R(\tau)
} where $L(\theta) \in G$ is arbitrary and the generators of $R(\tau)$ commute with $A$. This implies that  $\Lambda=R^{-1} \dot{R}$ commutes with $A$. 

In the presence of diffeomorphisms the solution is modified to
\al{ \label{gLMR}
g( \theta , \tau ) = L (\theta) M (\theta , \tau ) R(\tau ) 
}
where $L$ and $R$ the same as above,  and 
\al{ \label{MTexp}
M (\theta ,\tau ) = \text{T} \exp \left( - e \int_{- \infty}^{\tau} dt \ \xi  A \right) 
}
with the boundary condition $g (\theta , \tau \rightarrow - \infty) = 1$. 

Now, let us consider the Virasoro equation $0= \delta D$. Using \eqref{gLMR},  we get 
\al{
\Lambda = g^{-1} \partial_{\tau} g= R^{-1} \dot{R} - e \xi A
}
Inserting this into $0=\delta D$ we obtain
\al{ \label{VirKMsemidirectEoM2}
0  = D' \xi + 2 D \xi' + q \xi''' + e \text{Tr} \ ( A \xi' A +  A \xi A' )
}
Introducing a new field 
\al{ \label{DTrAA}
\tilde{D} \equiv D + \frac{e}{2} \ \text{Tr} \ ( A A)
}
 equation \eqref{VirKMsemidirectEoM2} simplifies to 
\al{ \label{newisotropy}
0 = \tilde{D} ' \xi + 2 \tilde{D} \xi' + q \xi'''
}
so that the KM variables $A,\Lambda$ disappear from the equation.  Thus $\tilde{D}$ is invariant under gauge transformations contrary to $D$ :
\al{
\delta_{\text{gauge}} D & = - \text{Tr} \ ( A \Lambda' ) \neq 0 \\ 
\delta_{\text{gauge}} \tilde{D} & = 0
}
We will use this result to obtain the gauge-invariant extension of the diff field $D_{\mu \nu}$.

\section{Direct Product of Two Virasoro Algebras }
\label{VirasoroDirectProduct}
The 2D algebra of generators of conformal transformations turns out to be (see e.g. \cite{ketov}) direct product of two copies of Witt algebras 
\al{
[\ell_m , \ell_n] & = (m-n) l_{m+n} \\
[\bar{\ell}_m , \bar{\ell}_n ] & = (m-n) \bar{\ell}_{m+n} \\
[\ell_m , \bar{\ell}_n ] & = 0
}

Upon quantization (of the underlying conformal field theory) these commutation relations pick up central extensions yielding  the direct product of two Virasoro algebras. Quantization, however, may put further restrictions on the generators.

Commutation relations for the direct product of two Virasoro algebras, with central charges left arbitrary, are given by
\al{ 
[L_m , L_n] & = (m-n) L_{m+n} + (c m^3 + h m ) I \delta_{m+n} \notag\\ [\bar{L}_m , \bar{L}_n] & = (m-n) \bar{L}_{m+n} + (\bar{c} m^3 + \bar{h} m ) \bar{I} \delta_{m+n}  \notag\\ [L_m , \bar{L}_n ] & = 0 
}
where $c, h , \bar{c}, \bar{h}$ are constants and $I$ and $\bar{I}$ are the generators of centers.

 Firstly, notice that, in the geometric actions constructed in the previous sections, whenever two parts of an algebra did commute (like the non-central and the central parts of the algebra) they yielded separate terms summed in the geometric action. The situation here is the same. Since the two copies of the Virasoro algebras commute with each other, the geometric action of the direct product algebra will be given by the sum of the geometric actions coming from each copy. 
 
  Secondly, the role of the LCCs $x^{\pm}$ are switched once we switch from $\{L\}$ algebra to $\{ \bar{L}\}$ algebra. Hence, we can get the geometric action for the latter, from the first  simply by $x^+ \leftrightarrow x^-$.  We will denote the noncentral coadjoint element coming from the first algebra by $D$ (with central charge $\mu$), and from the second one by $\bar{D}$ (with central charge $\bar{\mu}$). Similarly, the corresponding Polyakov fields (circle diffeomorphisms) will be denoted by $f$ and $\bar{f}$, respectively.

Recall the geometric action obtained from a single Virasoro algebra
\al{ 
S = \int d^2 x \ D(x^-) \ \frac{\partial_+f}{\partial_- f}   -\frac{\mu}{2} \int d^2 x \ \left[ \frac{(\partial^2_- f) ( \partial_+ \partial_- f )}{(\partial_- f)^2} - \frac{(\partial^2_- f)^2 (\partial_+ f)}{(\partial_- f)^3} \right] 
}
where the coadjoint vector used in building the geometric action is $\text{Ad}^*_f (D , \mu)$. 

Then, the geometric action for the direct-product algebra becomes
\al{ \label{DirectProductVirasoroGeoAction}
S & = \int d^2 x \ D(x^-) \ \frac{\partial_+f}{\partial_- f}   -\frac{\mu}{2} \int d^2 x \ \left[ \frac{(\partial^2_- f) ( \partial_+ \partial_- f )}{(\partial_- f)^2} - \frac{(\partial^2_- f)^2 (\partial_+ f)}{(\partial_- f)^3} \right] 
\notag\\ 
& +  \int d^2 x \ \bar{D}(x^+) \ \frac{\partial_- \bar{f}}{\partial_+ \bar{f}}   -\frac{\bar{\mu}}{2}\int d^2 x \ \left[ \frac{(\partial^2_+ \bar{f}) ( \partial_- \partial_+ \bar{f} )}{(\partial_+ \bar{f})^2} - \frac{(\partial^2_+ \bar{f})^2 (\partial_- \bar{f})}{(\partial_+ \bar{f})^3} \right] 
}
Notice that at this point there is no relationship between $D, \bar{D}$, their central charges $\mu , \bar{\mu}$ or the diffeomorphisms $f, \bar{f}$ acting on the coadjoint elements. It is just denotation of distinct variables. 

Upon quantization of a conformal theory with classical algebra given by direct product of two copies of Witt algebras $\{ l_n \} , \{ \bar{l}_m \}$ a complication arises (see  \cite{GSWsuperstring} Section 2.2). The energy momentum tensor of the 2D conformal theory does vanish classically (constraint equations of the classical conformal field theory). Since Virasoro generators are the Laurent modes of the energy momentum tensor they do need to vanish classically as well. At the quantum level, however, vanishing of $L_n$ for all $n \in \mathbb{Z}$ lead to inconsistencies (due to the central extension terms). Fortunately, by correspondence principle one only needs to have the expectation values of the quantum operators corresponding to Virasoro generators to vanish which can be achieved without setting all of them to zero, only half is sufficient. So one imposes the following conditions to be satisfied by the physical states of the theory
\al{
L_n \left| \phi \right> = 0 = \bar{L}_n \left| \phi \right> \ \ \text{for $n > 0$} \ \ \ \text{and} \ \ \ L_0 = \bar{L}_0 
}

How do these translate into a condition in terms of the noncentral coadjoint elements $D, \bar{D}$? The Virasoro generators (adjoint elements) interact with the coadjoint elements through the pairing between them and the coadjoint action. Since our main case of interest is the simplest type orbits, namely Diff$S^1/S^1$ whose isotropy algebra is generated by $L_0$ (and $\bar{L}_0$ for the second copy) the latter quantization condition implies in this case that the coadjoint elements $D, \bar{D}$ are fixed by the same element in their corresponding orbits. 

In particular in \eqref{DirectProductVirasoroGeoAction} we know that $D$ and $\bar{D}$ will be identified with $D_{--}$ and $D_{++}$, respectively. Moreover,  $\partial_+ f / \partial_- f$ and $\partial_- \bar{f} / \partial_+ \bar{f}$ will be identified with $h_{++}$ and $h_{--}$, respectively. Then we can rewrite \eqref{DirectProductVirasoroGeoAction} as 
\al{
S & = \int D_{--} h_{++} + \int D_{++} h_{--}+ W_+ [f]  + W_-[f] \\ & = \int D_{\mu \nu} h^{\mu \nu} + W[f] - \Lambda(f, \bar{f}) \ \ \ \text{with} \ \ \ D_{-+} = 0 
}
where we used 
 \eqref{P2DGLCCFullActionWithCounterterm}. Hence, we see that this is the extension of \eqref{P2DGLCCFullActionWithCounterterm} with $D_{++}$ turned on. The question that follows then is whether we can recover the counterterm $\Lambda$ in \eqref{P2DGLCCFullActionWithCounterterm} from \eqref{DirectProductVirasoroGeoAction}  using the restrictions placed by quantization  discussed above. We  will not continue this analysis here.

\chapter{TRANSVERSE ACTIONS}
\label{Transverse Actions}
\section{Introduction}
In the previous chapter we  obtained  geometric actions on  the coadjoint orbits of  Kac-Moody (KM) and Virasoro algebras. In each case, the non-central coadjoint element was seen as a background field coupled to group-valued bosonic fields.

In this chapter we are going to lift each of these background fields to  a dynamical field by providing a Hamiltonian for each and lifting the isotropy equation of coadjoint orbits to a  constraint field equation.

\begin{figure}[H]
\begin{center}
\includegraphics[scale=.5]{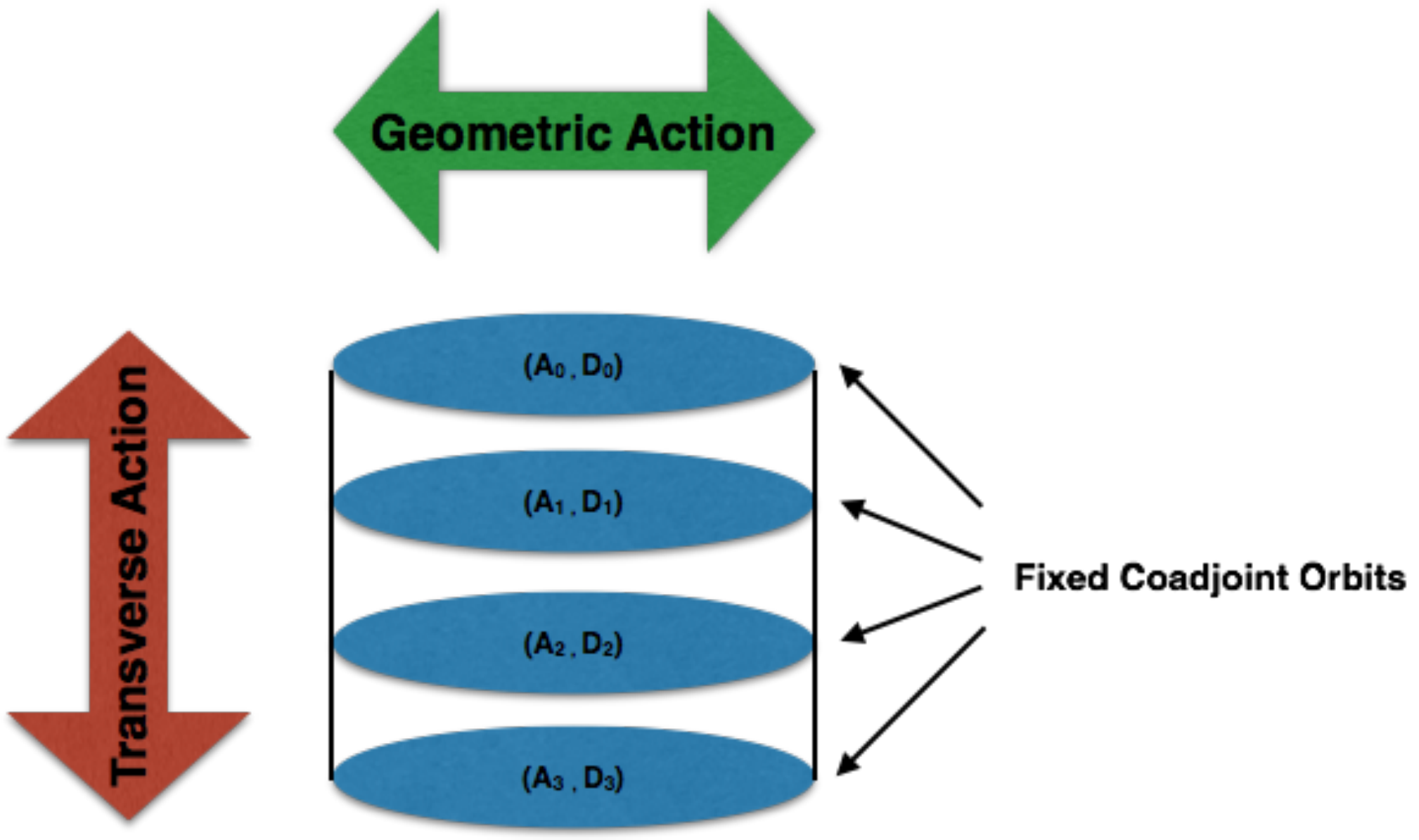}
\end{center}
  \caption{Geometric vs Transverse Actions}
  \label{geovstrans}
\end{figure}

 The isotropy equation \eqref{isotropyalgebrageneral} defines the set of adjoint vectors that fix the coadjoint element on its orbit, i.e.  that do not move it along the orbit. We then visualize infinitely many copies of  coadjoint orbits  stacked as sheets and interpret the isotropy equation as defining motion transverse to each sheet, i.e. from one sheet to the other, but not along the sheet itself (see Figure \eqref{geovstrans}).  This is the motivation for the term transverse. Coadjoint transformations on the orbits lift to local transformations of the constructed field theory.
 
We are going to apply this method first to the KM coadjoint element $A$ reaching the Yang-Mills (YM) theory of a gauge  field $A_{\mu}$, then to the Virasoro coadjoint element $D$, reaching the diffeomorphism field theory, the dynamical theory of  $D_{\mu  \nu}$.

The method of transverse actions provided here is originally from \cite{BLR97}, \cite{BLR00}, \cite{takeshithesis}. However, at the step where one picks up an ansatz for the momentum to construct the action, the equation \eqref{Xansatz}, we diverge. 
The reason for changing the momentum ansatz was the observation that the action for the diff field, obtained in \cite{BLR00} does not yield the same momentum as in the ansatz taken. In \cite{BLR00}, the YM form for the momentum is directly assumed in the KM case. Here, we do not assume the YM form. Rather, we introduce a fairly relaxed ansatz for the momentum. Then the YM  form arises automatically from the constructed action. This seems puzzling at first, but is essentially due to the Gauss law constraint being introduced into the action. Applying the same line of reasoning to the diff field we reached a different action than the one obtained in \cite{BLR00}. Moreover, the structures of the actions for the YM field and the diff field become similar upon this modification.

In this chapter, $x$ denotes the spatial coordinate unless otherwise stated. 
\section{Yang-Mills Action from Kac-Moody Algebra } 
\label{YMfromKM}
\subsection{Kac-Moody Gauge-Fixing }
\label{KMgaugefixing}
Recall the coadjoint action  \eqref{semidirectiso} of the adjoint vector $\mathcal{F}=(\xi(x),\Lambda(x),a)$ on the coadjoint vector $B=(D(x),A(x),\mu)$ of the semi-direct product of Virasoro and KM algebras, \  $\text{ad}^*_{\mathcal{F}}(B) = (\delta D(x), \delta A(x),0)$ 
with
\al{ \label{Virpartofcoad}
\delta D & = 2 \xi' D + D' \xi + q\xi''' + 2\beta \xi' - \text{Tr} \ (A \Lambda') \\ \delta A & =  A' \xi + \xi' A - [\Lambda , A ] + e^{-1}\Lambda' \label{KMpartofcoad}
}
where we introduced $q \equiv c \mu  / (2 \pi) $, $\beta \equiv h \mu / (4 \pi)$ and $e \equiv (k \mu)^{-1}$ for simplicity. $\delta D=0$ and $\delta A=0$ are the isotropy equations.

We can separate $\delta A$ and $\delta D$  into their pure Virasoro and  KM sectors. 
The pure KM sector ($\xi=0$) of $\delta A$ is
\al{\label{coadjointKMpureKM}
\delta A (x) & = -[ \Lambda(x) , A (x)] + e^{-1} \partial \Lambda(x) \notag\\ & = (-f_{abc} \Lambda_b (x) A_c(x) + e^{-1} \partial \Lambda_a (x))T_a
}
where we introduced 
$\Lambda(x) = \Lambda_b(x)T_b $ and $ A (x) = A_c (x)T_c $. Here $\{ T_a \}$ forms a basis for the Lie algebra of the base group of KM group. 
If we identify the coadjoint field $A$ with the space component $A_1$ of the YM field $A_{\mu}$ in 2D, then the transformation  above corresponds to a time-independent gauge transformation of $A_1$ : 
\al{\label{resgaugetrasformation}
A_1(t,x)  \rightarrow A^g_1(t,x) = g(x) A_1(t,x) g^{-1} (x) - e^{-1} ( \partial_1 g(x) ) g^{-1} (x)  
} 
For an infinitesimal transformation, $g(x) = 1 -  \Lambda_a (x) T_a$, \eqref{resgaugetrasformation} yields
\al{ \label{delAU}
\delta A_1 = \delta A_{a1} T_a \equiv A^g_1- A_1    =  (  -  [ \Lambda (x) , A_1 (t,x) ]_a +  e^{-1} \partial_1 \Lambda_a (x)  ) T_a
}
This  is the same as \eqref{coadjointKMpureKM} except that $A_1=A$ now also depends on time.

\subsection{Gauss Law }
We introduce the  operator 
\al{ \label{GaussQ}
Q = \int dx \,  e^{-1} G_a (x) \, \Lambda_a (x) 
}
(the coupling constant $e^{-1}$ is  introduced for convenience) that generates $\delta A_1$ :
\al{ \label{consdelA}
\delta A_{1a} (x) = \{  A_{1a} (x) ,Q \}  
}
Here $\{  \ ,  \ \}$ is the standard Poisson bracket (PB) and introducing the conjugate momentum $\pi^1_a$ to $A_{1a}$ it is explicitly given by the spatial integral
\al{
\{ F , G \} & = \int d x \left[   \frac{\delta F}{\delta A_{1c} (x)} \frac{\delta G}{\delta \pi^1_c (x)} - \frac{\delta F}{\delta \pi^1_c (x)} \frac{\delta G}{\delta A_{1c} (x)}\right]
}
In particular, we have 
\al{
\{ A_{1a} (x) , \pi^1_b (y) \} = \delta_{ab} \delta (x-y)
}
Combining \eqref{GaussQ}, \eqref{delAU} and \eqref{delAai} equation \eqref{consdelA} yields
\al{ \label{delAai}
\delta A_{1a} (x) &  = \frac{\delta Q}{\delta \pi^1_a (x) } = f_{abc} \Lambda_b(x) A_{1c}(x) - e^{-1} \partial_1 \Lambda_a (x) 
}
From this we deduce 
\al{
G_a (x) & = e f_{abc}  A_{1b}(x)  \pi^1_c (x) +  \partial_1 \pi^1_a(x) \\ & = ( e[A_1 (x) , \pi^1 (x) ] +   \partial_1 \pi^1(x)  )_a
}
$G_a=0$ is the well-known  Gauss law constraint. In the context of YM theory, Gauss law is the field equation for $A_0$ which is non-dynamical since $\pi^0$ vanishes. It is encountered in Dirac's constrained formalism\footnote{Dirac's formalism is reviewed in  Section \ref{DiracConstraintSummary}. } as the secondary constraint that follows from the consistency condition of the primary constraint $\pi^0=0$.

We also calculate $\delta \pi^1_a (x) =  \{ \pi^1_a (x) , Q \}$:
\al{ \label{delpiia}
\delta \pi^1_a (x)  = -\frac{\delta Q}{\delta A_{1a} (x)}   = e f_{abe}  \Lambda_b(x) \pi^1_e (x) = e [ \Lambda (x)  , \pi^1 (x)]_a 
}
This shows that $\pi^1$ is gauge-covariant as \eqref{delpiia} is the infinitesimal reduction of 
\al{
\pi^1_g = g^{-1} \pi^1 g
}
for a time-independent gauge transformation $g$. 
\subsection{Kac-Moody Transverse Action }
 Notice that the momentum has a hidden time index by its definition via the action functional (to be constructed) 
\al{ 
 \pi^1 = \frac{\delta S}{\delta (\partial_0 A_1 )} (= \pi^{01} )
 }
 
  We take the ansatz
\al{\label{piAansatz}
\pi^1 = -\partial^0 A^1 + L^{10}
}
where  $L$ is a rank-two object that does not involve $\partial^0 A^1$, but may involve the fields $A_1, A_0$ and their spatial derivatives. 

Transverse action is constructed using the isotropy equation $
\delta A_{1a}  =0$.  This enforces the dynamics to be transverse to the orbits, i.e., evolution of the field does not move the field on the orbit. 
We would like to define an action functional that enforces this condition. Since $\delta A_1$ is generated by $G_a$, the isotropy equation can be  enforced by the Gauss law constraint, $G_a = 0$. Hence, we introduce the following prescription for the transverse Lagrangian
\al{ \label{transverseactionprescription}
\mathcal{L} = \text{ST} - \mathcal{H} + \lambda C
}
where ST stands for the symplectic term,  $\mathcal{H}$ for Hamiltonian density, $C$ for the constraint and $\lambda$ for the Lagrange multiplier of $C$.  

Since $G_a$ is linear in the momentum, it has a hidden upper  time index. Then, for the Lagrangian to be a GCT-scalar, we need to contract $G_a$ with an object having a lower time index. Again if the only fields at hand are $A_0$, $A_1$ and their derivatives then the simplest choice is a constant times $A_0$. This is indeed a requirement to obtain the YM action.   

For convenience, let us define $B_{\mu \nu} \equiv \partial_{\nu} A_{\mu}$. Then the pieces  become
\al{
\text{ST} & =  (\partial_0 A_{1a} )   \pi^1_a = B_{10a}  (-B^{10}_a + L^{10}_a)  \notag\\ \mathcal{H} & =  (1/2) \pi_{1a} \pi^1_a =  (1/2)  (-B_{10a} + L_{10a} )(-B^{10}_a + L^{10}_a)  \notag\\ \lambda C & = c A_{0a} ( e f_{abc} A_{1b} (- B^{10}_c + L^{10}_c )  +  \partial_1 ( - B^{10}_a + L^{10}_a ) )
}
Combining we get
\al{
\mathcal{L} & = \text{ST} - \mathcal{H} + \lambda C \notag\\ & = -\frac{1}{2} B_{10a} B^{10}_a + \frac{1}{2} L_{10a} L^{10}_a -cA_{0a} (\partial_1 B^{10}_a) + cA_{0a} (\partial_1 L^{10}_a) \notag\\ & \hspace{0.86 in} - ce f_{abc} A_{0a} A_{1b} B^{10}_c + ce f_{abc} A_{0a} A_{1b} L^{10}_c 
}
Then we recompute the momentum 
\al{
\pi^1_a & \equiv \frac{\delta S}{\delta(\partial_0 A_{1a})} = \frac{\delta S}{\delta B_{10a}} = \frac{\partial \mathcal{L}}{\partial B_{10a}} - \partial_1 \left( \frac{\partial \mathcal{L}}{\partial(\partial_1 B_{10a} )} \right) \notag\\ & = - \partial^0 A^1_a + c \partial^1 A^0_a - c e f_{abc} A^0_b A^1_c
}
We see that this yields the YM momentum only when $c=1$ :
\al{
\pi^1_a =  \partial^1 A^0_a - \partial^0 A^1_a +  e f_{abc} A^1_b A^0_c = F^{10}_a \label{mompia}
}
Setting $c$ to any real constant is legitimate as it amounts to rescaling $A_0$ which does not affect the dynamical characteristics of the theory since $A_0$ is nondynamical. The meaning of setting $c=1$, on the other hand, is that the Gauss law is the constraint associated with $A_0$, not with an arbitrary multiple of it.

Hence, the field equations force $L^{10}$ (thereby the momentum $\pi^{1}$) to take the desired form;  we did not need to enforce it (as in \cite{BLR00}). The rest is straightforward. 
Inserting the momentum \eqref{mompia} back into $\mathcal{L}$,  and writing $\pi^1 = F^{10}$ we get
\al{
\mathcal{L} & = \partial_0 A_{1a} F^{10}_a - \frac{1}{2} F_{10a} F^{10}_a + A_{0a} ( \partial_1 F^{10}_a + e f_{abc} A_{1b} F^{10}_c )  \notag\\ & = \frac{1}{2} F_{10a} F^{10}_a + \partial_1 ( A_{0a} F^{10}_a ) 
}
where we applied partial integrations   and rearranged the indices. Let $y$ denote the spacetime coordinate $(t,x)$.
Then, the action reads
\al{
S = \int d^2 y \ \mathcal{L} =  \frac{1}{2} \int d^2 y \  F_{10a} F^{10}_a + \int d^2 y \ \partial_1 (A_{0a} F^{10}_a ) 
}
Using Stoke's theorem the second term is converted to an integral on the boundary :
\al{
\int_V d^2 y \ \partial_1 (A_{0a} F^{10}_a ) = \int_{\partial V} ds_1   A_{0a} F^{10}_a 
}
We will assume the boundary conditions (such as $A_{0a}=0$ or $F^{10}_a=0$ on $\partial V$) that make this term vanish. Hence, we get
\al{
S =   \frac{1}{2} \int d^2 y \  F_{10a} F^{10}_a 
}

Since in 2D we have 
$F_{\mu \nu a} F^{\mu \nu}_a  = 2 F_{10a} F^{10}_a$
the action can be written as
\al{
S =  \frac{1}{4} \int d^2 y \ F_{ \mu \nu a} F^{\mu \nu}_a 
}
Using the normalization $\text{Tr} \ ( T_a T_b ) =  \delta_{ab}$ we get
\al{ \text{Tr} \ (F_{\mu \nu} F^{\mu \nu} )  = F_{\mu \nu a} F^{\mu \nu}_b \ \text{Tr} \ ( T_a T_b ) = F_{\mu \nu a} F^{\mu \nu}_a
}
Thus, the action becomes
\al{
S =  \frac{1}{4} \int d^2 y \ \text{Tr} \ ( F_{\mu \nu} F^{\mu \nu} )
}
We can covariantly extend this action  to higher dimensions as
\al{
S =  \frac{1}{4} \int  d^N y \ \sqrt{g} \ \text{Tr} \ ( F_{\mu \nu} F^{\mu \nu} )
}
Since $F_{\mu \nu}$ are the components of a two-form they are not affected by covariantization. 

\subsection{Virasoro Sector of Kac-Moody Coadjoint Transformation }
Now, let us examine the Virasoro sector of  \eqref{KMpartofcoad} :
\al{ \label{deltaxiA}
\delta A = A' \xi + \xi' A
}
where  $A=A(x)$,  $\xi=\xi(x)$  and $'\equiv d/dx$. As we have  seen, $A$ can be identified with the space component of the YM vector field $A_{\alpha}$ in 2D. Below, we will reach an argument further supporting this and one that will be helpful in the construction of the diff transverse action. 

The statement that  $A_{\alpha}$ is a (covariant) vector field  means that under an infinitesimal coordinate transformation generated by a (contravariant) vector field $\xi^{\beta}$, $A_{\alpha}$ must transform according to
\al{ \label{deltaxiAmu}
\delta_{\xi} A_{\alpha} = \xi^{\beta} (\partial_{\beta} A_{\alpha}) + A_{\beta} (\partial_{\alpha} \xi^{\beta})
}
In 2D this constitutes two transformations, $\delta_{\xi} A_0$ and $\delta_{\xi} A_1$.   
The latter becomes
\al{
\delta_{\xi} A_1 & = \xi^0 \partial_0 A_1 + \xi^1 \partial_1 A_1 + A_0 \partial_1 \xi^0 + A_1 \partial_1 \xi^1 \notag\\ & = ( \xi^1 \partial_1 A_1 + A_1 \partial_1 \xi^1 ) + (\xi^0 \partial_0 A_1 + A_0 \partial_1 \xi^0)
}
For this to match up with \eqref{deltaxiA} we only  need $\xi^0 = 0$. Then for the remaining component we have 
\al{
\delta_{\xi} A_0 |_{\xi^0 = 0} = \xi^1 \partial_1 A_0 + A_1 \partial_0 \xi^1
}
$A_0$ is expected to transform as a scalar under spatial transformations (since it has no spatial index) so we also  need $\partial_0 \xi^1=0$. Thus together we have
\al{
\xi^0& =0 =\partial_0 \xi^1 \notag\\
\delta_{\xi} A_1 & = \xi^1 \partial_1 A_1 + A_1 \partial_1 \xi^1 \notag\\ \delta_{\xi} A_0 & = \xi^1 \partial_1 A_0
}

This is compatible with the temporal gauge (i.e. $A_0 =0$ implies $\delta A_0=0$), but does not require it. 

Similarly consider a  gauge tranformation 
\al{ \label{gaugetransformationgeneric}
A^g_{\mu}  = g A_{\mu} g^{-1} - e^{-1} g \partial_{\mu} g^{-1}
}
Since the infinitesimal generator $\Lambda$ on coadjoint orbits is only space-dependent, it can only generate time-independent gauge transformations. For an infinitesimal time-independent gauge transformation $g \approx 1 - \Lambda$,  \eqref{gaugetransformationgeneric} yields
\al{
\delta_{\Lambda} A_1=  A^g_1 - A_1 \approx e^{-1} \Lambda' - [\Lambda , A_1] 
}
Thus correspondence with KM element requires only the time-independence of the gauge transformations.
Under such transformations \eqref{gaugetransformationgeneric} yields
\al{
A^g_0 & = g A_0 g^{-1} - e^{-1} g \partial_0 g^{-1} \notag\\ & = g A_0 g^{-1}
}
This, as well, is compatible with the temporal gauge $A_0=0$, but does not require it.

Apart from the fact that $A_0$ is invisible on orbits, then comes the question "Why do we require the temporal gauge?" The answer, as shown in the previous chapter, comes from the geometric action. The term involving the noncentral Kac-Moody element $A$ is identified as a background gauge field interacting with WZNW field  only in temporal gauge, \eqref{diVecchia1}, \eqref{diVecchia2}.  

Thus, we have reached the result that the coadjoint transformation given in \eqref{deltaxiA} corresponds to the residual ($A_0=0$), time-independent ($\partial_0 \xi^{\mu} = 0$), spatial ($\xi^0=0$) infinitesimal coordinate transformation (or Lie derivative) of a rank-one1 tensor $A_{\alpha}$ ($A=A_1$). 
A similar analysis in the case of the diff field  will play a crucial role in constructing its dynamical theory.
\section{Diffeomorphism Field Action } \label{Diffaction}
\subsection{Virasoro Gauge-Fixing }
\label{Virasoro Gauge-Fixing}
Consider the pure Virasoro sector ($\Lambda=0$) of  \eqref{Virpartofcoad}
\al{ \label{coadjDch4}
\delta_{\xi} D = 2 \xi' D + D' \xi + q \xi''' + 2\beta \xi'
}
In analogy with the YM case, we claim that the field $D$ can be identified with the space-space component of a rank-two tensor $D_{\mu \nu}$ in 2D apart from central extensions and under certain assumptions.  Let us find out the assumptions needed. 

The Lie derivative of a rank-two tensor $D_{\mu \nu}$ along a vector field $\xi^{\lambda}$ is given by
\al{ \label{Liederranktwo}
\delta_{\xi} D_{\mu \nu} = (\partial_{\mu} \xi^{\lambda})D_{\lambda \nu} + (\partial_{\nu} \xi^{\lambda}) D_{\mu \lambda} + \xi^{\lambda} (\partial_{\lambda} D_{\mu \nu} ) 
}
which, including the central extension terms is modified to
\al{ \label{Liderranktwowithcentral}
\delta_{\xi} D_{\mu \nu} = (\partial_{\mu} \xi^{\lambda})D_{\lambda \nu} + (\partial_{\nu} \xi^{\lambda}) D_{\mu \lambda} + \xi^{\lambda} (\partial_{\lambda} D_{\mu \nu} )  +  q \partial_{\mu} \partial_{\nu} \partial_{\lambda} \xi^{\lambda} +\beta (\partial_{\mu} \xi_{\nu} + \partial_{\nu} \xi_{\mu} )
}

We would like to match up (11) component of \eqref{Liderranktwowithcentral} with \eqref{coadjDch4}. 
In 2D, $D_{\mu \nu}$ has four independent components. We first assume that $D_{\mu \nu}$ is symmetric\footnote{Note that for the antisymmetric part of the diff field,  the central extension terms vanish.}. We then have for the (11) component 
\al{
\delta D_{11}  & = \ \xi^1 \partial_1 D_{11}  + 2 \partial_1 \xi^{1} D_{11} + q \partial_1  \partial_1  \partial_1 \xi^1 + 2\beta \partial_1 \xi_1  \notag\\ & \hspace{0.8in} \ + ( \xi^0 \partial_0 D_{11} + 2 D_{01} \partial_1 \xi^0 + q  \partial_1  \partial_1 \partial_0 \xi^0)
}
We see that the necessary and sufficient condition is 
\al{ \label{gaugefix1}
\xi^0 = 0
}

Next, let us consider $D_{01}$. We expect it to behave as a (covariant) vector under time independent, spatial transformations as it has a single spatial index. Its transformation, using \eqref{gaugefix1}, becomes
\al{ \label{deltaD01}
\delta D_{01}   =  \xi^1  \partial_1 D_{01} + D_{01}  \partial_1 \xi^1 + (  D_{11} \partial_0 \xi^1 + q  \partial_0 \partial_1  \partial_1   \xi^1 + \beta  \partial_0 \xi_1)
}
Hence, we need the additional condition that
\al{ \label{gaugefix2}
\partial_0 \xi^1 =0
}

The last component $D_{00}$, is expected to be a spatial scalar and this holds without any additional conditions. Its transformation using \eqref{gaugefix1} and \eqref{gaugefix2} becomes
\al{
\delta D_{00}   =  \xi^1 \partial_1 D_{00} 
}
To summarize, so far we have 
\begin{subequations}
\label{nogaugefixingtrfs}
\al{
\xi^0& =0 = \partial_0 \xi^1 \\
\delta D_{11} & = \xi^1 \partial_1 D_{11} + 2 \partial_1 \xi^1 D_{11} + q \partial_1^3 \xi^1 +2 \beta \partial_1 \xi_1 \\  \delta D_{01} & = \xi^1  \partial_1 D_{01} + D_{01}  \partial_1 \xi^1  \\  \delta D_{00} & = \xi^1 \partial_1 D_{00} 
}
\end{subequations}

Now, recall that the identification \eqref{lightconediffmetric} of diff-metric coupling in the geometric action for the Virasoro algebra required the temporal gauge $D_{01}=0$. Moreover, since $D_{01}$ has a spatial index, if it were nonzero it should have shown itself on the coadjoint orbit. 
 Thus we must take the temporal gauge, $D_{01}=0$ for correspondence. This is consistent with \eqref{deltaD01}, \eqref{nogaugefixingtrfs} as setting $D_{01}=0$ implies $\delta D_{01}=0$. 

  We can not repeat this argument for $D_{00}$ since even if it is non-vanishing in higher dimensions it would project to zero on the spatial hypersurface. And the reason we haven't encountered it as part of the diff-Polyakov interaction was that the  P2DG action, central part of the Virasoro geometric action, assumes the lightcone gauge $h_{--}=0$ and this component multiplies $D_{00} \leftrightarrow D_{++}$ as discussed. 
  
  However, since it is invisible on the orbits we will "choose" to take it zero in constructing the action, although this is not a requirement. We shall call this the full temporal gauge, and in this gauge transformations become
  \begin{subequations} \label{lastset}
\al{
\xi^0& =0 =\xi^1  \\
\delta D_{11} & = \xi^1 \partial_1 D_{11} + 2 \partial_1 \xi^1 D_{11} + q \partial_1^3 \xi^1 + 2 \beta \partial_1 \xi_1  \label{deltaD11lastset} \\  \delta D_{01} & = 0 =D_{01}\\  \delta D_{00} & = 0 = D_{00} 
}
\end{subequations}

\subsection{Diff-Gauss Law }
\label{DiffGaussLaw}
Although we are in 2D (where we have a single spatial index) we are going to keep the spatial indices of the diff field for higher dimensions i.e. $D_{11} \rightarrow D_{ij}$. This does not affect the analysis though simplifies the  generalization to higher dimensions later on. The transformation \eqref{deltaD11lastset}  is then lifted to 
\al{ \label{deltaDijD11}
\delta D_{ij} = \xi^k \partial_k D_{ij} + \partial_i \xi^k D_{kj} + \partial_j \xi^k D_{ik} + q \partial_i \partial_j \partial_k \xi^k + \beta ( \partial_i \xi_j + \partial_j \xi_i)
}

In analogy with the analysis in the previous section, we propose that $\delta D$ given in \eqref{lastset} (as $\delta D_{11} \rightarrow \delta D_{ij}$) is  generated by the operator
\al{ \label{QGausssmeared}
Q=\int d x \ G_k (x) \xi^k (x)
}
through
\al{ \label{diffQPB}
\delta_{\xi} D_{lm} (x) = \{  D_{lm} (x) , Q \}
}

The Poisson bracket is, as before,  assumed to be standard :
\al{ \label{diffPB}
\{ F , G \} : = \int d x \left[   \frac{\delta F}{\delta D_{ij} (x)} \frac{\delta G}{\delta X^{ij} (x)} - \frac{\delta F}{\delta X^{ij} (x)} \frac{\delta G}{\delta D_{ij} (x)}\right]
}
where the integral is over a spatial hypersurface and we introduced the conjugate momentum $X$ to $D$. Since we took $D$ to be symmetric, $X$ also must  be symmetric by its definition through the Lagrangian. Then, in particular, we have 
\al{
\{ D_{ij} (x) , X^{lm} (y) \} = \frac{1}{2} (\delta^l_i \delta^m_j + \delta^l_j \delta^m_i ) \delta(x-y)
}
where the delta function is the spatial one. 
$G_k$ is the analog of the Gauss law operator of YM. We  call it the diff-Gauss law operator. 
Evaluating \eqref{diffQPB}  we get
\al{
 \frac{\delta Q }{\delta X^{lm}(x)}    = (\partial_l \xi^k) D_{km} & + (\partial_m \xi^k) D_{lk} + \xi^k (\partial_k D_{lm}) \notag\\ & + q \partial_l \partial_m \partial_k \xi^k + \beta ( \partial_l \xi_m + \partial_m \xi_l)
}
From this we deduce the diff-Gauss law operator
\al{ \label{DiffGaussLawOriginal}
G_k &  = X^{ij}  \partial_k D_{ij}  -  \partial_i ( X^{ij}  D_{kj} ) -\partial_j(D_{ik}  X^{ij})- q \partial_k \partial_j \partial_i X^{ij}  - 2 \beta \partial_i X^i_{\  k}
\notag\\ 
& = X^{ij}  \partial_k D_{ij}  -  2\partial_i ( X^{ij}  D_{kj} ) - q \partial_k \partial_j \partial_i X^{ij}  - 2 \beta  g_{k j} \partial_i X^{ij}
}
where we used the symmetry of $X$. 
The shortcut prescription to find the diff-Gauss law is given by 
\al{ \label{diffGaussshortcutpres}
G_k = ( - \delta_{\xi} D_{ij} )_{\xi^k \rightarrow X^{ij0} }
}
where $\delta_{\xi} D_{ij}=\delta D_{ij}$ is given by \eqref{deltaDijD11}. 

We also compute $\delta X$ : 
\al{ \label{deltaXQ}
\delta_{\xi} X^{lm}  & =  \{  X^{lm} , Q  \}  \notag\\ & =   - (\partial_k \xi^k  ) X^{lm}  - (\partial_k X^{lm}  ) \xi^k + (\partial_k \xi^l  ) X^{km}  + (\partial_k \xi^m  ) X^{kl}  \notag\\ & = - \Big( \xi^k \partial_k X^{lm} - \partial_k \xi^l X^{ k m} - \partial_k \xi^m X^{lk} + \partial_k \xi^k X^{lm} \Big) 
}
The sum of the first three terms corresponds to the Lie derivative of a rank-two spatial tensor and the last term shows that $X^{lm}$ is a rank-two spatial tensor density of weight one \cite{takeshithesis}. 
In fact  in 1D \eqref{deltaXQ} reduces to 
\al{\label{deltaXQ1D}
\delta X = - \xi' X - X' \xi + \xi' X + \xi' X = \xi' X - X' \xi
}
which is not the transformation of a rank-two tensor; rather that  of a generator $\xi$ of the coordinate transformation i.e. of a vector field.

\subsection{Diff Field Action }
\label{TheDiffeomorphismACtion}
Now, we are ready to construct the Lagrangian according to the prescription \eqref{transverseactionprescription}
\al{ \label{transverseactionprescriptionrevisit}
\mathcal{L} = \text{ST} - \mathcal{H} + \lambda C
}
 The diff-Gauss law has a lower space index and a hidden upper time index (since it is linear in the momentum) so its Lagrange multiplier is of the form $\lambda_0^k$. The obvious choice is  $D_0^{ \ k}$. Let us write down the pieces
\al{ \label{transverseLprescappliedtodiff}
\text{ST} & = (\partial_0 D_{ij}) X^{ij} \\ \mathcal{H} & = (1/2) X_{ij} X^{ij}  \\ 
\lambda (C) & = D_0^{ \ k}  ( X^{ij} \partial_k D_{ij} -  2\partial_i ( X^{ij} D_{kj} ) - q \partial_k \partial_j \partial_i X^{ij}  - 2 \beta  g_{k j} \partial_i X^{ij} )
}
where again we assumed the standard Hamiltonian and the symplectic term is fixed by the PB \eqref{diffPB}. Combining the pieces we get
\al{ \label{diffLag1}
\mathcal{L}_D =  D_0^{ \ k}   X^{ij} & \partial_k D_{ij} - 2 D_0^{ \ k} \partial_i (X^{ij} D_{kj})  - q D_0^{ \ k}  \partial_k \partial_j \partial_i X^{ij}- 2 \beta D_{0j} \partial_i X^{ij}  \notag\\  & + (\partial_0 D_{ij}) X^{ij} - (1/2) X_{ij} X^{ij}  
}

At this point let us mention an ambiguity. We can continue with this form of the action or we can partially integrate the higher order $q$ term to get a simpler action. This doesn't make a difference in flat space, in which \eqref{diffLag1} is written, since partial derivatives commute and boundary terms  can be made to vanish with appropriate boundary conditions. In the original theory the spatial hypersurface was a circle whose boundary is empty, thus anything defined on it vanishes anyway. Moreover, the momentum (and the field equations) are unaffected by this  partial  integration. However, after covariantization the two actions, partially integrated and the original, differ. Covariant derivatives do not commute and the difference between two distinct choices yield a factor involving the Riemann tensor components. 

We shall continue with the partially-integrated version.  
We then get 
\al{\label{Ldifftrans}
\mathcal{L}_D = X^{ij} ( D^k_{\ 0}  \partial_k D_{ij} & + 2 \partial_i D^k_{\ 0}  D_{kj} + q \partial_i \partial_j \partial_k D^k_{\ 0}  +2 \beta \partial_i D_{0j} + \partial_0 D_{ij} ) \notag\\ & - (1/2) X^{ij} X_{ij}  
}

For convenience, we shall introduce the notation $\partial_k D_{ij} \equiv D_{ijk}$, $\partial_l \partial_k D_{ij} \equiv D_{ijkl}$ etc. In analogy with the YM case, we declare the ansatz 
\al{ \label{Xansatz}
X^{ij} = D^{ij0} + Y^{ij0}
} 
where we require that $Y$ is a functional of the diff field  components and their space derivatives, but not time derivatives. Then, we get
\al{
\mathcal{L}_D   &= (D^{ij0} + Y^{ij0} ) \Big( D^k_{\ 0}   D_{ijk} + 2  D^k_{\ 0 i}  D_{kj} + q  D^k_{\ 0kji} +2 \beta  D_{0ji} \Big)  \notag\\ & \hspace{0.5in} + (1/2)  D^{ij0}D_{ij0} - (1/2) Y^{ij0} Y_{ij0} 
} 
Only the part involving velocities will be relevant for the momentum calculation :
\al{
\mathcal{L}_{\text{rel}} = D^{ij0} \Big( D^k_{\ 0}   D_{ijk} + 2  D^k_{\ 0 i}  D_{kj} + q  D^k_{\ 0kji} +2 \beta D_{0ji}  \Big) +(1/2)  D^{ij0} D_{ij0} 
}
Now, let us recompute the momentum from the constructed Lagrangian:
\al{ \label{diffmomflat}
X^{mn0} \equiv \frac{\delta S_{\text{rel}}}{\delta D_{mn0}}  = D_k^{ \ 0} D^{mnk} + 2 D_k^{ \ 0 m} D^{kn} + q D_k^{\ 0 k n m } + 2 \beta D^{0nm} + D^{mn0} 
}
Since diff momentum needs to be symmetric in its first two indices, we take 
\al{
X^{mn0}= D^{mn0} &+ D_k^{ \ 0} D^{mnk} +  D_k^{ \ 0 m} D^{kn} +  D_k^{ \ 0 n} D^{km}  \notag\\ &+ q D_k^{\ 0 k n m }  +  \beta ( D^{0nm} + D^{0mn})
}
The shortcut prescription to find the diff momentum is the following :
\al{ \label{shortcutmomentumprescription}
X_{ij0} = \partial_0 D_{ij} + (  \delta_{\xi} D_{ij} )_{\xi^k \rightarrow D_0^{ \ k} }
}

Comparing \eqref{diffmomflat} with \eqref{Ldifftrans} we see that
\al{ \label{LXsquare}
\mathcal{L}_D & = X^{ij0}  X_{ij0} - (1/2) X^{ij0} X_{ij0} \notag\\ & = (1/2)  X^{ij0} X_{ij0} 
}
Finally we covariantize 
\al{ \label{LD=Xsquare}
\mathcal{L}_D = (1/2) \sqrt{g} X^{\mu \nu \lambda} X_{\mu \nu \lambda}
}
This is  analogous to the YM Lagrangian, $ \propto \sqrt{g} F^{\mu \nu}_a F_{\mu \nu a}  $. 
We will introduce a constant $\alpha$ for the terms involving two diff fields since the (physical) dimension of these terms will, in general,  be different, and from here on  $\nabla_{\lambda} D_{\mu \nu} \equiv D_{\mu \nu \lambda}$, $\nabla_{\sigma} \nabla_{\lambda} D_{\mu \nu} \equiv D_{\mu \nu \lambda \sigma}$ etc. Then the covariantized momentum in \eqref{LD=Xsquare} reads
\al{ \label{covariantXmom}
X^{\mu \nu \lambda}  = D^{\mu \nu \lambda} & + \alpha D_{\sigma}^{ \ \lambda} D^{\mu \nu \sigma} +  \alpha  D_{\sigma}^{ \ \lambda \mu} D^{\sigma \nu}+\alpha  D_{\sigma}^{ \ \lambda \nu} D^{\sigma \mu}  \notag\\ &  +  \beta D^{\mu \lambda \nu} + \beta D^{\nu \lambda \mu} +q D_{\sigma}^{ \ \lambda \sigma \mu \nu}
}

The $q$-term in the starting expression \eqref{Liderranktwowithcentral} was invariant under the change of derivative indices since the expression was written in flat space. Now, however, we covariantized the theory and covariant derivatives do not commute. That's why we need to symmetrize this term. Then the momentum becomes  
\al{ \label{covariantXmom2}
X^{\mu \nu \lambda}  = D^{\mu \nu \lambda} & + \alpha D_{\sigma}^{ \ \lambda} D^{\mu \nu \sigma} +  \alpha  D_{\sigma}^{ \ \lambda \mu} D^{\sigma \nu}+\alpha  D_{\sigma}^{ \ \lambda \nu} D^{\sigma \mu}  \notag\\ &  +  \beta D^{\mu \lambda \nu} + \beta D^{\nu \lambda \mu} + q D_{\sigma}^{ \ \lambda (\sigma \mu \nu)}
}
where $( \mu \nu \cdots )$ denotes symmetrization of the indices inside.  

The $\beta$ term in \eqref{coadjDch4} could have been made to vanish at the very beginning by a constant shift of $L_0$ in \eqref{stringVirasorocommutation} (see Section \ref{vanishingbeta}).  So, here we will take $\beta=0$ for simplicity. Let us mention, however, that this does not affect the validity of the arguments below about the structure of the Lagrangian. Nonzero $\beta$ terms simply add more corrections to the propagator, three-point and four-point vertices; they do not change the general structure. With $\beta$ set to zero the Lagrangian simplifies to 
\al{
2 \mathcal{L}_D & = D^{\mu \nu \lambda} D_{\mu \nu \lambda} \notag\\ &  \hspace{0.1in}
+ 2 \alpha (  D^{\mu \nu \lambda} D^{\sigma}_{\ \lambda} D_{\mu \nu \sigma} + 2 D^{\mu \nu \lambda} D^{\sigma}_{\ \nu} D_{\sigma \lambda \mu}  ) 
\notag\\ &  \hspace{0.1in}
+ \alpha^2 ( D^{\mu \nu \lambda} D_{\mu \nu \sigma} D^{\rho}_{\ \lambda} D^{\sigma}_{\ \rho} +2 D^{\mu \nu \lambda} D_{\sigma \mu \lambda} D^{\sigma}_{\ \rho} D^{\rho}_{\ \nu} \notag\\ & \hspace{0.6in} + 2 D^{\mu \nu \lambda} D_{\sigma \mu \rho} D^{\rho}_{\ \nu} D^{\sigma}_{\ \lambda} +4 D^{\mu \nu \lambda} D_{\sigma \rho \mu } D^{\rho}_{\ \nu} D^{\sigma}_{\ \lambda} ) 
\notag\\ & \hspace{0.3in} 
+ 2q D^{\mu \nu \lambda} D^{\sigma}_{ \ \lambda (\mu \nu \sigma)} 
\notag\\ & \hspace{0.3in} 
+ 4\alpha q (  D^{\mu \nu \lambda} D^{\sigma}_{ \ \lambda} D^{\rho}_{\ \sigma (\mu \nu \rho)} + 2  D^{\mu \nu \lambda} D^{\rho}_{ \ \nu} D^{\sigma}_{\ \mu (\rho \lambda \sigma)} ) 
\notag\\ & \hspace{0.4in} 
+ q^2 D_{\sigma}^{ \ \lambda (\sigma \mu \nu)} D^{\rho}_{\ \lambda (\rho \mu \nu)}
}

\begin{figure}[H]
\begin{center}
\includegraphics[scale=.5]{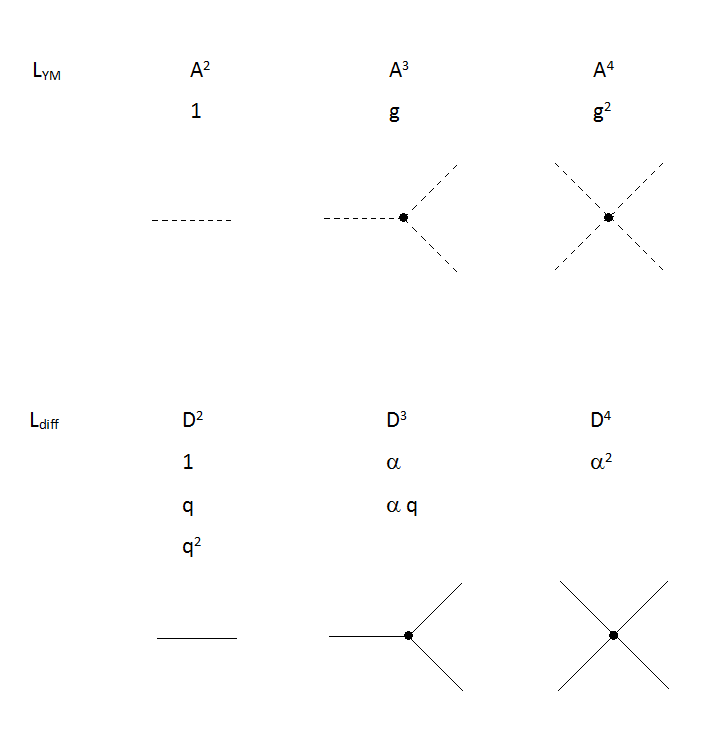}
\end{center}
  \caption{Yang-Mills vs Diff Field - Structure Comparison}
  \label{YMvsdiff}
\end{figure}

To see the contributions to the propagator and interactions better we may rearrange the terms as 
\al{
\mathcal{L}_D = (L_0 +L_q + L_{q^2}) + (L_{\alpha} + L_{\alpha q} ) + L_{\alpha^2}
}
where $L_0$ is the no coefficient term, $L_q$ is the term with coefficient $q$, etc. Then the structure is as  shown in  Figure \eqref{YMvsdiff}. 

We have an ordinary kinetic term and two corrections to the propagator with coefficients $q^2$ and $2q$. We have a three-point vertex with coefficient $2 \alpha$ and a correction to it with coefficient $2 \alpha q$. Finally we have a four-point vertex with coefficient $\alpha^2$. 
So in addition to the Lagrangian being in the form of momentum square as of YM theory, the expanded expression has the same structure of interaction vertices as in  YM theory apart from the higher order corrections (which is inevitable since the Virasoro central extension is of higher order contrary to those of KM algebra).

The previously proposed \cite{BLR00} Lagrangian  is obtained from this one if we get rid of terms with $q^2, \alpha q$ and $\alpha^2$ coefficients and rescale the $\alpha$ and $q$ terms by $1/2$. This is equivalent to taking only the symplectic part  of \eqref{LD=Xsquare}.  We shall call it the BLRY Lagrangian : 
\al{ \label{BLRYLagrangian}
\mathcal{L}_{\text{BLRY}} = (1/2) D_{\mu \nu \lambda} X^{\mu \nu \lambda} 
} 
As mentioned before this Lagrangian is built using the ansatz $X_{\mu \nu \lambda} = D_{\mu \nu \lambda}$ for the momentum.  However, it is easy to see that the BLRY action does not yield back the same momentum.\footnote{Note that this is a requirement only before covariantization. It is easy to see that after the covariantization neither action yields back the same momentum. This is due to the loss of the assumption that the correction $Y$ to the velocity term in \eqref{Xansatz} is independent of the time derivative upon covariantization.} This was the observation that led to the momentum ansatz \eqref{Xansatz} and eventually to the modified action.  
Moreover,  the stuctural correspondence with  YM theory is lost as the four-point vertex is not included in \eqref{BLRYLagrangian}. Finally, in \cite{BLR97}, \cite{BLR00}  only one possible $q$-term was taken into account rather than using the symmetrized expression involving all possibilities that reduce to the same expression in flat spacetime.

We have thus obtained a covariant action for the diff field assuming the diff field is a tensor\footnote{Although our gauge fixing arguments at the beginning were made for a non-tensor, for covariantization to make sense we need to assume at this point that the diff field is a tensor.  In Section \ref{CoordGaugeTrfoftheDiffField}, we are going to review problems of the transverse formalism, and look for alternative ways  to come up with a diffeomorphism invariant.}. Our gauge fixing arguments  were made in flat space.  Covariantization, however, changes the gauge structure of the theory. In particular, upon covariantization $D_{i 0}$ components of the diff tensor become dynamical and in order to obtain diff-Gauss law as a constraint  we must enforce $D_{i0}=0=\partial_0 D_{i0}$. The additional conditions for time derivatives are needed, since the action is second-order in $D_{i0}$ and in higher order theories time derivatives of the fields need to be treated as independent variables. Moreover, the invisible component $D_{00}$ showed up and the Lagrangian is third-order in it.  

None of these issues are seen in YM theory; covariantization does not affect the gauge structure; $A_0$ remains nondynamical. This is expected since in YM theory  gauge fixing refers to an internal gauge group, not to the coordinate transformations, so are unaffected by covariantization.  

Finally, let us note down the field equations. 
To obtain the field equations we need to apply variation with respect to $D^{\mu \nu}$. This amounts to computing 
\al{
\delta \mathcal{L} = X_{\mu \nu \lambda} \delta X^{\mu \nu \lambda}
} 
Then we partially integrate all the derivatives of the  variations $\delta D$s in $\delta X^{\mu \nu \lambda}$ and  rearrange the indices to express the result in the form $\delta D^{\mu \nu} ( \cdots )$. The expression $( \cdots )$ is the field equation $0=\delta S / \delta D^{\mu \nu}$.  With this method we get
\al{
0 = - \frac{\delta S}{\delta D^{\mu \nu}} & =  X_{\mu \nu \lambda}^{ \ \ \ \ \lambda} + \beta ( X_{\mu \lambda \nu}^{ \ \ \ \ \lambda}  + X_{\nu \lambda \mu}^{ \ \ \ \ \lambda}  ) + q X^{\sigma \lambda}_{ \ \ \ \nu (\mu \sigma \lambda)} \notag\\ & + \alpha \left[ D^{\sigma \lambda}_{\ \ \ \mu} X_{\sigma \lambda \nu} + 2 D_{\mu}^{\ \, \lambda \sigma} X_{\sigma \nu \lambda} - \partial_{\sigma} \Big(   D^{\sigma \lambda} X_{\mu \nu \lambda} + 2 X^{\sigma \lambda}_{\  \ \ \nu} D_{\mu \lambda} \Big)             \right] 
}

 We will analyze the covariantized theory in 2D Minkowski spacetime in Sections \ref{TransverseAfter} and \ref{BLRYTotalHamiltonian}.

\section{Interactions of  Diffeomorphism Field} \label{DiffInteractions}
The prescription to obtain the interactions of the diff field with the matter and gauge fields have been proposed in \cite{BLR00}. 
\subsection{General Prescription to Obtain Interactions of  Diff Field }

The self-interaction of the diff field has the structure 
\al{  \label{diffselfinter}
\mathcal{L}_{\text{int}}  = X^{\lambda \mu \rho} Z_{\lambda \mu \rho}
}
where $X^{\lambda \mu \rho}$ is the covariantized conjugate momentum of the diff field and $Z_{\lambda \mu \rho}$ is the covariantized (but not centrally extended) Lie derivative of the diff field. We can build the interaction Lagrangian of the diff field with other fields imitating the  structure of the self-interaction of the diff field. Namely, for a field $\phi$ with covariant conjugate momentum $\pi_{\phi}$, we introduce
\al{\label{interactionsgeneralprescr}
\mathcal{L}_{\text{int}} = \pi_{\phi} \, ( \delta_D \phi)
}
where $\delta_D \phi$ is the "diff variation" of $\phi$. Explicitly the construction is carried out in the following steps \cite{BLR00} :  
\begin{itemize}
\item Contract the conjugate momentum of the field with the Lie derivative of the matter field with respect to a vector field $\xi^k$. 
\item Replace the vector field $\xi^{k}$ with $D^{\ k}_0$ so that the Lie derivative is extended to a diff variation.
\item Fully covariantize the action.\index{covariantization}
\end{itemize}
Applying this prescription to  the diff
field itself we recover \eqref{diffselfinter}
\al{
\mathcal{L}_{0th} & = X^{ij0} ( \xi^k \partial_k D_{ij} + D_{kj} \partial_i \xi^k + D_{ik} \partial_j \xi^k ) \rightarrow 
\notag\\ 
\mathcal{L}_{1st} & = X^{\lambda \mu 0} (  D_0^{ \ \alpha} \nabla_{\alpha} D_{\lambda \mu} +  D_{\alpha \mu} \nabla_{\lambda} D_0^{ \ \alpha} +  D_{\lambda \alpha} \nabla_{\mu} D_0^{ \ \alpha}) \rightarrow 
\notag\\ 
\mathcal{L}_{\text{int}} & =  X^{\lambda \mu \rho} (D_{\rho}^{ \ \sigma} \nabla_{\sigma} D_{\lambda \mu} + D_{\sigma \mu} \nabla_{\lambda} D_{\rho}^{\ \sigma} + D_{\lambda \sigma} \nabla_{\mu} D_{\rho}^{ \ \sigma} ) 
}

 $\mathcal{L}_{\text{int}}$ corresponds to the part of the total diff Lagrangian  involving the constant $\alpha$ i.e. terms with coefficients $\alpha, q \alpha, \beta \alpha$ and $\alpha^2$. 
Notice that even though $D_{\mu \nu}$ is taken as a tensor in the transverse method, one does not simply use tensoriality to find its interactions  with other matter. 

Below we are going to show application of this method for interactions of the diff field with the point particle, spin one half and spin one fields.

\subsection{Diff Field Interacting with  Point Particle }
Components of the velocity four-vector of a point particle are given by 
\al{
u^0 = \frac{dt}{d \tau} = \gamma \hspace{0.2in} \text{and} \hspace{0.2in} u^i = \frac{dx^i}{d \tau} = \gamma \frac{dx^i}{d t} = \gamma v^i
}
where $\gamma = (\sqrt{1-v^2})^{-1}$. Then the four-momentum components become 
\al{
p^{\mu} = m u^{\mu} = ( \gamma , \gamma v^i)
}

From the transformation, $x^{\mu} \rightarrow x^{\mu} + \xi^{\mu}$, of the coordinate four-vector we can read the diffeomorphism shift of the coordinate three-vector, $\delta x^i = \xi^i$. 

Then, the prescription \eqref{interactionsgeneralprescr} yields the interaction term in following steps
\al{
S_0 & =\int dt \, p_i \delta x^i = \frac{1}{m} \int m \gamma\,d \tau  \, p_i \xi^i   = \frac{1}{m} \int d \tau  \,  p_i \xi^i p^0 \notag\\ S_1 & = \frac{1}{m} \int d \tau  \,  p_i D_0^{ \ i}  p^0 \notag\\ S_{\text{int}} & = \frac{1}{m} \int d \tau  \,  p_{\mu} D_{\nu}^{ \ \mu}  p^{\nu} = \frac{1}{m} \int d \tau \, p^{\mu} D_{\mu \nu} p^{\nu} 
}

We are going to introduce a factor of $\lambda/2$ for convenience where $\lambda$ is a dimensionful constant that makes this interaction  dimensionless. So we take 
\al{ \label{Spp-D}
S_{\text{int}} = \frac{\lambda}{2m} \int d \tau \, p^{\mu} D_{\mu \nu} p^{\nu} & = \frac{\lambda m}{2} \int d \tau \, \dot{z}^{\mu} \dot{z}^{\nu} D_{\mu \nu} 
}

The variation of $S_{\text{int}}$ with respect to the  coordinate four-vector is 
\al{\label{intvarz}
\delta_z S_{\text{int}} =- \lambda  m \int d \tau \, \delta z^{\rho} \ [ \ddot{z}^{\nu} D_{\rho \nu} + (1/2) ( \partial_{\mu} D_{\rho \nu } + \partial_{\mu} D_{\nu \rho} - \partial_{\rho} D_{\mu \nu} ) \dot{z}^{\mu} \dot{z}^{\nu} ] 
}

 The action of a free point particle is \cite{MTW}
\al{
S_{pp} & = \frac{m}{2} \int d \tau \,  \frac{d z^{\mu}}{d \tau} \frac{d z_{\mu}}{d \tau} = \frac{m}{2} \int d \tau \, g_{\mu \nu}  \frac{d z^{\mu}}{d \tau} \frac{d z^{\nu}}{d \tau} 
}
Comparing this with \eqref{Spp-D} we see that the diff field  adds to the metric i.e. 
\al{ 
S_{pp} + S_{\text{int}} = \int d \tau \, \dot{z}^{\mu} \dot{z}^{\nu} (g_{\mu \nu}  + \lambda D_{\mu \nu} ) 
}
 The variation of $S_{pp}$ with respect to the coordinate four-vector is
\al{ \label{ppvarz}
\delta_z S_{pp} = - m \int d \tau \, \delta z^{\rho} \, [\,  g_{\mu \rho} \ddot{z}^{\mu} + (1/2) ( \partial_{\nu} g_{\mu \rho} + \partial_{\nu} g_{\rho \mu} - \partial_{\rho} g_{\mu \nu} ) \dot{z}^{\mu} \dot{z}^{\nu} ] 
}
If there are no other fields present the two variations \eqref{intvarz} and \eqref{ppvarz} combine to give the perturbed geodesic equation 
\al{
\lambda m [ \, \ddot{z}^{\nu} D_{\rho \nu} & +  (1/2)  ( \partial_{\mu} D_{\rho \nu } + \partial_{\mu} D_{\nu \rho} - \partial_{\rho} D_{\mu \nu} ) \dot{z}^{\mu} \dot{z}^{\nu} \, ] \notag\\ 
& + m [\,  g_{\nu \rho} \ddot{z}^{\nu} + (1/2) ( \partial_{\mu} g_{\nu \rho} + \partial_{\mu} g_{\rho \nu} - \partial_{\rho} g_{\mu \nu} ) \dot{z}^{\mu} \dot{z}^{\nu} \, ] = 0
}
Contracting with $g^{\rho \alpha}$ and rearranging we get 
\al{ \label{diffmat}
(\ddot{z}^{\alpha}  + \lambda D^{\alpha}_{\ \nu} \ddot{z}^{\nu}) + (1/2) [\, \lambda    g^{\rho \alpha} (\partial_{\mu} D_{\rho \nu } + \partial_{\mu} D_{\nu \rho} - \partial_{\rho} D_{\mu \nu} ) + \Gamma^{\alpha}_{\mu \nu}  ]  \dot{z}^{\mu}  \dot{z}^{\nu} = 0
}
Hence, point particle interaction suggests that the diff field may be a linear perturbation to the spacetime metric. 

\subsection{Diff Field Interacting with  Dirac Fermion }
\label{DiffFermionInt}
Next, let us apply the transverse interaction prescription \eqref{interactionsgeneralprescr} to the coupling of the diff field to a Dirac fermion. From the covariant Dirac Lagrangian follows the covariant generalized momentum for the Dirac fermion  
\al{
\pi^{\beta} = \sqrt{g}  \, \overline{\psi}  \, \gamma^{\beta}
}
The Lie derivative of $\psi$ with respect to a vector field $\xi$ is given by \cite{Kosmann}
\al{
\xi^{\alpha} \nabla_{\alpha} \psi - \frac{1}{4} \nabla_{[\alpha} \xi_{\beta]} \gamma^{\alpha} \gamma^{\beta} \psi
}
where $\nabla$ is the spin covariant derivative. Extending this to a diff variation we get
\al{
D^{\ \alpha}_{ \lambda} \nabla_{\alpha} \psi - \frac{1}{4} \nabla_{[\alpha} D_{\beta] \lambda} \gamma^{\alpha} \gamma^{\beta} \psi
}
With this we write the interaction Lagrangian as
\al{ \label{difffermionbranson}
\mathcal{L}_{\psi D} = \sqrt{g} \ \overline{\psi} \ \gamma^{\lambda} \left( D^{\ \alpha}_{ \lambda} \nabla_{\alpha} \psi  - \frac{1}{4} \nabla_{[\alpha} D_{\beta] \lambda} \gamma^{\alpha} \gamma^{\beta} \psi \right) 
}
In flat spacetime $\nabla_{[\alpha} D_{\beta]\lambda} \gamma^{\alpha} \gamma^{\beta} $ vanishes and this reduces to
\al{ 
\mathcal{L}^{\text{flat}}_{\psi D} & = \overline{\psi} \gamma^{\lambda}  D_{ \lambda \alpha} \eta^{\alpha \beta}  \partial_{\beta} \psi \notag\\ 
& = (\overline{\psi} \gamma^{\alpha} \partial_{\beta} \psi ) D_{\alpha}^{\ \beta} \label{diffspinorint1}
}

It is stated in \cite{BLR00} that the factor $\bar{\psi} \gamma^{\alpha} \partial_{\beta} \psi $ multiplying the diff field corresponds to $\partial_{\tau} s / \partial_{\theta} s \leftrightarrow \partial_+ f / \partial_- f = h_{++}$ in the bosonization of fermions. We find, however, that this expression corresponds to the energy momentum (EM) tensor, not to the metric that couples to the fermion.  Let us show this. 

The Dirac Lagrangian in a curved spacetime reads 
\al{
\mathcal{L} = \sqrt{g} \, \overline{\psi} \gamma^{\mu} \nabla_{\mu} \psi
}
where $\nabla_{\mu}$ includes a spin connection term. Since we will consider the flat space EM tensor at the end we can drop the spin connection term here. We then get 
\al{
\mathcal{L} = \sqrt{g} \, \overline{\psi} \gamma^{\mu} \partial_{\mu} \psi
}
Note that in this expression $\gamma^{\mu} = e^{\mu}_a \gamma^a$  where $e^{\mu}_a$ are the inverse vielbein and $\gamma_a$ are the true Dirac matrices satisfying the Clifford algebra. It is straightforward to vary the action to obtain the EM tensor  as
\al{ \label{TmunucurvedDirac}
T^{\mu \nu} = \frac{1}{\sqrt{g}} \frac{\delta S}{\delta g_{\mu \nu}} = \overline{\psi} \gamma^{\mu} \partial^{\nu} \psi + \frac{1}{2} g^{\mu \nu} \overline{\psi} \gamma^{\mu} \partial_{\mu} \psi
}
where we used 
\al{
\delta \sqrt{g} = \frac{1}{2} \sqrt{g} g^{\mu \nu} \delta g_{\mu \nu} 
}
The second term in \eqref{TmunucurvedDirac} vanishes on-shell by the field equation $\gamma^{\mu} \partial_{\mu} \psi = 0$ so that we are left with 
\al{
T^{\mu \nu} =  \overline{\psi} \gamma^{\mu} \partial^{\nu} \psi 
}
This shows that the diff-spinor interaction \eqref{diffspinorint1} in  flat spacetime reads 
\al{
\mathcal{L}_{\psi D}^{\text{flat}} = T^{\mu \nu} D_{\mu \nu}
}
Just as matter fields couple to the gauge field via the current, $J^{\mu} A_{\mu}$, they couple to the spacetime metric via their EM tensor $T^{\mu \nu} g_{\mu \nu}$. 
Hence, we again see that the diff field emerges as a perturbation to the spacetime metric if the transverse interaction prescription  \eqref{interactionsgeneralprescr} is used to obtain its interaction. 

This is interesting since the diff field is originally, i.e. before considering it as a field on its own right coupling to other fields, is nothing but the EM tensor. In fact, the EM tensor of a 2D conformal field theory transforms as in \eqref{coadpassive}, \eqref{coadinfinitesimal}, i.e.  as a Virasoro coadjoint element. However, once we treat it as a distinct dynamical field, and  mimick its self-interaction to obtain its interactions with other fields, its role changes from the EM tensor to a perturbation to the spacetime metric.

\subsection{Gauge-Invariant Diff Field  }

Recall the coadjoint transformations \eqref{semidirectiso} of the semidirect product algebra
\begin{subequations}
\label{semidirectisorevisited}
\al{
\label{Virasoroisoequation}
\delta D (\theta) & = 2 \xi' D + D' \xi + \frac{c \mu}{2 \pi} \xi''' + \frac{ h \mu}{2 \pi} \xi' - \text{Tr}\  (A \Lambda') \\ \delta A(\theta) & = A' \xi + \xi' A - [\Lambda , A ] + k \mu \Lambda' 
\label{KMisoequation}
}
\end{subequations}

Let us first analyze the second transformation \eqref{KMisoequation}. $A$ is treated as a YM field in 1D, or space component of a YM field, in 2D, in temporal gauge $A_0=0$. In both cases terms involving $\xi$ represent the Lie derivative of the gauge field. The terms involving $\Lambda$ represent an  infinitesimal gauge transformation of the YM field in 1D, or  time-independent gauge transformation of the space component of the YM field in 2D. 

Now, let us analyze the first transformation \eqref{Virasoroisoequation}. In analogy with above the terms involving $\xi$ should represent the Lie derivative of the diff field $D$. However, the constructed Lagrangian is a GCT invariant only if the diff field is a tensor. In fact, if it is not a tensor then its covariant derivative  is not defined. This is an important flaw of the transverse method which we will try to cure later in the thesis. Again, in analogy with the YM case, the term involving $\Lambda$ should represent a gauge transformation of the diff field. However, in building the interaction of the diff field with the gauge field, diff field was treated as a gauge-invariant object \cite{BLR00}. We will handle this here since its solution is simpler.
 
Recall that in \eqref{DTrAA} we obtained a gauge-invariant object $\tilde{D}$ from the Virasoro coadjoint element $D$. Since, in 2D, we identified $D$ with the $D_{11}$ component of the diff field in \ref{Virasoro Gauge-Fixing}, and $A$ with the $A_1$ component of the YM field we can rewrite \eqref{DTrAA} as
\al{
\tilde{D}_{11} = D_{11} + \frac{e}{2} \text{Tr} (A_1 A_1) 
}
We can easily extend this to higher dimensions as
\al{ 
\tilde{D}_{\mu \nu}  \equiv  D_{\mu \nu} + \frac{e}{2} \, \text{Tr} \, (A_{\mu} A_{\nu} ) \label{gaugeinvdiff}
}
Note that in this extension,  temporal gauge $A_0=0$ of  YM theory corresponds to the full temporal gauge $D_{\mu 0}=0$ of the diff field theory. So the gauge-invariant extension  \eqref{gaugeinvdiff} is compatible with the gauge fixing arguments of the transverse formalism. 

Recall that the point particle and spinor interactions of the diff field suggested the diff field may be a perturbation to the spacetime metric. Motivated with this, we are going to use the gauge-invariant extension \eqref{gaugeinvdiff} in a setting where the metric and gauge field meet, namely the Born-Infeld action. 
A mass term included in the YM action or the Born-Infeld action breaks the gauge invariance, yet through \eqref{gaugeinvdiff}, inclusion of the diff field into the picture may yield a mass term for the gauge field without breaking  gauge invariance. We are going to investigate this possibility by incorporating the diff field via the nonabelian extension of the Born-Infeld action \cite{tseytlinnonabelianborninfeld}. 

The nonabelian Born-Infeld action reads 
\al{ \label{NABornInfeld}
\mathcal{L}  =\text{STr} \sqrt{ \det (g_{\mu \nu} + c F_{\mu \nu})} - \sqrt{g} 
}
In this expression, the determinant is taken only in the indices $\mu ,\nu$  whereas  the symmetric trace is defined  as  
\al{
\text{STr} \, (A_1 \cdots A_n) = \frac{1}{n!} \text{Tr} \,( A_1 \cdots A_n + \text{all permutations} ) 
}
where the trace Tr is taken in the fundamental representation over the Lie algebra indices of the gauge field. The curvature tensor is given by 
\al{
F_{\mu \nu} = \partial_{\mu} A_{\nu} - \partial_{\nu} A_{\mu} + e [A_{\mu} , A_{\nu} ] 
}

Introducing $\kappa \tilde{D}_{\mu \nu}$ into the determinant \eqref{NABornInfeld} for some constant $\kappa$,  the Lagrangian becomes 
\al{ \label{NABornInfelddiff0}
\mathcal{L} + \sqrt{g}  & = \text{STr} \, \sqrt{ \det (g_{\mu \nu} + c F_{\mu \nu} + \kappa \tilde{D}_{\mu \nu})} 
\notag\\ & = \sqrt{g} \,  \text{STr} \, \sqrt{ \det ( \delta^{\mu}_{\ \nu} + c F^{\mu}_{\ \nu} + \kappa \tilde{D}^{\mu}_{\ \nu} )} 
\notag\\ & = \sqrt{g} \, \text{STr} \, \exp\left(\frac{1}{2} \, \text{tr} \, \ln(\delta^{\mu}_{\ \nu} + c F^{\mu}_{\ \nu} + \kappa \tilde{D}^{\mu}_{\ \nu} ) \right) 
}
Using the Taylor series 
\al{
\ln (1+x) = x - \frac{x^2}{2} + \frac{x^3}{3} + \cdots 
}
for matrices, the logarithm can be expanded to second-order in $c$ and $\kappa$  
\al{\label{NABornInfelddiff}
\mathcal{L} + \sqrt{g}  & = \sqrt{g} \, \text{STr} \, \exp \left( \frac{1}{2} \, \text{tr}  \left[  c F^{\mu}_{\ \nu} + \kappa \tilde{D}^{\mu}_{\ \nu} - \frac{1}{2} (c F^{\mu}_{\ \nu} + \kappa \tilde{D}^{\mu}_{\ \nu} )^2 \right] \right) 
\notag\\ & = \sqrt{g} \, \text{STr} \, \exp \left( \frac{1}{2}    \kappa \tilde{D}^{\mu}_{\ \mu} - \frac{1}{4} \text{tr} \, (c F^{\mu}_{\ \nu} + \kappa \tilde{D}^{\mu}_{\ \nu} )^2 \right)  
\notag\\ & = \sqrt{g} \, \text{STr} \, \exp \left( \frac{1}{2}    \kappa \tilde{D}^{\mu}_{\ \mu} - \frac{1}{4} c^2 F^{\mu}_{\ \nu} F^{\mu}_{\ \nu}  - \frac{1}{4} \kappa^2  \tilde{D}^{\mu}_{\ \nu} \tilde{D}^{\mu}_{\ \nu} \right) 
}
where we used  antisymmetry of $F$ and symmetry of $\tilde{D}$. 
Expanding the exponential to second-order in $c$ and $\kappa$ we get 
\al{
\mathcal{L}  = \sqrt{g} \, \text{STr} \left(  \frac{1}{2}    \kappa \tilde{D}^{\mu}_{\ \mu} - \frac{1}{4} c^2 F^{\mu \nu} F_{\mu \nu}  - \frac{1}{4} \kappa^2  \Big(\tilde{D}^{\mu \nu} \tilde{D}_{\mu \nu} - \frac{1}{2} (\tilde{D}^{\mu}_{\ \mu} )^2\Big) \right) 
}

Now, if we treat the diff field part of $\tilde{D}_{\mu \nu}$ as a linear perturbation to the metric, to second-order in $A$ and to first-order in $D$,  we get 
\al{
\frac{\mathcal{L}}{\sqrt{g}} & = \frac{\kappa}{2}\,   D^{\mu}_{\ \mu} + \frac{e \kappa}{4} \,  \text{Tr}\,  (A^{\mu} A_{\mu}) - \frac{e\kappa^2}{4} \, D^{\mu \nu} \, \text{Tr} \, (A_{\mu} A_{\nu}) 
\notag\\ 
& + \frac{e \kappa^2}{8} \, D^{\mu}_{\ \mu} \,  \text{Tr} \, (A^{\nu}  A_{\nu} ) - \frac{c^2}{4}\,  \text{STr} \, (F^{(0)}_{\mu \nu} F^{(0)\mu \nu} ) 
}
where $F^{(0)}_{\mu \nu} \equiv \partial_{\mu} A_{\nu} - \partial_{\nu} A_{\mu}$. We can rewrite this result more symmetrically as 
\al{ \label{diffborninfeldA2D1}
\frac{\mathcal{L}}{\sqrt{g}}  = \frac{\kappa}{2} \,  D^{\mu}_{\ \mu} & +\frac{e \kappa}{4}  \left( g^{\mu \nu} - \kappa D^{\mu \nu} + \frac{e \kappa}{2} g^{\mu \nu} D^{\lambda}_{\ \lambda} \right) \text{Tr} \, (A_{\mu} A_{\nu} ) \notag\\ & - \frac{c^2}{4} \, \text{STr} \, (F^{(0)}_{\mu \nu} F^{(0)\mu \nu} ) 
}
The first term on the right is a correction to the metric determinant as can be shown using the expansion
\al{
\det (B + \epsilon C) = \det (B) + \epsilon \, \text{Tr} \, (B^{-1} C)  + O(\epsilon^2)
}
For $B_{\mu \nu} = g_{\mu \nu}$, $\epsilon= \kappa$ and $C_{\mu \nu} = D_{\mu \nu}$ this yields
\al{
\sqrt{g+\kappa D} & = \sqrt{g + \kappa \, g \, g^{\mu \nu} D_{\mu \nu}  + O ( \kappa^2) } \notag\\ & = \sqrt{g} \left( 1 + \frac{\kappa}{2} \, D^{\mu}_{\ \mu} + O( \kappa^2) \right) 
}
verifying the claim. 
Finally, the second term on the right of \eqref{diffborninfeldA2D1} is a mass term for the gauge field.

 \subsection{Diff Field Interacting with  Spin One Field  } \label{DiffSpinoneinteraction}

It is stated in \cite{BLR00} that  spin-one coupling of the diff field is a good test for the prescription \eqref{interactionsgeneralprescr}  since it should have non-trivial contributions to the isotropy equations\index{isotropy equation} for both $A$ and $D$ fields.  Upon including this interaction term, the isotropy equations for both KM and Virasoro algebras should be reproduced from the field theory upon reduction to the 1D spatial hypersurface of a 2D flat spacetime as field equations of $A_0$ and $D_{\mu 0}$ followed by the background conditions $A_0=0=D_{\mu 0}$.

Let us apply \eqref{interactionsgeneralprescr} to spin-one coupling.
The covariant YM action (with $e=1$) yields the following generalized momentum
\al{ \label{gencovmomYM}
\pi^{\rho \lambda} = \sqrt{g} F^{\rho \lambda} = \sqrt{g} \ ( \partial^{\rho} A^{\lambda} - \partial^{\lambda} A^{\rho} + [A^{\rho} , A^{\lambda} ] )
}
Spatial Lie derivative of the gauge field lifts to the covariant diff variation as
\al{
\delta A_i & = \xi^k \partial_k A_i + A_k \partial_i \xi^k \notag\\  & \rightarrow   D^k_{\ 0} \partial_k A_i + A_k \partial_i D^k_{\ 0} \notag\\ 
\ & \rightarrow D^{\alpha}_{\ \rho} \nabla_{\alpha} A_{\lambda} + A_{\alpha} \nabla_{\lambda} D^{\alpha}_{\ \rho} \notag\\ & \ \ \ \ =D^{\alpha}_{\ \rho} \partial_{\alpha} A_{\lambda} + A_{\alpha} \partial_{\lambda} D^{\alpha}_{\ \rho} \equiv (\delta_D A)^{(0)}_{\rho \lambda} \label{standardBLRYAmuinteraction}
}
Then the interaction term becomes 
\al{ \label{unmodifiedADinteraction}
\mathcal{L}_{\text{int}} = \text{Tr} \, \Big( \pi^{\rho \lambda}  (\delta_D A)_{[\rho \lambda]} \Big)
}
where we antisymmetrized the diff variation $(\delta_D A)_{\rho \lambda}$ since $\pi^{\rho \lambda}$ is antisymmetric. 

In \cite{BLR00} the suggested covariant diff-variation is different from \eqref{standardBLRYAmuinteraction}: 
\al{\label{BLRYdiffvariation}
(\tilde{\delta}_D A)_{\rho \lambda}=    D^{\alpha}_{\ \rho} \partial_{\alpha} A_{\lambda} + A_{\alpha} \partial_{\lambda} D^{\alpha}_{\ \rho} - \partial_{\rho} ( D^{\alpha}_{\ \lambda} A_{\alpha} )
}
We shall call the interaction obtained from the modified diff variation the BLRY interaction
\al{ \label{BLRYinteractionterm}
\tilde{\mathcal{L}}_{\text{int}} = \text{Tr} \, \Big( \pi^{\rho \lambda}  (\delta_D A)_{[\rho \lambda]} \Big)
}

We will investigate the BLRY and unmodified interactions below. First, let us review the arguments in \cite{BLR00}. 
It is stated that  when $D_{10}=0$, $A_0$ has no conjugate momentum even in the presence of the diff field.  This is true as we show below. 
The argument continues as follows. This interaction term is  still not gauge invariant. One may preserve gauge invariance by introducing a group-valued  scalar field $V$, transforming under right multiplication by a group element $h$ as $V \rightarrow Vh$. Then the interaction Lagrangian is modified to
\al{
\tilde{\mathcal{L}}_{\text{int}} =\sqrt{g} \ \text{Tr}  \, \Big(  F^{\rho \lambda} ( D^{\alpha}_{\ \rho} \partial_{\alpha} \tilde{A}_{\lambda} + \tilde{A}_{\alpha} \partial_{\lambda} D^{\alpha}_{\ \rho} - \partial_{\rho} ( D^{\alpha}_{ \ \lambda} \tilde{A}_{\alpha} ) ) \Big)
}
with
\al{ \label{VshiftofAmu}
\tilde{A}_{\mu} = A_{\mu} - V^{-1} \partial_{\mu} V
}
The following is suggested as the Lagrangian of the $V$ field 
\al{
\mathcal{L}_V = m^2_A \int d^n x \ (V^{-1} \partial_{\mu} V - A_{\mu} ) ( V^{-1} \partial_{\mu} V - A_{\nu} ) (g^{\mu \nu} + D^{\mu \nu} ) 
}
It is claimed that the variation of the interaction Lagrangian with respect to $D_{10}$, followed by  the conditions $V=1$ and $A_0 = 0=D_{0 \nu}$  needed for reduction to the  coadjoint orbits, yields the expected contribution  $\text{Tr} \, (A E')$ (which is the field lift of the term $\text{Tr} \, (A \Lambda')$ in the isotropy equation) where $A = A_1$ and $E=F^{01} = \pi^1$.  It is also claimed that the  variation with respect to $A_0$ yields the expected contribution $(XA)'$ (which is the field lift of the term $\xi A' + \xi' A$ in the isotropy equation).

Let us investigate the arguments provided above in 2D Minkowski spacetime with the metric convention $(+t, -x)$. 
In 2D the only nonvanishing momentum component is $\pi^{01} = - \pi^{10}$ as the momentum $\pi^{00}$ for $A_0$ vanishes by the definition \eqref{gencovmomYM}.  So the  BLRY interaction term \eqref{BLRYinteractionterm} reduces to 
\al{
\tilde{\mathcal{L}}_{\text{int}} = 2 \,\text{Tr} \, \Big( \pi^{01} (\tilde{\delta}_D A)_{[01]} \Big)
}

We introduce
\al{
A \equiv A_ 1 \ \ , \ \  A_0 \equiv M  \ \ , \ \  D_{11} \equiv D  \ \ , \ \  D_{01} \equiv N \ \ , \ \  D_{00} \equiv \varphi
}
Note that $A$ and $M$ are Lie algebra valued i.e. $A = A^a T^a$, $M= M^a T^a$ in a basis $\{T^a\}$ for the Lie algebra with the convention $\text{Tr}\, (T^a T^b) = \delta^{ab}$.
We will also denote time derivative with a dot and space derivative with a prime. In this notation, the momentum reads 
\al{
\pi^{01} = F^{01} & = \partial^0 A^1 - \partial^1 A^0 + [A^0 , A^1] \notag\\ & =  - \dot{A} + M' - [M, A]
}
Similarly the BLRY diff variation reads 
\al{
(\tilde{\delta}_D A)_{[01]} & = \varphi \dot{A} - 2 (NA)' - 2 \partial_t ( NM ) + DM' \notag\\ & + 2 \varphi' M + \varphi M' + 2 \dot{D} A + D \dot{A} 
}
and the unmodified diff variation reads 
\al{
(\delta_D A)_{[01]} = \varphi \dot{A} -  (NA)' -  \partial_t ( NM ) + DM' +  \varphi' M + \dot{D} A 
}
So it is true for both the BLRY and the unmodified diff variations that $A_0=M$ do not receive momentum contributions when $D_{10}=N$ (and its time derivative) vanishes. 

Denoting $E \equiv \pi_{01} = - \pi^{01}$ the variation of the BLRY interaction  with respect to $N=D_{01}$ yields 
\al{
\frac{\delta \tilde{S}_{\text{int}}}{\delta N } = -2 \, \text{Tr} \, ( A E' + M \dot{E}) = -2 \, \text{Tr} \, (A^{\mu} \partial_{\mu} E)
}
So in the temporal gauge $A_0=0$ this expression yields the desired contribution  up to a factor of $-2$. The same contribution follows from the unmodified interaction term \eqref{standardBLRYAmuinteraction} up to a factor of $-1$ i.e. 
\al{ \label{unmodifiedNvariation}
\frac{\delta S_{\text{int}}}{\delta N } = - \text{Tr} \, ( A E' + M \dot{E}) = - \, \text{Tr} \, (A^{\mu} \partial_{\mu} E)
}

The variation with respect to $M=A_0$ evaluated at $D_{0\mu}=0=A_0$ yields 
\al{  \label{modifiedMvariation}
\frac{\delta \tilde{S}_{\text{int}}}{\delta M }\bigg|_{D_{0\mu}=0=A_0} =-2 (\dot{D} A)' + [A, \dot{A}] D 
} 
The diff momentum evaluated at the background values $D_{0\mu}=0$ becomes $X=\dot{D}$ for both BLRY and full diff theories. Also for the temporal gauge $A_0=0$, YM momentum reduces to $E=\dot{A}$. So \eqref{modifiedMvariation} becomes
\al{
\frac{\delta S_{\text{int}}}{\delta M }\bigg|_{D_{0\mu}=0=A_0} =-2 (XA)' + [A,E] D
}
So at this step, our calculation yields an inconsistent result with the claim in \cite{BLR00} due to the presence of the last term. 

 The unmodified interaction  yields the contribution
\al{ \label{unmodifiedMvariation}
\frac{\delta S_{\text{int}}}{\delta M }\bigg|_{D_{0\mu}=0=A_0} =- (XA)' + (DE)'
} 
so it also contains an undesired term $-(DE)'$. 

Finally, the shift \eqref{VshiftofAmu} yields a gauge invariant interaction only if the diff field is treated as a gauge invariant object, as  the shift $V^{-1} \partial_{\mu} V$ is nullifying the gauge transformation of $A_{\mu}$. Since, in our interpretation the diff field has a nontrivial gauge transformation 
\al{
\delta_{\text{gauge}} = - \text{Tr} \, (AE')
}
the proposed interaction Lagrangian \eqref{BLRYinteractionterm} is not gauge invariant in our conventions. However, we have shown in the previous section that one can introduce the gauge-invariant combination \eqref{gaugeinvdiff}
\al{ \label{tildeDAA}
\tilde{D}_{\mu \nu} \equiv  D_{\mu \nu} + (e/2) \, \text{Tr}\, ( A_{\mu} A_{\nu})
}
built from the diff and gauge fields. So, we need to  simultaneously shift the gauge field $A \rightarrow \tilde{A}$ as in \eqref{VshiftofAmu}  and the diff field $D \rightarrow \tilde{D}$ as in \eqref{tildeDAA}. 

This, however, does affect the variations with respect to $M$ performed above. So the problem at hand is more complicated than it looks. Let us state the results for the unmodified interaction \eqref{unmodifiedADinteraction} upon the shift \eqref{tildeDAA}. The result \eqref{unmodifiedNvariation} of the variation with respect to $D_{01}=N$ does not change, which is good. The variation with respect to $A_0=M$, \eqref{unmodifiedMvariation} changes to 
\al{
-(\tilde{X}A)' + (\tilde{D}E)' - (1/2) A \, \text{Tr}\, (AE')
}
where  ($e$ is set to $1$) 
\al{
\tilde{D} & = D + (1/2)\, \text{Tr} \, (AA)
\\
\tilde{X} & = \dot{D} + \text{Tr}\, (A \dot{A}) = X + \text{Tr} \, (AE) \label{XEmomformula}
}
where the second equation holds in the temporal gauge $D_{\mu 0}= 0 = A_0$. So the case is unsolved. This suggests that the interaction prescription \eqref{interactionsgeneralprescr} may not applicable to the gauge field. 
Otherwise, 
the interaction prescription \eqref{interactionsgeneralprescr} and the gauge-invariant extension \eqref{gaugeinvdiff} should  be inspected carefully, especially the last steps, namely the covariantization. We will not continue to this analysis.

\subsection{Cocyle Motivated Spinor Interaction }
 
In this section we are going to introduce an alternative diff-spinor interaction. For this purpose we will use the calculation in Section \ref{Covariantcocyle} as a guide. 
Consider the following  third-order covariant differential expression
 \al{
I_{\mu \nu} = (\nabla_{\mu} \nabla_{\nu} + k S_{\mu \nu} ) \nabla_{\rho} \xi^{\rho}
 } 
 where $\xi^{\rho}$ is a contravariant vector field, $S_{\mu \nu}$ is an arbitrary rank-two tensor, $k$ is a constant and $\nabla_{\mu}$ is the Levi-Civita connection.
Expansion of the covariant derivatives in terms of the ordinary derivatives and connection coefficients yields 
 \al{ \label{JacIND}
 I_{\mu \nu} & = \partial_{\mu} \partial_{\nu} \partial_{\rho} \xi^{\rho} + ( \Gamma^{\rho}_{\rho \lambda} \partial_{\mu} \partial_{\nu} \xi^{\lambda} - \Gamma^{\lambda}_{\mu \nu} \partial_{\lambda} \partial_{\rho} \xi^{\rho} ) \notag\\ & + (\partial_{\nu} \Gamma^{\rho}_{\rho \lambda} \partial_{\mu} \xi^{\lambda} + \partial_{\mu} \Gamma^{\rho}_{\rho \lambda} \partial_{\nu} \xi^{\lambda} - \Gamma^{\lambda}_{\mu \nu} \Gamma^{\rho}_{\rho \sigma} \partial_{\lambda} \xi^{\sigma} + k S_{\mu \nu} \partial_{\rho} \xi^{\rho} ) 
 \notag\\ & + ( \partial_{\mu} \partial_{\nu} \Gamma^{\rho}_{\rho \lambda} \xi^{\lambda}- \Gamma^{\lambda}_{\mu \nu} \partial_{\lambda} \Gamma^{\rho}_{\rho \sigma} \xi^{\sigma} + k S_{\mu \nu} \Gamma^{\rho}_{\rho \lambda} \xi^{\lambda} )  
 }
 where the terms are ordered as $\xi''', \xi'', \xi', \xi$. 
In 1D, this reduces to
 \al{ \label{JacI1D}
 I_{\text{1D}} = \xi''' + ( 2 \Sigma + k S) \xi' + (\Sigma' + k S \Gamma) \xi
 }
 where we defined 
 \al{
 \Sigma \equiv \Gamma' - \Gamma^2/2
 }
 
We showed in Section \ref{connectionVircoadjointelement} that $\Sigma$ Lie-transforms as a Virasoro coadjoint element with central charge one. We also showed that  adding a rank-two tensor  to a  coadjoint element yields another coadjoint element with the same central charge, so the combination $k D \equiv 2\Sigma + k S$ is a coadjoint element with central charge two. Then 
 \al{
 D = (2/k) \Sigma + S
 }
 is a Virasoro coadjoint element with central charge $2/k$. 

  The last term in \eqref{JacI1D} is problematic since it does not transform in any good way. In the  calculation of the cocyle in Section \ref{Covariantcocyle}, however,  this term was not present since $I_{\mu \nu}$ is just part of the cocyle \eqref{covcocyle}; we have to contract it with $\eta^{\mu} dx^{\nu}$ and antisymmetrize the expression in $\eta$ and $\xi$. Namely, 
 \al{ \label{jacobiatorantisymmetrization}
 J = \int  dx^{\mu} \, \eta^{\nu} (\nabla_{\mu} \nabla_{\nu} + k S_{\mu \nu} ) \nabla_{\rho} \xi^{\rho} - \int  dx^{\mu} \, \xi^{\nu} (\nabla_{\mu} \nabla_{\nu} + k S_{\mu \nu} ) \nabla_{\rho} \eta^{\rho}
 }
 Indeed the 1D reduction of the cocyle $J$ has no $\xi \eta$ term 
 \al{
 J_{\text{1D}} & = \int dx \, \Big [ ( \eta \xi''' - \xi \eta''') 
 + ( 2 \Lambda + k S ) ( \eta \xi' - \xi \eta')  \Big] \notag\\  & = \int dx \, \Big [ ( \eta \xi''' - \xi \eta''') 
 + k D ( \eta \xi' - \xi \eta') \Big]
 }

Now, if we can form a GCT-vector out of the spinor field then we can use the analysis above to obtain a diff-spinor interaction. For this purpose, consider the spacetime Dirac matrix 
\al{
\gamma^{\mu}(x) = e^{\mu}_a (x) \gamma^a
} 
so it does transform as a vector under GCT.  Then the field   
\al{
\varphi^{\mu}(x) = \gamma^{\mu}(x) \psi(x) 
} 
also transforms as a vector under GCT. 
 Under a Lorentz transformation  $e^{\mu}_a \rightarrow e^{\mu}_b \Lambda^b_{\ a}$ of the frame fields we have the following 
\al{
\psi \rightarrow \rho \psi \ \ \ , \ \ \ \gamma^{\mu} \rightarrow \rho \gamma^{\mu}  \rho^{-1}
}
where $\rho \equiv \rho(\Lambda)$ is the Lorentz transformation  matrix for the spinor representation (see e.g. \cite{weldon}).  Then $\varphi^{\mu}$ transforms as 
\al{
\varphi^{\mu} \equiv \gamma^{\mu} \psi \rightarrow \rho \gamma^{\mu} \rho^{-1} \rho \psi = \rho \gamma^{\mu} \psi = \rho \varphi^{\mu} 
}
So, $\varphi^{\mu}$ is also a Lorentz spinor.
Similarly we have 
\al{
\overline{\varphi}^{\mu} \equiv \overline{\psi} \gamma^{\mu} \rightarrow \overline{\psi} \rho^{-1} \rho \gamma^{\mu} \rho^{-1} = \overline{\psi} \gamma^{\mu} \rho^{-1} = \overline{\varphi}^{\mu} \rho^{-1}
}
This also shows that objects of the form $\overline{\varphi}^{\mu}  \varphi^{\nu}$ are local Lorentz invariant. 
 
Now we are ready to introduce the interaction,
\al{ \label{alternative3rdorderint}
\mathcal{L} = \overline{\psi} \gamma^{\mu} \gamma^{\nu} ( \nabla_{\mu} \nabla_{\nu} + S_{\mu \nu} ) \nabla_{\rho} \varphi^{\rho}
}
Consider just the differential expression
\al{
I_{\mu \nu} =  (\nabla_{\mu} \nabla_{\nu} + S_{\mu \nu} ) \nabla_{\rho} \varphi^{\rho}
}
It is expanded as 
 \al{ 
 I_{\mu \nu} & = \partial_{\mu} \partial_{\nu} \partial_{\rho} \varphi^{\rho} + ( \Gamma^{\rho}_{\rho \lambda} \partial_{\mu} \partial_{\nu} \varphi^{\lambda} - \Gamma^{\lambda}_{\mu \nu} \partial_{\lambda} \partial_{\rho}  \varphi^{\rho} ) \notag\\ & + (\partial_{\nu} \Gamma^{\rho}_{\rho \lambda} \partial_{\mu}  \varphi^{\lambda} + \partial_{\mu} \Gamma^{\rho}_{\rho \lambda} \partial_{\nu}  \varphi^{\lambda} - \Gamma^{\lambda}_{\mu \nu} \Gamma^{\rho}_{\rho \sigma} \partial_{\lambda}  \varphi^{\sigma} + k S_{\mu \nu} \partial_{\rho}  \varphi^{\rho} ) 
 \notag\\ & + ( \partial_{\mu} \partial_{\nu} \Gamma^{\rho}_{\rho \lambda} \varphi^{\lambda}- \Gamma^{\lambda}_{\mu \nu} \partial_{\lambda} \Gamma^{\rho}_{\rho \sigma}  \varphi^{\sigma} + k S_{\mu \nu} \Gamma^{\rho}_{\rho \lambda}  \varphi^{\lambda} )  
 }
So, in 1D it reduces to 
 \al{ \label{1Daltspinordiffint}
 I_{\text{1D}} & = \varphi''' + ( 2 \Sigma + k S) \varphi' + (\Sigma' + k S \Gamma) \varphi 
 }

 We can write the interaction Lagrangian \eqref{alternative3rdorderint} in a more symmetric way as follows. The hermitian conjugate of $\varphi^{\mu}$ becomes 
\al{
(\varphi^{\mu})^{\dagger} = \psi^{\dagger} (\gamma^{\mu})^{\dagger} = \psi^{\dagger} \gamma^0 \gamma^{\mu} \gamma^0 = \overline{\psi} \gamma^{\mu} \gamma^0 
}
which implies 
\al{
\overline{\varphi}^{\mu} : = (\varphi^{\mu})^{\dagger} \gamma^0 =  \overline{\psi} \gamma^{\mu} 
}
Thus \eqref{alternative3rdorderint} can be simplifed to (with the added symmetrization condition)
\al{
\mathcal{L} = \overline{\varphi}^{(\mu} \gamma^{\nu)} ( \nabla_{\mu} \nabla_{\nu} + S_{\mu \nu} ) \nabla_{\lambda} \varphi^{\lambda} 
}
The field $\varphi^{\mu}$ satisfies the following 
\al{
\overline{\varphi}^{\mu} \varphi^{\nu} = g^{\mu \nu} \overline{\psi} \psi + 2 e^{\mu}_a e^{\nu}_b \overline{\psi} \sigma^{ab} \psi
}
which implies 
\al{
\overline{\varphi}^{\mu} g_{\mu \nu} \varphi^{\nu} = n \overline{\psi} \psi
}
where $n$ is the spacetime dimension. Since the diff field $D_{\mu \nu}$ is symmetric we also have 
\al{
\overline{\varphi}^{\mu} D_{\mu \nu} \varphi^{\nu} =  \overline{\psi} \mathcal{D} \psi
}
where $\mathcal{D} = g_{\mu \nu} D^{\mu \nu}$. 

What would be the analog of \eqref{jacobiatorantisymmetrization} in order to get rid of the unwanted last term in \eqref{1Daltspinordiffint}? One may employ a modification used in 2D covariant Dirac theory for showing the vanishing of the spin connection (see e.g. \cite{nakahara}, Section 7.10.3), namely, hermitianizing the action. For this purpose, we may simply define the interaction Lagrangian as $\tilde{\mathcal{L}} = \mathcal{L} + \mathcal{L}^{*}$.

\section{Supersymmetric Extension }
\label{DiffSUSY}
This section is based on the analysis  in \cite{delius90} and \cite{GR01}. Also \cite{friedan1986conformal} is used as a main reference.  The supersymmetric expressions for the pieces of the superdiffeomorphism action are the same though the resulting action is modified in accordance with the modified momentum ansatz in Section \ref{Diffaction}. 

\subsection{Superdiffeomorphism Field}
The superVirasoro algebra contains (in the NS sector) the bosonic Virasoro generators $L_m, m \in \mathbb{Z}$, the fermionic generators $G_{\mu}, \mu \in \mathbb{Z}+1/2$, and the center generator $I$. The commutation relations are given by
\al{
[L_m , L_n] & = (m-n) L_{m+n}  + \frac{1}{8} \hat{c} (m^3 -m) \delta_{m+n} I\notag\\ [L_m, G_{\mu} ] & = \left(\frac{1}{2} m  - \mu\right) G_{m+\mu} \notag\\ \{ G_{\mu} , G_{\nu} \} & = -  4 L_{\mu + \nu} - \frac{1}{2} \hat{c} \left( \mu^2 - \frac{1}{4} \right) \delta_{\mu + \nu} I
}
A generic algebra (or adjoint) element takes the form 
\al{
\hat{A} =  A^m L_m +  A^{\mu} G_{\mu} + \frac{1}{8} a \hat{c} I
}
We can introduce a superfield $A(z, \theta)$ corresponding to the non-central part of $\hat{A}$ as 
\al{
A(z, \theta) =  A^m z^{m+1} + 2 \theta  A^{\mu} z^{\mu + 1/2} \ \  \ \leftrightarrow \ \ \  A^m L_m +  A^{\mu} G_{\mu} 
}
where $\theta$ is the Grassmann variable. 
Then the complete generic element $\hat{A}$ corresponds to the doublet 
\al{ \label{AhatAz}
\hat{A} \ \leftrightarrow \ ( A(z, \theta) , a) 
}
We also introduce fields corresponding to the generators $L_n$ and $G_{\mu}$ as  
\al{
L(z) =  z^{-n-2} L_n \hspace{0.2in} \text{and} \hspace{0.2in} G(z) =  z^{- \mu - 3/2} G_{\mu} 
}
We can combine these into a superfield 
\al{
T(z, \theta) = \frac{1}{2} G(z) + \theta L(z) 
}
With all these, the correspondence \eqref{AhatAz} solidifies by the equality
\al{
\oint \frac{dz}{2 \pi i} \, d \theta \, A(z, \theta) \, T(z, \theta) + \frac{1}{8} a \hat{c} I = \hat{A}
}
For convenience let us abbreviate $(z, \theta) \equiv Z$ so that we have $A(z, \theta) = A(Z)$ and also 
\al{
dZ \equiv \frac{dz}{2 \pi i} d \theta 
}
The commutator of two generic elements $(A(Z),a)$ and $(B(Z),b)$ becomes 
\al{ \label{superVirgen}
[ (A,a) , (B,b)] = \left( (\partial A) B - A \partial B - \frac{1}{2} (\mathcal{D} A) (\mathcal{D} B) \ , \ \oint dZ \ (\partial^2 \mathcal{D} A) B \right) 
}
where $\partial \equiv \partial_z$ and we introduced the superderivative 
\al{\label{superder}
\mathcal{D} \equiv \partial_{\theta} +  \theta \partial
}
The equality \eqref{superVirgen} can be directly verified by performing the $z$ and $\theta$ integrations. The definition \eqref{superder} of the superderivative implies $\mathcal{D}^2 =  \partial$. 

Elements of superDiff($S^1$) are diffeomorphisms $(z, \theta) \mapsto (\tilde{z} (z, \theta) , \tilde{\theta} (z, \theta))$ such that the supersymmetric line element scales by a superfield 
\al{ \label{DNR904.7}
dz + \theta d \theta \mapsto d \tilde{z} + \tilde{\theta} d \tilde{\theta} = \phi(z, \theta) ( dz + \theta d \theta ) 
}
This is the generalization of the case of Diff($S^1$) where $dz \mapsto d \tilde{z} = f(z) dz$. Necessary and sufficient condition for \eqref{DNR904.7} is 
\al{
\mathcal{D} \tilde{z} - \tilde{\theta} \mathcal{D} \tilde{\theta} = 0 
}
This condition implies that the superderivative $\mathcal{D}$ transforms as
\al{ \label{DNR904.8b}
\tilde{\mathcal{D}} = ( \mathcal{D} \tilde{\theta})^{-1} \mathcal{D}
}
and $dz d \theta$ transforms as
\al{\label{dzdtheta}
d \tilde{z}  d \tilde{\theta} = (\mathcal{D} \tilde{\theta})  dz   d \theta
}

One can define an $h$-differential $A$ as 
\al{
A = A(z, \theta) (dz   d \theta)^{2h} 
}
This induces the transformation property of the superfield $A$ as  
\al{ \label{hdifffinite}
\tilde{A} (\tilde{z} , \tilde{\theta} ) = A(z, \theta) ( \mathcal{D} \tilde{\theta})^{-2 h} 
}
An infinitesimal transformation generated by an adjoint vector $F$ becomes
\al{\label{hdiffinfini}
 \delta_F A = - F \partial A - \frac{1}{2} \mathcal{D} F \mathcal{D} A -h (\partial F) A 
}
Comparison  with \eqref{superVirgen} suggests that adjoint elements transform as $-1$ differentials. 

The adjoint representation of the centrally extended group is given by 
\al{ \label{DNR904.13}
(A (Z) , a) \overset{g}{\mapsto} \left( A_g (Z) , a + 2 \oint dZ  S(Z, \tilde{Z}) A(Z) \right)
}
where 
\al{ \label{DNR904.13b}
A_g (\tilde{Z} ) = A(Z) ( \mathcal{D} \tilde{\theta} (Z))^2 
}
and the superSchwarzian $S(Z, \tilde{Z})$ is given by 
\al{
S (Z, \tilde{Z} ) = \frac{\mathcal{D}^4 \tilde{\theta}}{\mathcal{D} \tilde{\theta}} - 2 \frac{\mathcal{D}^3 \tilde{\theta}\  \mathcal{D}^2 \tilde{\theta}}{ ( \mathcal{D} \tilde{\theta})^2} 
}
An infinitesimal transformation $g$ generated by an adjoint vector $F$ becomes
\al{
\delta_F A & = - F \mathcal{D}^2 A - \frac{1}{2} \mathcal{D} F \mathcal{D} A + (\mathcal{D}^2 F) A \notag\\ & =  - F \partial A - \frac{1}{2} \mathcal{D} F \mathcal{D} A + (\partial F) A 
}
Thus the finite adjoint action agrees with the infinitesimal adjoint transformation given in \eqref{superVirgen}. It is also straightforward to show the representation property  
\al{
(A_g , a_g) \overset{h}{\mapsto} ((A_g)_h, (a_g)_h) = (A_{gh} , a_{gh}) 
}

Coadjoint vectors $(B^* , b^*)$ can be introduced via the following pairing 
\al{
\left< (B^* , b^* ) | (A,a) \right> \equiv  b^* a + \oint dZ \ B^*(Z) A(Z) 
} 
and the coadjoint transformation is defined by requiring the group invariance of the pairing 
\al{
\left< (B^*_g , b^*_g ) | (A_g,a_g) \right> \overset{!}{=} \left< (B^* , b^* ) | (A,a) \right>
}
Using \eqref{DNR904.13}, \eqref{DNR904.13b} and \eqref{dzdtheta} we get  
\al{
(B^*(Z), b^*) \overset{g}{\mapsto} (B^*_g(Z) , b^*_g) 
}
where 
\al{
B^*_g (\tilde{Z})  = \Big( B^*(Z) -2 b^* S(Z , \tilde{Z} ) \Big) (\mathcal{D} \tilde{\theta})^{-3} \ \ \ \text{and} \ \ \  b^*_g  = b^*
}
Comparing this with \eqref{hdifffinite} we see that for $b^* = 0$,  $B^*$ transforms as a $3/2$-differential.
 
Performing an infinitesimal transformation generated by $F$ we get  
\al{ \label{infinisupercoad}
\delta_F B^* & = -F \mathcal{D}^2 B^* - \frac{1}{2} \mathcal{D} F \mathcal{D} B^* - \frac{3}{2} \mathcal{D}^2 F B^* - b^* \mathcal{D}^5 F \notag\\ & = - F \partial B^* - \frac{1}{2} \mathcal{D} F \mathcal{D} B^* - \frac{3}{2} \partial F B^* - b^* \mathcal{D} \partial^2 F 
}
Comparing this with the infinitesimal transformation \eqref{hdiffinfini} we again see that for $b^* = 0$,  $B^*$ transforms as a $3/2$-differential. If we introduce the decomposition $F(z,\theta) = \xi(z) + \theta \epsilon(z)$ with bosonic $\xi$, Grassmann $\epsilon$, and $B^*(z,\theta) = u(z) + \theta D(z)$ with bosonic $D$ and Grassmann $u$, the infinitesimal transformation reduces to\footnote{The second equation here, is different from the equations in  \cite{GR01} and \cite{delius90}. In each source the second equation was
\al{
\delta_F D & = - \xi \partial D - 2 \partial \xi D - \frac{1}{2} \epsilon \partial u- \frac{3}{2} \partial \epsilon u + b^* \partial^3 \xi 
} } 
\al{
\delta_F u & = - \xi \partial u - \frac{3}{2} \partial \xi u- \frac{1}{2} \epsilon D  + b^* \partial^2 \epsilon \notag\\ \delta_F D & = - \xi \partial D - 2 \partial \xi D - \frac{3}{2} \partial (\epsilon u) + b^* \partial^3 \xi 
}
Note that  reduction of the second equation to the Virasoro algebra (i.e. turning off supersymmetry) is exactly the coadjoint transformation of the diff field as desired. Also we introduced $D$ as the component multiplying $\theta$ so as to keep it bosonic. 
\subsection{2D Majorana Action} 
To extend the formalism developed so far to a field theory we lift the Grassmann variable $\theta$ to a  2D Majorana spinor $\theta^{\alpha}$ and the supersymmetric derivative to the operator
\al{
\mathcal{D} \rightarrow \mathcal{D}_{\mu} = \frac{\partial}{\partial \theta^{\mu}} - \frac{i}{2} \gamma^N_{\nu \mu} \theta^{\nu} \frac{\partial}{\partial z^N} \equiv \partial_{\mu} - \frac{i}{2} \gamma^N_{\nu \mu} \theta^{\nu} \partial_N
}
One should be careful in calculations since Greek letters refer to the spinor indices not to the spacetime indices; capital Latin letters refer to the spacetime indices. $\mathcal{D}_{\mu}$ satisfies
\al{
\{ \mathcal{D}_{\mu} , \mathcal{D}_{\nu} \} = - i \gamma^M_{\mu \nu} \partial_M
}
where we used $\{ \partial_{\mu} , \theta^{\nu} \} = \delta^{\nu}_{\mu} =\{ \theta^{\nu} , \partial_{\mu} \}$. 

The Dirac-Gamma matrices  satisfy 
\al{
\{ \gamma^A , \gamma^B \} = 2 \eta^{AB} 
}
We also introduce 
\al{
[ \gamma^A , \gamma^B ] = 2 \Sigma^{AB} 
}
Combining these equations we get 
\al{
\gamma^A \gamma^B = \eta^{AB} + \Sigma^{AB} 
}
or, explicitly in  spinor components
\al{
\gamma^A_{\alpha \beta} \gamma^{B \beta \lambda} = \delta^{\lambda}_{\alpha} \eta^{AB} + \Sigma_{\alpha}^{AB \lambda}
}

An adjoint element $F$ is promoted to a vector superfield $F^M$ 
\al{
F \rightarrow F^M = \xi^M + \theta^{\alpha} \gamma^M_{\alpha \beta} \epsilon^{\beta}
}
and a coadjoint element $B^*$ is promoted to a spin 3/2 superfield $B_{\mu M}$
\al{
B^* \rightarrow B_{\mu M} = \Upsilon_{\mu M} +D_{MN}  \theta^{\alpha} \gamma^N_{\alpha \beta} + \theta^{\alpha} \theta^{\beta} \gamma^N_{\mu [ \alpha} A_{\beta ] MN}
}

To deduce the extension of the coadjoint action \eqref{infinisupercoad} to higher dimensional case, it is instructive to recall how this was done in the Virasoro case. Ignoring the central extension the quadratic differential transformation 
\al{
\delta_{\xi} D =  \xi \partial D + 2 \partial \xi D
}
was lifted to the Lie derivative of a rank-two tensor as
\al{
\delta_{\xi} D_{MN} =  \xi^A \partial_A D_{MN} + \partial_M \xi^A D_{A N} + \partial_N \xi^A D_{M A}
}
Hence, upon lifting the term with coefficient two (the term specifying $D$ as a quadratic differential) was split into two terms, one for each spacetime index of a rank two tensor. Applying the same rule for the tensoral part of \eqref{infinisupercoad} we deduce
\al{
\delta^{\text{tens}}_F B_{\mu M} = F^N \partial_N B_{\mu M} + (\partial_M F^N) B_{\mu N} + \frac{1}{2} (\partial_N F^N) B_{\mu M} 
}
The first two terms are what we expect from a field with one lower spacetime index (covariant vector) and the last one is a correction telling us that $B_{\mu M}$ carries a density of weight 1/2. 
Now keeping track of the indices we can fully extend  \eqref{infinisupercoad} 
\al{
\delta_F B_{\mu M} & = F^N \partial_N B_{\mu M} + (\partial_M F^N) B_{\mu N} + \frac{1}{2} (\partial_N F^N) B_{\mu M} \notag\\ & \ \ + i (\mathcal{D}_{\lambda} F^N) \gamma^{\lambda \nu}_N (\mathcal{D}_{\nu} B_{\mu M} ) + q \ \mathcal{D}_{\mu} \partial_N \partial_M F^N
}

At this point let us recall the general structure of the diff field Lagrangian.  The diff field Lagrangian is of the form 
\al{
2L_{\text{diff}} =  X^{LMR} ( \nabla_R D_{LM} +  Y_{LMR} ) 
}
where $Y_{LMR}$ is the covariantized and centrally extended  Lie derivative of the diff field 'with respect to itself' in analogy with equation \eqref{shortcutmomentumprescription}. Just as in the non-super case, it turns out that $ \nabla_R D_{LM} + Y_{LMR}= X_{LMR}$ thus we get 
\al{
2L_{\text{diff}} =  X^{LMR} X_{LMR}
}

Hence, if we get the super extension of $X_{LMR}$ we get the super extension of the diff Lagrangian. For this purpose we first need to find the analog of $D_0^{ \ N}$, the Lagrange multiplier of the diff-Gauss law. Consider the following superfield
\al{
F_A^{ \ N} = E^{1/2} \gamma^{\alpha \beta}_A \mathcal{D}_{\alpha} B_{\beta}^N
}
where $E$ is the superdeterminant and 
 the superderivative $\mathcal{D}_{\mu}$ is covariantized 
\al{
\mathcal{D}_{\mu} = \partial_{\mu} - \frac{i}{2} \gamma^N_{\nu \mu} \theta^{\nu} \nabla_N
}
A straightforward calculation shows
\al{
E^{-1/2} F^{\ N}_0  = D_0^N & -i \theta^{\beta} \nabla_0 \Upsilon^N_{\beta} - i ( \Sigma_0^{ \ N})^{\beta}_{\nu} \theta^{\nu} \nabla_M \Upsilon^N_{\beta} \notag\\ & \ - i \gamma^M_{\lambda \nu} \theta^{\nu} \theta^{\lambda} \nabla_0 D_M^{ \ N} - i (\Sigma_0^{ \ L})^{\beta}_{\nu} \gamma^M_{\lambda \beta} \theta^{\nu} \theta^{\lambda} \nabla_L D_M^{ \ N}
}
so that the leading order term matches up with the Virasoro case.

With $F^N_A$ found, the rest is straightfoward. We introduce the analog of the covariantized Lie derivative of diff field as the superfield $Y_{A \mu M}$  \al{ 
Y_{A \mu M} & = F^N_A \nabla_N B_{\mu M} + \nabla_M F^N_A B_{\mu N} + \frac{1}{2} (\nabla_N F^N_A) B_{\mu M} \notag\\ & + i ( \mathcal{D}_{\lambda} F^N_A) \gamma^{\lambda \nu}_N (\mathcal{D}_{\nu}  B_{\mu M}) + q \ \mathcal{D}_{\mu} \nabla_N \nabla_M F^N_A
} 
Then the superfield corresponding to the diff momentum becomes
\al{
X_{A \mu M} = \nabla_A B_{\mu M} +  Y_{A \mu M} 
}
Finally the superdiff action reads  
\al{
S = -\frac{1}{2} \int d^2 x \,  d \theta^{\mu } \, d \theta^{\nu} \, X_{A \mu M} \, X_{B \nu N} \, \eta^{AB}\,  \eta^{MN} 
}

Just as in the ordinary case, the previously found superdiff action \cite{GR01}
 corresponds to the symplectic part of this action :
\al{
S_{\text{superBLRY}} = - \frac{1}{2} \int d^2 x \, d \theta^{\mu} \, d \theta{\nu} \, X_{A \mu M} \, \nabla_{A} B_{\mu M} \, \eta^{AB}\,  \eta^{MN} 
}

It is straightforward to modify the remaining fermion actions discussed in \cite{GR01}, namely, when the Grassman variable $\theta$ is lifted to a 3D Majorana spinor or a 2D, 4D chiral spinor. We shall not continue this analysis.

\chapter{DIFFEOMORPHISM FIELD IN 2D MINKOWSKI SPACETIME}
\label{Diff Field in 2D Minkowski}

\section{Introduction }
Before analyzing the transverse action introduced in the previous section in 2D Minkowski spacetime  we will first go over a slightly different approach held in order to obtain the diff field theory, that predates the transverse method, and can be said to be the origin of it. 

In search of a covariant theory for the diff field, in \cite{LR95}  authors applied the methods of \cite{rajeev88}. This  is different from the transverse action in that the diff-Gauss law constraint is implicitly introduced, i.e. it is introduced as a constraint generating an equivalence relation on the phase space, not explicitly as a term in the Lagrangian or the Hamiltonian. We shall call this field theory the DX theory. Applying the methods of \cite{rajeev88} they also obtained a reduction of the field theory to a finite-dimensional theory.

Here, we will first review the paper \cite{rajeev88}. Then we will  go over DX theory \cite{LR95} and  correct a mistake (Equation (4.10) of \cite{LR95}) changing some of the results in the subsequent analysis. However, application of Dirac's constraint Hamiltonian formalism shows that even with this correction the theory is invalid. This will be shown in the next chapter. 

We will, then, go back to the analysis of the transverse action in 2D. There are two cases to consider, before or after the covariantization step. The former leads to the DXN theory, to be studied  in the next chapter, and the latter leads to a complicated higher order theory, even in the 2D Minkowski spacetime. We will introduce a new gauge-fixing condition for the diff field components called the chiral gauge in which every  aspect of the covariantized theory simplifies. Moreover, in this gauge the theory is not constrained.

\section{Finite Reduction of YM Theory on a Cylinder }
\label{Rajeev88Summary}
Rajeev \cite{rajeev88} discusses solving YM theory on a cylinder in the Hamiltonian formalism, without using a gauge-fixing condition, reducing the field theory to a finite-dimensional theory and quantizing it. Let us note the main steps.

The curvature tensor and the  YM equation  are given in covariant form by
\al{\label{Raj881}
F_{\mu \nu} & = \partial_{\mu} A_{\nu} - \partial_{\nu} A_{\mu} + [ A_{\mu} , A_{\nu} ] \\ 0 & = \partial^{\mu} F_{\mu \nu} + [ A^{\mu} , F_{\mu \nu} ]  \label{Raj882}
}
Introducing $E_1 \equiv F_{01}$, the YM equation yields
\al{\label{Raj885}
0 & = \partial_x E_1 + [A_1,E_1] \\
0 & = \partial_t E_1 + [A_0 , E_1 ]  \label{Raj884}
}

$A_0$ is nondynamical (its momentum $F^{00}$ identically vanishes) and can be  eliminated from the equations by introducing a variable $T(t,x)$ valued in $G$ as the solution to the equation
$\partial_t T = T A_0$ 
with the boundary condition 
$T (t=0 ) = 1$. 
This results in new variables $A , E$ given by
\al{ 
A  = T A_1 T^{-1} + T \partial_x T^{-1} \ \ \ , \ \ \ E  = TE_1 T^{-1} 
}
Inverting these for $A_1, E_1$
and inserting  back into \eqref{Raj881}, \eqref{Raj885} and \eqref{Raj884}  one gets 
\begin{subequations} 
\label{Raj88eqnsinnew}
\al{ 
E  &= \partial_t A \\ \label{EequalAdot}
 0  &= \partial_t E \\
    0  &= \partial_x E + [A , E ]  \label{RajeevGausslaw}
} 
\end{subequations}
Thus $A_0$ is eliminated and  the new theory involves only $A$ and $E$.
  
The equations \eqref{Raj88eqnsinnew} follow from a canonical formalism where the configuration space is the space of functions
$Q = \{ A : S^1 \rightarrow \mathcal{G} \}$
and  $E$ is canonically conjugate to $A$. The unconstrained phase space is $\Gamma = Q \oplus Q$ consisting of all $(A, E)$ and the Hamiltonian yielding the first two equations is
\al{ \label{Raj8814}
H = \frac{1}{2} \int  \left< E , E \right> \ dx
}
where $\left< \ , \ \right> $ is a bilinear form on $\Gamma$ which can be taken as the Killing form (practically the trace) on $\mathcal{G}$. The third equation will be a first-class constraint. It can be introduced by defining the true phase space, $\tilde{\Gamma}$, to be the space of pairs $(A , E)$ satisfying this constraint. Then one can show that $\tilde{\Gamma}$ can be defined as the quotient of $\Gamma$ with respect to the  equivalence relation (the gauge equivalence)
\al{ \label{Raj8815}
& (A , E ) \sim ( g A g^{-1} +g  \partial_x g^{-1}  , g E g^{-1} ) \equiv ( A^g , E^g) 
}
The Hamiltonian is gauge-invariant, 
$H(A,E) = H(A^g . E^g)$,
so that $H$ is well-defined on $\tilde{\Gamma}$. 

The constraint \eqref{RajeevGausslaw} can be formally solved introducing the Wilson line  $S$ satisfying 
\al{
\partial_x S + A S = 0
} with the boundary condition 
$S(x=0) =1$. 
Then  
\al{ \label{YMwilsonloopsolved}
E (x) = S(x) E(0) S^{-1} (x)
} solves the constraint. Now we define a map $\phi : \Gamma \rightarrow G \times \mathcal{G}$
\al{ \label{Raj8821}
\phi(A , E )= (S(2 \pi) , E (0)) \equiv (q,p)
}
Here, $S(2 \pi)$ is the Wilson loop. Geometrically it is the parallel transport operator around a loop. $\phi$ satisfies 
(is said to be equivariant under gauge transformations) 
\al{ \label{Raj8822}
\phi (A^g, E^g) = ( g(0) \ q \ g(0)^{-1} , g(0) \ p  \ g (0)^{-1} ) 
} 
Therefore, $\phi$ on $\Gamma$ induces $
\phi : \tilde{\Gamma} \rightarrow G \times \mathcal{G} / G_{\text{adj}}$ 
i.e. the real phase space defined by the gauge constraint yields $G \times \mathcal{G}$ up to the adjoint transformation given above. This map is a bijection.

 The equations in the new  variables $(q,p)$ read 
\al{ \label{Raj88particleeqns}
\dot{p}  = 0 \ \ \ , \ \ \ q^{-1} \dot{q}  = -2 \pi p  
}
These equations follow from a canonical formalism with the  canonical one-form,
\al{ \label{Raj88oneform}
\theta = - \text{Tr} \ ( p q^{-1} dq )
}
and the Hamiltonian,
\al{ \label{Raj8832}
H = \pi \text{Tr} \  ( pp)
}
on the space  $G \times \mathcal{G}$.
Note that \eqref{Raj8832} is just the projection of the Hamiltonian \eqref{Raj8814} of the field theory. \eqref{Raj88particleeqns}, \eqref{Raj88oneform} and \eqref{Raj8832} define  a theory with a finite number of degrees. For quantization and the spectrum of this theory see \cite{rajeev88}.

\section{DX Gravity Theory and Its Finite Reduction }
\label{ReviewDXGravity}
\subsection{DX Field Theory}
Let us recall the basics about the gravity theory proposed in  \cite{LR95}. The  Virasoro coadjoint element $D$ is lifted to a dynamical field, i.e., $D(\theta) \rightarrow D(\theta, \tau)$. In analogy with \cite{rajeev88} the authors proposed the action
\al{
S = \frac{1}{\lambda} \int d \sigma \ d \tau \  X \partial_{\tau} D - \frac{1}{2 \lambda} \int d \sigma \ d \tau \ X^2
}
where $\lambda$ is a parameter introduced for dimensional reasons and $X$ is the momentum conjugate to $D$. Recall equations \eqref{Raj88eqnsinnew}. In analogy  $X$ is taken as
\al{ \label{XDdot}
X = \dot{D}
} 
so that the action becomes 
\al{
S = \frac{1}{2 \lambda} \int d \sigma \ d \tau \  X^2 = \frac{1}{2 \lambda} \int d \sigma \ d \tau \  \dot{D}^2
}
where dot denotes $\tau$-derivative. Hamilton's equations yield  \eqref{XDdot} and  
\al{
\dot{X}=0
}

The Virasoro analog of the Gauss law \eqref{RajeevGausslaw} is obtained as follows. Notice that \eqref{RajeevGausslaw} is the pure KM isotropy equation $\Lambda' - [\Lambda, A]=0$ with $\Lambda \rightarrow E$. In other words the isotropy generator $\Lambda$ lifts to the momentum $E$. This is not so by chance, but follows by $\delta A = \{A, Q\}$ with $Q = \int \Lambda G$ where $G$ is the Gauss law operator. This ensures that the KM coadjoint transformation is a gauge transformation generated by the Gauss law. We already used this correspondence in Section \ref{YMfromKM} to obtain YM theory from KM algebra. 

 In analogy, we lift the isotropy equation \eqref{coadinfinitesimal} on the coadjoint orbits of Virasoro algebra to the constraint equation, the diff-Gauss law 
\al{ \label{Gauss}
2 X' D + D' X + c X''' = 0
}
where prime denotes $\theta$-derivative and $c$ is a constant.  This equation ensures transversality of  dynamics to the orbits, and turns  the Virasoro coadjoint transformation into a local symmetry of the theory. This equation will be enforced on the phase space as in the previous section.  
\subsection{Finite Reduction}
\label{DXFiniteReduction}
The Wilson line $v(\theta)$ for the diff field is defined through
\al{ \label{Dv}
D(\theta)\equiv  c \, S ( \theta , v) = c \left( \frac{v'''}{v'} - \frac{3}{2} \left[ \frac{v''}{v'} \right]^2 \right)
}
so that the Wilson loop becomes\footnote{We investigate whether this is indeed the Wilson loop in Section \ref{DiffWilsonLR95}.} 
\al{
Q \equiv v (2 \pi )
}
Plugging  \eqref{Dv} in the diff-Gauss law \eqref{Gauss} we get
\al{
X (\theta) = \frac{X(0)}{\partial v (\theta)} \equiv \frac{P}{v'}
}
Taking a time derivative of \eqref{Dv} and using $X = \dot{D}$ this yields
\al{ \label{Xvtau}
\frac{P}{c(v')^3} = \partial_v^3 ( \dot{v})
}
where 
\al{ \label{partialcubev}
\partial_v = \frac{1}{v'} \partial_{\theta}
}
Equation \eqref{Xvtau} is solved by
\al{
\partial_{\tau} v (\theta) = \frac{P}{c} & \left(  \frac{1}{2} \int_0^{\theta} d \phi \ \frac{v^2 (\phi)}{ (\partial v (\phi) )^2} \right. \notag\\  & \ \ \ \left. - v(\theta) \int_0^{\theta} d \phi \ \frac{v (\phi)}{ (\partial v (\phi) )^2}  + \frac{1}{2} v^2 (\theta) \int_0^{\theta} d \phi \ \frac{1}{ (\partial v (\phi) )^2}\right)  \label{partauv}
}
where $\tau$-dependence of $v$ is suppressed. For simplification we define
\al{ \label{fth}
f( \theta )   \equiv   \int_0^{\theta} d \phi  \frac{v^2 (\phi)}{ 2(\partial v (\phi) )^2} 
 \ , \   g( \theta )   \equiv  \int_0^{\theta} d \phi  \frac{v (\phi)}{ (\partial v (\phi) )^2}   \ , \ 
h(\theta)   \equiv  \int_0^{\theta} d \phi  \frac{1}{2 (\partial v (\phi) )^2}   
}
We also have
\al{
\dot{v} (2 \pi ) = \partial_{\tau} \Big( v (2 \pi , \tau ) \Big)  = \partial_{\tau} \Big( Q (\tau ) \Big) = \dot{Q}
}
Thus evaluating \eqref{partauv} at $\theta = 2 \pi$ we get 
\al{ \label{partauv2}
\dot{Q} = \frac{P}{c} \Big(  f(2 \pi) - Q g(2 \pi) +  Q^2 h(2 \pi)  \Big)
} 

For orbits\footnote{We examine such orbits  in Section \ref{firsttypeorbits}.} in which $D$ is diffeomorphic to a constant $D_0$ one can solve \eqref{Dv} :
\al{ \label{spectificsolutiondiffWilson}
v ( \theta) = \exp \left( \pm i \alpha \theta \right)
}
where 
$\alpha \equiv \sqrt{2D_0/c}$. 
If  $f$, $g$, $h$ are evaluated for the solution $v = \exp(i \alpha \theta)$ one gets 
\al{ \label{partauv3}
\dot{Q} = \frac{2 \pi^3 P}{c(\ln Q)^3} \left( 3  - 4Q  +Q^2  + 2 \ln Q\right)
}

The momentum equation $\dot{X}=0$ reads 
\al{ \label{Pdot0}
\dot{P}= \frac{P \dot{v}'}{v'}
}
Taking a $\tau$-derivative of \eqref{partauv2} and evaluating at $\theta= 2 \pi$ one can compute $\dot{v}'(2 \pi)$ in terms of $Q$ and $P$. Then  equation \eqref{Pdot0} yields
\al{ \label{Pdot2}
\dot{P}= \frac{ (4 \pi)^3 P^2 (Q-1)^2}{c \, Q (\ln Q)^3}   
}

This result differs from the one given in \cite{LR95} :
\al{
\dot{P}  = \frac{2 \pi^3 P^2}{Q (\ln Q)^4} (3- 4 Q + Q^2 + 2 \ln Q ) ( \ln Q + 1 ) 
} 
Therefore we diverge in the analysis in the rest of the paper (the Hamiltonian, the symplectic structure and the quantization of the system). 
\subsection{Underlying Symplectic Theory}
We need to find a Hamiltonian $H$ and a symplectic structure $\omega$ which yield $\dot{Q}$ and $\dot{P}$ equations through the central equation 
\al{ \label{sympcent}
\omega ( \zeta_H , Y) = - dH (Y)
}
Here $\zeta_H$ is the Hamiltonian vector field associated with $H$, and $Y$ is an arbitrary vector field. Explicitly, if we denote the phase space coordinates as $z^1 = P$ and $z^2 = Q$ then we have $\zeta_H = \dot{z} = ( \dot{P} , \dot{Q})$. Denoting the single component  of $\omega$  with the same letter, equation \eqref{sympcent} yields 
\al{ \label{PomH}
\dot{P} = - \frac{1}{\omega} \frac{\partial H}{\partial Q}
\hspace{0.3in}
\text{and} 
\hspace{0.3in}
\dot{Q} =  \frac{1}{\omega} \frac{\partial H}{\partial P}
}
The Hamiltonian 
\al{ \label{HamDX}
H = \frac{\pi^3}{c} \frac{P^2}{ (3 + 2 \ln Q - 4 Q + Q^2)^2}
}
and the symplectic two-form
\al{ \label{omDX}
\omega = \frac{(\ln Q)^3}{(3 + 2 \ln Q - 4 Q + Q^2)^3}
}
yield the equations \eqref{partauv3} and \eqref{Pdot2} through \eqref{PomH}. Also the field to particle projection argument noted after the equation \eqref{Raj8832} in the YM case seems to roughly hold in the $DX$ theory: 
\all{
H(X,D) \sim \int X^2 \hspace{0.2in} \text{with}  \hspace{0.2in} X \sim \frac{P}{G(Q)}  \hspace{0.2in} \rightarrow  \hspace{0.2in} H(P,Q) \sim \left( \frac{P}{F(Q)} \right)^2
}

\subsection{Comparison of  Symplectic Theories}

Reconsider the Hamiltonian $H$ given in \eqref{HamDX} and the symplectic structure $\omega$ given in \eqref{omDX}.  
Those proposed in \cite{LR95} are given by 
\al{
\tilde{H} = \ln P + \ln Q + \ln ( \ln Q )
}
and 
\al{
\tilde{\omega} = \frac{( \ln Q)^3}{2 \pi^3 P^2 (3 + 2 \ln Q - 4 Q + Q^2) }
}
In this section we will compare these as functions. 

First the easy part, the momentum $P$ dependence. $H$ is well-defined for any $P$. In particular, for all $P$, we have $H \geq 0$ and $H = 0$ if and only if $P=0$. 

On the other hand, $\tilde{H}$ is not positive definite for all $P$. For $P>1$ we have $\tilde{H} > 0$, for $P=1$ we have $\tilde{H}=0$ and for $P<1$ we have $\tilde{H}<0$. Also as $P \rightarrow 0$ we have $\tilde{H} \rightarrow - \infty$. Furthermore $\tilde{\omega}$ is singular at $P=0$, whereas $\omega$ does not depend on $P$. 
 
 Now, the $Q$-dependence.   $H$ is well-defined for any $Q \geq 0$ except at $Q=1$ where $H \rightarrow + \infty$. In particular $H(0,P)=0$. Also $H \geq 0$ for all $Q$ and as $Q \rightarrow \infty$ we have  $H \rightarrow 0$. 
 
 $\tilde{H}$ is well-defined for only $Q >1$, i.e. it is not defined for $Q \leq 1$. It is $0$ at $Q_0=1.4215299358831166'$ (numerical solver) and $\tilde{H}(Q<Q_0) <0$ and $\tilde{H}(Q>Q_0)>0$. Also as $Q \rightarrow \infty$ it diverges. 
 
  $\omega$ is well-defined for all $Q \geq 0$ except at $Q=1$ and $Q=0$ However, it has a well defined limit at each of these points
\al{
\lim_{Q \rightarrow 0} \omega = \frac{1}{8} \hspace{0.3in} \text{and} \hspace{0.3in} \lim_{Q \rightarrow 1} \omega = 1
}
$\tilde{\omega}$ is well-defined everywhere except at the same points $Q=0$ and $Q=1$. However, it has a limit only at $Q=1$ :
\al{
\lim_{Q \rightarrow 0} \tilde{\omega} = + \infty \hspace{0.3in} \text{and} \hspace{0.3in} \lim_{Q \rightarrow 1} \tilde{\omega} = 0
}

Therefore, it seems that in both $Q$ and $P$ dependence $H$ and $\omega$ behave in a much nicer way than $\tilde{H}$ and $\tilde{\omega}$. 

\subsection{Quantization} 
For convenience let us rewrite the basic ingredients. We had the equations 
\al{
\dot{Q} = \frac{2 \pi^3}{c} \frac{ P Z(Q)}{(\ln Q)^3} 
\hspace{0.3in} \text{and} \hspace{0.3in} \dot{P} = \frac{4 \pi^3 }{c}\frac{ P^2 (Q-1)^2}{ Q (\ln Q)^3}
}
where 
\al{
Z(Q) \equiv 3- 4 Q + Q^2 + 2 \ln Q
}
These equations follow from
\al{
H = \frac{\pi^3}{c} \frac{P^2}{Z(Q)^2} \hspace{0.3in} \text{and} \hspace{0.3in} \omega = \frac{(\ln Q)^3}{Z(Q)^3}
}

The Poisson bracket (PB)  $\{  \ ,  \ \}$ is given by $\omega^{-1}$ and we apply Dirac's rule $\{  \ , \  \} \rightarrow -i [  \ ,  \ ]$ to quantize. For a 2D phase space with coordinates $ (z^1 , z^2) = (P,Q)$, the PB of two dynamical variables $F(P,Q)$ and $G(P,Q)$ becomes 
\al{
\{F  , G \} & = (\omega^{-1})^{ij} \frac{\partial F}{\partial z^i} \frac{\partial G}{\partial z^j} \notag\\ & = (\omega^{-1})^{12} \left( \frac{\partial F}{\partial P} \frac{\partial G}{\partial Q} -  \frac{\partial F}{\partial Q} \frac{\partial G}{\partial P} \right) 
}
For our theory this reads
\al{
\{F  , G \} = \frac{Z(Q)^3}{(\ln Q)^3} \left( \frac{\partial F}{\partial P} \frac{\partial G}{\partial Q} -  \frac{\partial F}{\partial Q} \frac{\partial G}{\partial P} \right)
}
Applying Dirac's rule we get 
\al{
[\hat{F}, \hat{G}] =  i \frac{Z(\hat{Q})^3}{(\ln \hat{Q})^3} \left( \frac{\partial \hat{F} }{\partial \hat{P}} \frac{\partial \hat{G}}{\partial \hat{Q}} -  \frac{\partial \hat{F}}{\partial \hat{Q}} \frac{\partial \hat{G}}{\partial \hat{P}} \right)
}
where hatted variables are corresponding quantum operators. 
In particular, we have 
\al{
[\hat{Q}, \hat{P} ] = - i \frac{Z(\hat{Q})^3}{(\ln \hat{Q})^3} 
}

Acting on a wavefunction $f$ we get 
\al{
[\hat{Q} , \hat{P} ] f = \hat{P} (Q) f
}
so that 
\al{
\hat{P} (Q) = - i \frac{Z(Q)^3}{(\ln Q)^3} 
}
Therefore we have 
\al{
\hat{P} = - i \frac{Z(Q)^3}{(\ln Q)^3}  \frac{\partial}{\partial Q}
}
We will suppress hats from now on, the distinction should be clear from the context.

The Schrodinger equation is 
\al{ 
H \psi = E \psi
}
There is an ambiguity in the ordering of operators in the Hamiltonian. We will choose the simplest ordering for which
\al{
H = \frac{\pi^3}{c} \frac{1}{Z(Q)^2} P^2
}
We need to find the action of $P^2$ on $\psi$
\al{
P^2 \psi & = - i \frac{Z(Q)^3}{(\ln Q)^3}  \frac{\partial}{\partial Q} \left( - i \frac{Z(Q)^3}{(\ln Q)^3}  \frac{\partial \psi }{\partial Q}  \right)
 \notag\\ & = - \frac{Z^6}{(\ln Q)^6} \left[3 \left( \frac{Z'}{Z} - \frac{1}{Q \ln Q} \right) \frac{\partial}{\partial Q} + \frac{\partial^2}{\partial Q^2} \right] \psi
 }
 where $Z' = dZ/dQ$. Inserting this into the Schrodinger equation we obtain 
 \al{
- \frac{\pi^3}{c} \frac{Z^4}{(\ln Q)^6} \left[ \frac{1}{Q} \left( \frac{2(Q-1)^2}{Z} - \frac{1}{ \ln Q} \right) \frac{\partial}{\partial Q} + \frac{\partial^2}{\partial Q^2} \right] \psi = E \psi
 }

This equation is not solvable in closed form for an arbitrary $E$. On the other hand, if we consider the simplest case of $E=0$ the equation simplifies to 
 \al{ \label{E0Schrodingereqn}
 \left[ \frac{1}{Q} \left( \frac{2(Q-1)^2}{3- 4 Q + Q^2 + 2 \ln Q} - \frac{1}{ \ln Q} \right) \frac{\partial}{\partial Q} + \frac{\partial^2}{\partial Q^2} \right] \psi = 0
 }
 which is of the form 
 \al{
 \psi'' + f(Q) \psi' = 0
 }
 with 
 \al{ \label{fQ}
 f(Q) = - \frac{1}{Q} \left( \frac{2(Q-1)^2}{3- 4 Q + Q^2 + 2 \ln Q} - \frac{1}{ \ln Q} \right)
 }
Then one gets the nonlocal solution 
 \al{ \label{E0psi}
 \psi(Q) = C_1 + C_2 \int_1^Q dq \ \frac{\ln q}{2 \ln q + q^2 - 4 q +3}
 }
 where $C_1$ and $C_2$ are constants. 
 
 The problem here is that the integrand of \eqref{E0psi} blows up at $Q=1$ which is due to $f(Q)$ being divergent at $Q=1$. However, 
 \al{\lim_{Q \rightarrow 1} (Q-1) f(Q-1)=-2
 } 
 so that the singularity is regular and we can apply Frobenius method to get a series solution. These problems seem to stem from  the singularity $Q=1$ of the Hamiltonian. 
  
 We will not continue this analysis since we later realized that the DX theory is  inconsistent as other constraints need to be provided on the phase space for consistency of the diff-Gauss law.   This will be shown in the next chapter where we introduce the DXN theory which is equivalent to the DX theory on the  constraint surface defined by the diff-Gauss law.  In fact the source of singularities encountered in the analysis above may be due to these missing contraints.

\section{Covariantized Transverse Action and Chiral Gauge}
\label{TransverseAfter}
In this section we will show that every aspect (momenta, field equations, etc.) of the  covariantized transverse action evaluated in 2D Minkowski spacetime simplifies in an ansatz that we call chiral ansatz (or chiral gauge). 

This suggests that the underlying gauge condition for the transverse action before covariantization is $D_{01}=N=0$ (which can be extended to $D_{0\mu}=0$ as $D_{00}=\varphi$ is completely invisible), but it becomes the chiral ansatz for the covariantized  transverse action. This ansatz soldifies the change of character of the theory after covariantization as we will see in a variety of aspects below. 
\subsection{Momenta}
After covariantization the nondynamical diff field component $N=D_{01}$ becomes dynamical, which implies that the diff-Gauss law is not a constraint any more. 
To remedy this problem we first set $N$ to zero. However, for this to be a consistent condition we need to make the conjugate momentum to $N$, $X^{010}$ vanish as well. It turns out, however, that there is a symmetry between $X^{010}$ and $X^{011}$ : 
\al{
& X^{010} = - \beta \varphi' - \alpha \varphi \varphi' - 6 q \ddot{\varphi}''\\
& X^{011} = +\beta \dot{D} - \alpha D \dot{D} - 6 q \dot{D}''
}
These are symmetric under $\beta \leftrightarrow -\beta$, $\varphi \leftrightarrow D$ and $t \leftrightarrow x$. So, instead of requiring  vanishing of just the conjugate momentum $X^{010}$ if we require vanishing of the generalized momentum $X^{01\mu}$ of $N$ we see that the required conditions also become symmetric: 
\al{ \label{chiralansatz}
N=0 \ \  , \ \ \frac{d}{dx} \varphi (x,t) = 0 = \frac{d}{dt} D(x,t)
}
We call this the chiral gauge. The motivation of the name is the following. Suppose we take  $x \leftrightarrow x^-$,  $t \leftrightarrow x^+$ as in the geometric actions. Then the chiral gauge reads 
\al{
D_{-+}=0 \ \ , \ \ \partial_- D_{++} =0 =  \partial_+ D_{--}
} 
These are precisely the conditions we used to build the geometric action on the coadjoint orbits of the direct product of two Virasoro algebras in Section \ref{VirasoroDirectProduct}. 
They represent the chiral splitting of the energy momentum tensor and field equations in 2D conformal field theories. 

What do these conditions imply for the momenta of the remaining fields? Since we made the diff component $D$ time-independent, its consistency requires the vanishing of its conjugate momentum. This is indeed the case, we automatically get $X^{110}=0$ in the chiral ansatz. The remaining component of its generalized momentum reads 
\al{
X^{111} = - D' - 2 \beta D' + 3 \alpha DD' + 6q D'''
}

As an interesting side note,  vanishing of $X^{111}$ yields the KdV equation (with free parameters $\beta, \alpha, q$)  as time derivative of $D$ vanishes by hypothesis. Alternatively, consider a generic KdV equation of the form 
\al{ \label{GenericKdV}
aD'+bDD'+cD'''=d \dot{D} 
}
for constants $a,b,c,d$. Introducing wave solutions $\tilde{D}(z) \equiv D(x,t)$ with $z = x+et$ with a constant $e$,  \eqref{GenericKdV} can be rewritten as 
\al{
(a-de) \partial_z \tilde{D} + b \tilde{D} \partial_z \tilde{D}+ c \partial_z^3 \tilde{D} = 0 
}

Similarly, the spatial component of the generalized momentum of $\varphi$ vanishes, $X^{001}=0$ and its conjugate momentum reads 
\al{
X^{000} = \dot{\varphi} + 2 \beta \dot{\varphi} + 3 \alpha \varphi \dot{\varphi} + 6 q \dddot{\varphi}
}

\subsection{Acyclicity }
We take the full diff Lagrangian as
\al{
\mathcal{L} = X_{abc} X^{abc}
}
where $X_{abc}$ is the diff momentum and is given by the covariant expression 
\al{
X_{abc}  = D_{abc} & + \beta ( D_{acb} + D_{bca} ) \notag\\
         & + g^{de} \Big(    \alpha ( D_{ec} D_{abd} + D_{db} D_{eca} + D_{da} D_{ecb} ) + q D_{ec(abd)} \Big) 
}
Here, $D_{abc} \equiv \nabla_c D_{ab}$ , $D_{abcd} \equiv \nabla_d \nabla_c D_{ab}$ etc and 
\al{
D_{(abd)} =  (  D_{abd} + D_{adb} + D_{bad} + D_{bda} + D_{dab} + D_{dba} )/6
} 

The acyclic Lagrangian is defined as  
\al{
\mathcal{L}_{\text{ac}} =  ( X_{abc} + X_{bca} + X_{cab} ) X^{abc} /3
}
The acyclicity condition can be stated as 
\al{
\Delta \equiv \mathcal{L} - \mathcal{L}_c \overset{!}{=}0
}
The first requirement is to set $\beta=1$ in the Lagrangian. Then the difference $\Delta$ at the covariant level reduces to terms with coefficients $\alpha q$, $\alpha^2$ and $q^2$.  If we ignore these terms we get back the BLRY Lagrangian. 
Hence, the BLRY action is the acyclic part of the diff action (when $\beta=1$) at the covariant level. 

We flat-reduce the difference from acyclicity, i.e. we replace covariant derivatives with ordinary ones,  then evaluate the expression in the temporal gauge $N=0$. As mentioned above, the temporal gauge  is necessary for obtaining the diff-Gauss law as a (constraint) field equation from the theory as $N$ is the Lagrange multiplier of the diff-Gauss law introduced in building the transverse action.

So far we have tried to analyze the action  with the  full temporal gauge 
\al{
N=0=\varphi
} 
For instance, in order to recover the diff-Gauss law as the field equation for $D_{01}=N$ we need $N=0$ and $\varphi$ is not at all present before covariantization.   In this gauge the difference from acycility does not vanish for the full diff Lagrangian.  

On the other hand, the difference from acyclicity vanishes for the full, covariantized theory in the chiral ansatz \eqref{chiralansatz}.

\subsection{Field Equations} 

For simplicity we are going to display equations only for the BLRY theory whereas for the full theory we are going to state the general features. 
 The field equations of the BLRY theory  evaluated in 2D Minkowski spacetime with the full temporal gauge $\varphi=N=0$ read
\al{
\text{FE}_{11} \ : \ \ 0 & = 2 \ddot{D} - 6 \alpha D'^{2} - 2 D'' - 8 \beta D'' - 12 \alpha D D'' - 24 q D'''' \\ 
\text{FE}_{12} \ : \ \ 0 & = 2 \alpha \dot{D} D' + 4 \beta \dot{D}' + 4 \alpha  D \dot{D}'  + 24 q \dot{D}'''
\\ 
\text{FE}_{22} \ : \ \ 0 & = 2 \alpha \dot{D}^2 -24 q \ddot{D}''
} 
On the other hand, in the chiral gauge we get
\al{
\text{FE}_{11} \ : \ \ 0 & = 24 q D'''' - 12 \alpha D D'' - 8 \beta D'' - 2 D'' - 6 \alpha D'^2  \\ 
\text{FE}_{12} \ : \ \ 0 & = 0  
\\ 
\text{FE}_{22} \ : \ \ 0 & = -24 q \ddddot{\varphi} - 12 \alpha \varphi \ddot{\varphi} + 8 \beta \ddot{\varphi} + 2 \ddot{\varphi} - 6 \alpha \dot{\varphi}^2 
} 
Now, this is remarkable situation. Firstly, the consistency condition for $N=0$ gauge is satisfied i.e. the field equation for $N=0$ is identically satisfied, instead of yielding the diff-Gauss law. Secondly, the field equations for $D(x)$ and $\varphi(t)$ are decoupled, and have the same structure, and become identical for  $\beta \leftrightarrow - \beta$ and $q \leftrightarrow -q$. These equations are not solvable in closed form for $\alpha \neq 0$ and $q \neq 0$. In the  case, $\alpha =0$, for instance, $D$ equation is solved by 
\al{
D(x) = c_1 + c_2 x +  2 q e^{- x/ \sqrt{2q}} (c_3 e^{2 x/ \sqrt{2q}} + c_4 ) 
}
and in $q=0$ case we have  
\al{
D_1(x) & = f(x) \equiv \frac{2 \beta +1}{3 \alpha} + \left(\frac{3}{4\alpha} -c_1^2 (x^2 - 2 c_2  x - c_2^2) \right)^{1/3} 
\notag\\ 
D_{2,3}(x) & = \frac{1 \pm i \sqrt{3}}{2}  f(x)
}
Since $\varphi$ equation is of the same form with independent variable $t$ and with slightly different coefficients, it is solved by similar expressions for the $\alpha=0$ or $q=0$.

The same features (i.e. decoupling and spatial $D$ vs temporal $\varphi$)  is seen in the full theory although the equations are of higher order (sixth) and more complicated. We still have the field equation for $N$ satisfied identically, i.e., $0=0$. Similar features hold for the energy momentum tensor, and the conservation equations. 

To summarize, in the chiral ansatz, there are no constraints, so there are no consistency conditions to check. The transverse action after covariantization evaluated in the 2D Minkowski spacetime reduces to the theory of two decoupled fields, satisfying the same structure of equations, one in space (so frozen), and the other in the time variable. Due to the decoupling, one can set one of the fields to zero and analyze the other one. As we will show below, the problem of symmetric criticality does not arise, so the consistency seen in momenta and the field equations is carried to the Lagrangian level. 

\subsection{Lagrangian and  Problem of Symmetric Criticality}
As a final note, let us mention another nice feature of the chiral ansatz. For this purpose let us introduce the problem of symmetric criticality (see, for instance, \cite{bojowald} Chapter 2). Namely, in general, the field equations obtained from a gauge-fixed  Lagrangian are not necessarily the same as the gauge-fixed field equations obtained from the non-gauge-fixed Lagrangian. In other words, variations and gauge fixing do not commute; at which step you apply the gauge fixing, on-shell or off-shell, does matter. 

For instance, if the temporal gauge $N=0$ is applied at the Lagrangian level, then the resulting Lagrangian will lead to $0=0$ as the field equation for $N$ as it does not depend on $N$. However, if the Lagrangian is first varied and the resulting field equations are subjected to $N=0$ gauge, then the field equation for $N$ no more yields $0=0$, it becomes a constraint equation (the diff-Gauss law). Therefore, by applying $N=0$ gauge at the Lagrangian level one loses information. 

Along the same line of reasoning in the full temporal gauge $N=0=\varphi$ symmetric criticality problem arises as (constraint) information about $N$ and $\varphi$ are completely lost, and moreover, $D$ field equation is altered. 

Now, the problem of symmetric criticality does not appear in the chiral gauge, since in this gauge, the field equation for $N$ becomes $0=0$ in either case and $D$ and $\varphi$ equations remain exactly the same (for both the full and the BLRY actions) as can be verified. So one can work with the gauge-fixed Lagrangian. For simplicity let us consider the chiral BLRY Lagrangian given by 
\al{
\mathcal{L}  =  (1+ 2 \beta + 3 \alpha \varphi ) \dot{\varphi}^2 + 6 q \dot{\varphi} \dddot{\varphi} -(1+ 2 \beta - 3 \alpha D) D'^2 + 6 q D' D''' 
}
Hence,  decoupling is apparent at the Lagrangian level and the chiral gauge is a consistent ansatz at the Lagrangian level as well.

\section{BLRY Hamiltonian in 2D Minkowski Spacetime}
\label{BLRYTotalHamiltonian}

Recall the BLRY Lagrangian,
\al{
\mathcal{L}=D_{abc} X^{abc}
}
In this form $\mathcal{L}$ seems to be singular and third-order in time for each diff field component, since the higher derivative terms are of the form $ q \, \nabla^c  D_{ab} \, \nabla^{(d} \nabla^b \nabla^{a)} D_{dc} $. This is an illusion since we can partially integrate one of the derivatives in the second factor, which is equivalent to the addition of a total divergence $\nabla_a f^a$ to $\mathcal{L}$, and it is well known that the field equations are invariant under this operation. Moreover, the resulting Lagrangian is second-order. 

 Carrying out this operation  we end up with the following Lagrangian  in  2D Minkowski spacetime : 
\al{ \label{2DMinkaftercov}
\mathcal{L} & = C + B_1 \dot{D} + B_2 \dot{N} + B_3 \dot{\varphi} + 8q D'' \dot{N}' + 16q \dot{D}' N'' \notag\\ & + A_1 \dot{D}^2 + A_2 \dot{N}^2 + A_3 \dot{\varphi}^2 \notag\\ & + - 4 \alpha N \dot{N} \dot{\varphi} - 16 q \dot{N}'^2 - 8 q \dot{D}' \dot{\varphi}' - 8 q N'' \ddot{N} - 4 q \ddot{D} \varphi'' \notag\\ & + 8 q \dot{N}' \ddot{\varphi} + 16 q \ddot{N} \dot{\varphi}' - 6 q \ddot{\varphi}^2 
}
The ordering  is such that the first line consists of zeroth and first-order terms in time, second and third lines consist of second-order terms in time and the last line consists of third and fourth-order terms in time. Here, 
\begin{subequations}
\al{
A_1 & = 1+ \alpha \varphi \ \ , \ \ A_2 = -2 -2 \beta + 2 \alpha (D -  \varphi) \ \ , \ \ A_3 = 1+ 2 \beta + 3 \alpha \varphi \\
B_1 & = - 2 \alpha N (D' - 2 \varphi') + (4 \beta  - 4 \alpha D) N' \\
B_2 & = 8 \alpha NN' -4 (\beta + \alpha \varphi) \varphi' \ \ , \ \ B_3 = - 2 \alpha N \varphi' \\
C&= C_1 D'^2 + C_2 N'^2 + C_3 \varphi'^2 + C_4 \\ 
C_1 &= -1 - 2 \beta +3 \alpha D \ \ , \ \ C_2 = 2 + 2 \beta -2 \alpha (D-\varphi) \\
C_3 & = -1 + \alpha D \ \ , \ \ C_4 = -4 \alpha D' N N' - 6 q D''^2
}
\end{subequations}

For a second-order field theory conjugate momenta can be defined through 
\al{\label{secondordermomentadefn}
\pi_{\phi} = \frac{\delta S}{ \delta \dot{\phi}} = \frac{\partial \mathcal{L}}{\partial \dot{\phi}} - \partial_{\mu}  \frac{\partial \mathcal{L}}{\partial  (\partial_{\mu} \dot{\phi})}  \ \ , \ \ \pi_{\dot{\phi}} = \frac{\delta S}{ \delta \ddot{\phi}} = \frac{\partial \mathcal{L}}{\partial \ddot{\phi}}
}
These yield 
\begin{subequations}
\al{
\pi_D &  = 2 A_1 \dot{D} + B_1 + 4q ( 3 \dot{\varphi}'' - 4 N''' )
\\
\pi_N & = 2 A_2 \dot{N} + B_2 - 4 \alpha N \dot{\varphi} + 8 q (- D''' + 5 \dot{N}'' - 3 \ddot{\varphi}')
\\ 
\pi_{\varphi} & = 2 A_3 \dot{\varphi} + B_3 - 4 \alpha N \dot{N} + 4q ( 2 \dot{D}'' - 6 \ddot{N}' + 3 \dddot{\varphi})
\\ 
\pi_{\dot{D}} & = - 4 q \varphi'' 
\\ 
\pi_{\dot{N}} & = - 8 q N'' + 16 q \dot{\varphi}' 
\\ 
\pi_{\dot{\varphi}} & = - 12 q \ddot{\varphi} + 8 q \dot{N}'
}
\end{subequations}
We see from the last three equations that the theory is singular in $D$ and $N$ whereas nonsingular in $\varphi$. 

We can convert the higher order theory to an ordinary first-order theory by a redefinition of variables as described in Section \ref{pons2ndorder}. 
For this purpose we introduce the new variables, 
\al{
Q_1 \equiv D , \ \ Q_2 \equiv \dot{D} , \ \ Q_3 \equiv N , \ \ Q_4 \equiv \dot{N} , \ \ Q_5 \equiv \varphi , \ \ Q_6 \equiv \dot{\varphi}
}
The main difference from the analysis in Section \ref{pons2ndorder} is that we now have three (infinite sets of) variables, and since they are fields on spacetime, in addition to the time derivatives there are spatial derivatives. Spatial derivatives have no significance in the canonical analysis so they can be treated as additional indices. 

We can then introduce the first-order (singular) Lagrangian
\al{
\mathcal{L}_T & = C + B_1 Q_2 + B_2 Q_4 + B_3 Q_6+ 8q Q_1'' Q_4' + 16q Q_2' Q_3'' \notag\\ & + A_1 Q_2^2 + A_2 Q_4^2 + A_3 Q_6^2 - 4 \alpha Q_3 Q_4 Q_6- 16 q Q_4'^2 - 8 q Q_2' Q_6' \notag\\ &   - 8 q Q_3'' \dot{Q}_4- 4 q \dot{Q}_2 Q_5''+ 8 q Q_4' \dot{Q}_6 + 16 q \dot{Q}_4 Q_6' - 6 q \dot{Q}_6^2 \notag\\ & + \lambda_1 (\dot{Q}_1 - Q_2) + \lambda_2 (\dot{Q}_3 - Q_4) + \lambda_3 (\dot{Q}_5 - Q_6 ) 
}
The constraints in the last line are added for correspondence with the original Lagrangian \eqref{2DMinkaftercov}. The factors $A_i$, $B_i$ and $C$ are spatial functionals of $Q_1, Q_3, Q_5$ only. 

We introduce the conjugate momenta 
\al{
P_1 \equiv \pi_D , \ \ P_2 \equiv \pi_{\dot{D}} , \ \ P_3 \equiv \pi_N , \ \ P_4 \equiv \pi_{\dot{N}} , \ \ P_5 \equiv \pi_{\varphi} , \ \ P_6 \equiv \pi_{\dot{\varphi}} \ \ , \ \ p_i \equiv \pi_{\lambda_i}
}
Then applying \eqref{secondordermomentadefn} to $S_T$ we can compute the momenta
\begin{subequations}
\label{momentaBLRYpons}
\al{
&P_{2i-1} = \lambda_i  , \ \ i=1,2,3 \\
&P_2  = - 4 q Q_5'' \\
&P_4 = - 8q Q_3'' + 16 q Q_6' \\
&P_6 = -12 q \dot{Q}_6 + 8q Q_4' \label{P6LTdiff} \\
&p_i  = 0 \ , \ \ i=1,2,3
}
\end{subequations}
The equation \eqref{P6LTdiff} can be inverted for $\dot{Q}_6$ yielding 
\al{
\dot{Q}_6 = - \frac{1}{12q} P_6 + \frac{2}{3} Q_4'
} 
so, the theory is nonsingular in $\varphi$. 

Except the $P_6$ equation \eqref{P6LTdiff}, all the equations \eqref{momentaBLRYpons} constitute primary constraints.  Out of these, only $P_2$ and $P_4$ equations are essential, signifying the singularity in $D$ and $N$. The  remaining ones arise due to the introduction of the Lagrange multipliers to reduce the second-order theory to a first-order theory, and can be eliminated using the field equations for $\lambda_i$. Let us denote the main primary constraints as
\begin{subequations}
\label{BLRYponsessentialconst}
\al{
\phi_1 & \equiv P_2 + 4 q Q_5'' \approx 0 \\
\phi_2 & \equiv P_4 + 8 q Q_3'' - 16 qQ_6' \approx 0
}
\end{subequations}

The canonical (or naive) Hamiltonian (density) is formed, as usual, as
\al{
\mathcal{H}_c = P_I \dot{Q}_I + p_i \dot{\lambda}_i - L_T \ \ \ , \ \  I=1, \cdots, 6\  \ ; \ \ i = 1, 2, 3 
}
This yields 
\al{
\mathcal{H}_c & = \lambda_1 Q_2 + \lambda_2 Q_4 + \lambda_3 Q_6 - A_1 Q_2^2 - A_2 Q_4^2 - A_3 Q_6^2 \notag\\ & \ \  - B_1 Q_2 - B_2 Q_4 - B_3 Q_6 - C - 8q Q_1'' Q_4' - 16 q Q_2' Q_3'' 
\notag\\ & \ \ + 4 \alpha Q_3 Q_4 Q_6  + 8q Q_2' Q_6' - (1/24q) (P_6 - 8 q Q_4' )^2 + 16 q Q_4'^2 
}

In order to form the total Hamiltonian density, $\mathcal{H}_T$ we need to add each primary constraint followed by the equations \eqref{momentaBLRYpons} with some multipliers. As mentioned above, however, some of these constraints are not essential. Hence, we will shortcut the procedure by eliminating these. For this purpose, we will not introduce the constraints $p_i=0$, $i=1,2,3 \ $ into the action and we will replace $\lambda_1, \lambda_2, \lambda_3$ by $P_1, P_3, P_5$, respectively. This gets rid of all the nonessential constraints. So we will introduce only the essential constraints $\phi_1, \phi_2$ given by  \eqref{BLRYponsessentialconst} into the Hamiltonian with multipliers $\mu_1, \mu_2$, respectively. The result is 
\al{
\mathcal{H}_T & = \mathcal{H}_c + \mu_1 \phi_1 + \mu_2 \phi_2 \notag\\ & = P_1 Q_2 + P_3 Q_4 + P_5 Q_6 - A_1 Q_2^2 - A_2 Q_4^2 - A_3 Q_6^2 
\notag\\ & \ \ - B_1 Q_2 - B_2 Q_4 - B_3 Q_6 - C - 8q Q_1'' Q_4' - 16 q Q_2' Q_3'' 
\notag\\ & \ \ + 4 \alpha Q_3 Q_4 Q_6  + 8q Q_2' Q_6' - (1/24q) (P_6 - 8 q Q_4' )^2 + 16 q Q_4'^2 
\notag\\ &  \ \ + \mu_1 ( P_2 + 4q Q_5'') + \mu_2 ( P_4 + 8q Q_3'' - 16q Q_6') 
}

Next step is checking  consistency conditions for the primary constraints to see whether we get secondary constraints or conditions on the functions $\mu_1, \mu_2$. We will not continue this analysis since the diff-Gauss law is lost as a constraint. $N$ became dynamical and the theory lost its connection to its origins in 1D. 

As a final note,  higher order analysis of the full diff Lagrangian is no different. Since the full theory is of higher order than the BLRY theory, one merely needs to define more variables to obtain a first-order reduction. 
\chapter{CONSTRAINT ANALYSIS IN 2D MINKOWSKI SPACETIME}
\label{2DMinkowskiConstraint}

\section{Introduction}
In this chapter we will first analyze the transverse action before covariantization. We call this theory the DXN theory. This theory is equivalent to the DX theory (Section \ref{ReviewDXGravity}) on-shell (i.e. on the constraint surface). The  diff-Gauss law arises as an explicit (secondary) constraint from Hamilton's equations rather than being implicitly enforced through an equivalence relation on the phase space. 

 We will apply Dirac's constrained Hamiltonian formalism to the DXN theory and show that consistency equation of the diff-Gauss law implies the existence of an independent constraint. The constraint algebra closes with one primary, three secondary constraints and a condition on the Lagrange multiplier. The diff-Gauss law turns out to be second-class unless the kinetic term itself is turned into a constraint.
 
Hence, we turn the kinetic term  to a constraint. This is good in one respect, that a theory of gravitation requires time reparametrization invariance \cite{DiracLectures} which, in turn, requires the Hamiltonian to consist purely  of first-class constraints (i.e. to vanish on-shell). 
However, since we chose the standard kinetic term of the form $\sim X^2$ (both for the finite reduction of the $DX$ theory and the transverse diff field theory) such a modification implies the theory is trivial. 
 
 In general relativity also  the Hamiltonian consists of constraints, yet the "kinetic term" is not standard. 
Indeed, in the diff field case the higher order terms suggest that the standard kinetic term may not be suitable to build the dynamical theory. An alternative kinetic term, which may lead to a nontrivial dynamical theory, will be introduced and analyzed.  

In the last section of this chapter, we review and summarize the tranverse diff theory, and its problems. Then we investigate the modifications required to maintain covariance of the diff Lagrangian and its interactions.   

\section{Transverse Action Before Covariantization -  DXN Theory }
\label{DXNTheoryPart1}
\subsection{Lagrangian  }
Recall that the momentum \eqref{diffmomflat} of the diff field in flat spacetime of arbitrary dimension, and in lower indices is given by
\al{
X_{ij0} = \partial_0 D_{ij} & - \Big[ D_{k0} \partial_k D_{ij} + D_{kj} \partial_i D_{k0} + D_{ik} \partial_j D_{k0} \notag\\ & \ \ \  \ \ \  + q \partial_i \partial_j \partial_k D_{k0} - \beta (\partial_i D_{0j} + \partial_j D_{i0})  \Big] 
}
In 2D, before covariantization, the diff field action involves only two degrees of freedom, $D_{01} \equiv N$ and $D_{11} \equiv D$. Only the latter is dynamical, having a conjugate momentum $X \equiv X^{110}$. For the Minkowski metric with the sign convention $(+t, -x)$ we have $X_{110}=X^{110}=X$. As usual we will  denote $\partial_1$ by a prime and $\partial_0$ by a dot.  With these we get
\al{
X = \dot{D}  -(   N D' +  2  D N'   -  2 \beta N' +  q N''')
} 
This suggests that we can also shift $D$ by a constant $ \beta$ and simplify this to 
\al{
X = \dot{D}  -(   N D' +  2  D N'    +  q N''')
}
We can write the momentum compactly as
\al{\label{XNDmom}
X = \dot{D} - \mathcal{G} [N;D]
}
where 
\al{ \label{diffGaussfunctional}
\mathcal{G} [ \xi ; D ] : = \xi D' + 2 \xi' D + q \xi''' 
}
is the "coadjoint action" of a vector field $\xi$  on the diff field. In the following we denote $\mathcal{G} \equiv \mathcal{G} [ N ; D]$. Notice that this is not the same as the diff-Gauss law operator, $G_1= \mathcal{G}[X;D]$.

Then the  Lagrangian before covariantization, \eqref{LXsquare}, reads
\al{
\mathcal{L} =\frac{1}{2} X^2 =  \frac{1}{2} ( \dot{D} - \mathcal{G} )^2
}
in 2D flat spacetime. We call the theory defined by this Lagrangian the DXN theory.  

\subsection{Full vs BLRY Lagrangians }
\label{FullvsBLRY2DMinkowski}
Recall the BLRY Lagrangian \eqref{BLRYLagrangian}
\al{
\mathcal{L}_{\text{BLRY}} = \frac{1}{2} D_{abc} X^{abc} 
}
Before covariantization, in 2D Minkowski spacetime the diff action reads 
\al{ \label{fulldiffbeforecov}
S = \frac{1}{2} \int ( \dot{D} -  \mathcal{G} )^2 
}
and the BLRY action reads 
\al{ \label{BLRYbeforecov}
S_{\text{BLRY}} = \frac{1}{2} \int \dot{D} ( \dot{D} - \mathcal{G})
}
where $\mathcal{G} \equiv \mathcal{G}[N;D]$ in \eqref{diffGaussfunctional}, and the measure $d^2x$ is  suppressed. 

We see from these expressions that it is not right to say that the BLRY Lagrangian is the $N=0$ limit of the diff Lagrangian.  Rather the $N=0$ limits of these theories do match up, as $\mathcal{G}$ vanishes. In particular, the field equation of $N$ in $N=0$ gauge yields the diff-Gauss law, in both cases. Let us show this. Varying \eqref{fulldiffbeforecov} with respect to $N$ we get 
\al{
\delta S & = \int  X \delta \mathcal{G} \notag\\  & =  \int X( \delta N D' + 2 \delta N' D+ q \delta N''') \notag\\ & = - \int \delta N( XD' + 2 X' D + q X''')
}
So the field equation of $N$ reads 
\al{ \label{diffGaussbeforecov}
 0 = XD' + 2X' D + q X'''
}
Similarly, for the BLRY action we get 
\al{
\delta S_{\text{BLRY}}  = &\frac{1}{2} \int \dot{D} (- \delta \mathcal{G} ) \notag\\  = - &\frac{1}{2} \int \dot{D}( \delta N D' + 2 \delta N' D+ q \delta N''') \notag\\  = &\frac{1}{2} \int \delta N( \dot{D} D' + 2 \dot{D}' D + q \dot{D}''')
}
So the field equation of $N$ reads 
\al{
0=( \dot{D} D' + 2 \dot{D}' D + q \dot{D}''')
}
This is the diff-Gauss law \eqref{diffGaussbeforecov} only for $N=0$.

\subsection{Hamiltonian } \label{DXNHamiltonian}
The conjugate momentum to $D$ is given by \eqref{XNDmom}. 
We also have the variable $N$, but the Lagrangian is independent of its velocity, so, denoting its conjugate momentum by $\pi$, we have the primary constraint 
\al{ \label{DXNprim}
\phi_1 \equiv \pi = \frac{\delta L}{\delta \dot{N}} = 0
}

The naive Hamiltonian density is defined as
\al{
\mathcal{H} & = \dot{D} X + \dot{N} \pi - \mathcal{L}   \notag\\ & = \frac{1}{2} X^2 + X ( N D' + 2 N' D  + q N''') 
}
Adding  the primary constraint \eqref{DXNprim} with a Lagrange multiplier $\lambda$ we get the total Hamiltonian density
\al{
\mathcal{H}_T& = \mathcal{H} + \lambda \pi_N \notag\\ & = \frac{1}{2} X^2 + X ( N D' + 2 N' D + q N''')  + \lambda \pi 
}

The total Hamiltonian is defined by the space integral 
\al{
H_T = \int dx \, \mathcal{H}_T
}
Hamilton's equations yield
\begin{subequations}
\al{ \label{diffADM1a}
\dot{D}  = &\frac{\delta H_T}{\delta X} = X + ND' + 2 N' D + q N''' \\ \label{diffADM2a} \dot{X}  = - &\frac{\delta H_T}{\delta D}  = X' N - X N' \\ \label{diffADM3a} \dot{N}  = & \frac{\delta H_T}{\delta \pi} = \lambda \\ \label{diffADM4a} \dot{\pi}  = - &\frac{\delta H_T}{\delta N} = XD ' + 2 X' D + q X'''
}
\end{subequations}
The first one is the momentum $X$ definition \eqref{XNDmom} reproduced. The second one is the main  dynamical equation for the diff field. The third one tells us that $N$ is arbitrary as its velocity is arbitrary. Finally the last one tells us that the diff-Gauss law  is obtained as a secondary constraint following from the consistency condition of the primary constraint (just as in the canonical analysis of the YM theory) i.e. 
\al{
\dot{\pi}_N = 0  \ \ \Rightarrow  \ \ X D' + 2 X' D + q X''' = 0
}

We will continue the constraint analysis in the next section. 
Let us remark here that the DXN theory, the transverse theory before  covariantization, is equivalent to the DX theory introduced in Section \ref{ReviewDXGravity}. The momentum reduces on-shell (i.e. for $N=0$) to $\dot{D}$ and the diff-Gauss law emerges as the field (constraint) equation of the nondynamical field $N$. In this equivalence, Rajeev's theory reviewed in Section \ref{Rajeev88Summary} corresponds to the DX theory and the original YM theory in 2D corresponds to the DXN theory.

Let us check the solutions to Hamilton's equations.
Consider the dynamical equation \eqref{diffADM2a}. To avoid confusion let us use $(s,t)$ instead of $(x,t)$ so that $X=X(s,t)$ and $N=N(s)$ (recall that $N$ is not dynamical). Then,
\al{
\frac{\partial X}{\partial t} =  N \frac{\partial X}{\partial s} - \frac{\partial N}{\partial s} X
}
The solution for $X$ is 
\al{
X(s,t)=N(s) F (z)
}
where 
\al{
z \equiv t + \int_1^s \frac{d \sigma}{N(\sigma)}
}
Plugging this solution into \eqref{diffADM1a} and solving for $D$ we get 
\al{
D(s,t) = \frac{1}{N^2(s)} \left[  G(z) - F(z) \int_1^s d \sigma \ N^2(\sigma) - q \int_1^s d \sigma \ N(\sigma) N'''(\sigma) \right]
}

\section{Constraint Analysis of DXN Theory } \label{ConstraintDXN}
Let us complete the constraint analysis started in the previous section according to Dirac's  formalism \cite{DiracLectures} which we reviewed in Appendix \ref{ConstraintAnalysis}. 

Using the primary constraint \eqref{DXNprim} we formed the total Hamiltonian and computed Hamilton's equations. There we saw that the consistency equation \eqref{diffADM4a} of the primary constraint $\phi_1$ implied the diff-Gauss law as a secondary constraint. In this section we will switch to the language of Poisson brackets (PB).

The standard PB  for a field theory with phase space variables $\varphi^a, \pi_b$   reads
\al{
\{ F (x) \ , \ G(y) \} =  \int dz \left( \frac{\delta F(x)}{\delta \varphi^a (z) }\frac{\delta G(y)}{\delta \pi_a (z) } - \frac{\delta F(x)}{\delta \pi^a (z) }\frac{\delta G(y)}{\delta \phi_a (z) } \right)
}
From this, for instance, follows 
\al{
\{ \varphi^c (x) \ , \ \pi_d (y) \} =  \int dz \, \Big( \delta_a^c \delta (x-z) \delta^a_d \delta (y-z) - 0 \Big) = \delta ^c_d \delta (x-y)
}
The  PBs for the DXN theory are taken to be in standard form
\al{
\{D(x) , X(y)\} = \delta (x-y) \ \ ; \ \ \{N(x), \pi(y)\}=\delta(x-y) \ \ ; \ \ \text{the rest vanish}
}
where $t$-dependence is suppressed as usual. This implies, for instance, 
\al{
\{D(y)  ,  X'''(x) \} = \int dz \, \delta (y-z) \, \partial_x^3 \delta (x-z) = \partial^3_x \delta (x-y)
}

Recall the total Hamiltonian density
\al{
\mathcal{H}_T = \frac{1}{2} X^2 + X \mathcal{G} + \lambda \pi
}
where
\al{
\mathcal{G} = ND'+ 2N'D+q N'''
}
For convenience below we shall denote space dependence by a subindex, e.g., $D_x \equiv D(x)$. In the language of PBs,  consistency condition for the primary constraint $\phi_1$ reads
\al{ \label{pricons}
0 \approx \dot{\phi_1} =  \{ \phi_1 , H_T \}
}
Let us show this explicitly  
\al{
0 & \approx \left\{ \pi_y \ ,  \int_x  \frac{1}{2} X^2 + X_x ( N_x D'_x + 2 N'_x D_x + q N'''_x ) + \lambda_x \pi_x  \right\} \notag\\ 
& = \int_x X_x \left\{ \pi_y , N_x D'_x + 2 N'_x D_x + q N'''_x  \right\} \notag\\ & = \int_x X_x \Big( \partial_x D_x (-\delta_{xy}) + 2 D_x \partial_x (-\delta_{xy}) + q \partial_x^3 (-\delta_{xy} ) \Big) \notag\\ & = - \int_x \delta_{xy} \Big( X_x \partial_x D_x - 2 \partial_x ( X_x D_x) - q \partial_x^3 X_x \Big)  \notag\\ & = X_y D'_y + 2 X'_y D_y + q X'''_y
}
Since this does not involve the Lagrange multiplier $\lambda$, it must be a secondary constraint. Thus we have
\al{
\phi_2  =  XD' + 2X'D + q X''' \approx 0 
}
i.e. the-diff Gauss law is obtained\footnote{This is in analogy with YM theory. Vanishing of the momentum $\pi^0 = F^{00}$ conjugate to $A_0$ implies the Gauss law constraint. The Gauss law is the field equation of $A_0$ (analog of \eqref{diffADM4a}) and it can also be derived by the PB equation $\dot{\pi}^0 = \{\pi^0 , H_{\text{YM}}\}$ (analog of \eqref{pricons}). }  as a secondary constraint that follows from the vanishing of the momentum of $N$.

Next, we need to check for the consistency condition of the  secondary constraint to see whether we get another secondary constraint. This yields 
\al{ \label{chi1HT}
0 & \approx \dot{\phi}_2 =  \{ \phi_2 , H_T\} \notag\\ & = \left\{ XD'+2X'D+qX''' , \int \frac{1}{2} X^2 + X \mathcal{G} + \lambda \pi \right\} 
}
After a long, but  straightforward calculation this yields 
\al{
 0  &  \approx 3 X' X + N \phi_2' + 2 N' \phi_2  \notag\\ & \approx 3X'X   
}
In the last (weak) equality we used the fact that in field theories  spatial derivatives of a constraint $(\phi_2)$ do not constitute  independent constraints. Thus we obtained a new secondary constraint 
\al{
\phi_3 \equiv  X' X 
}
where we rescaled it by a factor of 3. Notice that this is the derivative of the kinetic term $T=X^2/2$. 

Computing the consistency condition of $\phi_3$ we get 
\al{
0 & \approx \dot{\phi}_3 = \{ \phi_3 , H_T \}  = N \phi_3' - N' \phi_3 - N'' X^2 \notag\\ & \approx - N'' X^2 
}
Hence, we obtain another secondary constraint 
\al{
\phi_4 \equiv N'' X^2 
}
Notice that this is linear in the kinetic term, though it does not necessarily trivialize the theory as we can set $N''=0$ to evade $X=0$. 

Finally, the consistency condition of $\phi_4$ reads 
\al{
0 & \approx \dot{\phi}_4 = \lambda'' X^2 - 2 N' \phi_4 + 2 NN'' \phi_3 \notag\\ & \approx \lambda'' X^2 \label{DXNmultipliercondition}
}
Since this involves the Lagrange multiplier $\lambda$ of the primary constraint, it does not constitute a constraint, rather a condition on $\lambda$.

Next step is checking the PBs of the constraint algebra to determine the first-class constraints. For this purpose we are going to introduce the following commonly used tool in higher order field theories. Namely, if $\phi_x$ is a local expression involving the fields and their derivatives then we  define  $\phi[\mu]$ 
\al{
\phi [ \mu ]  = \int_x \mu_x \phi_x 
} 
In fact, we already used it in \eqref{QGausssmeared} to get the diff-Gauss law operator;  $Q = G[\xi]$ in \eqref{QGausssmeared}, in the notation we use here.   Calculations involving smeared out fields are essentially the same as in the previous section.

Here are the results for the constraint algebra
\al{
& \{ \phi_1 , \phi_{1,2,3} \} = 0 \\
& \{ \phi_1 , \phi_4[\lambda] \} = -    \lambda'' X^2 + 4 \lambda' \phi_3 + \lambda \phi_3'  \approx 0 \label{phi1phi4DXN}
}
where we used \eqref{DXNmultipliercondition}. This calculation shows that $\phi_1$ is a first-class constraint. 

Similarly, we compute 
\al{
& \{ \phi_4 , \phi_{3,4} \} = 0 \\
& \{ \phi_4 , \phi_2[\mu] \} = 2 \mu' \phi_4 - 2 N'' \mu \phi_3 \approx 0 
}
With the additional result \eqref{phi1phi4DXN} we see that $\phi_4$ is also a first-class constraint. 

The remaining two constraints (in particular, $\phi_2$, the diff-Gauss law) are second-class due to the following bracket 
\al{
\{ \phi_3 , \phi_2[\mu] \} = \mu'' X^2 + \mu' \phi_3 - \mu \phi_3' \approx \mu'' X^2
}
or, alternatively, 
\al{
\{ \phi_2 , \phi_3[\mu] \} = - 3 \mu' \phi_3 - \mu'' X^2 \approx - \mu'' X^2 
}
Both brackets tell us that in order for $\phi_2$ and $\phi_3$ to be first-class we need kinetic term $X^2$ to be a constraint since $\mu$ is an arbitrary function that we introduced, and we can not impose a condition on it.

\section{Consistent Constraint Algebra, Yet Trivial Theory }
\label{FreezingDiffTheory}
The constraint analysis in the previous section suggested that the kinetic term 
\al{
T = X^2 /2 
}
be turned into a constraint. Therefore, let us consider the following theory 
\al{ \label{weaklyvanishingH}
H_T = \int  ( \mu_1 \phi_1 + \mu_2 \phi_2 ) 
}
where 
\al{
\phi_1 \equiv X^2 / 2  \ \ \ \ , \ \ \ \ \phi_2 \equiv  XD' + 2 X' D + q X'''
}
and $\mu_1,\mu_2$ are Lagrange multipliers. 
The unconstrained phase space consists of field configurations $X, D$ and the nondynamical variable $N$ is discarded. The Hamiltonian vanishes weakly. 

We need to show that the constraint algebra closes, no new constraints arise from the consistency conditions. Moreover, for time reparametrization invariance we also need to make sure that the constraints are first-class, generating gauge transformations. 
We have 
\al{ \label{smearedHam}
H_T = \phi_1[\mu_1]+\phi_2 [\mu_2]
}

Here are the results for the constraint algebra defined by \eqref{weaklyvanishingH} 
\al{ \label{constraintsweaklyvanishingH}
\{ \phi_1[ \mu] , \phi_1 [\lambda] \} & = 0 \notag\\ 
\{ \phi_1[ \mu] , \phi_2 [\lambda] \} & = - \int ( 2 \lambda \mu' \phi_1 + 3 \mu \lambda \phi_2') \approx 0 \notag\\ 
\{ \phi_2[ \mu] , \phi_2 [\lambda] \} & = \int ( \mu \lambda' - \mu' \lambda) \phi_2 \approx 0
} 
Hence, $\phi_1$ and $\phi_2$ are each first-class constraints. This also implies that no new constraints arise as follows. We compute 
\al{
\{ \phi_1 [\lambda] , H_T \} &  =\{ \phi_1 [\lambda] , \phi_1 [\mu_1] \} + \{ \phi_1 [\lambda] , \phi_2 [\mu_2] \} \approx 0  \notag\\ 
\{ \phi_2 [\lambda] , H_T \} &  =\{ \phi_2 [\lambda] , \phi_1 [\mu_1] \} + \{ \phi_2 [\lambda] , \phi_2 [\mu_2] \} \approx 0 
}
where we used \eqref{smearedHam} and \eqref{constraintsweaklyvanishingH}. 

Finally we compute the gauge transformations generated by the constraints $\phi_1, \phi_2$. If $\phi_i$ is a first-class constraint, a gauge transformation of an arbitrary dynamical variable $F$ generated by $\phi_i$ is computed as 
\al{
\delta_i  F = \{ F , \phi_i[\xi_i]  \}
}
It is enough to compute the gauge transformation of the basic fields $D$ and $X$. $\phi_1$ generates the following gauge transformations 
\al{
\delta_1 D = \xi_1 X \ \ \ , \ \ \ \delta_1 X = 0
}
For $X$ in the isotropy algebra this would describe motion transverse to orbits. It is interesting to obtain the transverse motion as a gauge transformation rather from time evolution. In fact, there is no time evolution as we discuss below. 

And finally the main gauge transformations we are interested in are generated by the diff-Gauss law $\phi_2$ : 
\al{
\delta_2 D = \xi_2 D' + 2 \xi'_2 D + q \xi_2''' \ \ \ , \ \ \ \delta_2 X = \xi_2 X' - \xi_2' X
}
This is simply  invariance under the   Lie-derivatives or Virasoro adjoint and coadjoint transformations. 

The  trouble here is that  the first constraint $\phi_1=0$ implies the vanishing of the diff momentum $X$, freezing the dynamics, i.e. trivializing the theory. 
This suggests that we need to look for alternative expressions quadratic in the diff momentum, whose PB with the diff-Gauss law does not yield new constraints, yet whose vanishing does not imply a trivial theory, i.e. $X\neq 0$. In the next section we are going to introduce a candidate expression.

\section{Alternative Hamiltonian }
\label{AlternativeTransverseTheoryTprimeG}
In this section, we are going to introduce an alternative kinetic term that transforms nicely under the gauge transformations generated by the diff-Gauss law. 

Consider the following  expression
\al{ \label{alternativekineticterm}
T = \frac{1}{2} D X^2 - \frac{q}{4} X'^2 + \frac{q}{2} X X''
}
where prime denotes spatial  derivative. 
It is a kinetic term because it is quadratic in the diff momentum $X$.  
The corresponding smeared out expression satisfies the  PBs
\al{
\{ D, T [\mu] \} = \mu D X  + \frac{q}{2} ( \mu'' X + 3 \mu' X' + 3 \mu X'' )
}
and 
\al{
\{ X , T[\mu]\}  = - \frac{1}{2} \mu X^2 
}
Using these it is straightforward to show that 
\al{
\{ T [\mu] , T[\lambda]\}  = 0 
}
This result holds identically not just weakly. 

$T$ is related to the diff-Gauss law operator $G = XD' + 2 X'D + q X'''$ 
as 
\al{\label{TprimeXG}
T' = XG 
}
Therefore, a theory consisting  of a single constraint $T=0$ implies the diff-Gauss law for a nontrivial theory, i.e. $X\neq 0$. Conversely, according to \eqref{TprimeXG} the diff-Gauss law constraint $G=0$ implies that the kinetic term is space-independent i.e. $T = A(t)$ for some function $A$.
Note that vanishing of $T$ implies 
\al{
\frac{D}{q}  =   \left(-\frac{X'}{X} \right)'  -  \frac{1}{2} \left(-\frac{X'}{X} \right)^2 
}
If we define a function $f(t,x)$ by
\al{ \label{fprimeoverfxprime}
\frac{f''}{f'}=-\frac{X'}{X}
}
then $T=0$ is solved by
\al{ \label{DSxf}
D(t,x) =  q\, (S_x f)(t,x) 
}
where $S_x$ is the Schwarzian derivative with respect to the variable $x$. 

Recall that in Section \ref{DXFiniteReduction}, equation \eqref{DSxf} was proposed as defining the diff-Wilson line\footnote{We will reexamine the diff-Wilson line and loop in Section \ref{diffWilson}.}, and  together with the ansatz
\al{ \label{XPfprime}
X(t,x) = \frac{P(t)}{f'(t,x)}
}
for the diff momentum,  the diff-Gauss law was solved. Moreover, the diff momentum given by \eqref{XPfprime} automatically satisfies the condition \eqref{fprimeoverfxprime}. This raises the question of whether we can apply finite reduction method to the alternative theory defined by the single constraint $T \approx 0$. The answer, however, turns out to be negative as follows. The position variable associated with the corresponding finite theory is defined as  
\al{
Q(t) = f(t, 2 \pi)
}
and the momentum variable is read from \eqref{XPfprime} as 
\al{
P(t) = X(t,x) f'(t,x) \ \ \ \forall x
}
The trouble is that  there is no  natural way to couple $Q(t)$ and $P(t)$ given by these equations as $f(t,2 \pi)$ and $f'(t, 2 \pi)=\partial_x f(t,2 \pi)$ are independent variables upon finite reduction. 

Now let us investigate the transformation properties of $T$. 
It is straightforward to compute 
\al{ \label{alternativekineticspatialscalar}
[T , G[\xi]] = - \xi T' 
}
so that $T$ is a spatial scalar. 

Let us consider the case $T\neq 0$, and get classical solutions for the Hamiltonian \eqref{alternativekineticterm}. First, consider a simpler theory defined  by the first term of $T$, 
\al{
\mathcal{H} = \frac{1}{2} DX^2
}
Assuming standard symplectic structure Hamilton's equations yield 
\al{ \label{alttheoryDeqm}
\dot{D} & = \frac{\delta H}{\delta X} = \frac{\partial \mathcal{H}}{\partial X}=DX \\ \dot{X} & = -\frac{\delta H}{\delta D} = - \frac{1}{2} X^2 \label{alttheoryXeqm}
}
Using the first equation we can obtain the Lagrangian 
\al{
\mathcal{L} = X \dot{D} - \mathcal{H} = \frac{\dot{D}^2}{2D}
}
Equation \eqref{alttheoryXeqm} is solved by 
\al{
X = \frac{2}{t + A(x)}
}
for an arbitrary function $A$ of $x$. 
Inserting this back into \eqref{alttheoryDeqm} we get 
\al{
D=  ( t+ A(x))^2 B(x)
}
where $B$ is another arbitrary function of $x$.

Now, let us analyze the full theory given by $T$. The Hamiltonian 
\al{
\mathcal{H} = \frac{1}{2} DX^2 + \frac{q}{4} ( 2 XX'' - X'^2)
} 
leads to the equations 
\begin{subequations} \label{alternativetransversefieldequations}
\al{ \label{Ddotequationalternativetransverse}
\dot{D} & = DX + \frac{3q}{2} X'' 
\\ 
\dot{X} & = - \frac{1}{2} X^2 
}
\end{subequations}
Hence, the momentum equation is unchanged, so is the momentum solution,
\al{
X = \frac{2}{t + A(x)}
}
The diff field solution, on the other hand, is modified to 
\al{
D = ( t+  A(x))^2 \left( B(x)-3q\left( \frac{A'(x)^2}{2(t+A(x))^4}-\frac{A''(x)}{3(t+A(x))^3}\right) \right)
}

There does not seem to be an obvious way to get the Lagrangian in local form from the Hamiltonian, as \eqref{Ddotequationalternativetransverse} can not be inverted for $X$ in terms of $\dot{D}$ in any obvious way.

\section{On Possible Routes to Fix  Transverse Formalism} 
\label{CoordGaugeTrfoftheDiffField}
\subsection{Review and Summary of  Problems }
Throughout the analysis up to this point we tried to make apparent  inconsistencies of the tranverse formalism. Let us review the transverse formalism to see the problems and their relation better. We started transverse formalism by lifting the transformation \eqref{coadjDch4} 
\al{ \label{Sec66VirCoad}
\delta D = \xi D' + 2 \xi' D + q \xi'''
} of a Virasoro coadjoint element  to higher dimensions as the Lie derivative \eqref{Liderranktwowithcentral} 
\al{ \label{Sec66VirCoadLiftLieDer}
\delta D_{\mu \nu} = \xi^{\lambda} \partial_{\lambda} D_{\mu \nu } + \partial_{\mu} \xi^{\lambda} D_{\lambda \nu} + \partial_{\nu} \xi^{\lambda} D_{\mu \lambda} + q \partial_{\mu} \partial_{\nu} \partial_{\lambda} \xi^{\lambda} 
} 
of a rank-two object $D_{\mu \nu}$. Due to the third-order central term in the Lie derivative this rank-two object is not a tensor. We recovered \eqref{Sec66VirCoad}  from this higher dimensional lift as $\delta D = \delta D_{11}$ under the gauge-fixing conditions 
\al{ \label{gaugefixtogether}
\xi^0 = 0 = \partial_0 \xi^1
}
which restricts the GCTs to spatial and time-independent coordinate transformations. Then we introduced a conjugate momentum $X^{ij0}$ and obtained the diff-Gauss law constraint \eqref{DiffGaussLawOriginal} that generates the spatial part $\delta D_{ij}$ of $\delta D_{\mu \nu}$. Using an ansatz for the momentum we constructed the action $\propto X^{ij0} X_{ij0}$. This action has the diff-Gauss law as a constraint \eqref{transverseLprescappliedtodiff} with Lagrange multiplier $D_{0}^{\ i}$. 

So far everything  looks good. However, this action is not covariant. So we chose the simplest way to deal with this problem, namely we covariantized the action. This operation has two parts: indices $i , j , 0 \rightarrow \mu , \nu , \lambda$  and derivatives $\partial \rightarrow \nabla$. The first part changed the characteristics of the theory : $D_{00}, D_{0i}$ components which did not project on the coadjoint orbits, and $D_{ij}$ component which is the field theory lift of the coadjoint element now became the same field $D_{\mu \nu}$. $D_{0 \mu}$ is now  dynamical so the diff-Gauss law is no more a constraint. 

The second part, namely replacing partial derivatives with covariant derivatives, is only meaningful for a tensor. Moreover, for a nontensor even the first part is problematic as contraction of covariant indices does not yield a tensor if the covariant indices are attached to a nontensor. So we required the diff field to be a tensor.

Then comes the question : What was the point of the first half of the transverse formalism? Why did we introduce $\delta D_{\mu \nu}$ that is nontensorial? Why did we introduce a diff-Gauss law constraint that generates it, if it were going to be lost  upon covariantization? And there is also the hidden inconsistency we discovered in Section \ref{ConstraintDXN}, that even before covariantization the diff-Gauss law is not first-class, so it does not generate the gauge transformation $\delta D_{ij}$, so it is not a "Gauss" law.

After this cruel critique of all the work we have laid down so far let us investigate possible solution routes that would save us from the trouble of altogether abandoning the transverse formalism. We are going to investigate entirely different approaches to  come up with a diff field theory in the next chapter. 

From now on, we treat \eqref{Sec66VirCoadLiftLieDer}
as the coordinate transformation of the diff field, so the diff field is not a tensor, and we abandon covariantization. 
 
\subsection{Spatial Covariance of Transverse Theory }

For now let us give up our hopes on finding a full diffeomorphism invariant action. Do we, at least, have spatial diffeomorphism invariance? We do have, but it requires reviving a hidden quantity,  the metric determinant. Recall the Gauss variation \eqref{deltaXQ} of the diff momentum $X^{ij}$. This calculation showed that the diff momentum is a spatial rank-two tensor density of weight one. 

A tensor density of weight $w$ multiplied by $g^{w/2}$ is a tensor, where $g$ is the metric determinant. Hence, we can form a true rank-two spatial tensor out of $X^{lm}$ as  
\al{
\tilde{X}^{lm} = \sqrt{h} X^{lm}
}
where $h$ is the determinant of the spatial hypersurface of the spacetime. In flat space $h=1$ and there seems to be no difference between $\tilde{X}$ and $X$. However, this is an illusion and another example of the problem of symmetric criticality. In order to show diffeomorphism invariance we need to compute the Lie variation of the Lagrangian. Although $\sqrt{h}$ is one, we can evaluate it before or after the variation. The results are not the same. In fact, using the Lie derivative of $X^{ij}$ obtained from \eqref{deltaXQ} we can show that 
\al{
\mathcal{L}_1 = \frac{1}{2} X^{ij} X_{ij}
}
is not a spatial scalar. However, the Lagrangian  density
\al{
\mathcal{L}_2 = \frac{1}{2} \tilde{X}^{ij} \tilde{X}_{ij}
}
leads to a spatial diffeomorphism invariant action. We have $\mathcal{L}_2 = \mathcal{L}_1$ for $h=1$. This is an example of the difficulty of recovering a covariant theory from its  background evaluated content. 

Hence, although the diff field is not a tensor,  the diff Lagrangian density  before covariantization in the form 
\al{
\tilde{\mathcal{L}} = \frac{1}{2} \tilde{X}^{ij} \tilde{X}_{ij} = \frac{1}{2} h X^{ij} X_{ij}
}
is a spatial scalar. And the spatially invariant action reads 
\al{ \label{SpatiallyDiffInvAction}
S = \int dt \int d^3x \, h^{3/2} \, X^{ij} X_{ij} 
}
Note that we do not know the expression for $X^{ij}$ in the action above when the metric is not flat. That is, when the gauge conditions $\xi^0 = \partial_0 \xi^i$ are turned off,  we do not know what happens to  $X^{ij}$ components.  We will introduce a candidate below in Section \ref{ModifiedTransverseAction}.

\subsection{Spatially Covariant Extension of  Diff Field in 2D }

Since we abandoned covariantization and confined ourselves to spatial diffeomorphism invariance, is everything okay now? Not really.  
We also formed interactions of diff field with other fields and for those interaction terms to make sense  diff field  should be a tensor. For instance, point particle and spinor couplings suggested that  diff field is a perturbation to the spacetime metric, which we used to extend the nonabelian Born-Infeld action. However, a nontensorial perturbation to the metric will destroy all the good things it brought with general relativity. 

How can we recover the spatial diffeomorphism invariance  of  diff field itself? A quick solution that comes to mind is to complement  diff field with a correction that turns it into a tensor, just as  diff field is complemented with a correction involving the gauge field in \eqref{gaugeinvdiff} to yield a gauge invariant object. 

In particular, diff field can be complemented with a correction involving the spacetime metric or Levi-Civita connection to yield a tensor. In fact, in 1D the solution of this problem is simple. As discussed in Section \ref{connectionVircoadjointelement} one may subtract a multiple of $\Sigma= \Gamma' - \Gamma^2/2$ from diff field $D$ to obtain a rank-two tensor. This result, however, does not nicely extend to higher dimensions. In Section \ref{HigherDimLiftofSigma} we searched for higher dimensional extensions $\Sigma_{\mu \nu}$ of $\Sigma$ that transform as \eqref{Liderranktwowithcentral}, but could not find a consistent one. 

However, for spatial diffeomorphism invariance we may not need such a higher dimensional extension of $\Sigma_{\mu \nu}$.  Namely, instead of finding a $\Sigma_{\mu \nu}$ that transforms as \eqref{Liderranktwowithcentral}, it is sufficient to find a $\Sigma_{\mu \nu}$ whose $(11)$ component reduces to a multiple of $\Sigma= \Gamma' -\Gamma^2/2$. In this reduction the gauge-fixing conditions \eqref{gaugefixtogether}  
should be used. Recall the object
\al{  \label{Sigmamunurevisited}
\Sigma_{\mu \nu}  \equiv a \partial_{\lambda} \Gamma^{\lambda}_{\mu \nu} +b \partial_{\mu} \Gamma^{\lambda}_{\lambda \nu}  + c  \partial_{\nu} \Gamma^{\lambda}_{\mu \lambda} + d \Gamma^{\lambda}_{\mu \nu} \Gamma^{\sigma}_{\sigma \lambda} + e \Gamma^{\lambda}_{\mu \sigma} \Gamma^{\sigma}_{\nu \lambda} + f \Gamma^{\lambda}_{\mu \lambda} \Gamma^{\sigma}_{\nu \sigma} 
}
subjected to the condition 
\al{  \label{SigmamunuVirasorocondition}
q\equiv a+b+c = -2 (d+e+f)
} 
Then the 1D reduction of $\Sigma_{\mu \nu}$ yields a coadjoint element $\Sigma$ with central charge $q$. Now, let us rewrite the result we found for the Lie variation of $\Sigma_{\mu \nu}$ in \eqref{Sigmamunurevisited}: 
\al{
\delta \Sigma_{\mu \nu} = \xi^{\lambda} \partial_{\lambda} \Sigma_{\mu \nu} + \partial_{\mu} \xi^{\lambda} \Sigma_{\lambda \nu} + \partial_{\nu} \xi^{\lambda} \Sigma_{\mu \lambda} + q \partial_{\mu} \partial_{\nu} \partial_{\lambda} \xi^{\lambda} + \Delta _{\mu \nu}
}
where 
\al{
\Delta_{\mu \nu} & = (-a +d) \Gamma^{\rho}_{\mu \nu} \partial_{\rho} \partial_{\sigma} \xi^{\sigma} + (a+e) ( \Gamma^{\rho}_{\sigma \nu}  \partial_{\mu} \partial_{\rho} \xi^{\sigma} +  \Gamma^{\rho}_{\sigma \mu} \partial_{\rho} \partial_{\nu} \xi^{\sigma}) \notag\\ &  + f ( \Gamma^{\rho}_{\rho \nu} \partial_{\mu} \partial_{\sigma} \xi^{\sigma} + \Gamma^{\rho}_{\rho \mu} \partial_{\nu} \partial_{\sigma} \xi^{\sigma}) + (b+c+d) \Gamma^{\rho}_{\rho \sigma} \partial_{\mu} \partial_{\nu} \xi^{\sigma} 
}
This expression is too complicated to deal with, so let us look at simpler subcases first. 
Not every subcase works. For instance, the subcase \eqref{simplecaseSigmamunu} noticed for its simplicity in Section \ref{HigherDimLiftofSigma}   does not work since the gauge conditions \eqref{gaugefixtogether}
and the condition \eqref{SigmamunuVirasorocondition} are not compatible in that case as can be seen after a straightforward calculation. 

Here is, a working subcase : $a=2, e=-1$. For this choice we get 
\al{ \label{workingcase}
\Sigma_{\mu \nu}  \equiv 2 \partial_{\lambda} \Gamma^{\lambda}_{\mu \nu} - \Gamma^{\lambda}_{\mu \sigma} \Gamma^{\sigma}_{\nu \lambda} = \Sigma_{\nu \mu} 
}
and the difference from the desired variation reduces to 
\al{
\Delta_{\mu \nu} & = -2 \Gamma^{\rho}_{\mu \nu} \partial_{\rho} \partial_{\sigma} \xi^{\sigma} +  \Gamma^{\rho}_{\sigma \nu}  \partial_{\mu} \partial_{\rho} \xi^{\sigma} +  \Gamma^{\rho}_{\sigma \mu} \partial_{\rho} \partial_{\nu} \xi^{\sigma}
}
Now, following the analysis in Section \ref{Virasoro Gauge-Fixing},  conditions \eqref{gaugefixtogether} imply 
\begin{subequations}
\al{ \label{deltaSigma11}
\delta \Sigma_{11} & = \xi^1 \partial_1 \Sigma_{11} + 2 \partial_1 \xi^1 \Sigma_{11} + 2 \partial_1^3 \xi^1 + \Delta_{11} \\  \delta \Sigma_{01} & = \xi^1  \partial_1 \Sigma_{01} + \Sigma_{01}  \partial_1 \xi^1 + \Delta_{01} \\  \delta \Sigma_{00} & = \xi^1 \partial_1 \Sigma_{00} +\Delta_{00}
}
\end{subequations}
where the central charge of \eqref{workingcase} is two. Our goal is to find the conditions that make $\Delta_{\mu \nu}=0$ without imposing any additional conditions on $\xi^{\mu}$. Remarkably for our choice \eqref{workingcase}, $\Delta_{11}$ automatically vanishes under \eqref{gaugefixtogether}. Vanishing of $\Delta_{01}$ requires $
\Gamma_{01}^1 = 0$ and finally vanishing of $\Delta_{00}$ requires $\Gamma^1_{00}=0$. So we can state the analog of \eqref{nogaugefixingtrfs} as 
\begin{subequations}\label{nongaugefixingSigmamunu}
\al{ 
\xi^0& =0 =\partial_0 \xi^1 \hspace{0.2in} \text{and} \hspace{0.2in} \Gamma^1_{01} = 0 = \Gamma^1_{00} \\
\delta \Sigma_{11} & = \xi^1 \partial_1 \Sigma_{11} + 2 \partial_1 \xi^1 \Sigma_{11} + 2 \partial_1^3 \xi^1  \\  \delta \Sigma_{01} & = \xi^1  \partial_1 \Sigma_{01} + \Sigma_{01}  \partial_1 \xi^1  \\  \delta \Sigma_{00} & = \xi^1 \partial_1 \Sigma_{00} 
}
\end{subequations}
with 
\begin{subequations} \label{nongaugefixingSigmamunu2}
\al{
\Sigma_{11} & = 2 \partial_0 \Gamma^0_{11} + 2 \partial_1 \Gamma^1_{11} - (\Gamma^1_{11})^2 - (\Gamma^0_{01})^2 \\
\Sigma_{01} & = 2 \partial_0 \Gamma^0_{01} - \Gamma^0_{00} \Gamma^0_{01} \\
\Sigma_{00} & = 2 \partial_0 \Gamma^0_{00} - (\Gamma^0_{00})^2
}
\end{subequations}

Can we further restrict these expressions analogous to the full-temporal gauge conditions \eqref{lastset}? If we take, in addition, $\Gamma^0_{00} = 0 = \Gamma^0_{01}$ we get $\Sigma_{00}=0=\Sigma_{01}$ which automatically yields the consistency conditions $\delta \Sigma_{00}=0=\delta \Sigma_{01}$. These imply  $\Sigma_{11}=2 \partial_0 \Gamma^0_{11} + 2 \partial_1 \Gamma^1_{11} - ( \Gamma^1_{11} )^2$. Now, to match the 1D reduction of $\Sigma_{11}$ with the Virasoro coadjoint element $2 \Gamma' - \Gamma^2$, we need the additional condition  $\partial_0 \Gamma^0_{11}=0$ Hence, the analog of \eqref{lastset} becomes 
\begin{subequations}\label{fulltemporalgaugefixSigmamunu}
\al{ 
\xi^0& =0 =\partial_0 \xi^1 \hspace{0.2in} \text{and} \hspace{0.2in} \Gamma^1_{01} = 0 = \Gamma^1_{00}  = \Gamma^0_{00} =  \Gamma^0_{01} = \partial_0 \Gamma^0_{11}
\\ \Sigma_{11} & =  2 \partial_1 \Gamma^1_{11} - ( \Gamma^1_{11} )^2 \\
\delta \Sigma_{11} & = \xi^1 \partial_1 \Sigma_{11} + 2 \partial_1 \xi^1 \Sigma_{11} + 2 \partial_1^3 \xi^1  
\\  \delta \Sigma_{01} & = 0 = \Sigma_{01} 
\\  \delta \Sigma_{00} & = 0= \Sigma_{00} 
}
\end{subequations}

With the  set of conditions \eqref{nongaugefixingSigmamunu} or \eqref{fulltemporalgaugefixSigmamunu}, we obtain a spatial tensor in 2D from the diff field by
\al{
T_{\mu \nu} \equiv D_{\mu \nu} - \frac{q}{2} \Sigma_{\mu \nu}
}
For \eqref{nongaugefixingSigmamunu}, \eqref{nongaugefixingSigmamunu2} we have 
\begin{subequations}
\al{
\xi^0& =0 =\partial_0 \xi^1 \hspace{0.2in} \text{and} \hspace{0.2in} \Gamma^1_{01} = 0 = \Gamma^1_{00} \\
\delta T_{11} & = \xi^1 \partial_1 T_{11} + 2 \partial_1 \xi^1 T_{11} \\  \delta T_{01} & = \xi^1  \partial_1 T_{01} + T_{01}  \partial_1 \xi^1  \\  \delta T_{00} & = \xi^1 \partial_1 T_{00} 
}
\end{subequations}
and for \eqref{fulltemporalgaugefixSigmamunu} we have 
\begin{subequations}
\al{
\xi^0& =0 =\partial_0 \xi^1 \hspace{0.2in} \text{and} \hspace{0.2in} \Gamma^1_{01} = 0 = \Gamma^1_{00}  = \Gamma^0_{00} =  \Gamma^0_{01} = \partial_0 \Gamma^0_{11} \\
T_{11} & = D_{11} - \frac{q}{2} \Big(  2 \partial_1 \Gamma^1_{11} - ( \Gamma^1_{11} )^2 \Big)
\\
\delta T_{11} & = \xi^1 \partial_1 T_{11} + 2 \partial_1 \xi^1 T_{11} 
\\  \delta T_{01} & = 0=  T_{01} 
\\  \delta T_{00} & = 0=T_{00}
}
\end{subequations}

Now, it is tempting to think that this result would nicely extend to higher dimensions, but it does not. The reason is simple : for \eqref{gaugefixtogether} it is easy to compute spatial part of $\Delta_{\mu \nu}$ in higher than two dimensions
\al{
\Delta_{ij} & = -2 \Gamma^k_{ij}  \partial_k \partial_m \xi^m + \Gamma^k_{m j} \partial_i \partial_k \xi^m + \Gamma^k_{m i} \partial_k \partial_j \xi^m \notag\\ & = \partial_k \partial_n \xi^m ( -2 \Gamma^k_{ij} \delta_m^n + \Gamma^k_{m j} \delta_i^n + \Gamma^k_{m i}  \delta_j^n) 
}
Only in the 2D case we automatically obtain $\Delta_{ij}=0$ since all the spatial indices above reduce to $1$. In higher dimensional case there is no obvious way to solve $\Delta_{ij}=0$. 

Now this calculation shows that the transverse theory that we took is spatially diffeomorphism invariant, but the diff field itself admits a correction to become a spatial tensor only in 2D spacetime. Extension of this result to higher dimensions requires the modification of the very first step of the transverse formalism. Namely, the higher dimensional lift  \eqref{Liderranktwowithcentral} of the coadjoint transformation 
\al{ \label{Liderranktwowithcentralrevisit}
\delta_{\xi} D_{\mu \nu} = (\partial_{\mu} \xi^{\lambda})D_{\lambda \nu} + (\partial_{\nu} \xi^{\lambda}) D_{\mu \lambda} + \xi^{\lambda} (\partial_{\lambda} D_{\mu \nu} )  +  q \partial_{\mu} \partial_{\nu} \partial_{\lambda} \xi^{\lambda}
}
We will investigate this in the next section. 
\subsection{Full Covariance Recovered in Interactions}
\label{DiffInvRecovered}
The analysis in the previous section suggests that the higher dimensional lift \eqref{Liderranktwowithcentralrevisit} of the coadjoint transformation 
is too strict for the diff field to be complemented with a correction making it into a spatial  tensor. Although \eqref{Liderranktwowithcentralrevisit} is the most straightforward lift, 
what if we change it to include second-order center terms as well?   In other words, we can introduce the modified Lie derivative
\al{ \label{NewVariationofDiffTensor}
\delta_{\xi} D_{\mu \nu} = (\partial_{\mu} \xi^{\lambda})D_{\lambda \nu} + (\partial_{\nu} \xi^{\lambda}) D_{\mu \lambda} + \xi^{\lambda} (\partial_{\lambda} D_{\mu \nu} )  +  q \partial_{\mu} \partial_{\nu} \partial_{\lambda} \xi^{\lambda}  + (q/2) \Delta_{\mu \nu}
}
with 
\al{
\Delta_{\mu \nu} & = -2 \Gamma^{\rho}_{\mu \nu} \partial_{\rho} \partial_{\sigma} \xi^{\sigma} +  \Gamma^{\rho}_{\sigma \nu}  \partial_{\mu} \partial_{\rho} \xi^{\sigma} +  \Gamma^{\rho}_{\sigma \mu} \partial_{\rho} \partial_{\nu} \xi^{\sigma}
}
The modified Lie derivative \eqref{NewVariationofDiffTensor} reduces to
\al{
\delta D = \xi D' + 2\xi' D + q \xi'''
} 
in 1D. In 2D curved spacetimes, and in flat spacetimes of any dimensions, it automatically yields $\delta D = \delta D_{ij}$. 

 Moreover, the object
\al{
T_{\mu \nu} \equiv D_{\mu \nu} - \frac{q}{2} \Sigma_{\mu \nu}
}
is now a spacetime tensor in any dimensions; not just a spatial tensor. Hence, with this new proposal we can recover full diffeomorphism invariance at least in  interactions. We will investigate the transverse action obtained from \eqref{NewVariationofDiffTensor} in the next section.

Now we apply the same procedure as in Section \eqref{Virasoro Gauge-Fixing} to obtain the spatial reduction of the field components\footnote{Linear central extension is ignored for simplicity, though its presence does not affect the arguments below.}. In addition to the conditions 
\al{
\xi^0 = 0 = \partial_0 \xi^k
}
we need $\Gamma^m_{k0}=0$ to make $D_{0i}$ a spatial covariant vector and $\Gamma^m_{00}=0$ to make $D_{00}$ a spatial scalar. Hence the analog of \eqref{nogaugefixingtrfs} is given by
\al{
 \xi^0 & = 0 = \partial_0 \xi^k \ \ , \ \ \Gamma^m_{k0}= 0 = \Gamma^m_{00} 
 \\
\delta D_{ij} & = \xi^k \partial_k D_{ij} + \partial_i \xi^k D_{kj} + \partial_j \xi^k D_{ik} + q \partial_i \partial_j \partial_k \xi^k 
\notag\\
& + (q/2) ( -2 \Gamma^m_{ij} \partial_m \partial_k \xi^k + \Gamma^m_{kj} \partial_i \partial_m \xi^k + \Gamma^m_{ki} \partial_j \partial_m \xi^k ) 
\\
  \delta D_{0i} & = \xi^k \partial_k D_{0i} + \partial_i \xi^k D_{0k} 
\\
 \delta D_{00} & = \xi^k \partial_k D_{00}
}

Finally, we can introduce the analog of the full temporal gauge set \eqref{lastset}
\al{
 \xi^0 & = 0 = \partial_0 \xi^k \ \ , \ \ \Gamma^m_{k0}= 0 = \Gamma^m_{00} 
 \\
\delta D_{ij} & = \xi^k \partial_k D_{ij} + \partial_i \xi^k D_{kj} + \partial_j \xi^k D_{ik} + q \partial_i \partial_j \partial_k \xi^k 
\notag\\
& + (q/2) ( -2 \Gamma^m_{ij} \partial_m \partial_k \xi^k + \Gamma^m_{kj} \partial_i \partial_m \xi^k + \Gamma^m_{ki} \partial_j \partial_m \xi^k ) 
\\
  \delta D_{0i} & =0 =  D_{0i} 
\\
 \delta D_{00} & = 0 =  D_{00}
}
The change in $\delta_{\xi} D_{\mu \nu}$ will change the expression of the diff-Gauss law and as a result the diff momentum. 

\subsection{Modified Transverse Action }
\label{ModifiedTransverseAction}
We introduce a conjugate momentum $X^{ij}$ to $D_{ij}$ and form the diff-Gauss law. We can use the shortcut prescription \eqref{diffGaussshortcutpres} to get the modified diff-Gauss law
\al{
G^0_k & = X^{ij} \partial_k D_{ij} + \partial_i X^{ij}  D_{kj} + \partial_j X^{ij}  D_{ik} + q \partial_i \partial_j \partial_k X^{ij} \notag\\ 
& + (q/2) ( -2 \Gamma^m_{ij} \partial_m \partial_k X^{ij} + \Gamma^m_{kj} \partial_i \partial_m X^{ij} + \Gamma^m_{ki} \partial_j \partial_m X^{ij}) 
}
Since the $D$ dependent terms of the diff-Gauss law are the same, the Lie derivative \eqref{deltaXQ} of $X^{ij}$ is unchanged. Hence, $X^{ij}$ is still a spatial tensor density and 
\al{
\tilde{X}^{ij} = \sqrt{h} X^{ij}
}
is a rank-two spatial tensor. 
We apply the prescription \eqref{transverseactionprescriptionrevisit} to find the diff-Lagrangian involving the diff momentum. We introduce the same ansatz \eqref{Xansatz} for the diff momentum. We insert this ansatz into the action and recompute the momentum. The result can be quickly found by \eqref{shortcutmomentumprescription}
which yields 
\al{
X_{ij0} & = \partial_0 D_{ij} +  D_0^{\ k} \partial_k D_{ij} + \partial_i D_0^{\ k} D_{kj} + \partial_j D_0^{\ k} D_{ik} + q \partial_i \partial_j \partial_k D_0^{\ k}
\notag\\
& + (q/2) ( -2 \Gamma^m_{ij} \partial_m \partial_k D_0^{\ k} + \Gamma^m_{kj} \partial_i \partial_m D_0^{\ k} + \Gamma^m_{ki} \partial_j \partial_m D_0^{\ k} ) 
}
Finally, we get a spatially covariant Lagrangian density
\al{
\mathcal{L} = \frac{1}{2} h X_{ij0} X^{ij0}
}

Could this be the most general spatially covariant action \eqref{SpatiallyDiffInvAction} that we were looking for? It may be, but we can not say for sure (again due to the problem of symmetric criticality). In other words, there would be other possibilities that would yield back the same flat spacetime content. However, at least we have a candidate at hand. Moreover, we obtained a full diffeomorphism invariant extension of the diff field itself to be used in  interactions, and consistency  requires the modified diff variation \eqref{NewVariationofDiffTensor}, thereby the modified action. 

We finalize our study of transverse formalism at this point. For future research on transverse theory and recovering full diffeomorphism invariance of the diff Lagrangian we can suggest three directions. The first is a careful investigation of ADM decomposition of general relativity with the intention of reversing it. Namely, one needs structures analogous to the extrinsic curvature and lapse and shift functions to lift the hypersurface theory to the full spacetime theory.  
The second is examining the covariant proposal made in \cite{lanophd}, and how it may be related to the tranverse formalism or how it can be modified to be compatible with the transverse formalism. 
Finally, the third is application of projective connections in the transverse formalism context. This may provide a mechanism to recover the spacetime diff field theory from the theory on the spatial hypersurface. 
In the next chapter we will investigate alternative geometric/topological approaches for the diff field including the projective connection proposal.

\chapter{ALTERNATIVE IDEAS AND FUTURE DIRECTIONS}

\section{Euler-Poincare Theory of  Diff Field }
\label{DiffEulerPoincare}

Here we are going to introduce a canonical formalism called the Euler-Poincare (EP) formalism associated with Lie algebras. The main references for this section are \cite{marsdenratiu}, \cite{marsdenratiupaper}.

Let $\mathfrak{g}$ be a  Lie algebra and let $\xi : \mathbb{R} \rightarrow \mathfrak{g}$, $\xi(t) \in \mathfrak{g}$. The Lagrangian is taken to be a function of $\xi$ and the variations to be considered are of the form 
\al{ \label{ep1}
\delta_{\eta} \xi = \dot{\eta} + [\eta , \xi] = \dot{\eta} + \text{ad}_{\eta}  \xi
}
Then the principle of least action 
\al{ \label{ep2}
0 = \delta S = \delta \int dt \ L
}
leads to the so called Euler-Poincare (EP) equations 
\al{ \label{EPgenericeqn}
\frac{d}{dt} \frac{\delta L}{\delta \xi} = \text{ad}^*_{\xi} \frac{\delta L}{\delta \xi} 
}
Let us prove this statement
\al{
0 & = \delta \int dt \  L \notag\\ & = \int dt \ \frac{\delta L}{\delta \xi} \delta \xi \notag\\ & = \int dt \ \frac{\delta L}{\delta \xi} ( \dot{\eta} + \text{ad}_{\eta} \xi ) \notag\\ & = \int dt \ \frac{\delta L}{\delta \xi} ( \dot{\eta} - \text{ad}_{\xi} \eta ) \notag\\ & = \int dt  \  \left( - \frac{d}{dt}  \frac{\delta L}{\delta \xi} \right) \eta - \int dt \ \frac{\delta L}{\delta \xi} \text{ad}_{\xi} \eta 
}
Notice that $\delta L / \delta \xi$ is a coadjoint vector so the integrand is the pairing between an adjoint and a coadjoint element. Then we can use invariance of the pairing to rewrite the second term
\al{
0 & = \int dt  \  \left( - \frac{d}{dt}  \frac{\delta L}{\delta \xi} \right) \eta + \int dt \ \text{ad}^*_{\xi} \left(\frac{\delta L}{\delta \xi}\right) \eta \notag\\ & = \int dt \  \left(  - \frac{d}{dt}  \frac{\delta L}{\delta \xi} + \text{ad}^*_{\xi} \frac{\delta L}{\delta \xi}\right) \eta
}
For arbitrary $\eta$ that vanishes at the end points we recover EP equations \eqref{EPgenericeqn}.

Now let us consider the Virasoro algebra and the diff field. What Lagrangian will we pick for the diff theory? Diff field is a Virasoro coadjoint element, so we need to form its pairing with an adjoint vector, as in the proof above. Which adjoint vector will we take? A clue comes from the rigid body theory where the conjugate momentum is obtained by varying the Lagrangian with respect to the main variable. Thus we may contract the diff field with its momentum $X$, which behaves in 1D as an adjoint vector, \eqref{deltaXQ1D}. Hence, we pick 
\al{
L = \left< D | X \right> 
}
Then the EP equation \eqref{EPgenericeqn} reads
\al{
\frac{d}{dt} \frac{\delta L}{\delta X} = \text{ad}^*_X \frac{\delta L}{\delta X}
}
which yields
\al{ \label{EP21} 
\dot{D} = \text{ad}^*_X D =  XD' + 2 X' D + q X'''
}

How to interpret \eqref{EP21}? Let us take $X$ to be in the isotropy algebra for $D$ so that 
\al{
0 = \text{ad}^*_X D=XD' + 2 X' D + q X'''
}
This implies $\dot{D}=0$. This is a strange situation from the perspective of transverse formalism. When the diff momentum is taken in isotropy algebra so that it represents an infinitesimal motion that is transverse to the orbit, diff field freezes. Hence,  dynamics described by the theory should be completely on the orbit. However, on the orbit $D$ is fixed by diffeomorphisms. Therefore such a theory describes evolution between distinct but diffeomorphic field configurations for $D$. So this is a theory on the coadjoint orbit rather than transverse to it. A coadjoint transformation is not obtained as a gauge transformation (generated by the diff-Gauss law). Rather, it corresponds to  the  dynamical evolution. 
 
If we take $X=D$ as functions, the field equation \eqref{EP21} becomes 
\al{
\dot{D}  - 3DD' - q D''' = 0
}
Rescaling time $\tau(t) = -t/2$, we reach KdV equation for $q=1/2$ 
\al{
D_{\tau} + 6 D D_{\sigma} + D_{\sigma \sigma \sigma} = 0
}

Finally, if instead of $X$ we paired $D$ with $N$, the Lagrange multiplier of the diff-Gauss law in the transverse formalism,  then the field equation would read 
\al{
\dot{D} = ND' + 2 N' D + q N'''
}
which corresponds to vanishing of diff momentum in the transverse theory. This is compatible with the argument above that the EP theory of diff field does not provide dynamics transverse to orbits.

\section{Diff-Wilson Loop}
\label{diffWilson}
Rajeev \cite{rajeev88} discusses the finite reduction of Yang-Mills (YM) theory on a cylinder ($\mathbb{S}^1\times\mathbb{R}$) using Wilson loop methods. This has been summarized in Section \ref{Rajeev88Summary}. In particular, the Wilson line is used  to solve the Gauss law (equation \eqref{YMwilsonloopsolved}). 

This method has been imitated in the case of Virasoro algebra in references \cite{LR95} and \cite{hendersonrajeev} in a different way. The former is reviewed in Section \ref{ReviewDXGravity}. Since  diff field is a Virasoro coadjoint element,  the Wilson loop associated with  Virasoro coadjoint representation will be called the diff-Wilson loop. In this section we are going to examine  diff-Wilson loop more carefully.

In the  Virasoro case things are  more complicated due to the higher order local transformation. Hence, we need to understand the properties of the Wilson loop in a form that is suitable for generalization to a field theory related to the Virasoro algebra. The main references for this section are \cite{rajeev88},\cite{NairQFT}, \cite{LR95} and \cite{hendersonrajeev}. 
\subsection{Understanding Wilson Loop } 
\label{UnderstandingtheWilsonloop}
In gauge theories, the configuration space is the space of gauge fields $\mathcal{A}$ modulo the space of gauge transformations $\mathcal{G}$. However, the quotient is not well-defined for the full gauge group $\mathcal{G}$ because, in general, it acts on $\mathcal{A}$ with fixed points, i.e. there are nonidentity gauge transformations that fix an arbitrary gauge field. Only the subgroup $\mathcal{G}_0 \subset \mathcal{G}$ of gauge transformations that are equal to identity at $x=0$ acts without fixed points. So the configuration space is taken as $\mathcal{A} / \mathcal{G}_0$ which is a smooth manifold and the wavefunctions of the gauge theory can be viewed as functions on this space. 

On circle $S^1$, this quotient space is finite-dimensional; the only gauge invariant observable is the Wilson loop $W[A]$ around the circle \cite{rajeev88}. To define this object we first need to solve the parallel transport equation 
\al{ \label{Wilsonloopequation}
\frac{d\psi}{dx} + e A(x) \psi(x) = 0 
}
or 
\al{ \label{gaugecovderfundrep}
\nabla_A \psi (x) = 0 
}
where $\nabla_A$ is the covariant derivative with the gauge connection $A$. 

This equation is invariant under the simultaneous transformations  
\begin{subequations} \label{Wilsoninvarianceequations}
\al{ \label{gaugetrfofpsi}
& \psi(x) \mapsto \psi^g(x) = g(x) \psi(x)  \\
& A(x) \mapsto A^g (x) = g(x) A(x) g^{-1} (x) + e^{-1} g(x) dg^{-1} (x)
\label{gauge trf vector potential}}
\end{subequations}
as follows 
\al{
\nabla_{A^g} \psi^g & = (\psi^g)' + e A^g \psi^g \notag\\ & = g' \psi + g \psi' + e \left( g A g^{-1} - e^{-1} g' g^{-1} \right) g \psi \notag\\ & = g ( \psi' + e A \psi) = 0 
}

Now, although $A(x)$ is periodic, the solution $\psi(x)$ of \eqref{gaugecovderfundrep} is, in general, not periodic (so is not a well-defined function\footnote{See \cite{NairQFT} pages 183-184 for path-dependence of the Wilson loop leading to the standard path-ordered exponential definition of the Wilson loop.} on $S^1$). Instead the solution satisfies 
\al{ \label{hendrajWilsonloopdefined}
\psi (2 \pi) = W[A] \psi(0) 
}
where $W: \mathcal{A} \rightarrow G$ is the parallel transport operator, or the Wilson loop. For $g \in \mathcal{G}_0$ we have $g(0) = g(2 \pi) = I$. Then using \eqref{gaugetrfofpsi} we get  $\psi^g(2 \pi) = \psi(2 \pi)$, and $\psi^g(0) = \psi(0)$. Together these imply that 
\al{
W[A^g] = W[A]
}
i.e. the Wilson loop is gauge-invariant.

Note that, in the notation of Section \ref{Rajeev88Summary}, $\psi = S$, $e=1$ and the boundary condition $S(0)=1$ implies $S(2 \pi) = W[A]$.

\subsection{Virasoro Covariant Derivatives }
\label{Virasorocovariantderivatives}

The main reference for this section is  \cite{scherer88}. 
First, we will introduce some more structure related to the Virasoro algebra.  
Let $\widehat{\text{G}}$ denote the central extension of the group  $\text{G}=\text{Diff}(S^1)$ of orientation-preserving diffeomorphisms of $S^1$. Let $\Omega_p (S^1)=\{ \psi = \psi(\theta) (d \theta)^p \} $ be the space of densities of weight $p \in \mathbb{R}$ on $S^1$. These densities form a representation $R^{(p)}$ of G, with  the action of a diffeomorphism $f \in \text{G}$ given by 
\al{\label{Scherer21}
R^{(p)}_{f} \psi = ( f')^p \ \psi \circ f
}

The algebra of $\widehat{\text{G}}$ is given by $\widehat{\mathbf{g}}=\text{Vect}(S^1) \oplus \mathbb{R}$, and is called  the Virasoro algebra and its (regular) dual is denoted by $\widehat{\mathbf{g}}^*$, which can be identified with $\Omega_2(S^1) \oplus \mathbb{R}$ i.e. quadratic differentials together with real center.

Recall that the (active) coadjoint action of the group is given by 
\al{ \label{Scherer11}
\text{Ad}^*_{f} (u, b)  = \Big( (f')^2 \ u \circ f + b \ S(f) \ , \ b \Big)
}
where $(u,b)\in \widehat{\mathbf{g}}^*$ , $f \in \text{G}$ and $S(f)$ is the Schwarzian of $f$.

We would like to construct covariant derivatives associated with the coadjoint element $u$. 
We may, without loss of generality, consider only elements of the form $\widehat{u} = ( u, 1) \in \widehat{\mathbf{g}}^*$. 
For $k \in \mathbb{N}$ we may tensor $R^{((1-k)/2)}$ with the coadjoint representation  to get the following representation on $\widehat{\mathbf{g}} \otimes \Omega_{(1-k)/2} (S^1)$,
\al{ \label{Scherer22}
T^{((1-k)/2)} \equiv \text{Ad}^* \otimes R^{((1-k)/2)}
}

The Hill operator, 
\al{ \label{Scherer23}
\nabla^{(2)}_{\hat{u}} \equiv \frac{d^2}{d \theta^2} + u (\theta) 
}
maps densities of weight $-1/2$ to densities of weight $3/2$ for each $\widehat{u} \in \widehat{\mathbf{g}}$. 
It can be viewed as part of the mapping $\nabla^{(2)}$ defined as 
\al{ \label{Scherer24}
\nabla^{(2)} : \widehat{\mathbf{g}} \otimes \Omega_{-1/2} (S^1)  \rightarrow \Omega_{3/2} (S^1) : \widehat{u} \otimes \psi  \mapsto \nabla^{(2)}_{\widehat{u}} \psi
}
Hence, the mapping $\nabla^{(2)}$, connecting the representations $T^{(-1/2)}=\text{Ad}^* \otimes  R^{(-1/2)}$ and $R^{(3/2)}$, is a covariant differential operator. Explicitly, we have 
\al{ \label{Scherer25}
\nabla^{(2)}_{\text{Ad}^*_{f} \widehat{u}} = (f')^{3/2} \left( \frac{1}{f'^2} \frac{d^2}{d \theta^2} - \frac{f''}{(f')^3} \frac{d}{d \theta} + u ( f(\theta)) \right) (f')^{1/2} 
}

In general, we have operators
\al{ \label{Scherer32}
\nabla^{(k)}_{\widehat{u}} : \Omega_{(1-k)/2} (S^1) \rightarrow \Omega_{(1+k)/2} (S^1)
}
which can be seen as part of the mappings 
\al{ \label{Scherer33}
\nabla^{(k)}  : \widehat{\mathbf{g}} \otimes \Omega_{(1-k)/2} (S^1) \rightarrow \Omega_{(1+k)/2} (S^1) : \widehat{u} \otimes \psi  \mapsto \nabla^{(k)}_{\widehat{u}} \psi
}

For our purposes the other interesting case is the operator for $k=3$ given by\footnote{For  generic $k\in \mathbb{N}$,  see \cite{scherer88}.} 
\al{ 
\nabla^{(3)}_{\widehat{u}}  = \frac{d^3}{d \theta^3} + 4 u \frac{d}{d \theta} + 2 u' \label{D3} \ \ : \Omega_{-1}(S^1) \rightarrow \Omega_2 (S^1)
}
mapping vectors to quadratic differentials. So the operator, $\nabla^{(3)}_{\widehat{u}} : \Omega_{-1} (S^1) \rightarrow \Omega_2 (S^1)$,  acts on a Virasoro adjoint element and yields a Virasoro coadjoint element. This is the analog of the covariant derivative acting on the gauge potential and used to build the YM action.

Note that, one can use two vector fields (adjoint vectors) $\xi, \eta \in \text{Vect}(S^1)$ to get an invariant using the pairing, 
\al{
(\xi, \eta)   \overset{u}{\rightarrow} \left< \xi, \nabla_u^{(3)} \eta \right> \in \mathbb{R}
} 
In fact the Kirillov form $\Omega$ can be rewritten in terms of $\nabla^{(3)}$ as
\al{
\frac{1}{2} \left( \left< \nabla^{(3)}_u \eta \, ,\, \xi \right> - \left< \nabla^{(3)}_u \xi \, ,\, \eta \right> \right) & = \int d\theta \, u ( \xi \eta' - \xi' \eta) + \frac{q}{2} \int d\theta \, (\xi \eta''' - \xi''' \eta) \notag\\ & = \left< (u,0) \, , \, [ (\xi,0),(\eta, 0) ] \right> - q c( \xi , \eta ) \notag\\ & = \left< (u, q)  \, , \, [ (\xi,0),(\eta, 0) ] \right> \notag\\ & = \Omega \left( (u,q) \, \, [ (\xi, 0) , (\eta, 0) ] \right) 
}

As a final note, let us state that the solutions of $\nabla_u^{(3)} g = 0$ can be written as a product of the solutions of $\nabla_u^{(2)}f=0$. In other words, if $f_1, f_2$ are independent solutions of $\nabla_u^{(2)}f=0$. Then the general solution for $\nabla_u^{(3)} g = 0$ is given by \cite{hendersonrajeev} 
\al{
g= a f_1^2 + b f_1 f_2 + c f_2^2
}

\subsection{Wilson Loop Associated with Hill Equation  }
In this section we will form the Virasoro analog of the Wilson loop operator \cite{hendersonrajeev} discussed in Section \ref{UnderstandingtheWilsonloop}.
The analogue of the gauge field space $\mathcal{A}$ is given by the space of coadjoint elements (or quadratic differentials) $\mathcal{U} = \{ u :S^1 \rightarrow \mathbb{R} \}$. The full gauge group $\mathcal{G}$ is replaced by the group Diff$(S^1)=\{f  : S^1 \rightarrow S^1 \}$ of diffeomorphisms of the circle. It acts on $\mathcal{U}$ by\footnote{In \cite{hendersonrajeev}  adjoint and coadjoint elements include linear center as well. We take  linear centers vanishing here i.e. we use Gelfand-Fuchs cocyle.} 
\al{ \label{hendrajcoadjoint}
u (\theta) \mapsto  f\circ u (\theta) = u (f(\theta)) f'^2 (\theta) +q Sf (\theta) 
}
where $q$ is the central charge of $u$. 

As in the case of the gauge field, the full gauge group acts with fixed points (i.e. there are non-identity elements that fix the coadjoint element or the isotropy group is nontrivial). Analogue of the true gauge group $\mathcal{G}_0$ is the subset Diff$_0(S^1) \subset \text{Diff}S^1$ containing diffeomorphisms satisfying 
\al{ \label{Diff0S1diffeoconditions}
f(0) = 0 \ , \ f'(0) = 1 \ , \ f'' (0) = 0
}
The proof for the infinitesimal diffeomorphisms is as follows \cite{hendersonrajeev}. For infinitesimal diffeomorphisms of the form $f(\theta) = \theta - \xi(\theta)$ invariance $f \circ u = u$ reduces to 
\al{ \label{infinitesimalfixingDiff0S1}
u' \xi + 2 u \xi' +q \xi''' = 0
}
Conditions \eqref{Diff0S1diffeoconditions} translate to the conditions $\xi(0)= \xi'(0) = 0 = \xi''(0)$ for the generator. Inserting these in \eqref{infinitesimalfixingDiff0S1} and evaluating at $\theta=0$ we get $\xi'''(0)=0$.  By taking derivatives of \eqref{infinitesimalfixingDiff0S1}, evaluating at $\theta=0$, and using the conditions obtained repeatedly, one reaches $\xi^{(n)} (0)=0$ for all $n \in \mathbb{Z}$. This implies $\xi(\theta) = 0$. Therefore, the only infinitesimal diffeomorphism $f$ fixing a generic element is the identity $f(\theta) = \theta$. 

Assuming this result can be generalized to finite diffeomorphisms,  analog of the configuration space $\mathcal{A} / \mathcal{G}_0$ of the gauge theory becomes $\mathcal{U}/\text{Diff}_0(S^1)$. 

Analog of the differential equation \eqref{Wilsonloopequation} defining the Wilson loop  can be taken as the Hill equation associated with the coadjoint element $(u, q)$
\al{ \label{hendrajHillequation}
\nabla^{(2)}_u \psi \equiv \psi '' +  \frac{1}{2q} u \psi =0
} 
Analogous to the invariance of  equation \eqref{Wilsonloopequation} under the combined transformations \eqref{Wilsoninvarianceequations} of the gauge field and the Wilson line, Hill equation \eqref{hendrajHillequation} is invariant if $u$ transforms as \eqref{hendrajcoadjoint} and $\psi$ transforms as a density of weight $-1/2$ (equation \eqref{Scherer21} with $p=-1/2$). This is implied by \eqref{Scherer25} when we set $\nabla_u^{(2)} \psi = 0$. 

We have shown in Section \ref{Virasorocovariantderivatives} that   Hill operator is the first nontrivial member of a sequence of covariant derivatives, and it can be shown \cite{hendersonrajeev} that the solutions of the equation $\nabla^{(2s+1)}_u \psi = 0$ are just products of the $2s$ solutions of $\nabla^{(2s)}_u  \psi = 0$, so $-1/2$ densities are the analogues of the fundamental representation and the  $\nabla^{(2)}_u$ is the analog of the covariant derivative for the fundamental representation.

Just as in the case of the ordinary Wilson loop,  solutions to the Hill's equation 
\al{
\psi'' + \frac{1}{2q} u \psi = 0
}
are, in general, not periodic. A basis $\varphi_1, \varphi_2$ of solutions will change by a linear transformation $M[u]$ as one goes from $\theta = 0 $ to $\theta = 2 \pi$ : 
\al{ \label{HillMonodromy}
\left( \begin{array}{cc} \varphi_1 (2 \pi) \\ \varphi_2 ( 2 \pi) \end{array} \right) =  M[u] \left( \begin{array}{cc} \varphi_1 (0) \\ \varphi_2 ( 0) \end{array} \right) 
}
Taking a standard basis satisfying the conditions 
\al{ \label{HendRajBasisInitialCond}
\varphi_1 ( 0) = 0, \ \varphi'_1 (0)=1  \ \ , \ \ \varphi_2 ( 0) = 1, \ \varphi'_2 (0)=0 
}
the matrix $M[u]$ becomes
\al{\label{Rajeevmonodromy}
M[u] = \left( \begin{array}{cc} \varphi'_1(2 \pi) & \varphi_1 (2 \pi) \\ \varphi'_2 (2 \pi) & \varphi_2 (2 \pi) \end{array} \right) 
}
and it is invariant under Diff$_0 (S^1)$, as we show in the Appendix \ref{InvarianceHillWilson}. This matrix, also called the monodromy matrix of the Hill operator,  is the analogue of the Wilson loop $W[A]$ which is invariant under the action of $\mathcal{G}_0$. 

\subsection{Diff-Wilson Loop for $\nabla^{(3)}$ } 
\label{DiffWilsonnabla3}

In the previous section we  obtained the analog of the Wilson loop associated with the Hill operator $\nabla^{(2)}_u$ which maps $-1/2$ densities to $3/2$ densities. Although $\nabla^{(2)}$ is considered as the analog of the covariant derivative in the fundamental representation, Hill operator is not what we are looking for to build a theory of the diff field.
In fact, the only operator that can be used to build a theory of the diff field is $\nabla^{(3)}_D$ as the diff-Gauss law  operator can be rewritten as  
\al{
G = \delta_X D = \nabla^{(3)}_D X
}
Hence, we need to generalize the analysis of the previous section to $\nabla^{(3)}_D$  and obtain its monodromy matrix. 

Consider the equation 
\al{ \label{C1}
\nabla^{(3)}_D \psi = D' \psi + 2 D \psi' +q \psi''' = 0 
}
associated with the Virasoro covariant derivative $\nabla^{(3)}_D$ mapping $-1$ density (vector or adjoint element) to $2$ density (quadratic differential or coadjoint element ). Explicitly, 
\al{
\nabla^{(3)}_D \overset{\phi}{\rightarrow} (\phi')^{-2} \,  \nabla^{(3)}_{D_{\phi}} \, ( \phi')^{-1}
}
where $\phi \in \text{Diff}_0(S^1)$, and  
\al{
D_{\phi} (x) = \phi'(x)^2 D(\phi(x)) + q \, S \phi (x)
}

Introduce a basis for the solution
\al{ \label{C2} 
\psi(x) = \left( \begin{array}{ccc} \varphi_1 (x) \\ \varphi_2 (x) \\ \varphi_3 (x)  \end{array} \right) 
}
satisfying the conditions
\begin{subequations} \label{C3}
\al{ 
\varphi_1 (0 ) = 1 \ , \ \varphi_1'(0) = 0 \ , \ \varphi'' (0) = 0 \\ 
\varphi_2 (0) = 0 \ , \ \varphi_2'(0)=1 \ , \ \varphi_2''(0)=0 \\ \varphi_3 (0) = 0 \ , \ \varphi_3'(0)=0 \ , \ \varphi_3''(0)=  1 
}
\end{subequations}

The analog of  \eqref{HillMonodromy}  can be written as 
\al{ \label{C5} 
\psi(2\pi) & = M[D] \psi(0) 
}
Using the initial conditions \eqref{C3} we recover
\al{ \label{C6}
M[D] = \left( \begin{array}{ccc} \varphi_1 (2 \pi) & \varphi_1'(2 \pi) & \varphi_1''(2 \pi) \\ \varphi_2 (2 \pi) & \varphi_2'(2 \pi) & \varphi_2''(2 \pi) \\ \varphi_3 (2 \pi) & \varphi_3'(2 \pi) & \varphi_3''(2 \pi)  \end{array} \right) 
}

Now, $\psi$ transforms in the adjoint 
\al{ \label{C7}
\phi \circ \psi (x) = \psi( \phi(x)) [\phi' (x) ]^{-1} 
}
and $\phi \in \text{Diff}_0 (S^1)$ so 
\al{ \label{C8}
\phi(0) = 0 \ , \ \phi' (0) = 1 \ , \ \phi'' (0) = 0 \ , \ \phi'''(0) = 0 
}
and 
\al{ \label{C9}
\phi(x+ 2\pi) = \phi(x) + 2 \pi \ \ , \ \ \ \phi^{(n)} (x+2\pi) = \phi^{(n)} (x)
}
Evaluating \eqref{C7} at $x=0$ and $x=2 \pi$ and using \eqref{C9} it is straightforward to show
\al{ \label{C10}
(\phi \circ \psi) (2 \pi) = M[D] (\phi \circ \psi) (0)
}
i.e. 
\al{ \label{C11}
M[D^{\phi}] = M[D]
}

\subsection{First Type Coadjoint Orbits of Virasoro Algebra  } \label{firsttypeorbits}
Classification of coadjoint orbits of the Virasoro algebra has been done in \cite{lazutkin},   \cite{segalunitary}, and \cite{witten88}. Here we will follow \cite{witten88}.  It turns out that  classification is achieved by solving the isotropy equation\footnote{This is because a coadjoint orbit is obtained as the quotient of the group by the isotropy group for the orbit.} for a coadjoint element $(D(\theta),c)$,
\al{\label{iso}
0& =\delta_f D = f D' + 2f' D + q f''' \notag\\ 
q & \equiv -c /24 \pi
}
given $D(\theta)$,  with the requirement of periodicity of $f(\theta)$ and its derivatives, and up to the action of diffeomorphisms for both $f$ and $D$. We will restrict our attention to the simplest types of orbits, namely, for $D$ diffeomorphic to a constant. These are sometimes called the first type orbits \cite{LR95}.

 For  $D= D_0=\text{constant} \neq 0 $, \eqref{iso} reduces to 
\al{\label{iso1}
0 = 2f' D_0 + q f'''
}
Defining $g \equiv f'$ and
\al{
\omega^2 \equiv \frac{2D_0}{q}  = - \frac{48 \pi D_0}{c} 
} 
\eqref{iso1} becomes 
\al{
g'' = - \omega^2 g
}
whose solution is 
\al{
g = a \cos (\omega \theta) + b \sin (\omega \theta)
}
This implies 
\al{
f (\theta) = c_0 + c_1 \cos (\omega \theta) + c_2 \sin (\omega \theta)
}
for arbitrary constants $c_0 , c_1$ and  $c_2$. 

Periodicity, $f(2 \pi) = f(0)$ implies the condition
\al{ \label{cond}
c_2 \sin (2 \pi \omega) = c_1 ( 1 - \cos (2 \pi \omega))
} 
This condition motivates the following distinct cases : 

i) $\omega = n \in \mathbb{Z}-0$. Then \eqref{cond} is automatically satisfied and we have 
\al{ \label{stabZ}
f (\theta) = c_0 + c_1 \cos (n \theta) + c_2 \sin (n \theta)
}
and the vector $f \, d/d\theta$ stabilizes the coadjoint element
\al{ \label{Dinst}
D_0 = - \frac{nc}{4 8 \pi} 
}
The stabilizer (or isotropy group) is generated by \eqref{stabZ} which is a linear combination of $L_0 , L_n, L_{-n}$ where $L_m \equiv i e^{im \theta} d/d\theta$. For any $n \in \mathbb{Z}-0$, these generate $\text{SL}(2, \mathbb{R})$. Hence, in this case the coadjoint orbit is given by 
\al{
\text{Orb}(D_0) \cong \text{Diff} \, S^1/\text{SL}(2, \mathbb{R})^{(n)}
}

ii) $\omega \notin \mathbb{Z}-0$. Then arbitrarines of $D_0$ up to diffeomorphisms requires $c_1 = c_2 = 0$ and we are left with 
\al{
f = c_0 
}
and 
\al{
D_0 = \text{constant} \neq  - \frac{nc}{4 8 \pi} 
}
Thus the stabilizer is generated by $L_0$. It is the group of  rigid rotations of $S^1$ which is isomorphic to $S^1$ itself. Hence, the coadjoint orbit for this case becomes
\al{
\text{Orb}(D_0) \cong \text{Diff} \, S^1/S^1
}

\subsection{Diff-Wilson Loop Proposed in \cite{LR95}}\label{DiffWilsonLR95} 
Recall the definition \eqref{hendrajWilsonloopdefined} of the Wilson loop. If instead of evaluating $\psi$ at $x=0$ and $x=2 \pi$, we evaluated it at two arbitrary locations $z,y$ on the circle we would get 
\al{ \label{rajhendWilsonlinedefined}
\psi(z) = W[z,y;A] \psi(y)
} 
$W[z,y;A]$ is the Wilson line. We can evaluate this equation at the gauge transformed configurations 
\al{
\psi^g(z) = W[z,y;A^g] \psi^g(y)
}
Now, using \eqref{gaugetrfofpsi} we can rewrite this equation as 
\al{
g(z) \psi(z) =  W[z,y;A^g] g(y) \psi(y)
}
or 
\al{
\psi(z) = g^{-1}(z)  W[z,y;A^g] g(y) \psi(y)
}
Comparing this with \eqref{rajhendWilsonlinedefined} we see that 
\al{
W[z,y,;A] = g^{-1}(z)  W[z,y;A^g] g(y)
}
or 
\al{
W[z,y;A^g] = g(z)  W[z,y;A] g^{-1}(y)
}
Hence, the Wilson line is gauge-covariant. 

In \cite{LR95} the diff-Wilson line is proposed to be $v(x)$ satisfying
\al{ \label{LR95diffWilsonrevisit}
D = c \, S(v)
}
The analog of the gauge transformation is the Virasoro coadjoint transformation 
\al{ \label{coadpassiverevisit}
D \overset{f}{\mapsto} D^f = f'^2 D \circ f + c \, S(f)
}
Although $v$ is not a diffeomorphism (nor is, in general, well-defined on the circle) we should be able to use the identity \eqref{Schwarzian id}, as long as $v$ is not the inverted map in the identity. With this assumption \eqref{coadpassiverevisit} becomes
\al{
D^f = c \, S(v \circ f)
}
Comparing this with \eqref{LR95diffWilsonrevisit} we see that 
\al{
v^f = v \circ f
}
where $v^f$ is the proposed  Wilson line for the transformed field $D^f$.

Now, the Wilson loop is obtained by evaluation at $x= 2\pi$, i.e., $Q \equiv v(2\pi)$. We see that invariance of the diff-Wilson loop, $Q^f(2 \pi) = Q(2 \pi)$, requires $f(2 \pi) = 1$. Since $f$ is a diffeomorphism, alternatively, we need $f(0)=1$. This is not satisfied for all the diffeomorphisms in Diff$S^1$ nor for those in the subset Diff$_0S^1$ defined in \eqref{Diff0S1diffeoconditions}. 

However, it is satisfied for all $f \in \text{Diff}S^1$ by the solutions \eqref{spectificsolutiondiffWilson} taken. These solutions are valid for the first-type orbits, with the additional trivial solution, $v(x)=1$.   The problem, however, is that  finite reduction of DX theory is obtained for only a single solution, namely, $v = \exp( i \alpha x)$, with $\alpha = \sqrt{2D_0 / c}$.  It is in this sense that application  of the Wilson loop method of \cite{rajeev88}   is incomplete in \cite{LR95}. 

In Section \ref{DiffWilsonnabla3} we obtained the complete diff-Wilson loop $M[D]$ that is invariant under Diff$_0 S^1$ extending the analysis of \cite{hendersonrajeev} for the Hill operator $\nabla^{(2)}$ to the differential operator $\nabla^{(3)}$. In the next section we evaluate $M[D]$ for the first type orbits. The analysis in Section  \ref{ReviewDXGravity} should be extended with the results of the next section. Namely, instead of a theory on a 2D phase space $(Q,P)$, one gets a theory on 6D a  phase space with $(Q_i, P_i)$, $i = 1,2,3$.

\subsection{Diff-Wilson Loop for $\nabla^{(3)}_u$ on First Type Orbits } 

Reconsider the Wilson loop equation 
\al{ \label{firsttypeWilsoneq}
\delta_{\psi} D = \nabla^{(3)}_D \psi = D' \psi + 2 D \psi' + q \psi''' = 0 
}
This is the same as the isotropy equation. Now, if we  restrict $D$ to first type orbits, discussed in Section \ref{firsttypeorbits}, the general solution reads
\al{
\psi(x) = c_0 + c_1 \cos(\omega x) + c_2 \sin(\omega x)
}
The solution basis satisfying the initial conditions \eqref{C3} is given by
\begin{subequations}
\al{
\varphi_1 (x) & = 1 \\
\varphi_2 (x) & = \sin (\omega x) / \omega \\ 
\varphi_3 (x) & = (1- \cos(\omega x)) / \omega^2 
}
\end{subequations}
Hence,  the Wilson loop \eqref{C6} can be evaluated as  
\al{  \label{FirsttypeorbitsCovariantThreeWilsonLoop}
M[D] = \left( \begin{array}{ccc} 1 & 0 & 0 \\ \frac{\sin(2 \pi \omega)}{\omega} & \cos(2 \pi \omega) & - \omega \sin (2 \pi \omega)  \\ \frac{1}{\omega^2} ( 1- \cos(2 \pi \omega)) & \frac{\sin (2\pi \omega)}{\omega} & \cos(2 \pi \omega)  \end{array} \right) 
}

Now consider the two subcases discussed in Section \ref{firsttypeorbits}. Given $\omega \neq 0$ : 

Case (i) : $\omega \in \mathbb{Z}$. In this case we get
\al{
M[D] = 1_{3 \times 3} \ \ \ \ \ \text{if} \ \  D \in \text{Diff}S^1/SL(2, \mathbb{R})^{(n)}
}
In other words, for Diff$S^1/SL(2, \mathbb{R})^{(n)}$ orbits the diff-Wilson loop becomes trivial. 

Case (ii) : $\omega \notin \mathbb{Z}$. In this case we get . 
\al{
M[D] \neq  1_{3 \times 3} \ \ \ \ \ \text{if} \ \ D \in \text{Diff}S^1/S^1 \hspace{0.7in} 
}
So for Diff$S^1/S^1$ orbits we have a nontrivial diff-Wilson loop. 

It is interesting to consider time evolution, say, restricted to constant $D$ orbits. In this case one can start with an $\omega \notin \mathbb{Z}$ and end up with $\omega \in \mathbb{Z}$ (as $\omega$ is determined by $D$). Then how should the  corresponding transition between the Wilson loops, namely, from a nontrivial matrix to the identity matrix be physically interpreted?

\section{Diff Field as a Projective Connection   } 
\label{TWdiffdynamicaltheory}
\subsection{The Proposal} 

In \cite{brensinger} authors proposed an alternative approach to obtain a dynamical theory of the diff field. By identifying the diff field 
with a component of a  Thomas-Whitehead (TW) projective connection \cite{thomas1}, \cite{thomas2}, \cite{whitehead}, the interaction term \eqref{Pol2Ddiffintterm}
\al{\label{Pol2DdiffTW}
S_{\text{int}} = \int d^2 x \, D_{--} \, h_{++}
}
appearing in the Virasoro geometric action \eqref{SDPolyakovf} can be recovered from
\al{
S_{\text{int}} = \int d^3 x\ \sqrt{\det(-G)} K_{\alpha \beta} G^{\alpha \beta}
}
where $G_{\alpha \beta}$ is the dimensionally-extended metric obtained from the 2D Polyakov metric, using the chiral Dirac matrix  $\gamma^3$  for defining the third dimension, and $K_{\alpha \beta} = K^{\rho}_{\ \alpha \rho \beta}$
is the Ricci tensor derived from the TW projective connection.  

The action governing the dynamics of the diff field is proposed to be
\al{
S_D = \int d^3 x \ \sqrt{\det(-G)} K^{\alpha}_{\ \beta \gamma \rho} K_{\alpha}^{\ \beta \gamma \rho} 
}

 Our focus in this thesis will be on the very first step, namely, examining the relationship between TW projective connections and the diff field (i.e. a Virasoro coadjoint element) in Section \ref{TWVirasororeln}.

\subsection{Projective Connections on $\mathbb{RP}^m$} 
\label{TWProjectiveSection}
We  follow the review \cite{saunders} for the Thomas-Whitehead (TW) projective connections \cite{thomas1}, \cite{thomas2}, \cite{whitehead} and the alternative version (based on volume bundle) \cite{robertsTW}.

Given a vector space $V$ the associated projective space $P(V)$ is the set of one-dimensional subspaces of $V$. $P(\mathbb{R}^{n+1})$ is  denoted $\mathbb{RP}^n$. 

To introduce TW projective connections we refer to an alternative description of $\mathbb{RP}^m$. 
As a manifold, $\mbb{RP}^m$ is the quotient of $\mbb{R}^{m+1}_0\equiv\mbb{R}^{m+1}-0$ under the multiplicative action of $\mbb{R}_0\equiv\mbb{R}-0$. The infinitesimal generator of this action is the radial vector field given in Cartesian coordinates by $x^{\alpha} \partial_{\alpha} = \Upsilon$. 

We may represent objects on $\mbb{RP}^m$ as objects on  $\mbb{R}^{m+1}_0$ transforming appropriately under the $\mbb{R}_0$ action;  this will be expressed in terms of the Lie derivative with respect to $\Upsilon$, together with invariance under the reflection map $j:x \mapsto -x$. 

Functions on $\mbb{RP}^m$ may be represented by functions $f$ on  $\mbb{R}^{m+1}_0$ satisfying 
\al{
\mathcal{L}_{\Upsilon} f = \Upsilon f = 0  \ \ \ \text{and} \ \ \ j^*(f) = f
}
Let us call the set of such functions $\mathcal{F}_{\Upsilon}$. 

Similarly, vector fields on $\mbb{RP}^m$ may be represented by equivalence classes of vector fields $X$ on  $\mbb{R}^{m+1}_0$ satisfying 
\al{
\mathcal{L}_{\Upsilon} X \propto \Upsilon \ \ \ \text{and} \ \ \ j_* (X) = X
}
with equivalence 
\al{
Y \sim X \ \ \ \text{if} \ \ \ Y-X \propto \Upsilon
}
Let $\mathfrak{X}_{\Upsilon}$ denote the set of such vector fields. 

Let $X_E$ denote the equivalence class of $X \in \mathfrak{X}_{\Upsilon}$. The set $\mathfrak{X}_{\Upsilon , E}$ of equivalence classes  is a Lie algebra over $\mathcal{F}_{\Upsilon}$, with 
\al{
[ \,X_E , Y_E  \,] =  [X,Y]_E
}
For any $f \in \mathcal{F}_{\Upsilon}$ we have $X f \in \mathcal{F}_{\Upsilon}$ if $X \in \mathfrak{X}_{\Upsilon}$ and $Yf = Xf$ if $Y\sim X$. Thus, $X_E f$ is well-defined (as  $Xf$). Hence, $\mathfrak{X}_{\Upsilon,E}$ elements act as derivations on $\mathcal{F}_{\Upsilon}$. 

We may define a covariant derivative operator on $ \mathfrak{X}_{\Upsilon,E}$ as a map $\nabla :  \mathfrak{X}_{\Upsilon,E} \times  \mathfrak{X}_{\Upsilon,E} \rightarrow  \mathfrak{X}_{\Upsilon,E}$ which is $\mbb{R}$-bilinear, $\mathcal{F}_{\Upsilon}$-linear in the first variable and satisfies 
\al{ \label{covderonTW1}
\nabla_{X_E} ( f Y_E) = f \nabla_{X_E} Y_E + ( X_E f) Y_E 
}
A covariant derivative is symmetric if 
\al{
\nabla_{X_E}  Y_E - \nabla_{Y_E} X_E = [ \, X_E , Y_E \, ]  
}

We now relate such operators to the standard covariant derivative $D$ on $\mbb{R}^{m+1}$ by choosing a representative of each equivalence class. For this purpose, we introduce a one form $\omega$ on $\mbb{R}^{m+1}_0$ such that 
\al{
\left< \Upsilon , \omega \right> = 1 \ \ \ \text{and} \ \ \ j^* (\omega) = \omega
} 
Then for any vector field $X$ we set 
\al{
\tilde{X} = X - \left< X , \omega \right> \Upsilon
}
$\tilde{X}$ satisfies $\left<\tilde{X} , \omega \right> = 0$. 
Then  $Y \sim X$ implies $\tilde{Y} = \tilde{X}$, and,  $X \in \mathfrak{X}_{\Upsilon}$ implies that $\tilde{X} \in \mathfrak{X}_{\Upsilon}$.
Thus such a one form $\omega$ enables us to select a representative of each equivalence class.

If, furthermore, we have 
\al{
\mathcal{L}_{\Upsilon} \omega = 0
}
it follows that $\mathcal{L}_{\Upsilon} \tilde{X} = 0$. $\Upsilon$ is an infinitesimal affine transformation of $D$. Hence, when $\mathcal{L}_{\Upsilon} \tilde{X} = 0 = \mathcal{L}_{\Upsilon} \tilde{Y}$ we get 
\al{
\mathcal{L}_{\Upsilon} (D_{\tilde{X}} \tilde{Y} ) = D_{\mathcal{L}_{\Upsilon} \tilde{X}} \tilde{Y} + D_{\tilde{X}} ( \mathcal{L}_{\Upsilon} \tilde{Y}) = 0
}
$j$ is an affine transformation, so when $j_* (\tilde{X}) = \tilde{X}$ and $j_* (\tilde{Y})=\tilde{Y}$ we have $j_*(D_{\tilde{X}} \tilde{Y})=D_{\tilde{X}} \tilde{Y}$.  So for any $\omega$ satisfying the conditions given above we may define a symmetric connection $\nabla^{\omega}$ on $\mathfrak{X}_{\Upsilon}$ from the standard covariant derivative on  $\mathbb{R}^{m+1}$ as 
\al{
\nabla^{\omega}_{X_E} Y_E = \Big( D_{\tilde{X}} \tilde{Y} \Big)_E
}

As a final remark, let us mention that $\Upsilon$ has the property that 
\al{
D \Upsilon = \text{id}
}
where id is the identity tensor; and this equation determines $\Upsilon$ up to the addition of a constant vector field. 

To generalize this construction from $\mathbb{RP}^m$ to an arbitrary smooth manifold $M$ we need the analog of $\mathbb{R}^{m+1}_0$ in the case of $M$ i.e. a smooth manifold of one higher dimension. Then we can define analog of $D$, a covariant derivative on this higher dimensional manifold, from which we can recover the analog of $\nabla$, the covariant derivative on $\mathbb{RP}^m$. It turns out that the larger manifold we are looking for is the volume bundle $V(M)$ of $M$.

 \subsection{TW Connections over a Smooth Manifold}
\label{volumeTW}

Over an $m$-dimensional manifold $M$ with coordinates $x^a \equiv (x^1,\cdots,x^m)$ one may build the volume bundle $V(M)$. An element of the fiber is a volume form i.e. an $m$-form
\al{
\omega = c(\omega)\,  dx^1 \wedge \cdots \wedge dx^m
}
Here, $c(\omega)$ is the coordinate of $\omega$ with respect to standard coordinate basis  $\{ \partial_a \}$ of $M$ . Let us, instead, coordinatize the fibers  as 
\al{
\lambda = | c |^{1/m+1}
}
So the coordinates for the volume bundle become $x^{\alpha} \equiv (x^0 , x^1, \cdots , x^m)\equiv (\lambda , x^1,\cdots , x^m) $. 

There is a natural $\mathbb{R}_+$ action ($\mathbb{R}_+ \equiv  \{x>0, \  x \in \mbb{R} \}$) on this bundle that scales only the $\lambda$ coordinate : 
\al{
R_s  : V(M) \times  \mathbb{R}_+ \rightarrow V(M) : (\omega , s) \mapsto s \omega
}
In coordinates this reads 
\al{
x^a \mapsto x^a \hspace{0.3in} \text{and} \hspace{0.3in} \lambda \mapsto s^{1/m+1} \lambda
}
This action is generated by 
\al{
\Upsilon = \lambda \frac{\partial}{\partial \lambda}
}

A TW connection is a special connection on $V(M)$, namely, one that satisfies 
\al{ \label{volumecovder}
\tilde{\nabla} \Upsilon = \text{id}
}
Operators with tilde refer to the $m+1$ dimensional space, and those  without tilde refer to the projected $m$ dimensional space. 

Equation \eqref{volumecovder} implies the following 
\al{ \label{TWconnectioncoefficients}
\tilde{\Gamma}^{\alpha}_{00}=0=\tilde{\Gamma}^0_{a0} \hspace{0.3in} , \hspace{0.3in} \tilde{\Gamma}^b_{a0}=\lambda^{-1} \delta^b_a \ (\lambda >0) \hspace{0.3in} , \hspace{0.3in} \tilde{\Gamma}^0_{ab} = \lambda \mathcal{D}_{ab}
}
Here $\mathcal{D}_{ab}$ is an object intrinsic to the manifold $M$. 

The geodesic equations then read 
\al{\label{TWtemporalgeo} 
& \ddot{\lambda} + \lambda \mathcal{D}_{ab} \dot{x}^a \dot{x}^b = 0  \\
& \ddot{x}^c + \tilde{\Gamma}^c_{ab} \dot{x}^a \dot{x}^b = \left( - 2 \dot{\lambda} /\lambda \right)  \dot{x}^c \label{TWspatialgeo}
}
Here $\dot{F}\equiv dF/d\tau$. 

We reparametrize the paths $\tau \mapsto f(\tau)$ such that \eqref{TWspatialgeo} becomes affine (i.e. right hand side of it  becomes zero or geodesic becomes geodetic). This requires the condition
\al{
\ddot{f}/\dot{f} = - 2 \dot{\lambda}/\lambda 
}
When this condition is inserted into \eqref{TWtemporalgeo} one gets
\al{ \label{Schwarziangeodetic}
Sf(\tau) = 2 \mathcal{D}_{ab} \dot{x}^a \dot{x}^b
}
where $Sf(\tau)$ is the Schwarzian \eqref{TheSchwarzian}.

Now, given a TW connection $\tilde{\nabla}$ on $V(M)$, with the aid of a any one-form $v$ on $V(M)$ which is $\mathbb{R}_+$ invariant and satisfies $\left< \Upsilon , v \right> = 1$ one can construct a symmetric affine connection $\nabla^v$ on $M$. Such a one-form $v$ is the connection form on the principal bundle $V(M) \rightarrow M$. In fact $\tilde{\nabla}$ gives rise in this way to a projective equivalence class $[\nabla]$ of symmetric affine connections on $M$, the different members of the class corresponding to different choices of $v$. The difference $v' - v$ of two members of the equivalence class is the pullback of a one-form on $M$, which determines the projective transformation relating the two corresponding connections $\nabla^v, \nabla^{v'}$ on $M$. For more details, see \cite{saunders} and \cite{robertsTW}.

\subsection{The Theory}

 To obtain a metric on $V(M)$ one can use the Dirac algebra on $M$,
\al{ \{\g^a, \g^b\} = 2 g^{a b}.
}
Then the chiral Dirac matrix, $\g^{m+1}$ is given, up to a factor $k(\lambda)$ involving the volume parameter, by
\al{
\gamma(\la)_{m+1} = \frac{k(\la)}{m!} i^{\frac{m-2}{2}} \epsilon_{a_1 \cdots a_m}\g^{a_1}\cdots \g^{a_m}.
}
The new Dirac algebra, obtained with the addition of $\gamma_{m+1}$, defines a metric $G$ on $V(M)$ through
\al{
\{ \g_{\alpha}, \g_{\beta} \} = 2 G_{\alpha \beta},
}
 where  $\alpha,\beta = 1, \cdots, m+1.$ 
 
  One can contract $G_{\alpha \beta}$  with the projective curvature to obtain the interaction term \eqref{Pol2DdiffTW} in the Virasoro geometric action.  The volume factor of the extended manifold is given by 
 \al{
 \sqrt{- \det (G_{\mu \nu})} = \sqrt{- \det (g_{ab})} k(\lambda) 
 }

The factor $k(\lambda)$ is proposed to be fixed by the condition $\det G =  k(\lambda)^2 \det g' =1$ under a conformal transformation $g'_{ab} = \exp(2 \lambda) g_{ab}$. 
     
Now,  this construction can be applied to a 2D manifold with the Polyakov metric 
\al{
  g_{a b} = \left( \begin{array}{cc} 0 & 1 \\ 1 & 2h(\theta,\tau) 
\end{array} \right) 
}
where  $h= \partial_{\theta} f/\partial_{\tau} f$. The extended 3D metric becomes
\al{
G_{\alpha \beta} = \left( \begin{array}{ccc} 0 & 1& 0 \\
1& 2h(\theta,\tau) & 0 \\
0 & 0 & k(\lambda) \end{array} \right) 
}

It is straightforward to compute the projective curvature components $K^{\rho}_{\ \alpha \beta \gamma}$. In particular, the projective Ricci tensor components become
\al{
K_{\alpha \beta }= \begin{cases} 
 -\la \, \partial_{\la} \mathcal{D}_{\theta \theta}    & {\a=1, \b=1} \\
  -\la \, \partial_{\la} \mathcal{D}_{\theta \tau} - \partial^2_{\theta} h_{\tau \tau},              & \a=1, \b=2 \\
0, & \text{otherwise}. \end{cases}
} 
Then, using the metric $G_{\alpha \beta}$ on the 3D space, the proposed interaction term reads
\al{
S_{\text{Diff Inter}}  & = \int d\lambda \, d\theta \, d\tau \, \sqrt{-G} \, G^{\mu \nu} K_{\mu \nu}  
\notag\\  &= \int d \lambda \, d\theta \,  d\tau \, \frac{k(\lambda)}{2} ( \partial_{\la} \mathcal{D}_{\theta \theta}\, h_{\tau \tau }- \partial_{\la} \mathcal{D}_{\theta \tau} -\partial^2_{\theta} h_{\tau \tau}). 
}

The arguments provided for recovering the diff-Polyakov interaction term \eqref{Pol2DdiffTW} from this integral are as follows. The last term is the scalar curvature and is a total derivative so can be dropped. The middle term is decoupled from the metric so integrates to a constant. And, using the identification $D_{ab} = \lambda \partial_{\lambda} \mathcal{D}_{ab}$ on the $\lambda$ boundary, the first term yields the Polyakov-diff interaction term. 

Finally, the proposed action for the diff field read
\al{
S_{\text{ Diff}}   =  
\int d\theta \, d\lambda \, d\tau \, \sqrt{-G}\,  K_{\;\;\alpha \beta \gamma}^{\rho} K_{\;\;\mu \nu \sigma }^{\delta} G_{\alpha \mu}G_{\beta \nu}G_{\gamma \sigma}G^{\rho \delta}.
}
The  action for $\G^a_{bc}=0$ will only involve the diff field components so describes the free theory. It becomes
\al{
S_{\text{ Diff free}} & = \int d\theta \, d\lambda\,  d \tau \, \sqrt{2}  \left( \frac{1}{\lambda^2 k(\lambda)} \left( (\Delta_{\lambda})^2 \mathcal{D}_{\theta \theta} - 2 (\Delta_{\lambda} \mathcal{D}_{\theta \tau})^2 + ( \Delta_{\lambda} \mathcal{D}_{\tau \tau})^2 \right) \right. \notag\\ 
& - \sqrt{2} \lambda^2 k(\lambda)^3 \left( (\partial_{\tau} \mathcal{D}_{\theta \theta})^2 - (\partial_{\tau} \mathcal{D}_{\theta \tau})^2 - 2 (\partial_{\tau} \mathcal{D}_{\theta \theta})(\partial_{\theta} \mathcal{D}_{\theta \tau})\right) \notag\\ 
& - \sqrt{2} \lambda^2 k(\lambda)^3 \left( (\partial_{\theta} \mathcal{D}_{\theta \tau})^2 + 2 (\partial_{\tau} \mathcal{D}_{\theta \tau})(\partial_{\theta} \mathcal{D}_{\tau \tau}) + ( \partial_{\theta} \mathcal{D}_{\tau \tau})^2 \right) \Big)  
}
where 
\al{
\Delta_{\lambda}^2 ( \mathcal{D}_{ij}) \equiv \Delta_{\lambda} (\Delta_{\lambda} \mathcal{D}_{ij}) \ \ , \ \ \Delta_{\lambda} (\mathcal{D}_{ij}) = \lambda^2 k(\lambda) \partial_{\lambda} \mathcal{D}_{ij} 
}
For the field equations, momenta and constraint analysis see \cite{brensinger}.

\section{TW Projective Connection and Virasoro Algebra }  
\label{TWVirasororeln}

In this section we are going to  review in detail the relationship between the coadjoint representation of Virasoro algebra and projective geometry.
Then we are going to move back to the Thomas-Whitehead (TW) theory. 
As we have seen in the previous section, in \cite{brensinger}, the object $\mathcal{D}$ appearing in \eqref{TWconnectioncoefficients} as a  projective  TW  connection component is identified with the diff field. 

For this identification to be justified, $\mathcal{D}$ should reduce  to a Virasoro coadjoint element on a TW projective space over circle. This was not shown in \cite{brensinger}, and the main clue  at hand is  the appearance of the  Schwarzian derivative in \eqref{Schwarziangeodetic} as related to $\mathcal{D}$ upon the reparametrization that turns the main geodesic equation \eqref{TWspatialgeo} into a geodetic equation. 
We are going to fill in this  gap  by showing that even in geodesic frames the  equations hide Virasoro coadjoint transformation under certain conditions. These conditions  require more thought.

\subsection{Virasoro Algebra and Projective Geometry} 

Here we are going to lay down the mathematical motivation for the proposal in \cite{brensinger}, namely, the relation of  Virasoro algebra  to projective geometry in 1D. 
The main references for this section are \cite{ovsienkobook}, \cite{kirillov82}. 

Recall the projective space $\mathbb{RP}^m$ introduced in the previous section. Local coordinates on $\mathbb{RP}^n$ come from $\mathbb{R}^{n+1}$. If $x^{\alpha} = ( x^0, \cdots , x^n)$ are  local coordinates in $\mathbb{R}^{n+1}$ then in a chart with $x^{\beta} \neq 0$ the $n$ affine coordinates on $\mathbb{RP}^n$ are defined as  $y^{\alpha} = x^{\alpha} / x^{\beta}$, $\alpha \neq \beta$.
In the following we restrict our attention to $\mathbb{R P}^1$.
 If $(x,y)$ are local coordinates on $\mathbb{R}^2$ then in the region $y \neq 0$ the affine coordinate on $\mathbb{RP}^1$ is $\xi^{(1)} = x/ y$, and in the region $x \neq 0$ the affine coordinate on $\mathbb{RP}^1$ is $\xi^{(2)} = y/ x$. In the intersection region, $x \neq 0 \neq y$ the transition map $\xi^{(1)} = 1/\xi^{(2)}$ is a diffeomorphism. 

If the affine coordinate on $\mathbb{RP}^1$ is $y$ then a  projective transformation $g \in PGL(2, \mathbb{R})$ is defined by 
\al{
y \mapsto g(y) = \frac{ay+b}{cy+d}
}

Consider a nondegenerate curve $\gamma(t)$ in $\mathbb{R P}^1$ i.e. $\gamma : \mathbb{R} \rightarrow \mathbb{RP}^1$, and  nondegeneracy  means $\dot{\gamma}(t)\neq 0 \ , \forall t$. Two curves $\gamma_1 (t)$ and $\gamma_2(t)$ are projectively equivalent if they are related by  a projective transformation $g \in PGL(2, \mathbb{R})$ : 
\al{ \label{ProjectiveEquivalenceofCurves}
\gamma_1 \sim \gamma_2 \ \ \Leftrightarrow \ \ \gamma_2 (t) = g \circ \gamma_1(t)
}
The following is from \cite{ovsienkobook} Theorem 1.3. 
\begin{thm}
\label{SturmTheorem}
There is a one to one correspondence between the equivalence classes of non-degenerate curves in $\mathbb{R P}^1$ and Hill operators  
\al{ \label{HillSturmoperator}
L = \frac{d^2}{dt^2} + u(t)
}
where $u(t)$ is a smooth function. In the affine coordinate on $\mathbb{RP}^1$  a curve in the equivalence class is given by a function $f(t)$.  Then the corresponding Hill operator has the potential
\al{ 
u(t) = \frac{1}{2} S(f(t))
}
\end{thm}
See \cite{ovsienkobook} for the proof.

For arbitrary diffeomorphisms $f, g \in \text{Diff}(\mathbb{R P}^1)$ the Schwarzian derivative satisfies \eqref{Schwarzian id}. 
As a result, the action of a diffeomorphism $g \in \text{Diff}(S^1)$  on the Hill operator becomes: 
\al{
T_{g{-1}} : u  \mapsto (g')^2 u(g) + \frac{1}{2} S(g)
}
This is the same as the transformation \eqref{Virasoroactivecoad} of a Virasoro coadjoint element of central charge $1/2$. Therefore, we have the following correspondence
\al{ \label{HillVirasorocorrespondence}
(u(x) , c) \leftrightarrow  2 c \frac{d^2}{dx^2} + u(x)
}
between a Virasoro coadjoint element of central charge $c$ on the left and the Hill operator on the right. This shows that \textit{there is a one to one correspondence between Virasoro coadjoint elements and the projective equivalence classes of curves in $\mathbb{RP}^1$}. 

The second main result is obtained by discussing the consequences of the above result for projective manifolds.
In \cite{brensinger} it is stated that the potential $u(x)$ in \eqref{HillVirasorocorrespondence} is a projective connection, referring to \cite{kirillov82}. 
In \cite{kirillov82} Kirillov simply states that every Virasoro coadjoint element  defines a projective structure on $S^1$.  Indeed, this result can be reached using the Theorem \ref{SturmTheorem} as we show below \cite{ovsienkobook}. To clarify the matter, let us first review the concept of a projective structure. 

A projective structure is the analog of a differentiable structure on a smooth manifold in the case of a projective space. Explicitly, in 1D, a projective structure on $\mathbb{R}$ is given by an atlas $\{ (U_i , \varphi_i )\}$ where $\{U_i\}$ is an open covering of $\mathbb{R}$ and the maps $\varphi_i : U_i \rightarrow \mathbb{RP}^1$ are local diffeomorphisms such that the transition maps $\varphi_i \circ \varphi_j^{-1}$ on $\mathbb{RP}^1$ are projective. 

A projective atlas defines a smooth immersion $\varphi : \mathbb{R} \rightarrow \mathbb{RP}^1$, and  a projective structure gives a projective equivalence class of such immersions, in the sense of \eqref{ProjectiveEquivalenceofCurves}. The immersion $\varphi$, modulo projective equivalence (i.e. $\varphi(t) \sim g \varphi(t)$ with $g \in PGL(2, \mathbb{R})$), is called the developing map. The maps $\varphi_i$  are nondegenerate  curves in $\mathbb{RP}^1$ so that the Theorem \ref{SturmTheorem} states that the developing map $\varphi$ gives rise to a Hill operator \eqref{HillSturmoperator}. Therefore, \textit{ the space of projective structures on $\mathbb{RP}^1$ is identified with the space of Hill operators. }

The definition of the projective structure extends to $S^1$, but a new feature is needed. Identifying $S^1$ with $\mathbb{R}/ 2 \pi \mathbb{Z}$, the developing map satisfies the additional condition $\varphi (t+2 \pi) = M ( \varphi(t))$ for some $M \in PGL(2, \mathbb{R})$. The projective map $M$ is called the monodromy. In this case projective equivalence of $\varphi$ extends to include $M$ as well i.e. $(\varphi(t) , M) \sim ( g \varphi(t) , g M g^{-1})$ for $g \in PGL(2, \mathbb{R})$. The monodromy condition implies that the potential $u(t)$ satisfies $u(t+2 \pi) = u(t)$, which is a requirement for $u(t)$ to be a Virasoro coadjoint element associated with $S^1$. 

Now using the correspondence \eqref{HillVirasorocorrespondence} between Hill operators and Virasoro coadjoint elements we can restate the finding above  : \textit{Every Virasoro coadjoint element defines a projective structure on $S^1$}. 
In fact, what Kirillov \cite{kirillov82} reached was this result which seems to have nothing to do with a connection at this level.

 However, it does, due to the following correspondence \cite{matveev2017}. Namely, that a projective structure can be equivalently given by a torsion-free, projectively-flat connection. Let us open this a bit. A projectively-flat connection is a connection that is projectively equivalent to a flat connection (i.e. one for which curvature vanishes). Two connections $\tilde{\nabla}$, $\nabla$ are projectively equivalent \cite{levicivitaprojective} if there exists a one-form $\phi$ such that, for arbitrary vector fields $X, Y$
 \al{
 \tilde{\nabla}_X Y = \nabla_X Y + \phi(Y) X + \phi(X) Y
 }
 The motivation behind this definition is that projectively equivalent connections  yield the same geodesics (considered as unparametrized curves). 
 
 Now, if  a connection is projectively-flat then it defines locally-flat geodesics in a neighborhood of any point. If one forms an atlas from these neighborhoods then the transition map between them will be a projective transformation mapping straight lines to straight lines. In particular, in the one-dimensional case, what we obtain is a family of projectively related maps $\varphi_i : U_i \rightarrow \mathbb{RP}^1$, i.e. a projective structure in the sense described above. 
 
 Therefore, we can now restate the correspondence above : \textit{Every Virasoro coadjoint element (or every Hill operator) defines a projectively-flat connection on $\mathbb{R}$}.   We leave it to the researcher to investigate how this result extends to the concept of projective connection as described by Thomas, Whitehead and Roberts \cite{thomas1}, \cite{thomas2}, \cite{whitehead}, \cite{robertsTW}.

In any case, it is a brilliant idea to extend a Virasoro coadjoint element (so the diff field) to higher dimensions through a projective connection. In particular, in \cite{brensinger} Thomas and Whitehead's formalism is used for this extension. In the next section, we are going to investigate how and under what the conditions this extension reduces back to Virasoro coadjoint orbits.

\subsection{TW Projective Connections and  Diff Field } 
Now, we are ready to investigate the relationship between a TW projective connection and the diff field. In the following whenever the argument of a function is suppressed it is $\theta$, and prime denotes $d/ d\theta$ as usual. 
Reconsider the TW projective geodesic equations  \eqref{TWtemporalgeo}, \eqref{TWspatialgeo} with the circle $S^1$ as the parameter space,
\al{
& x''^c + \Gamma^c_{ab} x'^a x'^b = - 2 (\lambda'/\lambda) x'^c \\
&\lambda'' + \lambda \mathcal{D}_{ab} x'^a x'^b = 0 
}
 Defining 
\al{
\Lambda \equiv (\ln \lambda)' =  \lambda'/\lambda
}
it is easy to compute 
\al{
 \lambda''/\lambda= \Lambda' + \Lambda^2
}
Then we can rewrite the geodesic equations as 
\al{
& x''^c + \Gamma^c_{ab} x'^a x'^b=-2\Lambda x'^c \\ 
& -\mathcal{D}_{ab} x'^a x'^b = \Lambda' + \Lambda^2
}

For a 1+1D projective space over a circle these equations reduce to 
\al{\label{1DGammaeqn}
& x'' + \Gamma x'^2 = - 2 \Lambda x' \\ 
&  - \mathcal{D} x'^2 = \Lambda' + \Lambda^2 \label{1Dprojeqn}
}
Note that in these equations $\Gamma$ and $\mathcal{D}$ (but not $\lambda$, $\Lambda$) depend on geodesic parameter $\theta$ through the coordinate $x$ of the curve i.e. $\Gamma  = \Gamma(x (\theta))= (\Gamma\circ x) (\theta)$, $\mathcal{D} = (\mathcal{D} \circ x)(\theta)$. So these equations can be properly rewritten as 
\al{\label{1DGammaeqn}
& x'' + x'^2\Gamma \circ x  = - 2 \Lambda x' \\ 
&  - x'^2 \mathcal{D}\circ x = \Lambda' + \Lambda^2 \label{1Dprojeqn}
}
We can solve the first equation for $\Lambda$,
\al{
\Lambda= - \frac{x''}{2x'} - \frac{x'\Gamma\circ x }{2} 
} 
Inserting this into the second equation  we get
\al{ \label{DSigmaCorrect}
2 \mathcal{D}\circ x = \frac{(\Gamma \circ x)'}{x'} - \frac{(\Gamma\circ x)^2}{2} + \frac{1}{x'^2}Sx
}
Using the chain rule formula 
\al{\label{chainrule}
(f \circ g)' = (f' \circ g) g' 
}
we can rewrite \eqref{DSigmaCorrect} as 
\al{
\label{DSigmaCorrect2}
2 \mathcal{D}\circ x = \partial_x (\Gamma \circ x) - \frac{(\Gamma\circ x)^2}{2} + \frac{1}{x'^2}Sx
}

Recall the transformation \eqref{gammatrf1d} of the connection coefficient under  $x \mapsto \bar{x}(x)$.
\al{
\bar{\Gamma}(\bar{x}) = \frac{d x}{d \bar{x}} \Gamma(x) - \left(\frac{d x}{d\bar{x}} \right)^2 \frac{d^2 \bar{x}}{d x^2} 
}
If we can take the geodesic parameter $\theta$ as a coordinate, so that the geodesic curve becomes a coordinate transformation (or a circle diffeomorphism), then under $x \mapsto \bar{x}=\theta$ this reads 
\al{
\bar{\Gamma} (\theta) = \frac{dx}{d\theta} \Gamma (x(\theta)) - \left( \frac{dx}{d \theta} \right)^2 \frac{d^2 \theta}{dx^2}
}
Using 
\al{
\frac{d^2 \theta}{dx^2}=\frac{d\theta}{dx} \frac{d}{d\theta} \frac{d\theta}{dx}  = \frac{1}{x'} \left( \frac{1}{x'} \right)'
}
we get  
\al{
\bar{\Gamma} = x' \Gamma \circ x + \frac{x''}{x'}
}
From this we directly compute 
\al{\label{sigmatildeequation2}
\bar{\Gamma}'-\frac{\bar{\Gamma}^2}{2}=x'^2 \left( \frac{(\Gamma \circ x)'}{x'} - \frac{(\Gamma \circ x)^2}{2} \right) + Sx
 }
 where $Sx$ is the Schwarzian. 
Using the chain rule \eqref{chainrule}, this becomes
 \al{\label{sigmatildeequation3}
\bar{\Gamma}'-\frac{\bar{\Gamma}^2}{2}=x'^2 \left( \partial_x (\Gamma \circ x) - \frac{(\Gamma \circ x)^2}{2} \right) + Sx
 }
 
 The object 
 \al{
 \Sigma= \Gamma'-\frac{\Gamma^2}{2}
 }
 is already familiar from section \ref{connectionVircoadjointelement}. 
Equation \eqref{sigmatildeequation3} is the active coadjoint transformation \eqref{Virasoroactivecoad}  of $(\Sigma , 1)$ : 
\al{
\text{Ad}^*_x \Sigma \equiv  x \circ \Sigma \equiv  \bar{\Sigma} = x'^2 \Sigma \circ x + Sx
} 
Note that equation \eqref{Sigmainfinitesimalcoad}  in Section \ref{connectionVircoadjointelement}  is the infinitesimal version  of  \eqref{sigmatildeequation3}.

Comparing  equations \eqref{DSigmaCorrect2} with \eqref{sigmatildeequation3} we get
\al{
2 \mathcal{D}(x) x'^2 = \bar{\Gamma}'- \frac{\bar{\Gamma}^2}{2} = \bar{\Sigma}= x'^2 \Sigma \circ x + Sx \label{TWSigmatrf}
}
This shows\footnote{The bars on $\Gamma$ are simply notational convention; they mean $\Gamma$ is evaluated in the coordinate frame $\bar{x} \equiv \theta$.} that $2 \mathcal{D}(x) x'^2$ is a Virasoro coadjoint element of central charge $1$. It is the coadjoint transform of $\Sigma$ under $\theta \mapsto x(\theta)$.

We can recover the same result in the geodetic frame as follows. Consider the geodesic to geodetic reparametrization $\theta \rightarrow f(\theta)$ i.e. one for which equation \eqref{1DGammaeqn} becomes
\al{
\frac{d^2 x}{df^2} + \Gamma(x(\theta)) \left(\frac{dx}{df}\right)^2 = 0
}
and the equation \eqref{1Dprojeqn} becomes
\al{
Sf(\theta) = 2 \mathcal{D}(x(\theta)) \left( \frac{dx}{d\theta} \right)^2
}
Now, once again if we can identify the geodesic curve as a circle reparametrization, and take $f(\theta)=x(\theta)$, then these equations imply
\al{
\Gamma(x) & = 0 \label{flatgeodeticTWSigma} \\ 
 2 \mathcal{D}(x) x'^2 & = Sx(\theta) 
} 
Since the Schwarzian $S(x)$ is the coadjoint transform of zero coadjoint element under $\theta \rightarrow x(\theta)$, \  $2 \mathcal{D}(x) x'^2$ \ is the coadjoint transform of zero coadjoint element.  
The equation \eqref{flatgeodeticTWSigma} implies  $\Sigma\circ x =0$. Hence, we again see that $2 (\mathcal{D}\circ x) x'^2$ is the coadjoint transform of $\Sigma(=0)$.

To summarize, for a 1+1D projective space over circle, \textit{if we can identify the geodesic curve as a circle diffeomorphism} then $(\mathcal{D}\circ x)x'^2$ is a Virasoro coadjoint element of central charge $1/2$. In any coordinate frame it is the coadjoint transform of $\Sigma/2$. In the geodetic frame $\Sigma=0$. 
Hence TW geodesic equations underlie Virasoro coadjoint transformation in 1+1D projective space over circle.

\appendix \label{sec:Appendix}



\chapter{Dirac-Ostrogradsky Formalism}
\section{Constraint Analysis}
\label{ConstraintAnalysis}

This note contains a summary of Dirac's constrained Hamiltonian formalism and its application to a field theory, namely Maxwell's theory. The main references are   \cite{DiracLectures} and \cite{henneaux1992quantization}. In this section  we denote Poisson brackets by $[ \ , \ ]$.  
\subsection{Summary of Dirac's Constrained Hamiltonian Formalism } \label{DiracConstraintSummary}
If the momentum definition 
\al{ \label{Dirac1}
p_n = \partial L / \partial \dot{q}_n
} 
 does not lead to an independent function of $\dot{q}_n$ then \eqref{Dirac1} constitutes a primary constraint. Let $\phi_m (q,p)=0 , \ m=1,\cdots,M$ denote all the primary constraints. 
 
  The (naive) Hamiltonian is defined as
\al{
H = p_n \dot{q}_n - L 
}
However, due to the relations $\phi_m (q,p) =0$ the function 
\al{
H^* = H + c_m (q,p) \phi_m (q,p)
}
is equally good as a Hamiltonian on the constraint surface. 

The equations of motion on the constraint surface become
\al{
\dot{q}_n & = \frac{\partial H}{\partial p_n} + u_m \frac{\partial \phi_m}{\partial p_n} \notag\\  \dot{p}_n & = -\frac{\partial H}{\partial q_n} - u_m \frac{\partial \phi_m}{\partial q_n} 
}
where $u_m$ can depend on $\dot{q}$ as well as $q$, $p$. Time evolution of an
arbitrary dynamical variable $g=g(q,p)$ is given by
\al{ \dot{g} = [ g , H ] + u_m [g , \phi_m ]} 
where $[ \ , \ ]$ is the Poisson bracket (PB). 

We extend the ordinary PB to include variables that may depend on $\dot{q}_n$. Then  the evolution equation becomes 
\al{ \dot{g} = [ g , H  + u_m \phi_m ] = : [g , H_T]
}
$H_T$ is called the total Hamiltonian.  We define a relation, called  weak equality, denoted $\approx$, meaning  equal up to constraints (i.e. on the constraint surface). When there is a weak equality  one should evaluate the PBs first, only then the constraints. 

Any constraint equation $\phi = 0$ must hold throughout  time. This requires the so-called consistency conditions : 
\al{ 
0 \approx \dot{\phi}_m & = [\phi_m , H_T] \notag\\ & = [\phi_m , H ] + u_n [\phi_m , \phi_n] \label{Diracconsistency}
}

Let us consider possible outcomes of these conditions: 
\begin{itemize}
\item Inconsistency: e.g. $L=q$ $\Rightarrow$ equation of motion : $0 \approx 1$. 
\item $0 \approx 0$ if the right-hand side of \eqref{Diracconsistency} is  linear in $\phi_m$'s. 
\item It may yield an expression independent of $u_m$ and not linear in $\phi_m$'s. Then the consistency condition is a new constraint $\xi (q, p) =0$ called a  secondary constraint. For each secondary constraint one checks the consistency condition again. One exhausts all secondaries and their consistencies following the same procedure.  Let  $\phi_k \approx 0$ , $ k = M+1 , \cdots , M+K $ , denote all (K) secondary constraints. Then the complete set of constraints become
$\phi_j \approx 0$ , $ j = 1 , \cdots , M+K $. 
\item It may yield an inhomogeneous linear set of equations  $ [ \phi_j , H ] + u_m [ \phi_j , \phi_m ] \approx 0 $  in unknowns $u_m$. These are not treated as constraints.
\end{itemize}

Consider the inhomogeneous equation in the final case. Let us  define $[\phi_j , \phi_m ] \equiv c_{jm} (q, p)$ and $[\phi_j , H ] \equiv  d_j (q, p)$. Then the inhomogeneous equation reads $
 u_m c_{jm} = d_j$. 
There must be a solution to this equation. Let $u_m = U_m (q, p)$ be a solution. This solution is not unique;  
if $V_m (q, p)$ is an arbitrary solution to the homogeneous equation, $ V_m c_{jm} = 0$, then $U_m + V_m$ is also a solution to the inhomogeneous equation. Let $V_{am}$, $a = 1 , \cdots , A $ denote all independent solutions of the homogeneous equation. Then the general solution of the inhomogeneous equation  is  $u_m = U_m + \sum_a v_a V_{am} $ where $v_a$ are arbitrary, possibly time-dependent. In general A $\leq$ M. 

With all these we have 
\al{
H_T = H+ U_m \phi_m + v_a V_{am} \phi_m
} 
Define $H' = H + U_m \phi_m$ and $\phi_a = V_{am} \phi_a$. Then, $H_T = H' + v_a \phi_a$. Here $H'$ is fixed (by the consistency equations) whereas $v_a \phi_a$ is arbitrary, since $v_a$ is an arbitrary time-dependent coefficient. This arbitrariness implies that the evolution of any dynamical variable will involve an arbitrary piece, leading to an indeterministic theory. 

Let us introduce a further decomposition of the constraints. A dynamical variable, $R(q, p)$,  is said to be first-class if 
\al{
[R, \phi_j] \approx 0 \   , \  \forall j \ \ \Leftrightarrow \ \  [R , \phi_j ] = r_{j j'} \phi_{j'}
}  
One can show that if $R_1$ , $R_2$ are first-class,  then $[R_1 , R_2]$ is also first-class using the Jacobi identity of the PB. 

The fixed part, $H'$, of the total Hamiltonian is first-class: 
\al{
[H' , \phi_j ] = [H , \phi_j ] + [ U_m \phi_m , \phi_j ] = [H , \phi_j ] +  U_m [ \phi_m , \phi_j ] \approx 0
}
since by definition, $U_m$ is a solution to the inhomogeneous equation. 
The arbitrary part, $v_a \phi_a$, of the total Hamiltonian is also first-class: 
 \al{
 [v_a \phi_a , \phi_j ] = v_a [ V_{am} \phi_m , \phi_j ] = v_a (V_{am} [ \phi_m , \phi_j ])
 }
 since by definition of $V_{am}$, the term on the right  weakly vanishes. Moreover, observe that $\phi_a$ being a linear combination of primary constraints is a primary constraint. Hence, we have a total Hamiltonian which is the sum of a first-class term and a primary, first-class term. The number of independent arbitrary functions $v_a$ is equal to the number of first-class primary constraints. Because of indeterminacy, several choices of $q , p$  correspond to the same state. 

Consider the infinitesimal time evolution of a dynamical variable $g$ 
\al{
g (\delta t) & = g_0 + \dot{g} \delta t \notag\\ & = g_0 + [g , H_T ] \delta t \notag\\ & = g_0 + \delta t ( [ g , H' ] + v_a [g , \phi_a ] )
}
The arbitrariness is in the last term :   $\Delta g (\delta t) = \delta t (v_a - v'_a) [g , \phi_a ] \equiv \epsilon^a [ g , \phi_a ] $. This then constitutes a gauge transformation. So the gauge transformations are generated by primary first-class generators. 

The Dirac conjecture is that secondary first-class constraints also do generate gauge transformations. Henneaux and Teitelboim \cite{henneaux1992quantization} disprove this providing a counterexample, yet they also show that such counterexamples  are of marginal interest so that for the  theories we consider  Dirac conjecture holds. See \cite{henneaux1992quantization} for details of this argument.  

One therefore extends the total Hamiltonian to the extended Hamiltonian by adding also the secondary first-class constraints i.e. all first-class constraints taken into account. 

Second-class constraints are not related to gauge transformations and can be eliminated by a redefinition of the PB leading to the Dirac bracket. The procedure is straightforward yet not needed for our purposes.

\subsection{Extended Action with Only First-Class Constraints } 

We are going to denote the first-class constraints (primary and secondary) by $\gamma_a$. Then the extended action reads 
\al{ \label{HT1}
S_E [ q^n(t), p_n(t) , u^a(t)] = \int dt \ ( p_n \dot{q}^n - H - u^a \gamma_a ) 
}
where $u^a$ are the Lagrange multipliers. $H$ is also first-class and we have  
\begin{subequations} \label{HT2}
\al{
[\gamma_a , \gamma_b ] & = C_{ab}^c \gamma_c \\
[H , \gamma_a ] & = V_a^b \gamma_b
}
\end{subequations} 

The time evolution of a dynamical variable $F[q,p]$ is given by 
\al{ \label{HT3}
\dot{F} \equiv \frac{dF}{dt} = [F , H_E ] = [F, H ] + u^a [F , \gamma_a] 
}

The gauge transformation of $F$ is given by 
\al{ \label{HT4}
\delta_{\epsilon} F= [ F , G] \equiv [F, \epsilon^a \gamma_a] = \epsilon^a [F, \gamma_a] 
}
where we introduced the gauge generator $G \equiv  \epsilon^a \gamma_a$ where $\epsilon^a = \epsilon^a (t)$ and $\epsilon^a$ are independent of phase space coordinates $q^n, p_n$. The meaning of the gauge invariance of $S_E$ is that
\al{ \label{HT5}
\delta_{\epsilon} S_E = 0 \hspace{0.3in}  \text{if} \hspace{0.3in} \epsilon^a(t_1) = 0 = \epsilon^a (t_2)
}

The gauge invariance of $S_E$ requires  \cite{henneaux1992quantization}
\al{ \label{HT6}
\delta_{\epsilon} u^a = \dot{\epsilon}^a + u^c \epsilon^b C_{bc}^a - \epsilon^b V_b^a 
}
for the gauge transformation of the Lagrange multipliers.

\subsection{Maxwell Theory } 

Here we are going to examine a field theory example of the formalism introduced above, namely Maxwell theory in 4D. The Lagrangian is 
\al{
L = -\frac{1}{4} \int d^3x \, F_{\mu \nu} F^{\mu \nu} 
}
where $F_{\mu \nu} = \partial_{\mu} A_{\nu} - \partial_{\nu} A_{\mu}$. 
The variation of $L$ with respect to $\partial_0 A_{\mu}$ becomes
\al{
\delta L = \int d^3x \, F^{\mu 0} \delta (\partial_0 A_{\mu}) 
}
from which we read the conjugate momenta $B^{\mu}$
\al{
B^{\mu} = F^{\mu 0} 
}
Since $F_{\mu \nu}$ is by definition antisymmetric, we immediately get a primary constraint 
\al{ \label{Maxwellmomconstraint}
\phi_1 \equiv B^0_{\mathbf{x}} \approx 0
}
where we  included the 3D space coordinate $\mathbf{x}$ as a subscript since \eqref{Maxwellmomconstraint} constitutes a three-fold infinity of primary constraints. 

Then we build the naive Hamiltonian 
\al{
H & = \int d^3x \, B^{\mu} (\partial_0 A_{\mu})  - L \notag\\ & =  \int d^3 x \left( \frac{1}{4} F^{ij} F_{ij} + \frac{1}{2} B^i B_i - A_0 (\partial_i B^i) \right)
}
where a partial integration is applied and boundary terms are ignored. 
The consistency equation, $[ B^0 , H ] \approx 0$ of the primary constraint yields the following secondary constraint
\al{
\phi_2 \equiv \partial_i B^i \approx 0
}
which is the Gauss Law. Its consistency condition is trivial, $0 = 0$; so there are no other constraints. 

It is straightforward to compute the  algebra of constraints and of the Hamiltonian 
\al{
[\phi_i(\mathbf{x}) , \phi_j (\mathbf{x}')]&=0, \ \ i,j = 1,2 \\ [H, \phi_1(\mathbf{x})]&=\phi_2(\mathbf{x}) \ , \ [H, \phi_2(\mathbf{x})]=0
}
so each constraint is first-class, generating gauge transformations. Hence,  we can form the extended action as 
\al{
S_E [ B^{\mu}, A_{\mu} , \mu_1 , \mu_2 ] = \int d^4 x \ ( B^i \dot{A}_i + B^0 \dot{A}_0 - H - \mu_1 \phi_1 - \mu_2 \phi_2 )
}
where $\mu_i$ are the Lagrange multipliers for the constraints. 
As discussed in the previous section, the extended action is invariant under the gauge transformations generated by the first-class constraints. The gauge generator would read 
\al{
Q = \int dx \ ( \epsilon_1 \phi_1 + \epsilon_2 \phi_2) 
}
with independent functions $\epsilon_i$. This generates the following gauge transformations 
\al{
\delta A_0 = \epsilon_1 \ \ , \ \ \delta B^0 = 0 \ \ , \ \ \delta A_i = \partial_i \epsilon_2 \ \ , \ \ \delta B^i = 0 \ \ , \ \ \delta \mu_1= \dot{\epsilon}_1 \ \ , \ \ \delta \mu_2 = \dot{\epsilon}_2 - \epsilon_1
}

In order to recover the standard form of the gauge transformations, one imposes the gauge condition $\mu_2=0$ under which the extended action reduces to 
\al{
S_E [ B^{\mu}, A_{\mu} , \mu_1 ] = \int d^4 x \ ( B^i \dot{A}_i + B^0 \dot{A}_0 - H - \mu_1 \phi_1  )
}
Consistency of  the gauge condition  $\delta \mu_2=0$ require $\epsilon_1 = \dot{\epsilon}_2$. Defining $\epsilon \equiv \epsilon_2$ the residual gauge transformations read 
\al{
\delta A_{\mu} = \partial_{\mu} \epsilon \ \ , \ \ \delta B^{\mu} = 0 \ \ , \ \  \delta \mu_1 = \ddot{\epsilon}
}

\section{Ostrogradsky Formalism}
\label{Ostrogradsky}

\subsection{Second-Order Nonsingular Lagrangian}
The main reference for this section is  \cite{woodard15}.

Consider a system governed by a second-order Lagrangian $L(q, \dot{q}, \ddot{q})$ depending nondegenerately on $\ddot{q}$. The Euler-Lagrange (EL) equation becomes
\al{ \label{2DEulerLagrange}
0 = \frac{\delta S}{\delta q} = \frac{\partial L}{\partial q} - \frac{d}{dt} \frac{\partial L}{\partial \dot{q}} + \frac{d^2}{dt^2} \frac{\partial L}{\partial \ddot{q}} 
}
Nondegeneracy means $ \partial^2 L/ \partial \ddot{q}^2 \neq 0 $ which implies that \eqref{2DEulerLagrange} can be cast in the form 
\al{
q^{(4)} = \mathcal{F} (q, \dot{q} , \ddot{q}, q^{(3)}) \ \ \Rightarrow \ \ q(t) = \mathcal{Q} (t, q_0, \dot{q}_0, \ddot{q}_0 , q^{(3)}_0)
}
Because solutions $\mathcal{Q}$ depend on four initial values, there must be four canonical coordinates. Ostrogradsky takes 
\al{ \label{2DMomenta}
Q_1 \equiv q \ \ , \ \ P_1 \equiv \frac{\partial L}{\partial \dot{q}} - \frac{d}{dt} \frac{\partial L}{\partial \ddot{q}} \ \ , \ \ Q_2 \equiv \dot{q} \ \ , \ \ P_2 \equiv \frac{\partial L}{\partial \ddot{q}}
}
The assumption of nondegeneracy implies that one can invert \eqref{2DMomenta} to solve for $\ddot{q}$ in terms of $Q_1 , Q_2$ and $P_2$. That is, there exists an acceleration $A(Q_1, Q_2 , P_2)$ such that 
\al{ \label{2Dnondeg}
\frac{\partial L}{\partial \ddot{q}} \bigg|_{q=Q_1 , \dot{q}=Q_2 , \ddot{q}=A} = P_2
}
Note that $A(Q_1,Q_2,P_2)$ does not depend on $P_1$. The momentum $P_1$ is only needed for the third time derivative. 

Ostrogradsky's Hamiltonian is obtained by Legendre transforming on $\dot{q}$ and $\ddot{q}$, 
\al{
H(Q_1, Q_2 , P_1 , P_2) & = P_1 \dot{q} + P_2 \ddot{q} - L \notag\\ & = P_1 Q_2 + P_2 A(Q_1, Q_2, P_2) - L( Q_1, Q_2, A(Q_1, Q_2,P_2))
}
Hamilton's equations are given by
\al{
\dot{Q}_i = \frac{\partial H}{\partial P_i} \ \ , \ \ \dot{P}_i = - \frac{\partial H}{\partial Q_i}
}
Let us check that these equations  generate time evolution.  Since $A$ is independent of $P_1$, the first equation becomes
\al{
\dot{Q}_1 = \frac{\partial H}{\partial P_1} = Q_2
}
Just as expected this equation reproduces the time evolution $\dot{q} =\ddot{q}$. Similarly, the equation for $Q_2$ yields 
\al{
\dot{Q}_2 = \frac{\partial H}{\partial P_2} = A + P_2 \frac{\partial A}{\partial P_2} - \frac{\partial L}{\partial \ddot{q}} \frac{\partial A}{\partial P_2} = A
}
where \eqref{2Dnondeg} is used. The momentum definition $P_1$ in \eqref{2DMomenta} comes from the evolution equation for $P_2$ : 
\al{
\dot{P}_2 = - \frac{\partial H}{\partial Q_2} = - P_1 - P_2 \frac{\partial A}{\partial Q_2} + \frac{\partial L}{\partial \dot{q}} + \frac{\partial L}{\partial \ddot{q}} \frac{\partial A}{\partial Q_2} = - P_1 + \frac{\partial L}{\partial \dot{q}}
}
where \eqref{2Dnondeg} is used again. Finally, using \eqref{2Dnondeg}, the evolution equation for $P_1$ read 
\al{
\dot{P}_1 = - \frac{\partial H}{\partial Q_1} = - P_2 \frac{\partial A}{\partial Q_1} + \frac{\partial L}{\partial q} + \frac{\partial L}{\partial \ddot{q}} \frac{\partial A}{\partial Q_1} = \frac{\partial L}{\partial q}
}
which, purely in terms of $q$ read
\al{
\frac{d}{dt} \left( \frac{\partial L}{\partial \dot{q}} - \frac{d}{dt} \frac{\partial L}{\partial \ddot{q}} \right) = \frac{\partial L}{\partial q}
}
Thus, the canonical equation for $P_1$ reproduces the EL equation \eqref{2DEulerLagrange}.

\subsection{Second-Order Singular Lagrangian}
\label{pons2ndorder}
The main reference for this section is \cite{pons}. 

Given a second-order singular Lagrangian $L(q, \dot{q} , \ddot{q})$ we define new variables 
\al{
q_1 \equiv q \ \ ,\  \ q_2 \equiv \dot{q} \ \ , \ \  \dot{q}_2 \equiv \ddot{q}
}
and introduce the Lagrangian constraint 
\al{ \label{singular2Dlagconst}
q_2 = \dot{q}_1 
}
Then we form the first-order Lagrangian $L_T$ 
\al{
L_T (q_1, q_2 , \dot{q}_1 , \dot{q}_2 , \lambda ) = L(q_1 ,q_2 , \dot{q}_2 ) + \lambda (\dot{q}_1 - q_2) 
}
where $\lambda$ is the Lagrange multiplier for the constraint \eqref{singular2Dlagconst}. The definition of momenta, then, yield 
\al{
P_1 & = \frac{\partial L_T}{\partial \dot{q}_1 } = \lambda \\ 
P_2 & = \frac{\partial L_T}{\partial \dot{q}_2} = \frac{\partial L}{\partial \dot{q}_2} \\ 
\pi & = \frac{\partial L_T}{\partial \dot{\lambda}} = 0
}
All of these constitute a primary constraint : $P_1$ is independent of $\dot{q}_1$, $\pi$ is independent of $\dot{\lambda}$ and since by hypothesis $L$ is singular, $P_2$ is independent of $\dot{q}_2$. 
Hence, we can identify the primary constraints 
\al{
&  \chi_1 \equiv P_1 - \lambda = 0  \label{singular2Dprim1}\\ 
& \chi_2 \equiv \pi = 0 \\
& \phi (q_1 , q_2 , P_2 )\equiv P_2 -  \frac{\partial L}{\partial \dot{q}_2} = 0 
}
Out of these only the last one is essential, the first two followed due to the prescription applied to obtain a first-order Lagrangian. These will be eliminated below. 

Let us show the equivalence of the EL equations for the first-order Lagrangian $L_T$ and the Ostrogradsky equations for $L$ : 
\all{
\frac{d}{dt} \frac{\partial L_T}{\partial \dot{q}_1 } -\frac{\partial L_T}{\partial q_1} = 0  \Leftrightarrow \dot{P}_1 = \frac{\partial L}{\partial q_1} 
}
where we used \eqref{singular2Dprim1}. Similarly, for $q_2$ we get 
\all{
\frac{d}{dt} \frac{\partial L_T}{\partial \dot{q}_2 } -\frac{\partial L_T}{\partial q_2} = 0  & \Leftrightarrow \dot{P}_2 = \frac{\partial L}{\partial q_2} - \lambda   \\ & \Rightarrow \dot{P}_2 =  \frac{\partial L}{\partial q_2} - P_1 \\ & \Leftrightarrow P_1 =  \frac{\partial L}{\partial q_2} - \dot{P}_2 
}
If we insert the second equation into the first we get 
\al{
0 & = \frac{\partial L}{\partial q_1} - \frac{dP_1}{dt} \notag\\ &=  \frac{\partial L}{\partial q_1} -  \frac{d}{dt} \frac{\partial L}{\partial q_2}  + \frac{d^2 P_2 }{dt^2} \notag\\ & = \frac{\partial L}{\partial q_1} -  \frac{d}{dt} \frac{\partial L}{\partial q_2}  + \frac{d^2  }{dt^2} \frac{\partial L}{\partial \dot{q}_2} \notag\\ & = \frac{\partial L}{\partial x} -  \frac{d}{dt} \frac{\partial L}{\partial \dot{x}}  + \frac{d^2  }{dt^2} \frac{\partial L}{\partial \ddot{x}} 
}

We build the naive Hamiltonian as usual. 
\al{
H_c  = P_1 q_2 + \frac{\partial L}{\partial \dot{q}_2} \dot{q}_2 - L ( q_1 ,q_2 , \dot{q}_2 ) 
}
We define
\al{
H_2 (q_1 , q_2 , P_2 ) \equiv \frac{\partial L}{\partial \dot{q}_2} \dot{q}_2 - L ( q_1 ,q_2 , \dot{q}_2 ) 
}
This corresponds to the energy functional for $q_2$ and $P_2$ with the dynamics of $q_1$ frozen.  Note that $H_2$ will not involve $\dot{q}_2$ since those terms involving it will cancel in the subtraction above.  Then the canonical Hamiltonian becomes 
\al{
H_c  = P_1 q_2 +H_2 (q_1,q_2, P_2)
} 

Then using the primary constraints we can build the Dirac Hamiltonian
\al{
H_D = H_c + \eta (P_1 - \lambda) + \gamma \pi + \omega \phi
}
The term $\gamma \pi$ affecting only the evolution of $\lambda$ can be ignored.  The consistency condition of $\pi=0$ reads 
\al{
0 = \{ \pi , H_D \} = \eta \{ \pi , \lambda \} = \eta \approx 0 
}
so we can set the second term $\eta ( P_1 - \lambda)$ to zero as well. Hence, we have the simplified Hamiltonian 
\al{
H & = H_c + \omega \phi (q_1 , q_2 , P_2) \notag\\ & = P_1 q_2 +H_2 (q_1,q_2, P_2)+ \omega \phi (q_1 , q_2 , P_2)
}
The rest of the analysis is the standard application of Dirac's method i.e. we check the consistency equations for the primary constraints to look for secondary constraints and build the extended Hamiltonian etc. We will skip that to avoid repetition.

\pagebreak{}



\chapter{Miscellaneous}

\section{Lightcone Coordinates }
\label{Lightcone Coordinates}
In this section we are going to introduce the conventions used in 2D fermionic theories. These mostly agree with the conventions in \cite{WZNW}. 

The lightcone coordinates (LCC) are defined through $x^{\pm} = (x^0 \pm x^1)/\sqrt{2}$. These are inverted to give $x^0 = (x^+ + x^-)/ \sqrt{2}$ and $x^1 = (x^+ - x^-)/ \sqrt{2}$.
(Recall that the metric signature adopted in this thesis is $(+t,-x)$.) Using the transformation of the metric   under a coordinate transformation
 we get $g_{++} = 0 = g_{--}$ and $g_{+-} = g_{-+}=1$. These imply  $g^{++}=0=g^{--}$ and $g^{+-}=1=g^{-+}$.
 
  Using the metric in LCC  we get $A_{\pm} = A^{\mp}$, and $A_{\mu} B^{\mu} = A^+ B^- + A^- B^+ = A_+ B_- + A_- B_+$. If we normalize the Levi-Civita symbol $\varepsilon_{\mu \nu}$ so that $\varepsilon_{01}=1$, we also get $\varepsilon_{+-}=-1=\epsilon^{-+}$.

The Dirac algebra is defined through $\{ \gamma_{\mu} , \gamma_{\nu} \} = 2 \eta_{\mu \nu}$, and we have $\overline{\psi} = \psi^* \gamma^0$. A convenient representation is given by
\al{ \label{2DDiracmatricesrep}
\gamma^0 = \left( \begin{array}{cc} 0 & 1 \\ 1 & 0 
\end{array} \right) \hspace{0.2in} \gamma^1 = \left( \begin{array}{cc} 0 & -1 \\ 1 & 0 
\end{array} \right) \hspace{0.2in} \gamma_c=\gamma^0\gamma^1 = \left( \begin{array}{cc} 1 & 0 \\ 0 & -1 
\end{array} \right) 
}
 so that the chirality matrix satisfies $\gamma^2_c = +1$. In LCC we have
\al{ \label{LCC2}
\gamma^{\pm} = \frac{1}{\sqrt{2}} ( \gamma^0 \pm \gamma^1)  \ \ \ , \ \ \ (\gamma^+)^2 = 0 = (\gamma^-)^2 \ \ \ , \ \ \ \gamma^+ \gamma^- + \gamma^- \gamma^+ =2 
}
Then for the representation \eqref{2DDiracmatricesrep} we find 
\al{ \label{LCC4}
\gamma^+ = \sqrt{2} \left( \begin{array}{cc} 0 & 0 \\ 1 & 0 \end{array}  \right) \ \ \ , \ \ \ \gamma^- = \sqrt{2} \left( \begin{array}{cc} 0 &  1\\ 0 & 0 \end{array} \right)   
}

We define the chiral components $\psi_{\pm}$ of $\psi$ by requiring $\gamma_c \psi_{\pm} = \mp \psi_{\pm}$. Thus $\psi = ( \psi_- \ \psi_+)^T$. 
We also have
\al{
\overline{\psi} = \psi^{\dagger} \gamma^0 = \psi^T \gamma^0 = (\psi_- \ \psi_+)\left( \begin{array}{cc} 0 & 1 \\ 1 & 0 \end{array}  \right)  = (\psi_+ \ \psi_-) 
}

The Dirac equation in flat spacetime reads $i \gamma^{\mu} \partial_{\mu} \psi=0$, which,  in LCC, become
\al{
\partial_{\pm} \psi_{\mp} =0
} 
Therefore, $\psi_-$ represent left movers and $\psi_+$ represent right movers.

For the massless Dirac particle, the vector current $\overline{\psi} \gamma^{\mu} \psi$ and the axial current $\overline{\psi} \gamma^{\mu}  \gamma_c \psi$ are both conserved. In 2D these are related by 
\al{ 
\overline{\psi} \gamma^{\mu} \gamma_c \psi = \varepsilon^{\mu \nu} \overline{\psi} \gamma^{\nu} \psi
} 
So, the current conservation equations combine, and can be written as
\al{
\partial_{\mu} J^{\mu} = 0 =  \varepsilon^{\mu \nu} \partial_{\mu} J_{\nu}
}
In LCC these yield
\al{
\partial_- J_+ = 0 = \partial_+ J_-
}

\section{Vanishing of the $\beta$ term } \label{vanishingbeta}
Consider the generic form of the Virasoro algebra 
\al{
[L_m, L_n] = (m-n) L_{m+n} + (cm^3 + hm) \delta_{m+n}
}
where $m, n \in \mathbb{Z}$. 
In particular this implies that 
\al{ \label{LmLminusm}
[L_m, L_{-m} ] = 2m L_0 + (cm^3 +hm) = 2m (L_0 + h/2) + cm^3
}
Hence, if we redefine the generators as 
\al{
L'_0 = L_0 + h/2 \ \ \ , \ \ \ L'_m = L_m \ , \ m \neq 0 
}
then the new generators satisfy 
\al{
[L'_m , L'_n ] = (m-n) L'_{m+n} + cm^3 \delta_{m+n} 
}
It is straightforward to show that the redefinition only affects the commutator \eqref{LmLminusm}. This is a standard trick in string theory and shows that only the cubic center has a geometric meaning (see \cite{GSWsuperstring}, Section 3.2.2). 

Then the pairing \eqref{Virasoropairing} between the Virasoro adjoint and coadjoint elements implies that the coadjoint transformation,
\al{
\delta D = \xi D' + 2 \xi' + \beta \xi' + q \xi'''
}
 of a Virasoro coadjoint element $D$ can be simplified to
\al{
\delta D = \xi D' + 2 \xi' D + q \xi'''
} 
Therefore, the $\beta$ central extension term is at our disposal; it can be turned on when needed, and off when not needed.

\section{Schwarzian Derivative}
The Schwarzian derivative is defined as 
\al{ \label{TheSchwarzian}
 Sf(x) & \equiv \frac{d}{dx} \left( \log \left( \frac{df}{dx} \right) \right) - \frac{1}{2} \left( \log \left( \frac{df}{dx} \right) \right)^2 
 \\ & =  \left( \frac{f''(x)}{f'(x)} \right)' - \frac{1}{2} \left( \frac{f''(x)}{f'(x)} \right)^2
 \\ & = \frac{f'''(x)}{f'(x)} - \frac{3}{2} \left( \frac{f''(x)}{f'(x)} \right)^2
}
where prime denotes derivative with respect to $x$, and $x$ may refer to a real or complex variable depending on the context. 

For an infinitesimal map $f(x) = x - \xi(x)$ the Schwarzian can be evaluated to be $Sf(x) = - \xi'''(x)$. We will frequently use the alternative notation $S(x, f) \equiv Sf(x)$. When we are not interested in the argument of the Schwarzian derivative, and need to consider compositions we  also use the notation $S(f) \equiv Sf(x)$.

Schwarzian derivative satisfies the identity
\al{ \label{Schwarzian id}
S(g\circ f) = (f')^2 \ S(g) \circ f  + S(f),
} 
which we can explicitly write as
\al{ \label{Schwarzian id2}
S ( x , g \circ f )=(f'(x))^2 \  S ( f(x) , g \circ f ) + S( x, f) 
}
Evaluating this for $g=f^{-1}$, and denoting $v=x$, $f^{-1}(v)=u$ we get another useful identity
\al{ \label{SchwarzianInverseIdentity}
(S w) (v) = - \left( \frac{dw}{dv} \right)^2 (Sv) (w) 
}

Fractional linear transformations
\al{ \label{Mobiustrf2}
x \mapsto \frac{ax+b}{cx+d}
}
with $ad-bc \neq0$ form the kernel of the Schwarzian derivative , i.e.
\al{ \label{KernelofSchwarzian}
Sf(z) = 0 \ \ \Leftrightarrow \ \ f(z) = \frac{az+b}{cz+d} \ , \ ad- bc \neq 0 
}
This holds true for real or complex  $x, a, b, c, d$. In the former case transformations \eqref{Mobiustrf2} form $PGL(2, \mathbb{R})$ and in the latter case $PGL(2, \mathbb{C})$. 

As a consequence of \eqref{KernelofSchwarzian}, for an arbitrary fractional linear transformation $f$ and an arbitrary smooth map $g$ we have 
\al{
S(f \circ g) = S(g)
}
So, Schwarzian is a projective invariant.

\section{Invariance of  $M^{(2)}[u]$}
\label{InvarianceHillWilson}
Under the action of $\phi \in \text{Diff}_0(S^1)$, a $-1/2$ density $\psi$  transforms as  
\al{ \label{B1}
\phi \circ \psi(x) = \psi ( \phi(x)) [\phi'(x)]^{-1/2}
}
Since $\phi \in \text{Diff}_0(S^1)$ it satisfies  
\al{ \label{B2}
\phi(0) = 0 \ , \ \phi'(0) = 1 \ , \ \phi''(0) = 0 
}
Since $\phi$ is a diffeomorphism on $S^1$ we also have 
\al{ \label{B3} 
\phi(x+2 \pi) = \phi(x) + 2 \pi 
}
which implies 
\al{ \label{B4} 
\phi^{(n)} (x+ 2 \pi) = \phi^{(n)} (x) \ , \ \text{for} \ n>0 
}
In particular we have 
\al{ \label{B5} 
\phi(2 \pi) = \phi(0) + 2 \pi \ \ \ , \ \  \ \phi^{(n)} (2 \pi) = \phi^{(n)} (0) \ \text{for} \ n>0
}

The Wilson loop equation reads 
\al{
\psi(2 \pi) = M[u] \psi(0) \ \ 
}
Evaluating this for the transformed fields we get 
\al{ \label{B6} 
(\phi \circ \psi) (2 \pi) = M[u^{\phi}] ( \phi \circ \psi)(0) 
}
We would like to show invariance of the Wilson loop i.e. 
\al{ \label{monodromyHillinvariance}
M[u^{\phi}] = M[u]
}

Evaluating \eqref{B1} at $x=0$, and using \eqref{B2} we get
\al{ 
\phi ( \psi (0)) & = \psi ( \phi (0)) [\phi' (0) ]^{-1/2}  = \psi(0) \label{B7}
}
Next, evaluating \eqref{B1} at $x=2 \pi$, and using \eqref{B2}, \eqref{B3}, \eqref{B5} we get 
\al{
\phi( \psi(2 \pi))  & = \psi( \phi(2 \pi)) [\phi'(2 \pi)]^{-1/2} = \psi(2 \pi) 
\label{B8}
}
Inserting \eqref{B7} and \eqref{B8}  in \eqref{B6} we get  the desired result \eqref{monodromyHillinvariance}.

\pagebreak{}


%

%
\biblio{bib.bib} 

\end{document}